\documentclass[a4paper,12pt]{article}

\usepackage[centertags]{amsmath}
\usepackage{amssymb}
\usepackage{enumerate}
\usepackage{epsfig}
\usepackage{array}
\usepackage{bbm}
\usepackage{color}
\usepackage{a4wide}
\usepackage{cite}
\usepackage{makeidx}
\usepackage{lscape}
\usepackage{longtable}
\usepackage{hhline}
\usepackage{fancyhdr}
\definecolor{jddred}{rgb}{.5,0,0}
\definecolor{jdblue}{rgb}{.1,0,.4}
\definecolor{jtur}{rgb}{.2,.45,.45}
\definecolor{jora}{rgb}{.9,.4,0}

\usepackage{caption}
\usepackage{titlesec}

\titleformat{\chapter}[display]
{\normalfont\huge\bfseries}{\chaptertitlename\ \thechapter}{20pt}{\Huge}
\titlespacing*{\chapter} {00pt}{0pt}{15pt}

\usepackage{makeidx} 
\makeindex

\title{ Local Grand Unification \\ 
in the Heterotic Landscape}

\author{Jonas Schmidt}
\date{\today}

\newcommand{\Z}[1]{\ensuremath{\mathbbm{Z}_{#1}}} 
\newcommand{\tZ}[1]{\ensuremath{\tilde{\mathbbm{Z}}_{#1}}}
\newcommand{\E}[1]{\ensuremath{\mathrm{E}_{#1}}} 
\newcommand{\G}[1]{\ensuremath{\mathrm{G}_{#1}}}
\newcommand{\SO}[1]{\ensuremath{\mathrm{SO}(#1)}}
\newcommand{\SU}[1]{\ensuremath{\mathrm{SU}(#1)}}
\newcommand{\U}[1]{\ensuremath{\mathrm{U}(#1)}}

\newcommand{\tr}{\mbox{tr}}

\newcommand{\rmd}{{\rm d}}

\newcommand{\T}{\ensuremath{\boldsymbol{10}}}
\newcommand{\F}{\ensuremath{\boldsymbol{5}}}
\newcommand{\Tb}{\ensuremath{\boldsymbol{\overline{10}}}}
\newcommand{\Fb}{\ensuremath{\boldsymbol{\bar{5}}}}

\newcommand{\ket}[1]{\ensuremath{\left|#1\right>}}

\newcommand{\jvs}{\rule[-7pt]{0.00pt}{20pt}}
\newcommand{\jvsb}{\rule[-6pt]{0.0pt}{17pt}}
\newcommand{\jmod}[2]{#1 \, {\rm mod} \, #2}

\begin{document}
\date{\mbox{ }}

\title{ 
{\normalsize     
June 2009\hfill\mbox{}\\}
\vspace{1cm}
\bf Local Grand Unification\\ 
in the Heterotic Landscape \\[8mm]}
%
\author{Jonas Schmidt\\[2mm]
{\normalsize\it Deutsches Elektronen-Synchrotron DESY, Notkestrasse 85, 22607 Hamburg, Germany}
}
\maketitle

\thispagestyle{empty}

\begin{abstract}
\noindent
We consider the possibility that the unification of the electroweak interactions and the strong force
arises from string theory, at energies significantly lower  than the string scale.
As a tool,  an effective grand unified field theory in six dimensions is derived from an anisotropic
orbifold
compactification of the heterotic string. It is explicitly shown that all anomalies cancel
in the model, though anomalous Abelian gauge symmetries are present locally at
the boundary singularities. In the supersymmetric vacuum additional interactions
arise from higher-dimensional operators. We develop methods that relate
the couplings of the effective theory to the location of the vacuum, and
find that unbroken discrete symmetries play an important role for the
phenomenology of orbifold models. An efficient algorithm for the calculation
of the superpotential to arbitrary order is developed, based on symmetry arguments.
We furthermore present a correspondence between bulk fields of the orbifold model
in six dimensions,
and the moduli fields that arise from compactifying four internal dimensions on a manifold
with non-trivial gauge background.
\end{abstract}

\newpage
\tableofcontents


\section{Introduction}

The vast majority of experimental observations at high-energy facilities in the pre-LHC era is described to very high accuracy by the standard model of particle physics \cite{pdb}. The latter is a phenomenological model whose 19 parameters are deduced from the best fit to the available data. Major drawbacks of the standard model are that it fails to describe neutrino oscillations and the observed abundance of dark matter and dark energy  in the universe. 
Further issues concern the Higgs sector. The sensitivity of the Higgs mass to the cutoff scale implies large fine-tuning and suggests the existence of beyond standard model physics, such as supersymmetry (SUSY). In addition, the triviality problem of the quartic Higgs self-interaction at infinite cutoff, i.e.~the expected vanishing of the corresponding coupling constant at low energies in that limit \cite{wk74,kls88}, 
 indicates that the standard model is not  fundamental. 
Attempts of treating gravity and the other forces of nature on the same footing, namely as local gauge symmetries, lead to the same conclusion\footnote{Asymptotic safety scenarios of gravity \cite{nr06} and the Higgs/Yukawa sector \cite{gs09} provide an alternative viewpoint, giving up the premise of perturbativity.}. This is due to incurable ultraviolet divergencies, which are also present for supergravity (SUGRA) extensions of the standard model, i.e.~for local supersymmetry. 

The most popular candidate for a fundamental theory is string theory  \cite{v68,gsw87,p98,bbs07}. It is based on replacing the idea of elementary point particles by the concept of fundamental strings. Due to their one-dimensional nature, string interactions are not point-like and therefore finite by construction. 
Consistency arguments then imply that strings have to propagate in ten space-time dimensions. Since they include a description of gravity, the sole scale $M_s$ of string theory is set by the analogue of Newton's constant in ten dimensions (up to the string coupling $g_s\sim\exp (\phi)$, where $\phi$ denotes the dilaton). Effective theories in four dimensions are obtained by assuming that six space-like coordinates describe a compact geometry. Its size, the string coupling, and the string scale then combine to the four-dimensional Planck scale $M_P\sim 10^{19}$ GeV.
 
As yet it remains an open question if string theory describes nature. 
 It is known to have ten to the hundreds different vacuum
 solutions, related to assumptions about the geometry of the predicted extra dimensions, but it is not known 
 whether 
  the standard model is accommodated in any of them.
 However, explicit constructions have shown that one can get indeed very close to the standard model and study quasi-realistic
 vacua. Examples include heterotic string compactifications on smooth \cite{byx05} or singular \cite{krz04,fnx04,bhx06,
 lx07,kk06,bls07,bs08} geometries (see \cite{s08} for recent reviews), 
 intersecting D-branes and orientifold models (cf.~\cite{bcx05,ho04,bbx08} and references therein),
 free fermionic models \cite{df95}, and F-theory 
 constructions  \cite{dw08}.
All models have in common that they do not fully match low-energy phenomenology, but almost, and it does not seem
 impossible to discover the missing link. 

A promising strategy for identifying a string theory vacuum whose low-energy effective field theory coincides with the standard model may be to require that the electroweak and the strong forces unify at an intermediate scale. The existence of a grand unified theory (GUT) \cite{pdb} in nature is an assumption, backed by the group theoretical simplicity of the early Georgi--Glashow $\SU5$ \cite{gg74} and $\SO{10}$ \cite{g75} GUT proposals, and a quantitative argument from the renormalization group running of the coupling constants. For the minimal supersymmetric standard model (MSSM), with SUSY partner masses of order of the electroweak scale $M_{\rm EW} \sim 3 \cdot 10^2$ GeV, unification appears at $M_{\rm GUT}\sim 3\cdot10^{16}$, up to $4\%$ threshold corrections (cf.~\cite{r09}). If a GUT is realized in nature, this result suggests that also supersymmetry is likely to exist, with a soft mass scale related to the electroweak scale. For GUTs in accord with string theory, the phenomenological value of the gauge coupling at the GUT scale $\alpha_{\rm GUT}\sim1/25$
implies a further relation between the string scale $M_s$, the volume of  the compact internal dimensions, and the string coupling \cite{k85}.

The simplest possibility for constructing
  supersymmetric grand unified theories  in higher dimensions 
 from string theory  \cite{w85} may be the compactification of the heterotic string \cite{ghx85}
 on orbifolds  \cite{dhx85}. One version of heterotic
 string theory is equipped with the gauge group $\E8 \times \E8$, where the exceptional Lie group $\E8$ 
 inherits standard GUT groups, 
  and may therefore provide a promising framework. 
  
 Orbifolds are simple examples for compact internal spaces. They are flat except
 for singular boundary points, called `fixed points', and thus technically easy to handle. 
 In field theoretical model building, orbifolds   became a popular 
 approach for  higher-dimensional GUTs more than five years ago
 \cite{k00,hmn02,hno02,abc01,abc03} (see \cite{hmz08,bms09} for recent work on 5d orbifolds).
 Standard problems of four-dimensional GUTs, like the breaking of the larger gauge group to the standard model or the presence of exotics in the larger Higgs representations, have simple solutions in orbifold models in terms of the boundary conditions of the bulk fields.
 Furthermore, orbifold GUTs lead to the idea of `local grand unification' at the fixed points  
  \cite{bhx05}. The fact that matter and Higgs fields may have different localization
 properties on the compact space can lead to interesting phenomenology, like
 the outcome of a heavy top-quark, or, more  generally, 
 non-trivial predictions concerning flavor physics. 

The phenomenological success of field theoretical orbifold SUSY GUTs in higher dimensions led to a renewed interest in heterotic orbifold models, twenty years after the first proposals of three-family orbifold models \cite{inq87,fix88,imx88}. The model \cite{bhx06} was found by requiring   local $\SO{10}$ unification, and this approach turned out to be very fruitful. For the first time it was possible to decouple all exotics in a consistent string compactification, with quasi-realistic Yukawa couplings and phenomenology. Once this model was found, a systematic scan of the `mini-landscape' of similar heterotic orbifold models could be carried out, and approximately 300 standard model-like string compactifications were identified \cite{lx07}. 

The motivation behind the model \cite{bhx06}, however, was the idea to realize an orbifold GUT in six dimensions,
 similar to the field theoretical model \cite{abc01}, in string theory. The corresponding anisotropic limit  \cite{w96} was worked out in full detail in \cite{bls07}, including the explicit cancelation of all anomalies. The construction relies on the assumption that some of the compact internal dimensions
 are considerably larger than others, reflecting the gap between the GUT and the string scale. 
 Note that anisotropic orbifolds seem to be preferred by the perturbative heterotic string
  \cite{ht05} (for work on precision gauge coupling unification in similar models see \cite{drw08}). 
An explicit mechanism to stabilize the large dimensions at the GUT scale was addressed recently in \cite{bcs08,bms09}. The effective six-dimensional orbifold GUT \cite{bls07} has local $\SU5\times \U1_X$ unification, a unique $\U1_{B-L}$, matter parity,  a heavy top-quark, and a simple decoupling mechanism for most of the exotics \cite{bls07,bs08}. This model will be reviewed and studied in detail in this paper, bearing in mind that it can be seen as a representative of a full class of quasi-realistic string compactifications \cite{lx07}.
 
Most work in the literature has been concerned with compactifications directly to four dimensions.  
 One typically obtains a large number of fields in heterotic orbifold
 models, and it is not straightforward to distinguish matter fields, Higgs fields
 and exotics. 
 In the favored models exotic fields are vector-like, and can therefore be decoupled
 by the generation of  mass terms, after the transition to the supersymmetric vacuum. The latter generically corresponds to non-trivial configurations in field space. This is the consequence of a tachyonic direction that is sourced by a loop-induced Fayet--Iliopoulos term (cf.~\cite{ads87,ccx98}).
  The requirement of non-vanishing vacuum expectation values (vevs) for some of the fields is an important point for orbifold model building, since it leads to the simultaneous generation of infinitely many interaction terms. The strength of the effective coupling then scales with the number of involved vevs, analogue to the Froggatt--Nielsen mechanism \cite{fn79}.
  The generated interactions may be favored, like Yukawa couplings, or disfavored,
 like proton decay operators or large $\mu$-terms. The phenomenology is therefore 
 largely determined
 by the   set of singlets that acquire non-zero
 vacuum expectation values.
 
 Up to now, finding quasi-realistic vacua in heterotic orbifold models seems to be
 mainly an issue of computing power. In practice, one truncates the superpotential 
 at a chosen finite order in the singlet vevs (typically this order is 6 or 8 \cite{bhx06, lx07, lx07c}),
 and scans for vacuum configurations for which the $F$-term and the $D$-term equations can be fulfilled, with satisfactory phenomenology.
  However, one may ask if the search for 
 vacua with phenomenologically preferred properties could be improved
 by detailed studies of the connection between the singlets which contribute
 to the vacuum, and the resulting effective superpotential. We shall address this
 problem in this work by considering symmetry arguments that may help to
 identify interesting candidates for physical vacua. This extends the results of \cite{bs08}, where the importance of unbroken discrete symmetries was noted first. Since the symmetries in non-trivial vacuum configurations refer to the full (perturbative) superpotential, they can be used to forbid disfavored terms to arbitrary order in the singlet vevs\footnote{Alternatively, a sufficient suppression may be obtained if the symmetries are broken at high order in the singlets, cf.~\cite{kx08}.}. The detailed discussion of this method and its application to the effective orbifold GUT model in six dimensions is the main purpose of this work.
 
 The transition to the supersymmetric vacuum is related to a consistency issue which  generically arises in orbifold compactifications. The back-reaction of the vacuum expectation values of localized twisted states on the geometry corresponds to `blowing-up' the orbifold singularities (cf.~\cite{chx85,fix88,a94,ht06,ngx07,lrx08} and \cite{nhx09} for a recent analysis of the $\Z{\rm 6-II}$ case). The general observation is that the resulting smooth compactification has less zero modes and reduced symmetries. The claim that one may ignore this back-reaction in the vicinity of the orbifold point, i.e.~that the orbifold description remains a good approximation nearby the origin in field space, still lacks a rigorous proof.  
 However, it seems plausible from the observation that orbifolds correspond to specific points of the moduli spaces associated with manifolds. The matching of spectra of the orbifold and a smooth background geometry may then give insights on the role of the orbifold states and thus lead to a better understanding of orbifold vacua. Although this work is mostly concerned with orbifolds, we also comment briefly on the interpretation of the bulk states in six dimensions as zero modes of a smooth compactification.

 This paper is organized as follows.
Chapter \ref{cpt:string} introduces the heterotic string and reviews its compactification
 on an orbifold, clarifying 
 the role of lattice translations for the orbifold projection conditions in Section \ref{sec:compatibility}.
  The spectrum of the effective orbifold GUT model in six dimensions is calculated in Chapter~\ref{cpt:lGUT},
 corresponding to an anisotropic limit of the compactification. Section~\ref{sec:anomalies} explicitly confirms
  that all appearing anomalies are canceled.
 In Chapter \ref{cpt:vacuum}, we discuss the model in a non-trivial vacuum configuration. The superpotential of standard $\SU5$ GUTs is reviewed in Section~\ref{sec:SU5},
 before we turn to the local GUTs with that gauge symmetry 
 at the fixed points in the extra dimensions.
 Section \ref{sec:SM} identifies unique Abelian symmetries $\U1_X$ and $\U1_{B-L}$ and the standard model families.
  A minimal vacuum that preserves matter parity and implies
 a universal decoupling of many exotics is defined in Section \ref{sec:decoupling}.
 The symmetries of general vacua are analyzed in Section \ref{sec:symms}, furthermore it is shown
 for the above minimal vacuum 
 symmetries and vanishing couplings are in direct correlation.
 Section \ref{sec:kernel} 
 describes an algorithm which allows one to enlarge the
 vacuum without generating disfavored couplings, and two examples are discussed. 
 Section \ref{sec:allorders} then provides a method   which can be used for the calculation of 
  superpotential interactions to arbitrary order in a given vacuum, with improved efficiency
  compared to order-by-order scans.
 An outlook on work in progress concerning a connection of the model with smooth
 geometries is given in Chapter \ref{cpt:K3}, before we conclude in Chapter \ref{cpt:conclusions}.
 Some technical details of orbifold compactifications are studied in Appendix~\ref{app:orbifolds},
  basic facts about Lie groups and Lie algebras are collected in Appendix~\ref{app:lie}.
    Appendix \ref{app:anomalies} gives details of the  calculation of the reducible anomalies
  of the model.
  The states at the hidden fixed points and the four-dimensional zero modes are listed
  in appendices \ref{app:n21} and \ref{app:4d}, respectively.
  Finally, Appendix \ref{app:discrete} summarizes the transformation behavior  of all fields at the GUT fixed
  points with respect to the unbroken symmetries of the three studied vacua. From this
  information it can be inferred whether a particular coupling may be present in the vacuum,
  or is forbidden to arbitrary order in the singlets.

\section{The heterotic string on orbifolds}
\label{cpt:string}

This chapter briefly introduces the heterotic string \cite{ghx85} and describes its compactification 
 on orbifolds \cite{dhx85}. The results and ideas presented here are neither new nor do they originate from
 the author, except for the ansatz for the superposition of physical states in Appendix~\ref{app:gammas}. Instead they can be found in much more detail in textbooks, e.g. \cite{gsw87,p98,bbs07}, or
 reviews \cite{bl99,s08}. Here we widely follow the conventions of 
 \cite{kx90,bhx06}. 
 The main purpose of this section is to fix the notation and to present the
 foundations of the model which will be discussed in detail in this work.

\subsection{The world sheet action}

The heterotic string can be understood as a field theory on a  two-dimensional surface,
called the `string world-sheet', which is parametrized by one time-like and one
space-like direction \index{World-sheet!coordinates}
 $\tau$ and $\sigma$,   respectively. 
The theory can be formulated in terms of 26 bosonic
fields $X^\mu, X^I, \mu=0,\dots,9, I=1,\dots,16$, and 10 fermionic fields $\psi^\mu$, 
which play the role of coordinates of a target space.
Up to specification of the metrics on the world-sheet and the target space, the 
action \index{World-sheet!action}
 reads\footnote{Throughout this paper we follow the Einstein sum convention, if not stated otherwise.}
\begin{align}
\label{eq:S10}
    S=-\frac{M_s^2}{2 \pi} \int \rmd^2 \sigma \left[
    \partial_\alpha X_\mu \partial^\alpha X^\mu + 4 i \psi^\mu
    \left( \partial_\tau-\partial_\sigma \right) \psi_\mu
    +\sum_{I=1}^{16} \partial_\alpha X^I \partial^\alpha X^I
    \right] \,.
\end{align}
Here $M_s$ is the string scale.
On-shell, the string fields appearing in this action can be written in terms of
 functions of the combinations $\sigma_\pm=\tau \pm \sigma$, \index{World-sheet!fields}
 \index{Equations of motion!solutions}
\begin{align}
\label{eq:Xpsivar}
    X^\mu(\tau,\sigma) &=
    X^\mu_L(\sigma_+)+X^\mu_R(\sigma_-)\,, &
    \psi^\mu(\tau,\sigma) &=\psi^\mu_R(\sigma_-)\,,&
    X^I(\tau,\sigma) &= X^I_L(\sigma_+)\,,
\end{align}
where subscripts $L,R$ were introduced for left- and right-moving fields \index{Left-movers}
\index{Right-movers}, respectively. Due to the 
asymmetry between left-movers and right-movers only closed heterotic strings exist, which
implies specific boundary conditions at $\sigma$ and $\sigma+\pi$ in our conventions.

The  target space  contains the
 physical four-dimensional Minkowski space $M_4$.  
A consistent quantum theory can be formulated if
 the latter is embedded into its ten-dimensional analogue $M_{10}$,
labeled by the index $\mu$.
Additionally, the index $I$ labels coordinates of a 16-dimensional  internal compact space, which 
has to coincide with the root lattice of either $E_8 \times E_8$ or ${\rm Spin}(32)/\Z2$, in order to obtain  a consistent
(quantum) theory in ten dimensions. The corresponding fields $X_L^I$ are then interpreted as
 gauge fields. Throughout the whole work we shall consider the case of a 
 gauge group
 \begin{equation}
\label{eq:G10}
G_{10}=E_8 \times E_8\,.
\end{equation}
\begin{figure}
\begin{center}
\includegraphics[width=8cm]{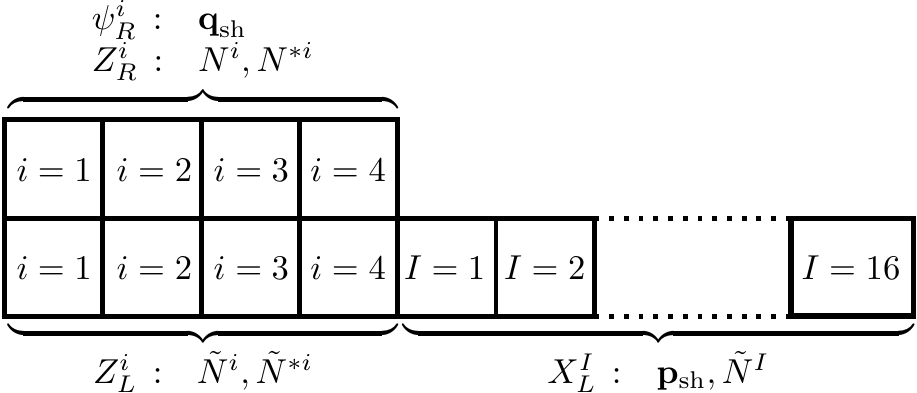}
\caption{The field content of the heterotic string. Instead of the 
fields $X_{L,R}^\mu, \psi_R^\mu, \mu=1,\dots,8$, the complex coordinates $Z^i_{L,R}=\frac{1}{\sqrt{2}}(
X_{L,R}^{2i-1}+iX_{L,R}^{2i}), i=1,\dots,4,$ and likewise for $\psi^i_R$,
are shown. After quantization, the fields give rise to states of a Hilbert space.
${\bf q}_{\rm sh}$ and $ {\bf p}_{\rm sh}$ represent
charges under the little  group and the gauge group of these states, respectively,
and $N^i,N^{*i},\tilde N^i,\tilde N^{*i},\tilde N^I$ denote oscillator numbers.}
\label{fig:hetboxes}
\end{center}
\end{figure}
The field content of heterotic string theory is sketched in Figure \ref{fig:hetboxes}.
The presence of the labels $\mu$ and $I$ is a consequence of the heterotic nature of the theory:
While $X^\mu_R$ and $\psi^\mu_R$ form a superstring in ten dimensions, $X_L^\mu, X_L^I$ are fields of a bosonic string in 26
dimensions. Thus supersymmetry charges are carried by right-moving fields, gauge charges by left-movers.
For quantization the product of left- and right-movers will be relevant and this information
is combined, yielding states which form full multiplets of $\mathcal N=1$ supergravity in ten dimensions
and of the gauge group (\ref{eq:G10}), at each mass level.

The theory which has been described so far
 does not describe nature as we observe it. First, it requires ten space-time dimensions,
and second, it has too much supersymmetry, namely 16 supercharges.
A common approach is to replace the ten-dimensional target space by the product space of a compact
six-dimensional geometry \index{Compact space} and the four-dimensional non-compact space,
\begin{align}
\label{eq:M10}
    \mathcal M_{10}&=\mathcal M_6 \times M_4\,.
\end{align}
If $\mathcal M_6$ has small volume, the effective theory at low energies is four-dimensional.
Furthermore, compactification on $\mathcal M_6$ can be responsible for supersymmetry
and gauge symmetry breaking, possibly yielding the standard model in four dimensions.
However, for non-flat manifolds $\mathcal M_6$ the quantization of the theory is
in general unknown. 
In that situation one usually considers the ten-dimensional supergravity on $\mathcal M_{10}$ as an effective 
theory.
Since in this work we are interested in compactifications on orbifolds, which are flat except for
singular points at the boundary, we are in the comfortable position to directly solve the dynamics
of the string fields and then quantize canonically.

The canonical momenta which follow from the action (\ref{eq:S10}) are not independent.
In light-cone gauge the unphysical coordinates are expressed
as combinations of  the time-like direction ($\mu=0$) and one of the spatial
directions (we choose $\mu=9$),
\begin{align}
\label{eq:Xpm}
    X^\pm(\tau,\sigma)&=\frac{1}{\sqrt 2}\left(X^0(\tau,\sigma) \pm X^9(\tau,\sigma)\right)\,, &
    \psi^\pm(\sigma_-)&=\frac{1}{\sqrt 2}\left(\psi^0(\sigma_-) \pm \psi^9(\sigma_-)\right)\,.
\end{align}
The independent directions are then the transverse coordinates $\mu=1,\dots,8$, with equations of motion
\index{Equations of motion}
\begin{align}
\label{eq:stringEOM}
    \partial_+ \psi^\mu &=0 \,,&
      \partial_+ \partial_- X^\mu&=0\,,&
      \partial_+ \partial_- X^I&=0\,,
\end{align}
where $\partial_\pm=\partial/\partial \sigma_\pm$ and $I=1,\dots,16$. Solutions to these equations will then allow 
straightforward canonical quantization. However, they depend on boundary conditions of the fields,
which eventually make the crucial difference between toroidal and orbifold compactifications.

\subsection{Classical strings on orbifolds}

\subsubsection{The orbifold geometry}

In the following we shall consider orbifold compactifications \index{Compact space} of the type \index{Orbifold!geometry}
\begin{align}
\label{eq:M6}
      \mathcal M_6 &= \frac{T^2\times T^2\times T^2}{\Z N} \,, 
\end{align}
where $T^2$ denotes a two-dimensional torus and $N$ is an integer number\footnote{For
$\Z N \times \Z M$ orbifolds see e.g. \cite{s08}.}.
The product form of $\mathcal M_6$ suggests to introduce three complex coordinates \index{Complex coordinates}
for the internal space\footnote{In this section we do not explicitly show the dependence of
coordinates $X^i$ on the world-sheet variables $(\tau,\sigma)$. 
All following transformation rules apply point-wise with respect to these variables.},
\begin{align}
\label{eq:Zi}
     Z^1 &=\frac{1}{\sqrt 2}\left(X^1+i X^2\right)\,,&
     Z^2 &=\frac{1}{\sqrt 2}\left(X^3+i X^4\right)\,,&
     Z^3 &=\frac{1}{\sqrt 2}\left(X^5+i X^6\right)\,.
\end{align}
Similarly, we introduce another complex coordinate for the two  transversal directions
in four-dimensional Minkowski space, $Z^4=(X^7+i X^8)/\sqrt{2}$.

\subsubsection*{The space group}

The action of a generating element of the `point group' $\Z N$ in (\ref{eq:M6}) on the toroidal coordinates
$Z^i$ can now be defined in terms of a
'twist vector' \index{Twist vector} $v$, with entries of order $N$,
\begin{align}
\label{eq:vN}
    {\bf v}&=\left(v^1,v^2,v^3;0\right) \,, &
    N v^j &\in \Z{ }\,, &
     j=1,2,3\,.
\end{align}
The zero in the last entry was included for later convenience. 
A coordinate $Z^i$ then transforms as
(here no summation over $i$ is implied)
 \begin{align}
\label{eq:thetas}
    Z^i &\rightarrow \vartheta_{(i)} Z^i\,,&
    \vartheta_{(i)} &\equiv e^{2 \pi i \, v^i}\,, & i&=1,\dots,4\,.
\end{align}
Thus the coordinates of the internal tori are rotated  by the point group \index{Point group} $\Z N$,
and the latter is embedded into internal Lorentz transformations. Since the on-shell degrees
of freedom transform under the little group \index{Little group} $\SO 8$, we have
\begin{align}
\label{eq:vSO8}
    \Z N &\subset \SO 8 \subset \SO{1,9}\,.  
\end{align}
As it turns out, there are additional phenomenological constraints on the twist vector, due to the 
requirement that the resulting low-energy spectrum should have $\mathcal N=1$ supersymmetry:
\index{Twist vector}
\begin{align}
\label{eq:condvsusy}
 	\sum_j v^j &=0 \,,    &
	v^j &\neq \jmod 01 \,, &
	j&=1,2,3\,.  
\end{align}
These conditions will become transparent during
 the discussion of the generic spectrum in Section \ref{sec:genericspectrum}.

Further isometries of the geometry (\ref{eq:M6}) are lattice translations of the three tori.
Each of them is defined as $T^2=\mathbbm C/ \pi \Lambda^j$ for a two-dimensional
lattice $\Lambda^j $, with lattice vectors \index{Lattice!vectors}
${\bf e}_{2j-1}, {\bf e}_{2j} \in \mathbbm C^4, j=1,2,3,$ of the form
\begin{align}
\label{eq:defea}
\Lambda^1\,&:&{\bf e}_1&=(e_1^1,0,0;0)\, &
{\bf e}_2&=(e_2^1,0,0;0)\, \\
\Lambda^2\,&:&{\bf e}_3&=(0,e_3^2,0;0)\, &
{\bf e}_4&=(0,e_4^2,0;0)\, \\
\Lambda^3\,&:&{\bf e}_5&=(0,0,e_5^3;0)\, &
{\bf e}_6&=(0,0,e_6^3;0)\, ,
\end{align}
with complex entries $e_a^i$. The product of the three tori can then be understood as $\mathbbm C^3 / \pi \Lambda$,
with $\Lambda = \sum_j \Lambda^j$.
The compactness of the internal space is expressed as invariance under translations \index{Lattice!translations}
\begin{align}
\label{eq:ztori}
    Z^i &\rightarrow Z^i+\pi m_a e_a^i\,,& m_a &\in \Z{ } \,,&i&=1,\dots,4\,.
\end{align}

In summary the `space group'\index{Space group}, which is the total isometry group of the space under consideration, 
can be written as 
\begin{align}
\label{eq:S}
   S&=\left\{\left. \left(\theta^k,m_a {\bf e}_a\right) \right| m_a \in \Z{ } \,, k=0,\dots,(N-1)
   \right\}  \,,
\end{align}
where $\theta={\rm diag}\,(\vartheta_{(1)},\vartheta_{(2)},\vartheta_{(3)},1)$.
$S$ is the discrete subgroup of the ten-dimensional Poincar\'e group
which corresponds to the crystallographic group
of the orbifold geometry.
The transformation of the coordinates ${\bf Z}=(Z^1,Z^2,Z^3,Z^4)$ \index{Complex coordinates}
under an element $g= \left(\theta^k,m_a {\bf e}_a\right) \in S$ is \index{Space group!transformation!coordinates}
\begin{align}
\label{eq:gZ}
    {\bf Z} \stackrel{g}{\mapsto} \theta^k {\bf Z} +\pi m_a {\bf e}_a\,.
\end{align}
The multiplication rule of the group (\ref{eq:S}) is
\begin{align}
\label{eq:multrule}
    \big(\theta^k,m_a {\bf e}_a\big)\big(\theta^{\tilde{k}},\tilde{m}_a {\bf e}_a\big)
    &=\big(
    \theta^{k+\tilde k},m_a {\bf e}_a+\theta^k \tilde m_a  {\bf e}_a
    \big) \in S\,,
\end{align}
and inverse elements 
 are given by
\begin{align}
\label{eq:inverse}
  \big(\theta^k,m_a {\bf e}_a\big)^{-1} &=
  \big(
  \theta^{-k},- \theta^{-k} m_a {\bf e}_a
  \big)\,.
\end{align}
For later usage we also introduce the `conjugacy class' \index{Conjugacy class} 
$[g]$ of an element $g=(\theta^k,m_a {\bf e}_a)$,
\begin{align}
\label{eq:gccdef}
    [g]&= \left\{\left.
    h g h^{-1} \right| h \in S
    \right\}=\left\{\left.
    \left(\theta^k,  \theta^{\tilde k}m_a {\bf e}_a+(\mathbbm 1-\theta^k) \tilde m_a {\bf e}_a
    \right)\right| \tilde k, \tilde m_a \in \Z{ }
    \right\}\,.
\end{align}
The set of vectors $(\mathbbm 1-\theta^k)m_a{\bf e}_a$
which appear in the bracket is denoted as
\begin{align}
\label{eq:sublattice}
    \Lambda_k &\equiv \left(\mathbbm1-\theta^k\right)\Lambda\,, 
\end{align}
and often referred to as the `sub-lattice' \index{Sub-lattice} associated with the twist $\theta^k$.

With the identification of the space group (\ref{eq:S}) the definition of the orbifold \index{Orbifold!geometry}
 in (\ref{eq:M6}) becomes
clear: It is a subset of $\mathbbm C^3$ onto which every point can be mapped 
by group transformations (\ref{eq:gZ}),  $\mathcal M_6=\mathbbm C^3/S$. 
Such a subset is also called the `fundamental domain', and all information
of the full space is contained in this region. The fundamental domain itself
is by definition invariant under all isometries.

\subsubsection*{Fixed points and fixed planes}

The crucial feature of orbifold constructions is the appearance of `fixed points' \index{Fixed point/plane/torus} and `fixed
planes', or `fixed tori'. They are   points ${\bf Z}_g$
 which are invariant under the action of a space group element $g=(\theta^k,m_a {\bf e}_a)\neq(\mathbbm 1,0)$,
\begin{align}
\label{eq:FPdef}
    {\bf Z}_g &= g {\bf Z}_g =   \theta^k {\bf Z}_g+\pi m_a {\bf e}_a\,.
\end{align}
Fixed points  correspond to localization in all six internal dimensions.
Fixed tori on the other hand describe invariance of a full torus $T^2$ under the twist, and
localization in the other two complex dimensions.
This situation will occur regularly in the example of relevance for this work. 
In general, the existence of invariant points or planes under isometry transformations
makes the crucial difference between toroidal and orbifold compactifications, as we will see in the following.

Any element $g=(\theta^k,m_a{\bf e}_a)$ which does not act like a pure translation
$Z^i\stackrel{g}{\mapsto} Z^i+\pi m_a e_a^i$ in one of the planes fulfills a fixed point equation (\ref{eq:FPdef}).
This is demonstrated in Appendix \ref{app:solveZ}, where the explicit solutions for ${\bf Z}_g$
are constructed.
 
However, not all space group elements generate independent fixed points. 
The latter are only defined up to the conjugacy class of the generating element $g$. This can be seen
by considering two fixed points ${\bf Z}_{g_1},{\bf Z}_{g_2}$ with generating elements $g_1,g_2$, respectively, and
$g_2=h^{-1} g_1 h$ for a $h \in S$. Then
\begin{align}
\label{eq:eqFP}
h {\bf Z}_{g_2} &= h g_2 {\bf Z}_{g_2} =g_1 h {\bf Z}_{g_2}\,,
\end{align}
which states that 
$h{\bf Z}_{g_2}$ fulfills the fixed point equation corresponding to $g_1$. 
 Therefore the two fixed points ${\bf Z}_{g_1}$ and ${\bf Z}_{g_2}$ are related by a symmetry transformation  and thus identical on the orbifold.
%
This observation will lead to crucial projection
conditions for physical states after quantization.

\subsubsection{Orbifold boundary conditions}
Heterotic string theory has only closed string solutions. This means that the string fields, which are functions of the
world-sheet coordinates $\tau$ and $\sigma$, must obey appropriate boundary conditions.
In our conventions they relate the field values at $\sigma$ to the ones at $\sigma+\pi$.

\subsubsection*{Bosonic coordinates}

Since the orbifold geometry identifies all points which are related by space group transformations (\ref{eq:gZ}),
the boundary conditions for the coordinate fields read \index{Boundary conditions!bosons}
\begin{align}
\label{eq:bcZ}
    {\bf Z}(\tau,\sigma+\pi)&=g  {\bf Z}(\tau,\sigma)
    =\theta^k  {\bf Z}(\tau,\sigma) +\pi m_a {\bf e}_a\,.
\end{align}
Here $g=(\theta^k,m_a {\bf e}_a) \in S$ labels different possible boundary conditions, corresponding to 
different localization properties of the string. It is already apparent that for non-trivial $g$ the constant
part of the solutions will be given by the corresponding fixed point coordinate, suggesting that
the string winds around that point and is trapped to its vicinity.

For later convenience, we can make the ambiguities of the phases in (\ref{eq:bcZ}) explicit.
We introduce a vector of positive frequencies ${\bf \omega}=(\omega^1,\omega^2,\omega^3;0)$,
\begin{align}
\label{eq:omega}
    \omega^i &=\jmod{k v^i}{1} \;,& &\text{$0<\omega^i \leqslant1$}, & i&=1,\dots,3\,,
\end{align}
and rewrite the boundary condition as
\begin{align}
	\label{eq:bcZi}
	Z^i(\tau,\sigma+\pi)&=e^{2 \pi i \left(
	\tilde n^i+\omega^i\right)}Z^i(\tau,\sigma) +\pi m_a e^i_a\,,
	\qquad \qquad {\bf \tilde n}=(\tilde n^1, \dots,\tilde n^4) \in \Z{}^4 \,.
\end{align}
The appearing  combinations $\tilde n^i+{\omega^i}$
will appear as fractional frequencies in the mode expansion 
of the on-shell fields ${\bf Z}$, and 
carry over to excitations after quantization.

\subsubsection*{Fermions and bosonized fermions}

For the fermionic fields in complex notation\index{Right-movers},
\begin{align}
\label{eq:psii}
      \psi^j(\sigma_-) &\equiv\frac{1}{\sqrt 2}\left(\psi^{2j-1}_R(\sigma_-)+i \psi^{2j}_R(\sigma_-)\right)\,,
      \quad j=1,\dots,4\,,
\end{align}
the transformation under a space group element $g=(\theta^k,m_a{\bf e}_a)$ is 
\index{Space group!transformation!fermions}
\begin{align}
\label{eq:gpsi}
	{\bf \psi} \stackrel{g}{\mapsto} \theta^k {\bf \psi} \,,
\end{align}
where ${\bf \psi}\equiv(\psi^1,\psi^2,\psi^3,\psi^4)$.
Note that invariance of the term $\psi^i \partial_+ \psi_i$ in the action (\ref{eq:S10}) forbids the appearance
of shifts for fermionic fields.
This leads to \index{Boundary conditions!fermions}
\begin{align}
\label{eq:bcpsiR}
    {\bf \psi} (\sigma_--\pi)&=\phantom{-}\theta^k    {\bf \psi} (\sigma_-) \qquad \qquad \text{(R)}\,,\\
    \label{eq:bcpsiNS}
     {\bf \psi} (\sigma_--\pi)&=-\theta^k    {\bf \psi} (\sigma_-) \qquad \qquad \text{(NS)}\,,
\end{align}
where (R) and (NS) denote Ramond 
and Neveu-Schwarz boundary conditions, respectively. 

It is instructive to represent the fermionic degrees of freedom $\psi^i$ by bosonic
fields 
$H^i$,
\begin{align}
\label{eq:defHi}
    \psi^i(\sigma_-)& \equiv e^{-2  i H^i(\sigma_-)}\,, & 
    \psi^{*i}(\sigma_-)& \equiv e^{2  i H^i(\sigma_-)}\,, 
\end{align}
which implies a transformation rule
\begin{align}
\label{eq:gH}
    H^i(\sigma_-) &\stackrel{g}{\mapsto} H^i(\sigma_-) - \pi k v^i\,,
\end{align}
and thus with ${\bf M}=(M^1,M^2,M^3,M^4)$ leads to the following boundary conditions:
\begin{align}
\label{eq:bcHR}
    H^i(\sigma_--\pi)&=H^i(\sigma_-)-\pi\left(
    k v^i+M^i \right)\phantom{ + m\frac12l} \,, \quad {\bf M} \in \Z{ }^4
     \qquad \qquad \text{(R)}\,, \\
 \label{eq:bcHNS}
    H^i(\sigma_--\pi)&=H^i(\sigma_-)-\pi\left(
    k v^i+M^i + \frac12\right)\,, \quad {\bf M} \in \Z{ }^4
     \qquad \qquad \text{(NS)}\,   .
\end{align}
These  conditions can be expressed by a single equation,
\begin{align}
\label{eq:bcH}
    {\bf H}_R(\sigma_--\pi)&={\bf H}_R(\sigma_-)-\pi\left(
     {\bf q}+k {\bf v}\right) \,, \qquad \qquad {\bf q} \in \Lambda^*_{\SO 8}\,,
\end{align}
where $\Lambda^*_{\SO 8}$ denotes the weight lattice of $\SO 8$\index{Weight lattice of $\SO8$}. 
In a canonical basis the latter is spanned by elements ${\bf q}$\index{Little group!charges},
\begin{align}
\label{eq:wSO8}
    \Lambda^*_{\SO 8} \,: \qquad
    {\bf q} &=(q^1,q^2,q^3,q^4) \,, \qquad
    \left\{\begin{array}{c}
    q^i \in \Z{ }\,, \quad \sum_i q^i \;\text{odd} \,,\\
    \text{or} \\
    q^i \in \Z{ }+\frac 12\,, \quad \sum_i q^i \;\text{even}\,.
    \end{array}\right.
\end{align}
After quantization, this will imprint information about the
transformation behavior under the unbroken subgroup of the little group
$\SO 8$ \index{Little group} onto the right-moving states.

\subsubsection*{Gauge fields}
For a consistent string theory it is generically required that the space group acts 
not only on the coordinate fields, but also on the
gauge degrees of freedom $X_L^I$. This can be understood as
a `gauge embedding' \index{Gauge embedding} $S \subset G_{10}$, expressed by a mapping
\begin{align}
\label{eq:SG}
   g= \left(\theta^k,m_a{\bf e}_a\right) &\mapsto \left(k V,m_a W_a\right)\,, 
 \end{align}
with `shift vector' \index{Shift vector} ${\bf V}=(V^1,\dots,V^{16})$ and `discrete Wilson lines' \index{Wilson lines}
\cite{hmn02}
    ${\bf W}_a=(W_a^1,\dots,W_a^{16})$. The transformation under the  space group
    element $g$ is then given by \index{Space group!transformation!gauge fields}
\begin{align}
\label{eq:gXI}
    X_L^I  &\stackrel{g}{\mapsto} X_L^I + \pi\left(k V^I+m_a W^I_a\right)\,, &
    I&=1,\dots,16\,.
\end{align}

The embedding (\ref{eq:SG}) is not arbitrary. First, ${\bf V}$ and ${\bf W}_a$ are
of finite order $N$ and $N_{(a)}$, respectively,
\begin{align}
\label{eq:VWorder}
    N {\bf V} &\in \Lambda_{E_8 \times E_8} \,, &  
   N_{(a)} {\bf W}_a &\in \Lambda_{E_8 \times E_8} \,.
\end{align}
Here $\Lambda_{E_8 \times E_8}$ denotes the root lattice of $E_8 \times E_8$ and
there is no summation over the index $a$. 
Note that the $X_L^I$ are only defined up to elements of $\Lambda_{\E8\times\E8}$,
by construction of the heterotic string, and therefore the right-hand side of each of the
conditions expresses trivial action on the fields $X_L^I$.

For each $a=1,\dots,6$, the integer $N_{(a)}$ denotes the `order of the lattice vector ${\bf e}_a$'\index{Lattice!vectors!order}. 
It is a 
divisor of the order $N$ of the orbifold which is determined
by  the equation $(\theta,{\bf e}_a)^{N_{(a)}}=(\theta^{N_{(a)}},0)$.
This states that for $N_{(a)}$ repeated transformations the effect of the 
lattice shift is trivial, as long as twists are involved.
 For the Wilson lines this translates into (\ref{eq:VWorder}).
 
Further conditions on the Wilson lines arise if two lattice shifts are equivalent
on the orbifold, $(\mathbbm 1,{\bf e}_a) \in [(\mathbbm 1,{\bf e}_b)]$. Then
also the two associated Wilson lines have to agree up to root lattice elements.
This restricts the number of distinct Wilson lines in a given orbifold geometry.

Second, there are consistency conditions from string theory,
namely  the requirement that the partition function
is modular invariant
\cite{dhx85,v86,imx88},
\begin{align}
	N \left( {\bf V}^2 -{\bf v}^2\right) &=\jmod 02\,.  
\end{align}
As a consequence, the gauge embedding is generically non-trivial.
Except for special situations, like for example a $\Z 3$ orbifold with ${\bf v}=(1/3,-2/3,1/3;0)$,
this condition implies a non-vanishing shift vector ${\bf V}$.
However, a large ambiguity remains in the definition of the gauge embeddings (\ref{eq:SG}),
especially for the choice of Wilson lines.
This gave rise to systematic scans over parts of the heterotic orbifold landscape 
\cite{lx07}.

With non-vanishing Wilson lines, the above conditions generalize.
 Here we state them in the form of
 `strong modular invariance conditions'\footnote{The counterparts
 are `weak modular invariance conditions', as discussed in
  \cite{s08}. Solutions of the weak modular invariance conditions
  also fulfill the strong modular invariance conditions, after
  adding appropriate elements of $\Lambda_{\E8\times\E8}$.}\index{Modular invariance conditions},
\begin{subequations}
\label{eq:smi}
\begin{eqnarray}
    \frac12\left({\bf V}^2-{\bf v}^2\right) &=& \jmod 01 \,, \\
    {\bf V}\cdot {\bf W}_a &=& \jmod 01\, ,\\
     {\bf W}_a \cdot {\bf W}_b &=& \jmod 01 \qquad \text{for} \, {\bf W}_a \neq {\bf W}_b\,,\\
     \frac 12 {\bf W}_a^2 &=&\jmod 01\,.
\end{eqnarray}
\end{subequations}

The resulting boundary conditions for the fields $X_L^I$ read \index{Boundary conditions!gauge fields}
\begin{align}
\label{eq:bcXLI}
    X_L^I(\sigma_+ +\pi)&=X_L^I(\sigma_+)+\pi \left(p^I +k V^I+m_a W_a^I\right)\,, &
    {\bf p}&=(p^1, \dots,p^{16})\in \Lambda_{\E 8 \times \E 8}\,.
\end{align}
The appearance of the vector ${\bf p}$ in the boundary conditions of the left-movers 
\index{Gauge symmetry!charges}
 is crucial for all what follows. After quantization it will
determine the transformation behavior of a state with respect to the present gauge symmetry.

\subsubsection{Classical solutions}
The equations of motion (\ref{eq:stringEOM}) have straightforward 
solutions\footnote{Contributions $X^I_R(\sigma_-)$ are forbidden by
 world sheet supersymmetry.} (\ref{eq:Xpsivar}),
\begin{align}
    Z^i(\tau,\sigma) &=
    Z^i_L(\sigma_+)+Z^i_R(\sigma_-)\,, &
    \psi^i(\tau,\sigma)&=e^{-2 i H^i(\sigma_-)} \,,&
    X^I(\tau,\sigma) &= X^I_L(\sigma_+)\,,
\end{align}
for $i=1,\dots,4, \, I=1,\dots,16$.
From the boundary conditions (\ref{eq:bcZi}), (\ref{eq:bcH}) 
and (\ref{eq:bcXLI})
one can read off the mode expansions (recall that $\omega^i=\jmod{k v^i}{1}\geqslant0$):
\index{Equations of motion!solutions}
\begin{subequations}
\label{eq:modeexp}
\begin{align}
\label{eq:modeexpZL}
    Z_L^i(\sigma_+)&=f^i_L(\sigma_+) 
    +\frac i2 \sum_{\tilde \nu^i_{\rm sh}\neq 0}
    \frac{1}{\tilde \nu^i_{\rm sh}} \tilde \alpha^i_{\tilde \nu^i_{\rm sh}} e^{-2 i \tilde \nu_{\rm sh}^i \sigma_+}  \,, &
    {\bf \tilde \nu}_{\rm sh} &\equiv {\bf \tilde n}- {\bf \omega}\,,  &
    {\bf \tilde n} &\in \Z{ }^4\,,\\
\label{eq:modeexpZR}
    Z_R^i(\sigma_-)&=f^i_R(\sigma_-)
    +\frac i2 \sum_{\nu^i_{\rm sh}\neq 0}
    \frac{1}{\nu^i_{\rm sh}} \alpha^i_{\nu^i_{\rm sh}} e^{-2 i \nu_{\rm sh}^i \sigma_-}  \,, &
    {\bf \nu}_{\rm sh} &\equiv {\bf n}+  {\bf \omega}\,,  &
    {\bf n} &\in \Z{ }^4\,,\\
 \label{eq:modeexpHR}
   H^i(\sigma_-) &=h^i+q_{\rm sh}^i \sigma_- + \frac i2 \sum_{ m \neq 0}\frac 1m  \beta_{m}^i
   e^{-2i m \sigma_-}\,, &
    {\bf q}_{\rm sh} &\equiv {\bf q}+ k {\bf v}\,,  &
    {\bf q} &\in \Lambda^*_{\SO 8}\,,\\
    \label{eq:modeexpXLI}
    X_L^I(\sigma_+)&=x^I+ p^I_{\rm sh} \sigma_+
    + \frac i2 \sum_{\tilde m\neq 0} \frac{1}{\tilde m} \tilde \alpha^I_{\tilde m} e^{-2 i \tilde m \sigma_+}\,, &
    {\bf p}_{\rm sh} &\equiv {\bf p}+ {\bf V}_g\,,&  {\bf p} &\in \Lambda_{\E 8 \times \E 8}\,,\\
        &&{\bf V}_g&\equiv k {\bf V}+m_a {\bf W}_a \,. \nonumber
\end{align}
\end{subequations}
Similar solutions apply for the conjugate coordinate fields $Z_L^{*i}, Z_R^{*i}, \psi_R^{*i}$.
The appearing constants $ \tilde \alpha^i_{\hat q^i_{\rm sh}},\tilde \alpha^{*i}_{\hat q^{*i}_{\rm sh}},
\alpha^i_{\hat r^i_{\rm sh}},\alpha^{*i}_{\hat r^{*i}_{\rm sh}},h^i,\beta^i_n,
x^I,\tilde \alpha^I_n$ 
depend on the space group element
$g=(\theta^k,m_a{\bf e}_a)$ that defines the  boundary condition.
This is also the case for the functions $f^i_L(\sigma_+)$, $f^{i}_R(\sigma_-)$, 
\begin{align}
\label{eq:fL}
    f^i_L(\sigma_+)&= \left\{ \begin{array}{l l}
    \frac 12z^i + \frac 12 \left( p^i+m_a  e_a^i \right)\sigma_+ \,, & \text{torus $T^i$ is invariant ($\vartheta_{(i)}^k=1$)} \\
    \frac 12z_g^i \,, &  \text{torus $T^i$ has a fixed point $z_g^i$ }
    \end{array}\right. \,, \\
    \label{eq:fR}
     f^i_R(\sigma_-)&= \left\{ \begin{array}{l l}
    \frac 12z^i + \frac 12 \left( p^i-m_a  e_a^i \right)\sigma_- \,, & \text{torus $T^i$ is invariant ($\vartheta_{(i)}^k=1$)}   \\
    \frac 12z_g^i \,, &  \text{torus $T^i$ has a fixed point $z_g^i$ }
    \end{array}\right. \,,
\end{align}
and their conjugates $f_L^{*i} = (f_L^i)^*$, $f_R^{*i} = (f_R^i)^*$.
Here $z^i$ and $p^i$ are arbitrary constants, they describe a free center-of-mass motion of the
string over the bulk of the torus $T^i$.

These solutions contain the localization properties of the string. 
We distinguish between `untwisted' strings\index{Sector!untwisted}, defined by $g=(\mathbbm 1,0)$, and
`twisted' strings\index{Sector!twisted}, which have  $\theta^k \neq \mathbbm 1$ and
 fulfill a non-trivial fixed point (or fixed torus) equation $g {\bf z}_g={\bf z}_g$.
 Equations (\ref{eq:fL}), (\ref{eq:fR}) show that
 untwisted strings are completely delocalized over the full geometry.
In contrast, twisted strings cannot move
freely, they are localized at least in one of the internal tori.

\subsection{Quantization and the low-energy spectrum}
\label{sec:quantization}

\subsubsection{The Hilbert space}

With the explicit mode expansions from the last section
canonical quantization in light-cone gauge is now straightforward. 
Here we first summarize the results before we discuss conditions
on physical states due to the requirement of compatibility with the underlying orbifold.

The states of the quantized theory follow from promoting the appearing coefficients 
 to operators which obey appropriate
canonical commutation 
relations. 
Each fixed point on the orbifold is associated to a class of space group elements $[g]$
and gives rise to distinct boundary conditions. Thus the coefficients of the corresponding
mode expansion and hence also the operators of the quantum theory depend
on this conjugacy class. 
They are defined on an associated Hilbert space $\mathcal H_{[g]}$,
and the total Hilbert space is of the form \index{Hilbert space}
\begin{align}
\label{eq:Hilbert}
   \mathcal H  &=\bigoplus_{[g]} \mathcal H_{[g]}\,,
\end{align}
where the direct sum is taken over all disjoint conjugacy classes $[g] \subset S$. Each contribution
$\mathcal H_{[g]}$ is referred to as a `sector'\index{Sector}. 
We stress that inequivalent solutions of the classical field equations will thus give rise to distinct states in the
effective field theory after quantization, with distinct localization properties and charges.

All states in a sector $\mathcal H_{[g]}$ are 
 subject to projections which define the
physical subset of states. They are `level matching conditions', which ensure that the masses
of left-moving states are equal to the ones of right-moving states, 
orbifold consistency conditions, which we will discuss in some detail later,
and `GSO projections'\index{GSO projection} \cite{gso76}
for the fermions. The latter guarantee that there are as many bosons as fermions in the theory.
In practice, it restricts the quantum numbers ${\bf q}$ to lie either in the integer lattice of $\Lambda^*_{\SO 8}$
(cf.~the upper line of (\ref{eq:wSO8})) or the half-integer lattice (cf.~the lower line of (\ref{eq:wSO8})).

Operators which correspond to positive (negative) frequencies are understood as creation (annihilation) 
operators\index{Hilbert space!operators}, 
\begin{align}
	&  \tilde \alpha^i_n,\tilde \alpha^{*i}_n,
	\alpha^i_n,\alpha^{*i}_n,\tilde \alpha^I_n,\beta^i_n
\;: \;
\left\{\begin{array}{cl}
n<0 & \text{creation operator},\\
n>0 & \text{annihilation operator},\\
\end{array}\right.
\end{align}
and ground states of the sector $\mathcal H_{[g]}$ are defined as states which vanish upon application
of any of the annihilation operators. They carry information about the internal discrete momenta 
${\bf p}_{\rm sh}$ and ${\bf q}_{\rm sh}$ of left- and right-movers, respectively:
\begin{align}
\label{eq:Omega}
    \ket {{\bf p}_{\rm sh},{\bf q}_{\rm sh}}_{[g]} &\equiv  
    \ket {{\bf p}_{\rm sh}}_L \otimes \ket{{\bf q}_{\rm sh}}_R  \otimes \ket{\chi}_{[g]}\,.
\end{align}
Here 
 $\ket{\chi}_{[g]}$ unifies a set of quantum numbers  which are
related to the localization of the state and will be specified shortly.

A general state in $\mathcal H_{[g]}$ \index{Sector!states} \index{Hilbert space!states|see{Sector states}} 
now takes the form ($n_i,m_i,p_i,q_i,r_i,u_i<0$)
\begin{align}
\label{eq:statesHg}
    \underbrace{\tilde \alpha^{i_1}_{n_1}\cdots \tilde \alpha^{i_N}_{n_N}}_{\text{from}\,Z_L^i}
    \underbrace{\tilde \alpha^{*j_1}_{m_1}\cdots \tilde \alpha^{*j_M}_{m_M}}_{\text{from}\,Z_L^{*i}}
	\underbrace{\alpha^{k_1}_{p_1}\cdots \alpha^{k_P}_{p_P}}_{\text{from}\,Z_R^i}
	\underbrace{\alpha^{*l_1}_{q_1}\cdots \alpha^{*l_Q}_{q_Q}}_{\text{from}\,Z_R^{*i}}
	\underbrace{\tilde \alpha^{I_1}_{r_1}\cdots \tilde \alpha^{I_R}_{r_R}}_{\text{from}\,Z_L^I}
	\underbrace{\beta^{s_1}_{u_1}\cdots \beta^{s_U}_{u_U}}_{\text{from}\,H_R^i}
	\ket {{\bf p}_{\rm sh},{\bf q}_{\rm sh}}_{[g]}  \,.
\end{align}
The oscillator numbers of excited states \index{Oscillator numbers}
are counted by the number operators\footnote{
Only the notion of those which are associated with the coordinate fields $Z$ will be needed later.} 
(here no $i$-summation)
\begin{align}
\label{eq:N}
    \tilde N^i &= \frac{1}{\omega^i} \sum_{\tilde \nu_{\rm sh}^i>0} \tilde \alpha_{-\tilde \nu^i_{\rm sh}}^i
     \tilde \alpha_{\tilde \nu^i_{\rm sh}}^{*i}   \,,&
   \tilde N^{*i} &= \frac{1}{\bar \omega^i} \sum_{\tilde \nu_{\rm sh}^{*i}>0} \tilde 
   \alpha_{-\tilde \nu^{*i}_{\rm sh}}^{*i} \tilde \alpha_{\tilde \nu^{*i}_{\rm sh}}^i   \,,
\end{align}
and similarly for $N^i, N^{*i}$. Here $\bar \omega^i,\tilde \nu_{\rm sh}^{*i}$ are defined similar to
$\omega^i,\tilde \nu_{\rm sh}^i$, with replacement $k \rightarrow -k$.
These number operators have non-negative integer eigenvalues, also denoted by
$N^i, N^{*i}, \tilde N^i ,\tilde N^{*i}$ in the following. 
The string theory origins of the appearing quantum numbers of a state
are summarized in Figure \ref{fig:hetboxes}.

\subsubsection{Orbifold compatibility}
\label{sec:compatibility}

Equivalent boundary conditions $g$ and $h g h^{-1}, g,h \in S,$ lead to the same Hilbert space $\mathcal H_{[g]}$,
since on the orbifold the corresponding fixed points are identical (cf.~Equation  (\ref{eq:eqFP})).
Explicitly, \index{Boundary conditions!equivalent}
\begin{subequations}
\begin{align}
\label{eq:bcZ2}
	{\bf Z}(\tau,\sigma + \pi ) &= g {\bf Z}(\tau,\sigma) \,,\\
\label{eq:bcZ3}	
	h {\bf Z}(\tau,\sigma + \pi ) &= (hgh^{-1})h {\bf Z}(\tau,\sigma) 
\end{align}
\end{subequations}
have to lead to the same physical states, and likewise for the other fields. This imposes important
consistency conditions on the physical states in $\mathcal H_{[g]}$, 
\index{Consistency conditions|see{Orbifold \\projection conditions}}
\index{Orbifold!projection conditions}
\begin{align}
	h \ket{\text{phys}} &=\ket{\text{phys}} \hspace{1cm}\text{for all $h\in S$}.
\end{align}
In the following, we will construct these states explicitly. For that we first
calculate a phase $\Phi_h$ which arises due to properties of the solutions (\ref{eq:modeexp}).
Then we will require that this phase is one for physical states, thereby defining
the orbifold projection. 

For each $g=(\theta^k,m_a{\bf e}_a)$ one can solve the equations of motion with boundary conditions
given by (\ref{eq:bcZ2}). After canonical quantization they give rise to Hilbert spaces $\mathcal H_g$,
with ground states \index{Sector!states}
\begin{align}
\label{eq:ketg}
\ket{{\bf p}_{\rm sh}, {\bf q}_{\rm sh}}_g& = 
\ket {{\bf p}_{\rm sh}}_L \otimes \ket{{\bf q}_{\rm sh}}_R  \otimes \ket{\chi_g}\,,
\end{align} 
where $\ket{\chi_g}=\ket{\chi_g^1}\otimes\dots \otimes \ket{\chi_g^4}$ collects quantum numbers which specify
the localization properties of the state: \index{Localization of states}
\begin{align}
\label{eq:chig}
\ket{\chi_g^i} &=\left\{\begin{array}{ll}
      	   \ket{p^i, m_a e_a^i} \,, & \text{$T^i$ is a fixed torus ($\vartheta_{(i)}^k=1$)} \,,\\
	   \ket{z_g^i} \,, &\text{$T^i$ has a fixed point $z_g^i$}\,.
\end{array} \right.
\end{align}
Thus the states in $\mathcal H_g$ either have center-of-mass momentum $p^i$ \index{Center-of-mass
momentum} and possibly
winding numbers $m_a$ \index{Winding modes/numbers}
along the directions $e_a^i$, or they are localized at the fixed point coordinate $z_g^i$
(compare to (\ref{eq:fL})).

In the former case, non-vanishing momenta and winding modes will induce massive states in the effective
low-energy theory. 
For example, 
the momenta $p^i$ take discrete values on the compact torus, which directly map to the Kaluza-Klein mass
of the state in the effective field theory after dimensional reduction. Assuming a compactification scale
much above the low-energy scale, we can ignore such contributions and 
restrict to 
$p^i=m_ae_a^i=0$ for bulk states of the plane $i$.

To keep the notation short in the following, we further introduce
the `local twist vector'\index{Twist vector!local} ${\bf v}_g$ and the `local shift vector'\index{Shift vector!local}
 ${\bf V}_g$, which already appeared in (\ref{eq:modeexp}),
\begin{align}
\label{eq:vgVg}
    {\bf v}_g &\equiv k {\bf v} \,, & {\bf V}_g &\equiv  k {\bf V}+m_a {\bf W}_a\,, 
\end{align}
such that  ${\bf q}_{\rm sh} = {\bf q} + {\bf v}_g$, ${\bf p}_{\rm sh} = {\bf p} + {\bf V}_g$.

\subsubsection*{Transformation rules for states}

A general space group element $h$ 
has non-trivial action on the states of $\mathcal H_g$. The constant parts of $H_R^i, X_L^I$
transform by shifts, cf.~(\ref{eq:gH}), (\ref{eq:gXI}),
\begin{align}
\label{eq:gzgh}
    h^i &\stackrel{h}{\mapsto} h^i - \pi v_h^i\,, &
    x^I &\stackrel{h}{\mapsto} x^I + \pi V_h^I \,. 
\end{align}
For momentum eigenstates 
this corresponds to the following transformation behavior:
\begin{align}
\label{eq:hpq}
   	\ket{ {\bf q}_{\rm sh} }_R &\stackrel{h}{\mapsto} e^{-2 \pi i {\bf q}_{\rm sh} \cdot {\bf v}_h}  \ket{ {\bf q}_{\rm sh} }_R \,, &
	\ket{ {\bf p}_{\rm sh} }_L &\stackrel{h}{\mapsto} e^{2 \pi i {\bf p}_{\rm sh} \cdot {\bf V}_h}  \ket{ {\bf p}_{\rm sh} }_L \,.
\end{align}
From the explicit mode expansions  (\ref{eq:modeexp})  one can infer the
transformation properties of the creation and annihilation operators, for example
\index{Space group!transformation!operators}
\begin{align}
\label{eq:halpha}
    \tilde \alpha^i_{\hat q_{\rm sh}^i} &\stackrel{h}{\mapsto} e^{2 \pi i v_h^i}\tilde \alpha^i_{\hat q_{\rm sh}^i} \,,&
   \tilde \alpha^{*i}_{\hat q_{\rm sh}^{*i}} &\stackrel{h}{\mapsto} e^{-2 \pi i v_h^i}\tilde \alpha^{*i}_{\hat q_{\rm sh}^{*i}} \,.
\end{align}
The corresponding excitations are counted by the integer vectors ${\bf \tilde N}=(\tilde N^1,\dots,\tilde N^4)$ and
 similarly ${\bf \tilde N}^*, {\bf N}, {\bf N}^*$.
 \index{Oscillator numbers}

However, as we will see the relevant states for the low-energy effective theory have ${\bf N}={\bf N}^*=0$,
 and we shall restrict to that case from now on.
It is then convenient to introduce a vector ${\bf R}=(R^1,\dots,R^4)$\index{R-symmetry@$R$-symmetry},
\begin{align}
\label{eq:Ri}
    R^i &\equiv q_{\rm sh}^i +\tilde N^{*i}-\tilde N^i\,,
\end{align}
where the $R^i$ differ for bosons and fermions in a multiplet, as we shall discuss in more detail later.
 A state \index{Sector!states} 
 with oscillator numbers can then be expressed as $\ket{{\bf p}_{\rm sh}, {\bf R} }\equiv
 \big|{\bf p}_{\rm sh},{\bf q}_{\rm sh}; {\bf \tilde N},{\bf \tilde N}^*\big\rangle$,
 \begin{align}
 \label{eq:stateNpq}
 \ket{{\bf p}_{\rm sh}, {\bf R} }&=
 \tilde \alpha_{n_1}^{i_1} \cdots \tilde \alpha_{m_1}^{*j_1}
 \cdots \ket{{\bf p}_{\rm sh}}_L \otimes \ket{{\bf q}_{\rm sh}}_R\,,
 \end{align}
 with transformation behavior \index{Space group!transformation!states}
\begin{align}
\label{eq:hstate}
   |{\bf p}_{\rm sh},{\bf R} \rangle
   &\stackrel{h}{\mapsto} e^{2 \pi i \left( {\bf p}_{\rm sh} \cdot {\bf V}_h
   -{\bf R} \cdot {\bf v}_h
   \right)}
   |{\bf p}_{\rm sh},{\bf R} \rangle\,.
\end{align}

Next we consider the transformation behavior of the center-of-mass contribution $\ket{\chi_g}$,
 defined in (\ref{eq:chig}). If $g$ has a fixed point ${\bf  z}_g$, then $h$ maps this to a fixed point
 of the conjugated element (cf.~(\ref{eq:eqFP})), \index{Space group!transformation!coordinates}
 \begin{align}
 \label{eq:zhz}
	\ket{{\bf z}_g}  
	&\stackrel{h}{\mapsto} \left| \vphantom{z_g}\right. {\bf z}_{h g h^{-1}}\big.\vphantom{z_g}\rangle
	\,.  
\end{align}
In general, this causes an inconsistency: For $[g,h]\neq 0$ the state on the right hand side
 is not contained in $\mathcal H_g$. This problem is solved in $\mathcal H_{[g]}$ by  considering a superposition
 of states\index{Sector!states}, 
 \begin{align}
\label{eq:chig2}
    \ket{\chi}_{[g]} &= \sum_{l \in [g]} \chi_l \ket{{\bf z}_l}\,,
\end{align}
with coefficients $\chi_l$. The states of interest are  eigenstates with respect to the action of all space group elements 
$h=(\theta^{\tilde k},\tilde m_a{\bf e}_a)$\index{Space group!eigenstate},
 \begin{align}
	  \ket{\chi}_{[g]}
	&\stackrel{h}{\mapsto} \tilde \Phi_h \ket{\chi}_{[g]}
	\,, &
	\tilde \Phi_{(\theta^{\tilde k},\tilde m_a{\bf e}_a)}&=\eta_{\tilde m_a{\bf e}_a}'\eta_{\tilde k}\,, 
\end{align}
where the multiplicative  form of the phase $\tilde \Phi_h$ \index{Space group!transformation!phase}
is a consequence of
the observation  $(\theta^{\tilde k},\tilde m_a{\bf e}_a) = (\mathbbm 1,\tilde m_a{\bf e}_a)(\theta^{\tilde k},0)$.
 Consistency with the group multiplication then further requires for arbitrary $k,\tilde k, m_a,\tilde m_a$
\begin{align}
	\eta_{(m_a+\tilde m_a){\bf e}_a}'&=\eta_{m_a{\bf e}_a}'\eta_{\tilde m_a{\bf e}_a}'\,,
	 &  \eta_{k+\tilde k}&=\eta_{k}\eta_{\tilde k}\,, &
	 \eta_{\theta^k m_a{\bf e}_a}'&=\eta_{m_a{\bf e}_a}'\,.
\end{align}
The latter relation follows from the explicit   inverse element (\ref{eq:inverse}).
Note that with (\ref{eq:gccdef}) it also implies that elements from the same conjugacy class
lead to the same geometrical phase, $\tilde \Phi_{\tilde g}=\tilde \Phi_{h\tilde gh^{-1}}$.

Every state is invariant under the unit element, $\tilde \Phi_{(\mathbbm 1,0)}=1$. However,
also for non-vanishing $\eta_{\tilde k}$ there is the possibility that one can find a translation
$m_a^{\tilde k}{\bf e}_a$ such that there is a cancelation between the two factors,
$\eta_{m_a^{\tilde k}{\bf e}_a}'\eta_{\tilde k}=1$. If $g$ itself is among these elements $(\theta^{\tilde k},m_a^{\tilde k}{\bf e}_a)$
  in a sector $\mathcal H_{[g]}$,  invariance of $\ket{{\bf z}_g}$ under
commuting elements (cf.~(\ref{eq:zhz})) is lifted to the superposition $\ket{\chi}_{[g]}$.

In Appendix \ref{app:gammas} we make an explicit ansatz for the superposition (\ref{eq:chig2}),
and find eigenstates $\ket{\chi}_{[g]}=\ket{ \gamma', \gamma }_{[g]} $
 of all $h=(\theta^{\tilde k},\tilde m_a{\bf e}_a)\in S$\index{Space group!eigenstate}\index{Space group!transformation!phase}\index{Space group!invariance|see{Orbifold\\ projection conditions}}\index{Superposition|see{Space group
 eigenstate}},
\begin{align}
\label{eq:eigenstate}
\ket{ \gamma', \gamma }_{[g]}  &=   \left( 
 \sum_{m_a'{\bf e}_a \in \Lambda}
      e^{2 \pi i k  \gamma_a' m_a'}\big(\mathbbm 1,m_a'{\bf e}_a\big) \right) 
     \left( \sum_{n=0}^{N -1} 
    e^{-2 \pi i n  \gamma_a' m_a} e^{-2 \pi i  \gamma n  } \big(\theta^n,0\big)\right)  \ket{{\bf z}_{g}}\,,	\\
     \label{eq:eigeneq}
    \ket{\gamma', \gamma }_{[g]}
     & \stackrel{h}{\mapsto} \tilde \Phi_h
       \ket{ \gamma', \gamma }_{[g]}\,, \quad \quad \quad \quad
       \tilde \Phi_h \equiv  e^{2 \pi i  \gamma_a' ( \tilde  k m_a-k \tilde m_a)}e^{2 \pi i \gamma \tilde k}\,,
\end{align}
where $\gamma'=(\gamma_1',\dots,\gamma_6')$ and $\gamma$ \index{Gamma-phases} are constants that characterize the state,
and $g=(\theta^k,m_a{\bf e}_a)$.
The presence of the transformation phase due to $\gamma'$ is a new result and 
 and extension of what is usually done in the literature, cf.~\cite{bhx06,lx07,lx07c}, but does not
 influence the particle spectra calculated there.
 We will comment more on this  in Sections \ref{sec:concon} and \ref{sec:genpc}.
 
The constants $\gamma',\gamma$
 are constrained by the conditions given in (\ref{eq:condglist0}) in Appendix \ref{app:gammas}.
One finds that consequently they must be of finite order. For example
  \begin{align}
  \label{eq:orderg}
	\gamma&=\frac{j}{N_{(g)}} \,, \hspace{2cm} j=1,\dots,N_{(g)}\,,
\end{align}
where $N_{(g)}$ is defined as the smallest positive integer with
\begin{align}
\label{eq:Ng}
   	\big(\mathbbm 1-\theta^{N_{(g)}}\big) m_a{\bf e}_a &
	\in \Lambda_k\,, \hspace{1cm}
	1 \leqslant N_{(g)} \leqslant k\,.
\end{align}
Note that $N_{(g)}$ is a divisor of $k$.
The $\gamma_a'$ are related to the geometry of the lattice. Their order corresponds to the order 
\index{Lattice!vectors!order} of the
associated lattice vector. An explicit example will be discussed in the next chapter.

The contributions $e^{2 \pi i  \gamma  \tilde k}$ to the phase $\tilde \Phi_h$
are conventionally referred to as `gamma-phases' \index{Gamma-phases} in the literature.
However, we stress that 
generically  a superposition of states from the whole relevant sub-lattice $\Lambda_k$
has to be considered in order to guarantee invariance under the full space group $S$.
This  induces an additional contribution $e^{2 \pi i  \gamma_a' ( \tilde  k m_a-k \tilde m_a)}$
to the   phase $\tilde \Phi_h$,
 which does not vanish for all choices of  $h=(\theta^{\tilde k},\tilde m_a{\bf e}_a)$.

In summary, in a sector $\mathcal H_{[g]}$, where $g=(\theta^k,m_a{\bf e}_a)$,
eigenstates with respect to the action of arbitrary space group elements read \index{Sector!states}
\begin{align}
\label{eq:physstates}
\ket{{\bf p}_{\rm sh}, {\bf R};  \gamma', \gamma}_{[g]}=\ket{{\bf p}_{\rm sh},{\bf R} } \otimes \ket{ \gamma',\gamma}_{[g]}\,.
\end{align}
Under transformation with an element $h=(\theta^{\tilde k},\tilde m{\bf e}_a)$ they pick up a phase $\Phi_h$,
\index{Space group!transformation!phase}
\begin{align}
\label{eq:Phi}
      \ket{{\bf p}_{\rm sh}, {\bf R};  \gamma', \gamma }_{[g]}
      &\stackrel{h}{\mapsto} \Phi_h
      \ket{{\bf p}_{\rm sh}, {\bf R}; \gamma', \gamma }_{[g]} \,, &
      \Phi_h &= e^{2 \pi i  \left( {\bf p}_{\rm sh} \cdot {\bf V}_h
   -{\bf R} \cdot {\bf v}_h
    \right)} \tilde \Phi_h\,,
\end{align}
where $\tilde \Phi_h$ was defined in (\ref{eq:eigeneq}).

\subsubsection*{Consistency conditions}
\label{sec:concon}

The transformation rule (\ref{eq:Phi}) describes the effect of a transition from boundary condition $g$ in (\ref{eq:bcZ2})
to boundary condition $h g h^{-1}$ in (\ref{eq:bcZ3}) on the states of the corresponding quantized theory.
Quantitatively, it is given  in form of a phase $\Phi_h$.
 However, on the underlying orbifold
geometry it is impossible to distinguish between conjugated space group elements. 
This translates into consistency conditions
\index{Orbifold!projection conditions} for physical states \index{Physical states|see{Orbifold\\
projection conditions}} in $\mathcal H_{[g]}$:
\begin{align}
\label{eq:Phi1}
   \mathcal H_{[g]}\;:\quad \quad \Phi_h&=1\qquad \text{for all $h\in S$.}
\end{align}
 
The phase $\Phi$ from (\ref{eq:Phi}) can be decomposed into a universal contribution for all superpositions
of states at equivalent fixed points, and a $  \gamma',\gamma$ \index{Gamma-phases}
 dependent geometrical factor $\tilde \Phi_h$.
If the latter phase is trivial, $\tilde \Phi_h=1$, the consistency condition is a non-trivial constraint
for the allowed Lorentz symmetry and gauge symmetry quantum
numbers, the familiar `orbifold projection'. For the superposition (\ref{eq:eigenstate}), this is the case
for all commuting elements $[h,g]=0$.
 
For $h\in S$ with $\tilde \Phi_h\neq1$,  additional phases  contribute
to the projection condition $\Phi_h=1$. The latter are then often
referred to as the `modified orbifold projection conditions'.
From all possible quantum numbers $({\bf p}_{\rm sh},{\bf R},\gamma', \gamma)$ they
project out the unphysical combinations.
Here the observation is that  for any pair of
charges $({\bf p}_{\rm sh},{\bf R})$, one can find an eigenstate of $h$ for which the  condition is
fulfilled, specified by a choice of the finitely many distinct $\gamma',\gamma$. Thus none of these pairs is absent in the
physical spectrum, but they are grouped into a finite number of subsets, characterized by labels $\gamma',\gamma$.
The latter respect gauge symmetry and supersymmetry, in the sense that all zero modes in a subset form
complete multiplets under these symmetries.
 Let $h\in S$ denote all elements which commute with $g$, and $r\in S$ the rest. Then
 the above statement can be sketched as
\begin{align}
\label{eq:setchphaseeffect}
&\begin{array}{r}
[h,g]=0\\
\text{$[r,g] \neq 0$}
\end{array}\;:&
&\big\{ \big( {\bf p}_{\rm sh},{\bf R}\big)\big\}\stackrel{ \Phi_{h} =1}{\longmapsto} 
\big\{\big({\bf p}_{\rm sh},{\bf R}\big)\big\}_{[g]}\stackrel{ \Phi_r=1}{\longmapsto} 
\big\{\big({\bf p}_{\rm sh},{\bf R};\gamma',\gamma\big) \big\}_{[g]}\,.
\end{align}
 Note that ignoring the existence of the translational gamma phases related
 to $\gamma'$ can only yield a spectrum with the correct charge vectors, if for localized states
 the last step in the above mapping 
 is skipped.  However, throughout this work
  we 
  take the viewpoint that  the information on the superposition of states at the infinitely many sub-lattice points
  is an additional quantum number for physical states,
   and thus the full set of projection conditions (\ref{eq:Phi1}) has to be fulfilled in all sectors.

\subsubsection{Conditions for massless states}

For the construction of an effective field theory as the low-energy limit of the compactified heterotic string
only massless states are of interest,
\begin{align}
\label{eq:m0}
	m_L^2 = m_R^2 = 0\,. 
\end{align}
A state (\ref{eq:statesHg}) from the sector $\mathcal H_{[g]}$ has three contributions to the above masses.
One is due to its internal momenta ${\bf p}_{\rm sh}, {\bf q}_{\rm sh}$. The second is the zero-point energy,
which has a universal
contribution due to the shifts $\omega^i$ 
in the twisted oscillator frequencies:
\begin{align}
\label{eq:deltac}
    \delta c &= \frac 12 \sum_{i=1}^3 \omega^i(1-\omega^i)\,.
\end{align}
Third, oscillator excitations lead to mass contributions, given by
the fractional oscillator numbers
\begin{align}
	\tilde N &= \sum_{i=1}^3 \left( \omega^i \tilde N^i + \bar \omega^i \tilde N^{*i} \right)\,, &
	 N &= \sum_{i=1}^3 \left( \omega^i  N^i + \bar \omega^i  N^{*i} \right)\,.
\end{align}

The resulting conditions for massless states (\ref{eq:m0}) are \index{Mass equations}
\begin{align}
\label{eq:masslessL}
    \frac 18 m_L^2 &= \frac 12 {\bf p}_{\rm sh}^2  -1+\delta c + \tilde N=0 \,, \\
    \label{eq:masslessR}
    \frac 18 m_R^2 &= \frac 12 {\bf q}_{\rm sh}^2 -\frac 12 +\delta c + N=0 \,.  
\end{align}
These equations formulate further conditions for states 
which appear in the low-energy effective field theory.
All other states will have masses of the order of the string scale, and will be neglected
in the analysis.

\subsubsection{The generic spectrum}
\label{sec:genericspectrum}

In general, the space group $S$ acts on all three complex internal dimensions $i=1,2,3$. 
The evaluation of the mass conditions (\ref{eq:masslessL}), (\ref{eq:masslessR}) and the 
orbifold projections (\ref{eq:Phi1}) for all sectors yields all zero modes which are fully compatible with
the compactification. Thus the resulting  spectrum  defines an effective
theory in four dimensions, assuming that the volume of all internal tori is small.
Here we give a brief overview over generic features of the low-energy spectrum of heterotic orbifold models.

\subsubsection* {The gravity sector} 
	Gravity \index{Sector!gravity} lives in the ten-dimensional bulk. Thus corresponding states arise from the untwisted
	sector \index{Sector!untwisted} $\mathcal H_{(\mathbbm 1,0)}$.  Furthermore, these states are uncharged under
	gauge symmetries, ${\bf p}=0$, do not involve superpositions, $\gamma'=\gamma=0$,
	 and include two-tensors with respect to Lorentz symmetry. 
	The relevant solutions of the mass equations (\ref{eq:masslessL}), (\ref{eq:masslessR})
	are
	\begin{align}
		\label{eq:gengravsecsol}
    		&\ket{\bf q}_R \otimes \tilde \alpha_{-1}^i \ket 0_L\,, &
		&\ket{\bf q}_R \otimes \tilde \alpha_{-1}^{*i} \ket 0_L\,,
	\end{align}
	with 
	\begin{align}
\label{eq:qSO8}
    	{\bf q} &= \underbrace{\left(\underline{\pm 1,0,0,0}\right)}_{{\bf 8}_v \; \text{of} \; \SO 8}
	\quad \text{or} \quad 
	{\bf q}= \underbrace{\pm \Big(\frac12,\frac12,\frac12,\frac12\Big), 
	\Big( \underline{\frac 12,\frac 12,-\frac 12,-\frac 12}\Big)}_{{\bf 8}_s \; \text{of} \; \SO 8}\,,
\end{align}
where underlines denote all possible permutations. The bosonic states lie in the integer lattice of $\SO 8$,
they transform as a vector. The fermions have right-moving momenta with half-integer entries, they
transform in the spinor representation ${\bf 8}_s$. The oscillator excitations transform like the coordinates $Z^i, Z^{*i}$,
\begin{align}
	\tilde \alpha_{-1} \equiv \left( \tilde \alpha_{-1}^1,\dots,  \tilde \alpha_{-1}^{4} ,
	\tilde \alpha_{-1}^{*1}, \dots,  \tilde \alpha_{-1}^{*4} \right) &= {\bf 8}_v \; \text{of} \; \SO 8\,.
\end{align}
Hence the uncharged solutions (\ref{eq:gengravsecsol})
of the untwisted mass equations can be identified with the ten-dimensional supergravity multiplet\index{Multiplet!supergravity!ten dimensions}:
\begin{align}
\label{eq:gravmul10}
     {\bf 8}_v \times {\bf 8}_v &= \underbrace{\bf 35}_{g_{MN}} + \underbrace{\bf 28}_{B_{MN}}  +\underbrace{\bf 1}_\phi\,, &
      {\bf 8}_s \times {\bf 8}_v &=\underbrace{{\bf 56}_c}_{\tilde g_{M}}+\underbrace{{\bf 8}_c}_{\tilde \phi} \,.
\end{align}
Here $g_{MN}$ ($\tilde g_{M}$) denotes the graviton (gravitino), $\phi$ ($\tilde \phi$) the dilaton (dilatino) and
$B_{MN}$ the antisymmetric two-form in ten dimensions.

The action of the space group on three of the four complex coordinates breaks the $\SO8 $ to the point group 
\index{Point group}
$\Z N \subset \SO 6\subset \SO 8$ and four dimensional helicity \index{Helicity} $\U 1_h \sim \SO 2 \subset \SO 8$,
\begin{align}
\label{eq:tildeLambda}
    \SO 8 &\rightarrow  \Z N \times \U 1_h \subset \SO 6 \times \U 1_h\,.
\end{align}
In our conventions the fourth entry of the vector ${\bf q}$ determines the transformation properties of a zero mode state
under helicity $\U 1_h$. Furthermore, we choose
 $q^4=0$ and $q^4=-1/2$ ($q^4=+1/2$) to correspond to a scalar and a left-handed (right-handed) Weyl fermion
 \index{Weyl fermion!four dimensions} in
four dimensions, respectively.

Thus without the orbifold projections, the solutions (\ref{eq:gengravsecsol}) give rise to four gravitinos,
 implying
$\mathcal N=4$ supersymmetry in four dimensions. The requirement that this should be broken to $\mathcal N=1$
results in constraints on the choice of viable twist vectors ${\bf v}$. First, we note that also in four dimensions
gravitinos involve a vector and a spinor index. In (\ref{eq:gengravsecsol}) the spinor arises from the
right-moving momentum and the vector from oscillators in flat space directions, $i=4$, which do not transform under the twist.
Thus the orbifold condition (\ref{eq:Phi1}) becomes
\begin{align}
\label{eq:cond4dgravitinos}
       {\bf q} \cdot {\bf v} &=\jmod 0 1\,.
\end{align}
If we choose the vectors ${\bf q}=\pm(1/2,1/2,1/2,1/2)$ from the possibilities in (\ref{eq:qSO8})
as representatives for the two helicity states \index{Helicity} of 
the four-dimensional gravitino\index{Gravitino}, we recover the condition \index{Twist vector!condition for
$\mathcal N=1$} (\ref{eq:condvsusy}) for $\mathcal N=1$ supersymmetry: 
\begin{align}
	 	\sum_j v^j &=0 \,,    &
	v^j &\neq \jmod 01 \,, &
	j&=1,2,3\,.
\end{align}
In summary, the four-dimensional supergravity multiplet \index{Multiplet!supergravity!four dimensions}
is given by the states (\ref{eq:gengravsecsol}), with
${\bf q}=(0,0,0,\pm1), \pm(1/2,1/2,1/2,1/2)$ and oscillator excitations in flat space directions, $i=4$.

Further zero modes arise from excitations in internal directions, $i=1,2,3$. They are again subject to the
orbifold projection conditions, and the number of solutions depends on the space group $\Z N$.
These states are `geometrical moduli'\index{Geometrical moduli}, they describe fluctuations of the internal metric
around the fixed background.

\subsubsection*{The gauge sector}

Further solutions of the mass equations (\ref{eq:masslessL}), (\ref{eq:masslessR}) in the untwisted sector are 
($I=1,\dots,16$) 
\begin{align}
\label{eq:cartan10}
    &\ket{{\bf q}}_R \otimes \tilde \alpha^I_{-1} \ket 0_L\,, &
    &\ket{{\bf q}}_R \otimes \ket{{\bf p}}_L \,, 
\end{align}
with ${\bf q}$ given by (\ref{eq:qSO8}),  and  ${\bf p}^2 =2$. The latter equation has
480 solutions, furthermore the oscillators above give rise to
16 uncharged Cartan generators\index{Cartan generators}. In total, (\ref{eq:cartan10}) 
 forms the gauge vector multiplet \index{Multiplet!vector!ten dimensions} in the adjoint representation of $\E 8 \times \E 8$ in ten dimensions.

Application of the orbifold projection conditions (\ref{eq:Phi1}) restricts the allowed combinations
of internal momenta ${\bf q}$ and ${\bf p}$ above. This never influences the number of Cartan generators,
and therefore the rank of the gauge group cannot be reduced by orbifolding. However, some of the
states with $q^4=\pm1$, which correspond to vectors in four dimensions, are projected out and thus 
the local gauge symmetry of the effective theory is reduced to  a subgroup of $\E8 \times \E8$.

\subsubsection*{The matter sector}

Matter arises from all twisted sectors $\mathcal H_{[g]}$, and also from the untwisted sector. It is given by all
charged massless states which fulfill the orbifold projection conditions and
 transform as chiral multiplets from the four-dimensional perspective ($q^4=0, q^4=-1/2)$, or as their
 conjugates. The matter spectrum is very model dependent; we will discuss a specific example in detail in the
 following chapter.
 
\subsection{Interactions}
\label{sec:selectionrules}

For realistic model building not only the spectrum is important, but also interactions between the various zero modes. 
They arise from correlation functions of string vertex operators, which  can in principle be calculated for orbifolds with
conformal field theory methods \cite{dfx87,krz04}. The results show that 
a superpotential coupling of the form \index{Couplings}
\begin{align}
\label{eq:Wterm}
    W &= \alpha \phi_1 \cdots \phi_M\,,
\end{align}
where $\alpha$ is a moduli dependent constant and $\phi_n, n=1,\dots,M$, are chiral multiplets of the low-energy effective theory,
can only be present if the associated quantum numbers fulfill a certain set of conditions, the `string selection rules'.
Some of them have straightforward interpretations in the effective field theory as gauge invariance and $R$-charge conservation.
Additionally,
it is required that the string boundary conditions are such that they are consistent with string interaction diagrams. 
The
resulting constraint is often referred to as the `space group selection rule', it is sensitive to the localization
properties of the involved states.

\subsubsection[Gauge invariance and discrete   $R$-symmetry]{Gauge invariance and discrete \boldmath $R$-symmetry}

Consider a term of $M$ fields, which arises from a superpotential contribution 
(\ref{eq:Wterm}) and couples  two fermions to $M-2$ bosons.
The explicit form of the string vertex operators for such an interaction then implies the following rules
for the bosonic components of the chiral multiplets $\phi_n$: \index{String selection rules}
\begin{align}
\label{eq:rulep}
   \sum_{n=1}^M {\bf p}_{\rm sh}^{(n)} &=0 \,, \\
   \label{eq:ruleR}
   \sum_{n=1}^M R^j_{(n)} &=\jmod{-1}{N_{(v^j)}} \,, \qquad \qquad j=1,2,3\,.  
\end{align}
Here the contributions from the multiplet $\phi_n$ are labeled by $(n)$, and $N_{(v^j)}$ is the order of
the sub-twist $\vartheta_{(j)}$ in the plane $j$.

Equation (\ref{eq:rulep}) states gauge invariance of the superpotential. 
Similarly, Equation (\ref{eq:ruleR}) describes charge conservation of three discrete $R$-symmetries,
acting as\footnote{Here no summation over $j$ is implied.}\index{R-symmetry@$R$-symmetry}
\begin{align}
\label{eq:Ract}
\phi(x,\theta) &\stackrel{R^j}{\mapsto} e^{2 \pi i R^j v^j} \phi (x,e^{\pi i v^j}\theta) \,, & j&=1,2,3\,,
\end{align}
on chiral multiplets $\phi(x,\theta)$, where the $R^j$ are the bosonic $R$-charges that appear in (\ref{eq:ruleR}),
and $\theta$ denotes the Grassmannian coordinates of four-dimensional superspace, transforming as 
$\theta \mapsto e^{\pi i v^j}\theta$.
The rule (\ref{eq:ruleR}) can then be understood as
\begin{align}
\label{eq:RW}
W &\stackrel{R^j}{\mapsto}  e^{2 \pi i \sum_n R^j_{(n)} v^j} W =  e^{-2 \pi i  v^j}W\,, & j&=1,2,3\,,
\end{align}
whence the action $\int \rmd^2 \theta W$ is invariant under the three discrete $R$-transformations.
Note that there is no continuous $R$-symmetry in the effective theory, as it is often assumed in
field theoretical model building.

As apparent from (\ref{eq:Ract}), the discrete $R$-symmetries do not commute with supersymmetry.
This can also be seen from their definition (\ref{eq:Ri}),
\begin{align}
	\label{eq:Rmod}
    R^j &\equiv q_{\rm sh}^i +\tilde N^{*i}-\tilde N^i=\jmod{k v^j}{1}\,, & j&=1,2,3\,,
\end{align}
 which implies a distinction between
bosons and fermions in a multiplet. Note that these $R$-charges also appear in the transformation phase (\ref{eq:Phi}).

The $N_{(v^j)}$-fold application of the transformation (\ref{eq:RW})  together with
 (\ref{eq:Rmod}) leads to a condition
on the twist quantum numbers $k_{(n)}$:
\begin{align}
	 e^{2 \pi i c^j \sum_n  k_{(n)} /N_{(v^j)} }&=1\,, & c^j &\equiv N_{(v^j)} v^j \in \Z{ }\,, & j&=1,2,3\,.
\end{align}
Since the appearing integers $c^j$ cannot be divisors of $N_{(v^j)}$, this implies the rule
\begin{align}
\label{eq:rulek}
    \sum_{n=1}^M k_{(n)} &= \jmod{0}{N} \,,
\end{align}
where $N$ is the order of the orbifold twist. The fact that this well-known sum rule is related
to the $R$-symmetries was first noted in \cite{bs08}.

\subsubsection{Space group selection rule}

A string interaction is only possible if the boundary conditions $[g_n]=[(\theta^{k_{(n)}},m_a^{(n)}{\bf e}_a)]$ 
associated to the $\phi_n$ in (\ref{eq:Wterm})
multiply to the identity: \index{String selection rules!space group}
\begin{align}
\label{eq:Pig}
    \prod_{n=1}^M [g_n] &= (\mathbbm 1,0) \,.  
\end{align}
The conjugacy classes \index{Conjugacy class} (\ref{eq:gccdef}) can be expressed as 
\begin{align}
\label{eq:cc2}
	[g_n] = (\theta^{k_{(n)}}, \Theta m^{(n)}_a{\bf e}_a+\Lambda_{k_{(n)}})\,,
\end{align}
 with
$\Theta=\{\theta^q|q \in \Z{ }\}$ and  sub-lattices $\Lambda_{k_{(n)}}=(\mathbbm 1-\theta^{k_{(n)}}) \Lambda$, 
which were introduced in (\ref{eq:sublattice}).
This gives 
\begin{align}
[g_n] [g_m]=(\theta^{k_{(n)}+k_{(m)}}, \Theta (m_a^{(n)}{\bf e}_a+m_a^{(m)}{\bf e}_a)+\Lambda_{k_{(n)}+k_{(m)}})\,, 
\end{align}
from which we infer that (\ref{eq:Pig}) 
again gives the condition (\ref{eq:rulek}),
and additionally\footnote{This 
follows from $\Lambda_{\sum k_{(n)}}=0$ and invariance of (\ref{eq:cc2}) under $m_a^{(n)}{\bf e}_a
\rightarrow m_a^{(n)}{\bf e}_a+\lambda$ for $\lambda \in \Lambda_{k_{(n)}}$.}
\begin{eqnarray}
\label{eq:Srules}
    \label{eq:summ0L}
    \sum_{n=1}^M  m_a^{(n)}{\bf e}_a &=& \jmod{0}{\sum_{n=1}^M  \Lambda_{k_{(n)}}} \,.
\end{eqnarray}
This condition restricts the allowed contributions from different fixed points to a superpotential coupling. 
It can thus be interpreted
as a constraint which is sensitive to localization properties of the involved states on the orbifold geometry.

\section{A local GUT in six dimensions}
\label{cpt:lGUT}

The phenomenological 
success of orbifold GUTs in five \cite{k00,hmn02}
 or six \cite{abc01,hno02} dimensions
was one of the main motivations to reconsider orbifold compactifications of the heterotic string \cite{krz04,fnx04,bhx06},
twenty years after their first proposal \cite{dhx85}.
Here an effective
orbifold GUT of co-dimension two is described, which is associated with an anisotropic orbifold
compactification of the heterotic string  \cite{bls07}.

The model which is presented here has the same geometry and gauge embedding as 
the model \cite{bhx06}. The latter is an example for an effective MSSM-like model in four dimensions,
which follows from compactifying the six internal dimensions of heterotic string theory on a specific
orbifold, with specific Wilson lines. It is a representative of a full class of models, as discussed in \cite{lx07},
its spectrum is reviewed in  Appendix \ref{app:4d}.
Here we focus on the possibility that such an orbifold compactification and the idea of supersymmetric
grand unification in the extra dimensions can be unified. This may be achieved by assuming that one
or two of the internal radii are considerably larger than the others, yielding an anisotropic
compactification \cite{w96}. For heterotic orbifolds this is an attractive possibility
to incorporate higher-dimensional GUTs as an intermediate step of the compactification \cite{krz04,fnx04,ht05,bhx06}.
In this picture, the GUT scale is linked to the compactification scale of some large internal compact
dimensions. At the string scale, there is no effective field theory limit and full string propagation
on the ten-dimensional geometry has to be considered.
 
In fact, there is some tension between the requirement that weakly coupled heterotic string theory is applicable
and precision gauge coupling unification \cite{ht05}. Here we shall assume that consistency with this
bound can be achieved in the present model\footnote{For related work 
on this issue see \cite{drw08}.},
once the vacuum structure and the stabilization mechanism are fully understood.

 Apart from the phenomenological interest in
the realization of a supersymmetric GUT in higher dimensions
within the framework of string theory, the main benefit of our approach is a better understanding
of the structure of the model. This will then lead to a simplification of the 
discussion of the vacuum (cf.~Chapter \ref{cpt:vacuum}), and it will provide a link towards
smooth Calabi--Yau compactifications without explicit blow-ups (cf.~Chapter \ref{cpt:K3}).

\subsection[The $\Z{\rm 6-II}$ orbifold]{\boldmath The $\Z{\rm 6-II}$ orbifold}
\label{sec:Z6II}

\begin{figure}[t]
\begin{center}
\includegraphics[width=16cm]{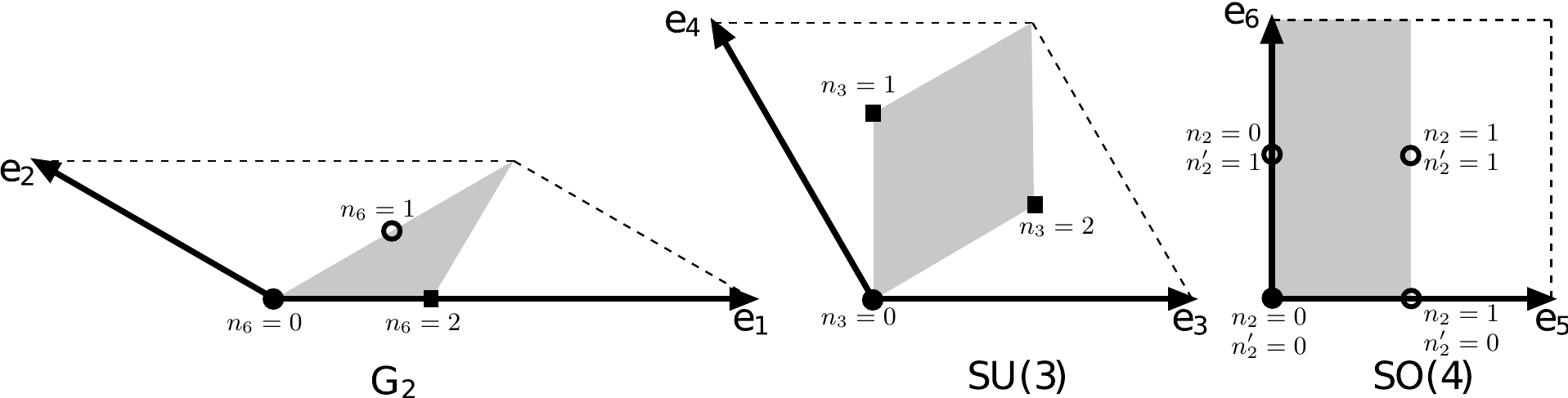}
\caption{The $\Z{\rm 6-II}$ orbifold geometry. 
The shaded area is the fundamental domain of the orbifold,
which by definition is invariant under the twist (\ref{eq:v6}). Note that in each plane the geometry
is that of the surface of a pillow.
Its boundary points are singularities, marked by circles, boxes and bold dots, denoting
fixed points with respect to a space group element with
 $k=3,2,1$, respectively. The fact that for $k=1$ the fixed points  in the $\SU3$- and the $\SO4$-plane
  are also invariant is not illustrated.
  The explicit coordinates and space group elements are collected in Table \ref{tab:fppos}.
}
\label{fig:z6II}
\end{center}
\end{figure}

We consider the compactification of six dimensions on the orbifold \index{Z6-II orbifold@$\Z{\rm 6-II}$ orbifold}
\index{Orbifold!geometry}
\begin{align}
\label{eq:GammaZ6II}
\mathcal M_6&=\frac{\mathbbm C^3}{\Gamma_{\Z{\rm 6-II}} \times \Z{\rm 6-II}}
\equiv\frac{\mathbbm C/\Gamma_{\G2} \times \mathbbm C/\Gamma_{\SU3} \times
 \mathbbm C/\Gamma_{\SO4}}{\Z{\rm 6-II}}
 \,, &
\end{align}
where 
 $\Gamma_G, G=\G2,\SU3,\SO4,$ denotes 
the two-dimensional root lattice\footnote{Throughout
the work we will frequently use the term `$G$-plane', either
 referring to  the two-torus $\mathbbm C/\Gamma_G$ or the corresponding
non-compact plane $\mathbbm C$, depending on  the context.}  of the Lie group $G$ and we omit factors of $\pi$. 
The action of the point group $\Z{\rm 6-II}$ is defined by the twist vector
\begin{align}
\label{eq:v6}
    {\bf v}_6&=\left(-\frac 16,- \frac 13,\frac 12;0 \right)\,.  
\end{align}
This orbifold was first suggested in \cite{krz04} and became a widely studied candidate for realistic model building
 \cite{fnx04,bhx06,lx07,fnw06,lx07c}. Its geometry is summarized in Figure \ref{fig:z6II}, the
 explicit lattice vectors are given in Table \ref{tab:es}.
 
\begin{table}[t]
  \centering
  \footnotesize
  \begin{tabular}{l|l || l|l || l|l}
	\jvsb Plane & Lattice vector &Plane & Lattice vector&Plane & Lattice vector \\
	\hline
	\jvsb $G_2$ & ${\bf e}_1=(1,0,0;0)$ & $\SU3$ & ${\bf e}_3=(0,1,0;0)$
	& $\SO4$ & ${\bf e}_5=(0,0,1;0)$\\
	\jvsb & ${\bf e}_2=\frac{1}{\sqrt 3}(e^{\frac{5 \pi}{6}i},0,0;0)$
	 & & ${\bf e}_4 = (0,e^{\frac{2 \pi}{3}i},0;0)$ 
	  & & ${\bf e}_6 = (0,0,i;0)$
\end{tabular} 
  \caption{Definition of the root lattice $\Gamma_{\Z{\rm 6-II}}$.}\label{tab:es}
\end{table}

The most apparent property of the $\Z{\rm 6-II}$ orbifold is the presence of invariant tori under the
sub-symmetries $\Z 2$ and $\Z 3$, generated by \index{Sub-twists}
 \begin{subequations}
\label{eq:v2v3}
\begin{eqnarray}
\label{eq:v2}
\Z 2\,: \quad&& {\bf v}_2 \equiv 3 {\bf v}_6 = \left(-\frac 12, -1,\frac 32 ;0\right)\,, \\
\label{eq:v3}
    \Z 3\,:\quad && {\bf v}_3 \equiv 2 {\bf v}_6 = \left(-\frac 13, -\frac 23,1;0 \right)\,. 
\end{eqnarray}
\end{subequations}
This shows that the $\SU3$- and the $\SO4$-plane
are invariant under the $\Z2$ and the $\Z3$ sub-twists,
respectively. Consequently, states which are localized in the $\SU3$-plane are bulk states in the
$\SO4$-plane, and vice-versa.

\subsubsection*{Fixed point structure}

The geometry has 12 fixed points and 14 fixed tori, related to 44 sectors $\mathcal H_{[g]}$. 
Each of them corresponds to one of the twisted sectors $T_k$ and
is specified by the following labels: \index{Z6-II orbifold@$\Z{\rm 6-II}$ orbifold!Fixed points/tori}
\begin{subequations}
\label{eq:fplabels}
\begin{align}
\label{eq:fplabels1}
    \text{12 fixed points in $T_1$:}&  & 
    n_6&=0\,, &  n_3&=0,1,2\,, & n_2&=0,1\,, & n_2'&=0,1\,,\\
 \label{eq:fplabels2}   
    \text{6 fixed tori in $T_2$:}&  & 
    n_6&=0,2\,, &  n_3&=0,1,2\,,\\
\label{eq:fplabels3}
    \text{8 fixed tori in $T_3$:}&  & 
    n_6&=0,1\,, &  & & n_2&=0,1\,, & n_2'&=0,1\,, \\
   \label{eq:fplabels4}   
    \text{6 fixed tori in $T_4$:}&  & 
    n_6&=0,2\,, &  n_3&=0,1,2\,,\\ 
 \label{eq:fplabels5}
    \text{12 fixed points in $T_5$:}&  & 
    n_6&=0\,, &  n_3&=0,1,2\,, & n_2&=0,1\,, & n_2'&=0,1\,.    
\end{align}
\end{subequations}
The localization of these fixed points is depicted in Figure \ref{fig:z6II} and quantified in
 Table \ref{tab:fppos}, from which also the associated space group elements can be inferred.
\begin{table}[t!!]
  \centering 
  \footnotesize
  \begin{tabular}{c|| l| l|| l| l|| c| c}
   	\jvsb Plane & \multicolumn{2}{c||}{Fixed Point} & \multicolumn{2}{c||}{Constructing element $g$}
	& \multicolumn{2}{c}{$\gamma$-phase}  \\
	\cline{2-7}
	\jvsb & Label & Coordinate & $k$ & $m_a{\bf e}_a$ & $N_{(g)}$ & $\gamma$\\
	\hline
	\hline
	\jvsb $G_2$ & $n_6=0$ & 0 & $1,2,3,4,5$ & 0 & $1$ &$0$\\
	\cline{2-7}
	\jvsb& $n_6=1$ & $\frac 12 \pi {\bf e}_1+\frac 12 \pi  {\bf e}_2$
	\jvsb&3 &$ {\bf e}_1+{\bf e}_2$ & $3$ &$\frac 13,\frac 23,1$\\
	\cline{2-4}
	\cline{6-7}
	\jvsb& $n_6=2$ & $\frac13 \pi  {\bf e}_1$ & 2 & & $2$ &$\frac 12,1$\\
	\cline{4-5}
	\jvsb& &   & 4 & $-{\bf e}_2$ & &\\	
	\hline
	\hline
	\jvsb$\SU3$ & $n_3=0$ & 0 &  $1,2,4,5$ & 0 \\
	\cline{2-5}
	\jvsb& $n_3=1$ & $\frac 13 \pi {\bf e}_3+\frac 23 \pi {\bf e}_4$&1,4&${\bf e}_4$ \\
	\cline{4-5}
	\jvsb&  &  &2,5&${\bf e}_3+{\bf e}_4$\\
	\cline{2-4}
	\jvsb& $n_3=2$ & $\frac 23 \pi {\bf e}_3+\frac 13 \pi {\bf e}_4$&1,4& \\
	\cline{4-5}
	\jvsb&  &  &2,5&${\bf e}_3$ \\
	\hhline{=====}
	\jvsb$\SO4$&$(n_2,n_2')=(0,0)$&0& $1,3,5$&0 \\
	\jvsb&$(n_2,n_2')=(1,0)$&$\frac 12 \pi {\bf e}_5$&1,3,5&${\bf e}_5$ \\
	\jvsb&$(n_2,n_2')=(0,1)$&$\frac 12 \pi {\bf e}_6$&1,3,5&${\bf e}_6$ \\
	\jvsb&$(n_2,n_2')=(1,1)$&$\frac 12 \pi {\bf e}_5+\frac 12 \pi {\bf e}_6$&1,3,5&${\bf e}_5+{\bf e}_6$ 
\end{tabular}
  \caption{List of the fixed points of the $\Z{\rm 6-II}$ orbifold. Note that
  the given coordinates and the translations of the constructing elements
  only refer to the given plane and have to be
  superposed.
  The planes $\SU3$ and $\SO4$ are invariant under the $\Z 2$ and $\Z 3$ sub-twists which 
  are generated by $\theta^3$ and $\theta^2$, respectively.
  This leads to the appearance of fixed planes, expressed by arbitrary coordinates in the invariant plane.
  The $\gamma$-phase is entirely determined by the localization in the
  $\G2$-plane. The $\gamma'$-phase is universally given by $\gamma'=\frac 16(0,0,2 l_3,2 l_3,3 l_5,3 l_6)$,
  for integers $l_3,l_5,l_6$.
  }\label{tab:fppos}
\end{table}

The number of different sectors \index{Sector} coincides with the number of disjoint conjugacy classes 
of the space group\footnote{We are grateful to Christoph L\"udeling for pointing this out.}.
This follows from the observation that the conjugacy class associated with an
element $(\theta^k,m_a{\bf e}_a)$ can be written as the product of all twists acting on 
the translation, and the conjugacy class \index{Conjugacy class} of the twist,
\begin{align}
	\left[ \left( \theta^k,m_a{\bf e_a} \right) \right] &= \left( \mathbbm 1, \Theta m_a{\bf e}_a \right)
	 \left( \theta^k,\Lambda_k\right)\,.
\end{align}
Here $\Theta=\{\theta^{\tilde k} | \tilde k =0,\dots,5\}$ collects all possible twists and 
$\Lambda_k$ denotes  the $k$-th sub-lattice,
defined in (\ref{eq:sublattice}) and explicitly given in Table \ref{tab:sublas}. 
For each $k$, one can  count the number of translations
$m_a{\bf e}_a$ that induce inequivalent sets and recover the above numbers.

Note that both the $\Z 2$ and the $\Z3$ sub-twist act non-trivially on the $\G2$-plane.
Fixed points of the covering space 
which are invariant under only one of these twists are then mapped onto conjugated fixed points
by the other, cf.~(\ref{eq:eqFP}),  (\ref{eq:zhz}).
Since the string mode expansions, which 
constitute the starting point of quantization, are formulated
in the covering space, conjugated fixed points give rise to different Hilbert spaces.
However, on the orbifold the two points agree and the Hilbert spaces have to be united,
as described in Section \ref{sec:quantization}.
Physical states 
are then superpositions of states at the different conjugated fixed points.
They are specified by quantum numbers $  \gamma', \gamma$ 
\index{Gamma-phases} and take the form (\ref{eq:eigenstate}).

We illustrate this with the example of the non-trivial $\Z3$ fixed point in the
$\G2$-plane, labeled by $n_6=2$. Table \ref{tab:fppos} specifies the generating space group element $g$
and the coordinate $z_g^1$ as \index{Space group!transformation!coordinates}
\begin{align}
	g&=(\theta^2,{\bf e}_1+{\bf e}_2)\,, &
	z_g^1&=\frac 13 \pi e^1_1\,,
\end{align}
where $e^i_a$ denotes the $i$-th entry of the vector ${\bf e}_a$ from Table \ref{tab:es}.
The action of $h=\theta^3$ is then
\begin{align}
z_g^1 \stackrel{h}{\mapsto}\tilde z^1 = \vartheta_{(1)}^3 z_g^1=-z_g^1\,,
\end{align}
which does not lie inside  the chosen fundamental domain of the orbifold. This coordinate is
the fixed point coordinate of the conjugated element $hgh^{-1}$:
\begin{align}
	hgh^{-1}&= (\theta^2,-{\bf e}_1-{\bf e}_2)\,, &
	\tilde z^1&\stackrel{\phantom{A}hgh^{-1}}{\longmapsto}
	-\left(\vartheta_{(1)}^2 z_g^1+e_1^1+e_2^1\right)=-z_g^1=\tilde z^1\,.
\end{align}
Physical states are superpositions
\begin{align}
\label{eq:gamma2}
n_6&=2\,: &
	\ket{\gamma} &=\ket{z_g^1}+e^{-2 \pi i \gamma}\ket{\tilde z^1}\,, & \gamma&=\gamma_{1,2}(\theta^3)=\frac 12,1\,,
\end{align}
up to normalization. They are eigenstates of $h$, \index{Space group!eigenstate}
\begin{align}
	\ket{\gamma} \stackrel{h}{\mapsto} \ket{\tilde z^1}+e^{-2 \pi i \gamma}\ket{z_g^1}
	=e^{2 \pi i \gamma}\ket{\gamma}\,,
\end{align}
as discussed in (\ref{eq:eigenstate}). Similarly the $\Z 2$ fixed point gives three eigenstates, with eigenvalues
\begin{align}
\label{eq:gamma1}
	n_6&=1\,: & &\gamma=\gamma_{1,2,3}(\theta^2)=\frac13, \frac 23,1\,. 
	& \phantom{\gamma}&\phantom{=\gamma_{1,2}(\theta^3)=\frac 12,1\,,}
\end{align}
In summary, states with the same quantum number $n_6$ are localized at the same point in the $\G2$-plane
of the orbifold.  Furthermore, they are grouped into two or three sub-classes of $\Z 6$ eigenstates,
specified by two or three possible eigenvalues 
$\gamma_j(\theta^{\tilde k})$ in (\ref{eq:gamma2}) or (\ref{eq:gamma1}), respectively.

The translational $ \gamma'$-phases under a transformation
by $h=(\theta^{\tilde k},\tilde m_a{\bf e}_a)$ for the $\Z{\rm 6-II}$ geometry are calculated in
Appendix \ref{app:gammas}, with result \index{Space group!transformation!phase}
\index{Z6-II orbifold@$\Z{\rm 6-II}$ orbifold!quantum numbers $\gamma'$}
\begin{align}
\label{eq:gamma'}
     \gamma'=\left(0,0,\frac {l_3}{3},\frac{l_3}{3} ,\frac{l_5}{2},\frac{l_6}{2}\right)\,,
\end{align}
where $l_3,l_5,l_6$ are independent integers.
This directly corresponds to the order of the lattice vector ${\bf e}_a$.
For directions which are related by symmetry transformations the entries coincide,
in the case of unit order of a lattice direction they have to be trivial.

 Recall that $\gamma'$ and $\gamma$ describe superpositions of states at fixed points
 which are related to the constructing element $g=(\theta^k,m_a{\bf e}_a)$ by translational and rotational conjugation,
 respectively, as apparent from (\ref{eq:eigenstate}). The phase which arises upon application
 of $h$ was given in (\ref{eq:eigeneq}),
 \begin{align}
\tilde \Phi_h \equiv  e^{2 \pi i  \gamma_a' ( \tilde k m_a- k \tilde m_a)}e^{2 \pi i \gamma \tilde k}\,. 
\end{align}
For elements $h$ which commute with $g$ one has 
\begin{align}
	\gamma \tilde k &=\jmod{0}{1} \,, & \tilde k \gamma_a' m_a &=\jmod{k \gamma_a' \tilde m_a}{1}\,,
\end{align}
such that the states $\ket{\gamma',\gamma}_{[g]}$ have the property
\begin{align}
[h,g]&=0\;: \hspace{2cm} \tilde \Phi_h=1\,. 
\end{align}

Also note that sub-lattice translations $\tilde m_a{\bf e}_a\in\Lambda_{\tilde k}$ from Table \ref{tab:sublas} 
do not contribute to the phase, $\gamma_a'\tilde m_a=\jmod 01$. This is an important property, since
it has the consequence that it is sufficient to evaluate the consistency condition for finitely many elements
 $h\in S$, which are representatives of disjoint conjugacy
classes.

As an example for the translational superposition, consider the fixed point at the origin of the $\SO4$-plane.
Its constructing element $(\theta,0)$ does not commute with lattice translations in that plane, for example
\begin{align}
	\big( \mathbbm 1, {\bf e}_5 \big) \ket{{\bf z}_{(\theta,0)}} &= \ket{{\bf z}_{(\theta,2 {\bf e_5})}}\,.
\end{align}
The   eigenstates of translations in the direction ${\bf e}_5$ are then given by the superpositions
\begin{align}
	\ket{\gamma'_5}_{[(\theta,0)]}=&\sum_{m\in \Z{ }} e^{-2 \pi i \gamma'_5 m} \ket{{\bf z}_{(\theta,2 m {\bf e_5})}}\,,
	& \gamma'_5&=\frac 12,1\,,
\end{align}
with eigenvalues $\mp 1$.

\subsubsection{The gauge embedding}

For the gauge embedding (\ref{eq:SG}) of the space group $S$ into the gauge group $\E8 \times \E8$,
the independent Wilson lines have to be identified. They correspond to the inequivalent lattice shifts
contained in $S$, which can be read off from the $\gamma'_a$ values in (\ref{eq:gamma'}).
%
There are no Wilson lines in the $\G2$-plane, one of order three in the $\SU3$-plane,
and two of order two in the $\SO4$-plane. We denote them by \index{Wilson lines}
\begin{align}
\label{eq:Ws}
{\bf W}_{(3)} &\equiv {\bf W}_3\,, &
{\bf W}_{(2)} &\equiv {\bf W}_5\,, &
{\bf W}_{(2)}' &\equiv {\bf W}_6\,,
\end{align}
where the ${\bf W}_a$ are associated to the lattice vector ${\bf e}_a$, cf.~(\ref{eq:SG}).
The new subscripts $(n)$ correspond to the order $n$ of the Wilson line.

Following \cite{bhx06}, we choose the gauge embedding \index{Gauge embedding}
\index{Z6-II orbifold@$\Z{\rm 6-II}$ orbifold!gauge embedding}
\begin{subequations}
\label{eq:gaugeemb}
\begin{eqnarray}
\label{eq:V6}
    {\bf V}_6 &=& \left( -\frac 12,-\frac12,\frac13,0,0,0,0,0\right)
    			\left(\frac{17}{6},-\frac 52,-\frac 52,-\frac 52,-\frac 52,-\frac 52,-\frac 52,\frac52\right) \,, \\
  {\bf W}_{(2)} &=& \left(-\frac 12,0,-\frac12,\frac12,\frac12,0,0,0\right)
  			\left(\frac{23}{4},-\frac{25}{4},-\frac{21}{4},-\frac{19}{4},-\frac{25}{4},-\frac{21}{4},-\frac{17}{4},
			\frac{17}{4}\right)\,,\phantom{AAAA} \\
  {\bf W}_{(2)}'&=&0\,,\\
  {\bf W}_{(3)} &=&\left( -\frac 16,\frac12,\frac12,-\frac16,-\frac16,-\frac16,-\frac16,-\frac16\right)
  			\left(0,-\frac23,\frac13,\frac43,-1,0,0,0\right)\,.  
\end{eqnarray}
\end{subequations}
With this choice   the strong modular invariance conditions
(\ref{eq:smi}) are fulfilled. 

The local shift vectors (\ref{eq:vgVg}) for a fixed point element $g$ with quantum numbers 
$k, n_2$ and $n_3$ can now be written in an economical form, \index{Shift vector!local}
\begin{align}
\label{eq:Vgn}
   {\bf V}_g \equiv {\bf V}_{(k,n_2,n_3)}&= k \left( {\bf V}_6+n_2 {\bf W}_{(2)} +n_3{\bf W}_{(3)} \right)\,.
\end{align}
If $g$ corresponds to a fixed $\SU3$- or $\SO4$-torus, the above formula applies
with $n_3=0$ or $n_2=0$, respectively.

The definition of the gauge embedding is the final input which is necessary for the calculation of the
spectrum. In summary, the model is specified by the tori lattices (Table \ref{tab:es}), from  which 
 the twist vector
(\ref{eq:v6}) can be inferred, 
and the gauge embedding (\ref{eq:gaugeemb}). 

\subsubsection{Space group selection rules for interactions}
\label{sec:sgrulesZ6II}

The string selection rules for allowed superpotential couplings \index{Couplings}
 in the general case were discussed  in section
\ref{sec:selectionrules}. While some of them have straightforward interpretations in the effective
field theory as gauge invariance and $R$-symmetry, this is not so clear
for the space group selection rule (\ref{eq:Srules}). For a coupling of $M$ chiral multiplets,
\begin{align}
\label{eq:W2}
    W &= \alpha \phi_1 \cdots \phi_M\,
\end{align} 
with quantum numbers labeled by the index $(l)$, it reads 
\begin{align}
\label{eq:summ2}
    \sum_l  m_a^{(l)}{\bf e}_a &= \jmod{0}{\sum_l  \Lambda_{k_{(l)}}} \,.
\end{align}
The sub-lattices \index{Sub-lattice} that appear on the right hand side of
 this equation are listed in Table \ref{tab:sublas} for the $\Z{\rm 6-II}$ 
geometry.
\begin{table}[t]
  \centering 
  \footnotesize
  \begin{tabular}{l|l|l|l}
	\jvsb Sub-lattice & $G_2$-plane &$\SU3$-plane &$\SO4$-plane \\
	\hline
	\jvsb$\Lambda_1=(\mathbbm 1-\theta^1)\Lambda$ &
	$a{\bf e}_1+b{\bf e}_2$ & $a{\bf e}_3+(3 b-a){\bf e}_4$ &$2a{\bf e}_5+2b{\bf e}_6$ \\
	\jvsb$\Lambda_2=(\mathbbm 1-\theta^2)\Lambda$ &
	$a{\bf e}_1+3b{\bf e}_2$ & $a{\bf e}_3+(3 b-a){\bf e}_4$ &0 \\
	\jvsb$\Lambda_3=(\mathbbm 1-\theta^3)\Lambda$ &
	$2a{\bf e}_1+2b{\bf e}_2$ & 0 &$2a{\bf e}_5+2b{\bf e}_6$ 	
\end{tabular}
  \caption{The sub-lattices of the $\Z{\rm 6-II}$ orbifold. $a$ and $b$ are independent integers
  in each plane. Note that $\Lambda_{6-k}=\Lambda_k$.
  }\label{tab:sublas}
\end{table}
The left hand side is determined by the number of states  at a fixed point $g$,
which can be represented by labels $(k,n_2,n_2',n_3,n_6)$, cf.~(\ref{eq:fplabels}).

For the $\G2$-plane, the sub-lattices $\Lambda_k$ do not coincide for all choices of $k$, cf.~Table \ref{tab:sublas}.
We thus consider each of them separately. For $k=1,5$, the sub-lattice $\Lambda_k=\Lambda_1$ and the root
lattice of $G_2$ are identical. Thus in that case there is no restriction from space group selection rules;
Equation (\ref{eq:summ2}) is always fulfilled.

If all states of a coupling are in the $T_2$ or $T_4$ sector, only the sub-lattice $\Lambda_2$ appears,
which implies a non-trivial condition for the allowed combinations of $m_a^{(l)}$.
Let $N_{k,n_6}$ denote the number of states that contribute to the coupling (\ref{eq:W2})
and which are localized at a fixed point of the
$\G2$-plane with
the corresponding quantum numbers. Inspection of Table \ref{tab:fppos} then gives
\begin{align}
	 \sum_l m_a^{(l)} {\bf e}_a &=
	  N_{2,2} {\bf e}_1+\left(  N_{2,2} -N_{4,2}\right) {\bf e}_2\,.
\end{align}
This vector lies in $\{a{\bf e}_1+3b{\bf e}_2|a,b\in \Z{ } \}\subset\Lambda_2$ if
\begin{align}
\label{eq:N22}
	N_{2,2}+2 N_{4,2}&=\jmod 03 \,.
\end{align}

Similarly one can evaluate (\ref{eq:summ2}) for the case that all contributions 
to a coupling come from the $T_3$ sector,
\begin{align}
	 \sum_l m_a^{(l)} {\bf e}_a &=
	  N_{3,1}\left( {\bf e}_1+  {\bf e}_2\right)\,,
\end{align}
which is in $\{ 2a{\bf e}_1+2b{\bf e}_2|a,b \in \Z{ } \}\subset\Lambda_3$ for
\begin{align}
\label{eq:N31}
	N_{3,1}&=\jmod 02 \,.
\end{align}

In terms of the quantum numbers of the individual states, (\ref{eq:N22}) and (\ref{eq:N31}) can
be unified by the following rule:
\begin{align}
\label{eq:Z6s}
    	\sum_{l} k^{(l)} n^{(l)}_6 &= \jmod 06\,.  
\end{align}
Note that states with $k=1,5$ give no contribution to the sum since they always have $n_6=0$.
We stress that the rule (\ref{eq:Z6s})
 only applies to couplings which have the property that the contributing
multiplets either all come from $T_1$ and $T_5$, or they all come from $T_2$ and $T_4$, or they
all arise from $T_3$. Any term with contributions from at least two of these three
branches will contain the sub-lattice $\Lambda_1$ in the space group selection rule
(\ref{eq:summ2}), which
completely covers the $G_2$ root lattice. In that case the rule (\ref{eq:summ2})
becomes trivial and (\ref{eq:Z6s}) does not apply.

A similar analysis is performed in Appendix \ref{app:sgselrules} for the $\SU3$- and the
$\SO4$-plane. In summary, the space group selection rules for a coupling of $M$ multiplets
 are \index{String selection rules!space group}
\begin{subequations}
\label{eq:srules}
\begin{eqnarray}
	G_2\,: \qquad \sum_{l=1}^M k^{(l)} n^{(l)}_6 &=& \jmod 06\,, \quad 
		\text{if} \;
		\left\{ \begin{array}{c}
			k^{(l)}-k^{(l')}=\jmod 06 \\
			\text{or}\\
			k^{(l)}+ k^{(l')}=\jmod 06
		\end{array}\right\} \; \text{for all $l,l'$}\,,\phantom{AAAA} \\
	%
	\SU3\,: \qquad \sum_{l=1}^M k^{(l)} n^{(l)}_3 &=& \jmod 03\,, \\
	\SO4\,: \qquad 
	\sum_{l=1}^M k^{(l)} n^{(l)}_2 &=& \jmod 02\,,	\\
	 \qquad \sum_{l=1}^M k^{(l)} {n_2'}^{(l)} &=& \jmod 02 \,.
\end{eqnarray}
\end{subequations}
These rules will be interpreted as discrete symmetries \index{Discrete symmetry} in the following chapter.

\subsection{The effective orbifold GUT}
\label{sec:6dgut}

Effective orbifold GUTs can be derived from the heterotic string by assuming that
 the six internal dimensions are compactified on an anisotropic geometry. 
 For the $\Z{\rm 6-II}$ orbifold this can be realized by taking some of the
 independent radii significantly larger than the others.
 %
 We now present the specific example which will then be studied for the remainder of this paper. 
The main motivation for the work is to understand more of the mechanisms behind the
zero mode spectra and couplings. If an intermediate effective GUT is realized in nature, the
knowledge of the details of a particular model can help to identify generic features of 
such compactifications. The hope is that   eventually these may be related to testable predictions. 
However, orbifold models appear to be rather ambiguous and first it is necessary to learn more
about their structure.
On a technical level,
the intermediate GUT approach simplifies the discussion, since the local GUT model is equipped with a larger symmetry
than the four-dimensional  effective theory. We shall profit from that when we study the vacua of the model
in the next chapter.

\subsubsection{Anisotropic limits of orbifolds}

The $\Z{\rm 6-II}$ orbifold  has four independent radii: One each for the $\G2$-plane and the $\SU3$-plane, and another
two for the $\SO4=\SU2\times \SU2$-plane. Thus in principle many effective GUT models can be derived from this
geometry, with any dimension between five and ten. 
In each case the anisotropy \index{Anisotropic geometry} allows one to split the
geometry into a small and a large part (here embedded into six real dimensions),
\begin{align}
\label{eq:Gammasl}
    \mathbbm R^6/\Gamma_{\Z{\rm 6-II}} &=
     \underbrace{\mathbbm R^m/\Gamma_{\rm small}}_{T^m_{\rm small}} \times 
    \underbrace{\mathbbm R^n/\Gamma_{\rm large}}_{T^n_{\rm large}}
 \,, &
    \mathcal M_{\rm eff} &=T^n_{\rm large}/\Z{N_{\rm eff}}\,,
\end{align}
with $m+n=6$.
This leads to a specific spectrum, gauge group and number of supersymmetry generators
for the effective field theory on the $\Z{N_{\rm eff}}$ orbifold $\mathcal M_{\rm eff}$.
The order $N_{\rm eff}$ of the effective orbifold
 is given by the order of the $\Z{\rm 6-II}$ twist in $T^n_{\rm large}$.
The full list of possibilities and the resulting bulk gauge groups
is given in table D.13 of \cite{bhx06}.

The massless spectrum of the string on a given compact orbifold geometry is independent of the
size of the compactification. However, the masses of the Kaluza-Klein modes will depend
on the moduli that describe the internal geometry. In the limit of infinite volume of $T^n_{\rm large}$,
some of these states become massless; they are zero modes of the compactification
of the small dimensions in $T^m_{\rm small}$, but not of
  the full geometry $\mathbbm R^6/\Gamma_{\Z{\rm 6-II}}$ at finite volume.
In the effective intermediate model $\mathcal M_{\rm eff}$, these states correspond to bulk states
with odd boundary conditions, which are projected out in the low-energy limit.
  
We derive the effective orbifold model \index{Effective orbifold}  in a two-step procedure:
\begin{enumerate}
  \item Treat $T^n_{\rm large}$ effectively as flat space
   and compactify only the dimensions in $T^m_{\rm small}$ on a compact geometry $\mathcal M_{\rm small}$.
   In  flat space all coordinates are independent, so the isometries of the latter coincide with the subset of the full space
   group $S$ which leaves $T^n_{\rm large}$ invariant,
   \begin{align}
\label{eq:Seff}
    	S_{\rm small}&=\left\{\left. g \in S \right| T^n_{\rm large} \; \text{is invariant under $g$} \right\}\,.  
\end{align}
  The resulting zero mode spectrum is the bulk
  spectrum of $\mathcal M_{\rm eff}$.
  \item For the local spectrum at a fixed point of $\mathcal M_{\rm eff}$,
  related to a space group element $g_f \in S$, calculate the zero modes of the 
  full $\Z{\rm 6-II}$ compactification to four dimensions which correspond to 
  such a localization.  These states should also be present
  in the anisotropic limit, and no new states should arise.
  For bulk states of the effective field theory, 
  the corresponding local phase $\Phi_{g_f}$ defines the behavior of 
  the associated field under the local $\Z{N_{\rm eff}}$ twist. 
  Zero modes are then required to be compatible with all local boundary conditions, $\Phi_{g_f}=1$.
\end{enumerate}  
  
The latter requirement is indeed equivalent to the  
  projection conditions in the case of a one-step
compactification to four dimensions. 
This follows from the observation that arbitrary elements $h \in S$ can be decomposed into elements  
which represent the different twisted sectors\index{Sector!twisted}. Each of them has a unique description in terms
of the labels $k,n_6,n_3,n_2,n_2'$, as given in (\ref{eq:fplabels}), and 
corresponds to a generating space group element, which we choose as in Table \ref{tab:fppos}.
With notation $g_{k,{\bf n}}$, where ${\bf n}=(n_6,n_3,n_2,n_2'),$ one can write
\begin{align}
\label{eq:decomph}
    h&=\prod_{f=(k,{\bf n})} g_f^{l_f} \,, \quad \quad  l_f \in \Z{ }\,, 
\end{align}
where $f=(k,{\bf n})$ runs over all 44 fixed point sectors. For a given $h\in S$
and the identification $g_f^0 \equiv (\mathbbm 1,0)$ it is always possible
to find an appropriate set of exponents $l_f$.
For example,
\begin{subequations}
\label{eq:motlocproj}
\begin{eqnarray}
	{\bf e}_1 = g_{3,(1,0,0,0)}g_{3,(0,0,0,0)}g_{4,(2,0,0,0)}g_{2,(0,0,0,0)}\,, & &
	{\bf e}_2 = g_{3,(0,0,0,0)}g_{4,(2,0,0,0)}g_{5,(0,0,0,0)}\,, \phantom{AAA}\\
	{\bf e}_3 = g_{2,(0,2,0,0)}g_{4,(0,0,0,0)}\,,\hspace{3.07cm}& &
	{\bf e}_4 = g_{4,(0,1,0,0)}g_{2,(0,0,0,0)}\,, \\
	{\bf e}_5 = g_{3,(0,0,1,0)}g_{3,(0,0,0,0)}\,,\hspace{3.07cm} & &
	{\bf e}_6 = g_{3,(0,0,0,1)}g_{3,(0,0,0,0)}\,.	
\end{eqnarray}
\end{subequations}
Thus for any sector, the conditions \index{Orbifold!projection conditions}
\begin{align}
\label{eq:proj2}
\Phi_{g_f}&=1\hspace{1cm} \text{for all fixed points elements $g_f\in S$}\,,
\end{align}
 are a reformulation of the projection conditions to the zero modes in four dimensions. 
For the above bulk fields in the effective field theory on $\mathcal M_{\rm eff}$, 
this statement 
coincides with the familiar orbifold boundary conditions.

At intermediate scales between the compactification scales of the large and the small dimensions
of the anisotropic orbifold,
the full string compactification   including all massive modes should be described by the effective
orbifold field theory on $\mathcal M_{\rm eff}$ to good approximation. 
In the following, we calculate the corresponding spectrum   by solving \index{Effective orbifold!projection conditions}
 \begin{subequations}
\label{eq:pceffgut}
\begin{eqnarray}
\label{eq:pceff1}
\text{Localized states}:   && \Phi_{g_{f'}}=1 \hspace{1cm} \text{for all fixed point elements}\;g_{f'}\in S \,,\\
\label{eq:pceff2}
\text{Bulk states}:  && \Phi_{g_{f'}}=1 \hspace{1cm} \text{for all}\;{g_{f'}}\,: \;
\text{no localization in $T^n_{\rm large}$}\,,\\
\label{eq:pceff3}
\text{Projection to ${\bf z}_{g_f}$}:  && \Phi_{g_{f'}}=1  \hspace{1cm} \text{for all}\;{g_{f'}}\,: \;
\text{localization at ${\bf z}_{g_f}$ in $T^n_{\rm large}$}\,, \phantom{AAAA}\\
\label{eq:pceff4}
\text{Zero modes}:  && \Phi_{g_{f}}\hspace{.1cm}=1 \hspace{1cm} \text{for all}\;{g_{f}}\,\hspace{.1cm}: \;\;
\text{localization in $T^n_{\rm large}$}\,.
\end{eqnarray}
\end{subequations} 
Here $f,f'=(k,{\bf n})$ correspond to one of the 44 fixed point sectors of the model and
$g_f,g_{f'}$ are some representatives of the associated conjugacy classes.
  Note that localized states as well as bulk states can arise from several inequivalent sectors.
 Each of them has different phases $\Phi_h$ and is   constrained by an
  independent set of  conditions.
  
 The last two conditions apply to bulk states and require that (\ref{eq:pceff2}) is fulfilled
 simultaneously. If the projection (\ref{eq:pceff3}) is evaluated
 for all fixed point elements in the plane, the solutions are
 zero modes of the orbifold.
 
Alternatively to the condition (\ref{eq:pceff1}) for   states localized at ${\bf z}_{g_f}$ in $\mathcal M_{\rm eff}$, 
which is the same as the projection
to zero modes in four dimensions, one could discuss the condition
\begin{align}
\label{eq:pceff1b}
 & \Phi_{g_{f'}}=1 \hspace{1cm}  \text{for all}\;{g_{f'}}\,: \;
\text{localization at ${\bf z}_{g_f}$ in $T^n_{\rm large}$}\,.
\end{align}
The difference to (\ref{eq:pceff1}) is that there,  compatibility with other fixed point elements with localization in
 $T^n_{\rm large}$ is required. This assigns the corresponding translational phases $\gamma'_a$
 \index{Gamma-phases}
 to the states. One observes that the assignment does commute with the four-dimensional gauge
 symmetry, but not with the larger local gauge symmetries at   the orbifold fixed points. The reason for that
 is that precisely the lattice translations which imply the difference between the local gauge symmetries
 and the effective gauge symmetry in the low-energy limit are also responsible for the choice of $\gamma'$,
 for a state with given charge vectors.
 However, what we are aiming at is an effective description for the compactification of the string on
 a six-dimensional orbifold, so the assignment cannot be avoided for a consistent description.
 We conclude that the $\gamma_a'$ values in the directions of $T^n_{\rm large}$ are properties
 of the four-dimensional zero mode limit, rather than quantum numbers of the local theory at a fixed point.

\subsubsection{Geometry of the effective GUT}

\begin{figure}[t!]
\begin{center}
\includegraphics[width=6cm]{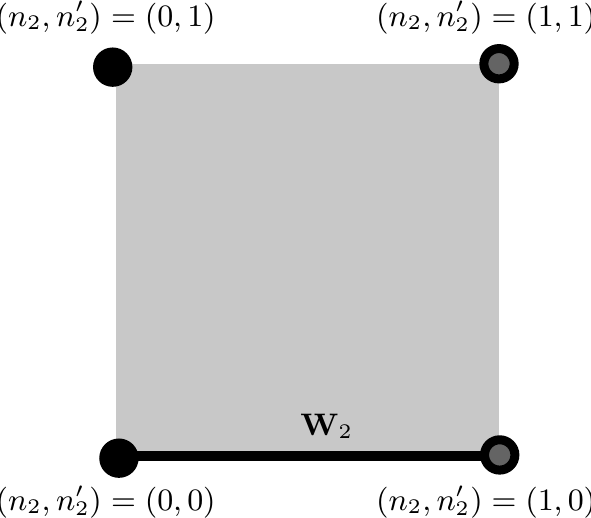}
\caption{Sketch of the geometry of the effective co-dimension two $\Z2$ orbifold. 
It corresponds to the surface of a pillow, which means that there is a back side to the
gray bulk region, cf.~Figure \ref{fig:z6II}. The model has one non-vanishing Wilson line ${\bf W}_2$,
which implies that the fixed points with the same label $n_2$ are equivalent,
the other ones are not.}
\label{fig:so4}
\end{center}
\end{figure}

Starting from the $\Z{\rm 6-II}$ geometry, described in Section \ref{sec:Z6II}, we now propose an effective co-dimension two
orbifold model. In the language of (\ref {eq:Gammasl}) it corresponds to the choice 
(we now switch back to complex coordinates, but still label tori by their real dimension)
\index{Effective orbifold!geometry}
\begin{align}
	 T^4_{\rm small}=\mathbbm C^2/\Gamma_{G_2\times\SU3} 
	  \,, \hspace{2cm} T^2_{\rm large} =\mathbbm C/\Gamma_{\SO 4}\,,
\end{align} 
where the subscripts `large' and `small'   refer to a scale comparable to the GUT scale, and a much
smaller scale, related to the string or the Planck scale, respectively.
 The exact choice should eventually be compatible with precision gauge coupling unification in the resulting model,
 and the perturbative string description.
 
 The $\Z{\rm 6-II}$ twist contains  a $\Z 2$ and a $\Z 3$ sub-twist\index{Sub-twists}, cf.~(\ref{eq:v2v3}),
  \begin{subequations}
\label{eq:v2v3b}
\begin{eqnarray}
\label{eq:v2b}
\Z 2\,: \quad&& {\bf v}_2  = \left(-\frac 12, -1,\frac 32 ;0\right)\,, \quad 
\text{transforms $T^4_{\rm small}$ and $T^2_{\rm large}$}\,, \\
\label{eq:v3b}
    \Z 3\,:\quad && {\bf v}_3   = \left(-\frac 13, -\frac 23,1;0 \right)\,, \quad
    \text{leaves $T^2_{\rm large}$ invariant}\,.
\end{eqnarray}
\end{subequations}
The entries $v_N^i, i=1,2,3,$ correspond to rotations in the $G_2$, $\SU3$ and $\SO4$-plane, respectively. 
The $\SO4$-plane is invariant under the $\Z3$ in $\Z{\rm 6-II}$, and
 we can thus understand the effective 6D orbifold GUT limit as
  \begin{align}
	\mathcal M_6 &= \frac{\mathbbm C^2/\left(\Gamma_{\G2\times\SU3}\times \Z3\right) \times 
	\mathbbm C/\Gamma_{\SO4}}{\Z2} 
	\hspace{.5cm}
	 \stackrel{T^4_{\rm small} \; \text{small}}{\longrightarrow} 
	\hspace{.5cm}
	\mathcal M_{\rm eff} =\mathbbm C/\left(\Gamma_{\SO4} \times \Z 2\right)\,. 
\end{align}
Everything in the effective orbifold model is fixed by the underlying $\Z{\rm 6-II}$ model,
specified by the twist (\ref{eq:v6}) and the gauge embedding (\ref{eq:gaugeemb}).
 Note that the states at the fixed points of $\mathcal M_{\rm eff}$ are invariant under
   $\Z6$, while bulk states arise from modding out  the $\Z3$ subgroup.
   
The effective $\Z2$ orbifold is sketched in Figure \ref{fig:so4}, where also the 
non-vanishing Wilson line \index{Wilson lines} in the ${\bf e}_5$ direction is indicated.
 This has the effect of shifting the local embedding of the $\Z2$ at
 the fixed points with quantum number $n_2=1$ against the one at $n_2=0$.
 Since there is no second Wilson line in the ${\bf e}_6$ direction, fixed points with the
 same label $n_2$ are equivalent, in the sense that their spectra are identical.

\subsubsection{Generic projection conditions}
\label{sec:genpc}

Every  sector $\mathcal H_{[g]}$ \index{Sector} is characterized by a set of labels 
\begin{align}
\text{states at ${\bf z}_g$}\,: \hspace{1cm} k, n_6,n_3,n_2,n_2',l_3,l_5,l_6,\gamma,
{\bf p}_{\rm sh},{\bf R}\,.
\end{align}
In the following we shall use the convention that $\gamma=0$ 
\index{Gamma-phases} either corresponds to the untwisted
sector, or to localization
at the origin of the $\G2$-plane, cf.~Figure \ref{fig:z6II}. We write
\begin{align}
	\gamma&=\frac{j}{N_{(g)}}\,, \hspace{1cm} j=0 \hspace{.5cm}\Leftrightarrow \hspace{.5cm} n_6=0\,,
\end{align}
and $j=1,\dots,N_{(g)}$ otherwise, which means if the localization in the $\G2$-plane is described by $n_6=2$
or $n_6=3$. Furthermore, states which are delocalized in the $\SU3$- or the $\SO4$-plane
have $l_3\equiv0$ or $l_5\equiv0,l_6\equiv 0$ in all formulae, respectively. This is a consequence of
the trivial 
sub-lattices for the states in these planes, which do not give rise to superpositions.

A general $g$ is related to  states which are characterized by
\begin{subequations}
\label{eq:genericprg}
\begin{eqnarray}
	{\bf p}_{\rm sh}&=& {\bf p}+{\bf V}_g = {\bf p}+ k \left( {\bf V}_6+n_2 {\bf W}_{(2)}+n_3 {\bf W}_{(3)} \right)\,, \\
	{\bf R}&=&{\bf q}_{\rm sh} +{\bf \tilde N}^*-{\bf \tilde N}=k {\bf v}_6 +{\bf \tilde N}^*-{\bf \tilde N}\,, \\
	\gamma'_a m_a&=&\frac{l_3}{3}k n_3+\frac{l_5}{2}k n_2+\frac{l_6}{2}k n_2'\,.
\end{eqnarray}
\end{subequations}

We now aim to evaluate the projection condition $\Phi_h=1$ for these states. This we
evaluate for all $h\in S$
or from a relevant subset, for example $h\in S_{\rm small}$, defined in (\ref{eq:Seff}). The 
elements $h$ also have a specific set of labels,
\begin{align}
h\,: \hspace{1cm} \tilde k,  \tilde n_6, \tilde n_3, \tilde n_2, \tilde n_2'\,,
\end{align}
associated with
\begin{subequations}
\label{eq:genericprh}
\begin{eqnarray}
	{\bf V}_{g_{f'}} &=&{\bf V}_g+\left(\tilde k-k\right){\bf V}_6
	+\left(\tilde k \tilde n_2-k n_2\right) {\bf W}_{(2)}+\left(\tilde k \tilde n_3-k n_3\right) {\bf W}_{(3)}\,, \\
	{\bf v}_{g_{f'}}&=&{\bf v}_g+\left(\tilde k-k\right) {\bf v}_6\,,\\
		\gamma'_a \tilde m_a&=&\frac{l_3}{3}\tilde k \tilde n_3+\frac{l_5}{2}\tilde k \tilde n_2
		+\frac{l_6}{2}\tilde k \tilde n_2'\,.
\end{eqnarray}
\end{subequations}

In a first step, one can solve the condition for $h=g$. This is a non-trivial projection of all available states
onto physical subsets of charge vectors,  \index{Orbifold!projection conditions}
\begin{align}
\label{eq:1ststep}
	\text{First step}\,: \hspace{1cm} {\bf p}_{\rm sh} \cdot {\bf V}_g
   -{\bf R} \cdot {\bf v}_g
     =\jmod 01 \hspace{.5cm} \Rightarrow \hspace{.5cm}
    \text{solution for ${\bf p}_{\rm sh},{\bf R}$} \,.
\end{align}
  
 Next, one has to consider $h\neq g$. In that case there may be a non-trivial geometrical phase,
 \index{Geometrical phase|see{Space group\\ transformation phase}}
 \index{Space group!transformation!phase}
 \begin{align}
	\tilde \Phi_h&=e^{2\pi i \frac{l_3}{3} \left( n_3-\tilde n_3 \right) k \tilde k}
	e^{2\pi i \frac{l_5}{2} \left( n_2-\tilde n_2 \right) k \tilde k}
	e^{2\pi i \frac{l_6}{2} \left( n_2'-\tilde n_2' \right) k \tilde k} 
	e^{2\pi i \gamma \tilde k}\,.
\end{align}
With (\ref{eq:genericprh}) one finds from $\Phi_h=1$ for $h$, taken from a subset of 
 $S$, 
 that the additional geometric projection conditions are \index{Orbifold!projection conditions}
\begin{subequations}
\label{eq:2ndstep}
\begin{eqnarray}
\text{Second step}\,: \hspace{5.9cm}
	\left(\tilde n_2'-n_2'\right)\frac{l_6k}{2} &=& \jmod 01\,,\\
	\left(\tilde n_2-n_2\right) \left({\bf p}_{\rm sh} \cdot {\bf W}_{(2)}-\frac{l_5 k}{2}\right)&=&\jmod 01
	\,,\\
	\left(\tilde n_3-n_3\right) \left({\bf p}_{\rm sh} \cdot {\bf W}_{(3)}-\frac{l_3 k}{3}\right)&=&\jmod 01
	\,,\\
	\left(\tilde k-k\right)\left( {\bf p}_{\rm sh} \cdot
	\left( {\bf V}_6+n_2 {\bf W}_{(2)}+n_3 {\bf W}_{(3)}\right)-{\bf R}\cdot {\bf v}_6+\frac{j}{N_{(g)}}\right)
	&=&\jmod 01 
	\,,\hspace{1cm}\phantom{a}
\end{eqnarray}
\end{subequations}
for all $\tilde k, \tilde n_3,\tilde n_2,\tilde n_2'$ that specify the considered elements $h$.
Invariance under the full space group then corresponds to  \index{Orbifold!projection conditions}
\begin{subequations}
\label{eq:2ndstepS}
\begin{eqnarray}
	\frac{l_6k}{2} &=& \jmod 01\,,\\
	\frac{l_5 k}{2}&=&\jmod{{\bf p}_{\rm sh} \cdot {\bf W}_{(2)}}{1}
	\,,\\
	 \frac{l_3 k}{3}&=&\jmod{{\bf p}_{\rm sh} \cdot {\bf W}_{(3)}}{1}
	\,,\\
	\frac{j}{N_{(g)}}&=&\jmod{\left[{\bf R}\cdot {\bf v}_6-{\bf p}_{\rm sh} \cdot
	\left( {\bf V}_6+n_2 {\bf W}_{(2)}+n_3 {\bf W}_{(3)}\right)\right]}{1}\,.
\end{eqnarray}
\end{subequations}
Note that for bulk states in a plane the corresponding geometrical quantum numbers $l_i$
are zero. This may imply further projection conditions for the charge vectors ${\bf p}_{\rm sh}$,
${\bf R}$. Since localization in the $\SO4$-plane requires $k=1,3,5$, one always finds $l_6=\jmod02$
in our model. We can thus completely ignore this quantum number in the following.
This is a consequence of the vanishing Wilson line ${\bf W}_{(2)}'$, associated with the
lattice vector  ${\bf e}_6$.

The geometrical projection conditions  fix the  phases $\gamma',\gamma$ 
\index{Gamma-phases} for physical states.  
A sketch of this two-step projection procedure was given in (\ref{eq:setchphaseeffect}).  
 The   result states that non-trivial phases in translational superpositions
only appear if Wilson lines are present. On first sight, the introduction of $\gamma',\gamma$
 enlarges the physical Hilbert space
since the phases correspond to new eigenstates of the space group operators, which are
overseen if one restricts to the case of unit coefficients.
This larger multiplicity is then reduced by the consistency conditions in the presence of
Wilson lines, and in the end the number of solutions is not increased due to the new phases.
However, we stress that for orbifolds with non-vanishing Wilson lines all geometrical
phases are present and take non-trivial values\footnote{If one is only interested in the 
results for the spectrum, one can decide to ignore the translational $\gamma'$-phases
for localized states from the start, and simultaneously also the projection conditions for elements $h\neq g$
in a sector $\mathcal H_{[g]}$. However, for bulk states  one still requires the
 rules ${\bf p}_{\rm sh}\cdot {\bf W}_a=\jmod 01$. This is indeed a popular approach in the
literature, cf.~\cite{bhx06,lx07,lx07c}.}.
  
\subsubsection{The gravity sector}

The gravity sector \index{Sector!gravity} of the six dimensional effective theory consists of the
supergravity multiplet, a dilaton multiplet (also called tensor multiplet) and gravitational moduli related to the
four small compact dimensions. All these states are uncharged under gauge
symmetries and arise from the untwisted sector. As discussed in section
\ref{sec:genericspectrum}, they are of the form
	\begin{align}
		\label{eq:gengravsecsol2}
		&\big|\big.{\bf p}=0,{\bf R}= {\bf q}+\underbrace{(\underline{1,0,0,0})}_{{\bf \tilde N}^*}
		\big.\big\rangle\,, &
		&\big|\big.{\bf p}=0,{\bf R}= {\bf q}-\underbrace{(\underline{1,0,0,0})}_{{\bf \tilde N}}
		\big.\big\rangle\,,
	\end{align}
	with $\gamma'=0,\gamma=0$, and oscillator numbers ${\bf \tilde N}^*, {\bf \tilde N}$. They
	correspond to excitations associated to operators $ \alpha_{-1}^{*i},\alpha_{-1}^i, i=1,\dots,4$, respectively.
The right-moving momenta ${\bf q}$ are constrained by the conditions (\ref{eq:masslessL}), (\ref{eq:masslessR}),
 which state that the states have to be massless. 
Solutions for ${\bf q}$ take the form (\ref{eq:qSO8}) and correspond to representations ${\bf 8}_v+{\bf 8}_s$ of $\SO 8$,
the oscillators form another ${\bf 8}_v$.

These states are now subject to the projection condition (\ref{eq:pceff2}).
It has to be evaluated for all fixed point elements $g_{f'}$ for which the $\SO4$-plane is a fixed torus.
These are the ones with $ \tilde k=2,4$, since that property is related to the $\Z3$ which leaves $T^2_{\rm large}$
invariant. Untwisted states have $k=0, l_3=0,l_5=0,j=0, \tilde n_2=n_2=0$, and the conditions (\ref{eq:2ndstep}) 
reduce to
\begin{align}
\label{eq:prgr}
	\left( {\bf q} +{\bf \tilde N}^*-{\bf \tilde N} \right) \cdot {\bf v}_3 &=\jmod 01\,.
\end{align}

First, consider solutions with excitations $\alpha_{-1}^i, \alpha_{-1}^{*i}$ in the large planes, $i=3,4$.
These operators will again transform in the vector representation of the little group of the
Lorentz group for the six large dimensions. \index{Little group}
The latter is given by $\SO4 = \SU2\times \SU2$,
\begin{align}
	\SO8 \rightarrow \Z3 \times \SO4 = \Z3 \times \SU2\times \SU2\,,
\end{align}
and the right-moving momenta are 
	\begin{align}
\label{eq:qSO4}
    	{\bf q} &= \underbrace{\left(0,0,\underline{\pm 1,0}\right)}_{({\bf 2},{\bf 2}) \; \text{of} \; \SU2\times \SU2}
	\quad \text{or} \quad 
	{\bf q}= \underbrace{\pm \Big(\frac12,\frac12,\frac12,\frac12\Big), 
	\pm\Big( \frac 12,\frac 12, -\frac 12,-\frac 12\Big)}_{2\cdot(\mathbbm 1, {\bf 2})
	 \; \text{of} \;\SU2 \times \SU2} \,,
\end{align}
where the fermionic states correspond to a left-handed Weyl fermion in six dimensions\index{Weyl fermion!six dimensions}.
Thus in total, one recovers the contents of the supergravity multiplet 
 and the dilaton multiplet 
in six dimensions   \cite{ns86,p98}:
\begin{subequations}
\label{eq:qgrav}
\begin{eqnarray}
	({\bf 2},{\bf 2})  \times ({\bf 2},{\bf 2})  &=& 
	\underbrace{({\bf 3},{\bf 3})}_{g_{MN}} + 
	\underbrace{({\bf 3},\mathbbm 1)}_{B_{MN}^+}+ \underbrace{(\mathbbm 1,{\bf 3})}_{B_{MN}^-}
	+\underbrace{(\mathbbm 1,\mathbbm 1)}_\phi
	\,, \\
	  ({\bf 2},{\bf 2})  \times \left[2 \cdot (\mathbbm 1, {\bf 2}) \right] &=&
	  \underbrace{2\cdot ({\bf 2},{\bf 3})}_{\tilde g_{M}}
	  +\underbrace{2\cdot({\bf 2},\mathbbm 1) }_{\tilde \phi}
	  \,.
\end{eqnarray}
\end{subequations}
Here $B_{MN}$ denotes the antisymmetric two-form\index{Antisymmetric tensor field},
 which can be split into an anti-self-dual part $B_{MN}^-$ and
a self-dual part $B_{MN}^+$. These contributions appear in the supergravity multiplet and the tensor multiplet, 
respectively, where the latter multiplet contains also the dilaton $\phi$. The fermions are given by the 
gravitino $\tilde g_M$ in six dimensions and the dilatino~$\tilde \phi$,
\index{Multiplet!supergravity!six dimensions}
\index{Multiplet!tensor (six dimensions)}
\begin{align}
	&\text{Gravity multiplet:} \quad \left( g_{MN}, B_{MN}^-, \tilde g_M \right)\,, &
	&\text{Tensor multiplet:} \quad \big( B_{MN}^+,\phi,\tilde \phi \big)\,.
\end{align}
Note that the number of bosonic degrees of freedom equals the number of
 fermionic ones within each of these multiplets.

Second, the projection condition (\ref{eq:prgr}) also has solutions with excitations in the internal directions,
\begin{align}
&\left\{\begin{array}{ll}
{\bf \tilde N}=(1,0,0,0), &{\bf \tilde N}^*=0, \\
{\bf \tilde N}=0, &{\bf \tilde N}^*=(1,0,0,0)
\end{array}\right\}& &\text{or}&
 &\left\{\begin{array}{ll}
{\bf \tilde N}=(0,1,0,0), &{\bf \tilde N}^*=0, \\
{\bf \tilde N}=0, &{\bf \tilde N}^*=(0,1,0,0)
\end{array}\right\}. \nonumber
%
\end{align}
For each of these pairs of choices for ${\bf \tilde N}, {\bf \tilde N}^*$,
  the right-moving momenta of four bosonic degrees of freedom and a right-handed Weyl fermion in six dimensions
  are given by \index{Weyl fermion!six dimensions}
	\begin{align}
\label{eq:qSO4b}
    	{\bf q} &=  \underbrace{\left(\underline{\pm1,0},0,0\right)
	 }_{4 \cdot(\mathbbm 1,\mathbbm 1) \; \text{of} \; \SU2\times \SU2}
	\quad \text{or} \quad 
	{\bf q}= 	\underbrace{\pm\Big( \frac 12,-\frac 12,\underline{\frac 12,-\frac 12}
	\Big)}_{2\cdot({\bf 2},\mathbbm 1) \; \text{of} \; \SU2\times \SU2}\,.
\end{align}
These ${\bf q}$ represent the eight degrees of freedom
 that are contained in  one hypermultiplet \index{Multiplet!hypermultiplet (six dimensions)}
  in six dimensions\footnote{This
 is twice the content of a half-multiplet $2\cdot (\mathbbm 1,\mathbbm 1)+({\bf 2},1)$, cf.~appendix
 B.7 of \cite{p98}.}. Note that the oscillator numbers of two states with opposite
 momenta 
 correspond to excitations in complex conjugated directions,
 $({\bf q},{\bf \tilde N}) \leftrightarrow (-{\bf q},{\bf \tilde N}^*)$.
The above solutions give rise to
 two hypermultiplets,
 \begin{align}
\label{eq:C1C2}
    C_1, \;\;C_2\,, 
\end{align}
  which are related to excitations in the planes $G_2$ and $\SU3$,
  respectively.   The associated eight bosonic degrees of freedom are the two corresponding volume moduli
  \index{Geometrical moduli}
  and off-diagonal fluctuations of the metric and the antisymmetric tensor.
   Note that there are no complex structure moduli, since the geometry of the model is fixed
  to the orbifold.

\subsubsection{Gauge vectors}

The gauge symmetry in the six-dimensional bulk is identified by analysis of the massless vector representations 
of the $\SO4$ Lorentz transformations,
which survive the orbifold projection. These states come from the untwisted sector and hence have $\gamma'=\gamma=0$.
There are 16 Cartan generators\index{Cartan generators} 
which survive the projection, as it was the case for $\E8\times \E8$,
but with  right-moving momenta restricted by (\ref{eq:prgr}),
\begin{align}
\label{eq:cartan6}
&\text{16 Cartan generators:}& &\alpha^I_{-1} \ket{{\bf p}=0,{\bf R}={\bf q}}\,,&I&=1,\dots,16\,, 
\end{align}
\begin{align}
\label{eq:qvec2}
{\bf q} &= \underbrace{\left(0,0,\underline{\pm 1,0}\right)}_{({\bf 2},{\bf 2}) \; \text{of} \; \SU2 \times \SU2}
	\quad \text{or} \quad 
{\bf q}= \underbrace{\pm \Big(\frac12,\frac12,\frac12,\frac12\Big), 
	\pm\Big( \frac 12,\frac 12, -\frac 12,-\frac 12\Big)}_{2\cdot
	(\mathbbm 1,{\bf 2}) \; \text{of} \; \SU2 \times \SU2}\,,
\end{align}
These ${\bf q}$ vectors 
describe a  vector multiplet in six dimensions \index{Multiplet!vector!six dimensions}
\cite{p98}.

Next, we find the states with the same right-moving momenta, but also non-vanishing
left-moving momenta ${\bf p}$,
\begin{align}
\label{eq:pVgf'}
&\ket{{\bf p},{\bf q} } \,: & {\bf q}\; &\text{from (\ref{eq:qvec2})}  \,, &
{\bf p}\cdot {\bf V}_{g_{f'}} &=\jmod 01\,.
\end{align}
The latter condition is the projection condition (\ref{eq:pceff2}), $\Phi_{g_{f'}}=1$,
evaluated for fixed point elements $g_{f'}$ which leave the $\SO4$-plane invariant.
Their local shift vectors were defined in (\ref{eq:Vgn}),
\begin{align}
\label{eq:Vgfp}
   {\bf V}_{g_{f'}}  &= \tilde k \left( {\bf V}_6 + \tilde n_3{\bf W}_{(3)} \right)\,, &
   \tilde k&=2,4\,, & \tilde n_3&=0,1,2\,.
\end{align}
Thus (\ref{eq:pVgf'}) is a set of six conditions on the vectors ${\bf p}$. They are equivalent to
two equations, ${\bf p}\cdot {\bf V}_3=\jmod 01$, ${\bf p}\cdot {\bf W}_{(3)}=\jmod 01$,
where ${\bf V}_3 \equiv 2 {\bf V}_6$.  The same conditions follow from
(\ref{eq:2ndstep}), with $k=0, l_3=0, l_5=0, j=0, n_2=\tilde n_2=0$. There are $30^2$ solutions,
\begin{align}
\label{eq:psolbulk}
    {\bf p}&\in 
    \underbrace{\left\{
    \pm(1,0^2,\underline{-1,0^4}),(0^3,\underline{1,-1,0^3})
    \right\}}_{\#=30}\otimes
    \underbrace{\left\{
     (0,\underline{1,-1,0},0^4),(0^4,\underline{\pm1,\pm1,0^2})
    \right\}}_{\#=30}
    \,,
\end{align}
where we introduced the notation $0^n$ for $n$ subsequent entries of zero.
    
The solutions (\ref{eq:psolbulk}) are a subset of the roots \index{Roots} of $\E8 \times \E8$. In fact, they are
the roots of a subgroup, whose embedding  is a consequence
of the specific form of the shift vector and the Wilson line. We now identify this subgroup 
by standard methods, as collected
in Appendix \ref{app:lie}. For that we isolate the simple roots \index{Simple roots}
${\bf p}_i$ among the above set of left-moving
momenta, 
 \begin{align}
\label{eq:srootsbulk}
   	{\bf p}_i \in&\big\{
	\underbrace{(1,0^2,-1,0^4),(0^3,1,-1,0^3),(0^4,1,-1,0^2),(0^5,1,-1,0),(0^6,1,-1)}_{\SU6}\big\} \nonumber \\
	\otimes &\big\{
	\underbrace{(0,1,-1,0^5),(0^2,1,-1,0^4)}_{\SU3},\underbrace{(0^4,1,-1,0^2),
	(0^5,1,-1,0),(0^6,1,-1),(0^6,1,1)}_{\SO8}\big\}\,,
\end{align}
 \begin{table}[t]
    \centering
    \footnotesize
    \begin{tabular}{c|l|c|c|c|c}
  $\U1$ &Generator Embedding into $E_8 \times E_8$ & Bulk & $n_2=0$ & $n_2=1$ & 4d \jvsb \\
  \hhline{======}
  $t_1 $ & $ \left(0, 1, 0, 0, 0, 0, 0, 0\right)\left(0, 0, 0, 0, 0, 0, 0,
    0\right)$ & $\surd$ & $\surd$ & $\surd$ & $\surd$ \jvsb \\ 
  \hhline{------}
  $t_2 $ & $ \left(0, 0, 1, 0, 0, 0, 0, 0\right)\left(0, 0, 0, 0, 0, 0, 0,
    0\right)$ & $\surd$ & $\surd$ & $\surd$ & $\surd$ \jvsb \\ 
  \hhline{------}  
  $t_3 $ & $ \left(1, 0, 0, 1, 1, 1, 1, 1\right)\left(0, 0, 0, 0, 0, 0, 0,
    0\right)$ & $\surd$ & $\surd$ & $\surd$ & $\surd$ \jvsb \\ 
  \hhline{------}  
  $t_4 $ & $ \left(0, 0, 0, 0, 0, 0, 0, 0\right)\left(1, 0, 0, 0, 0, 0, 0,
    0\right)$ & $\surd$ & $\surd$ & $\surd$  & $\surd$\jvsb \\ 
  \hhline{------}  
  $t_5 $ & $ \left(0, 0, 0, 0, 0, 0, 0, 0\right)\left(0, 1, 1, 1, 0, 0, 0,
    0\right)$ & $\surd$ & $\surd$ & $\surd$ & $\surd$ \jvsb \\ 
  \hhline{------}  
  $t^0_6 $ & $ \left(5, 0, 0, -1, -1, -1, -1, -1\right)\left(0, 0, 0, 0, 0, 0, 0, 0\right)$ &
  $\times$ & $\surd$ & $\times$ & $\surd$ \jvsb \\  
  \hhline{------}  
   $t^1_6 $ & $ \left(5, 0, 0, -10, -10, 5, 5, 5\right)\left(0, 0, 0, 0, 0,
    0, 0, 0\right)$ & $\times$ & $\times$ & $\surd$  & $\surd$\jvsb \\ 
  \hhline{------}
  $t_7 $ & $ \left(0, 0, 0, 0, 0, 0, 0, 0\right)\left(0, 1, 1, -2, 0, 0, 0,
    0\right)$ & $\times$ & $\times$ & $\surd$  & $\surd$\jvsb \\ 
  \hhline{------}  
  $t_8 $ & $ \left(0, 0, 0, 0, 0, 0, 0, 0\right)\left(0, 0, 0, 0, -1, -1, -1,
    1\right)$ & $\times$ & $\times$ & $\surd$ & $\surd$ \jvsb \\ 
  \hhline{------}
      $t_Y  $ & $ \left(0,0,0,\frac{1}{2},\frac{1}{2} , -\frac{1}{3},
  -\frac{1}{3},-\frac{1}{3}\right)\left(0,
 0, 0,0, 0, 0, 0, 0\right)$ & $\times$ & $\times$ & $\times$ & $\surd$ \jvsb  \\
   \hhline{------}
  $t_{\rm an}^0 $ & $ \left(5, 0, -4, -1, -1, -1, -1, -1\right)\left(5, -1, -1, -1, 0, 0, 0,
  0\right)$ & $\times$ & $\surd$ & $\times$ & $\surd$ \jvsb \\ 
  \hhline{------}
  $t_{\rm an}^1 $ & $ \left(1, 3, -1, 1, 1, 1, 1, 1\right)\left(-4, 4, 4, 4,
    0, 0, 0, 0\right)$ & $\times$ & $\times$ & $\surd$ & $\surd$ \jvsb \\
  \hhline{------}
  $t_{\rm an}^{(\text{4d})} $ & $ \left(\frac{11}{6},\frac{1}{2} , -\frac{3}{2},
  -\frac{1}{6},-\frac{1}{6}, -\frac{1}{6}, -\frac{1}{6}, -\frac{1}{6}\right)\left(1,
  \frac{1}{3}, \frac{1}{3}, \frac{1}{3}, 0, 0, 0, 0\right)$ & $\times$ & $\times$ & $\times$ & $\surd$ \jvsb  \\
  \hhline{------}
    $t_X  $ & $(0,1,1,-\frac{2}{5},-\frac{2}{5},-\frac{2}{5},-\frac{2}{5},-\frac{2}{5})(\frac{1}{2},\frac{1}{6},\frac{1}{6},\frac{1}{6},
   0,0,0,0)$ & $\times$ & $\surd$ & $\times$ & $\surd$ \jvsb  \\
   \hhline{------}
    $t_{B-L} $ & $(0,1,1,0,0,-\frac{2}{3},-\frac{2}{3},-\frac{2}{3})(\frac{1}{2},\frac{1}{6},\frac{1}{6},\frac{1}{6},0,0,0,0)$ &  $\times$& $\times$ &$\times$  & $\surd$ \jvsb  \\
\end{tabular}
\caption{Definition of the $\U1$ generators. The last four columns
  indicate whether the generator is orthogonal  to all generators
   of the semi-simple group  ($\surd$) in the bulk, at the fixed 
points or in four dimensions, or not ($\times$). 
The anomalous $\U1$'s are linear combinations of the commuting $\U1$'s 
at the fixed point specified by the superscript
or in four dimensions; they are denoted by $t_{\rm an}^0$,
  $t_{\rm an}^1$ and $t_{\rm an}^{(\text{4d})}$, respectively.} 
\label{tab:t}
\end{table}
and read off the group from the corresponding Cartan matrix \index{Cartan matrix}
 $A_{ij}=2 {\bf p}_i \cdot {\bf p}_j/|{\bf p}_i|^2$:
\begin{align}
	A_{ij}&=\underbrace{\left(\begin{array}{rrrrr}
	2 & -1 & 0 & 0 & 0 \\
	-1 & 2 & -1 & 0 & 0\\
	0 & -1 & 2 & -1 & 0\\
	0 & 0 & -1 & 2 & -1\\
	0 & 0 & 0 & -1 & 2
	\end{array}\right)}_{\SU6} \oplus
	\underbrace{\left(\begin{array}{rrrrrr}
	2 & -1 &   \\
	-1 & 2 & \\
	 &  & 2 & -1 & 0 &0\\
	 &  & -1 & 2 & -1&-1\\
	 &  & 0 & -1 & 2&0\\
	 & & 0 &-1 & 0 & 2
	\end{array}\right)}_{\SU3 \times \SO8}	\,.
\end{align}
Here the blank off-diagonal parts have zero entries. The Cartan matrices can be understood as definitions of
the corresponding Lie groups. We conclude that the six-dimensional bulk theory has the gauge group
\index{Gauge symmetry!bulk}
\begin{align}
\label{eq:bulkgaugegroup}
    G_{\rm bulk}&=\SU6 \times \U1^3 \times \left[ \SU3 \times \SO8 \times \U1^2 \right]\,.  
\end{align}
The $\U1$ factors \index{U1@$\U1$!factors} 
have to be present since the total number of uncharged Cartan generators is unchanged
by the orbifold projection condition, cf.~(\ref{eq:cartan6}). For example, $\SU6$ has rank five,
and five Cartan generators
together with the 30 states from (\ref{eq:psolbulk}) form the ${\bf 35}$ of $\SU6$, the adjoint representation.
The remaining three Cartan generators can only generate three additional $\U1$'s in the visible sector 
(which means with origin in the first $\E8$). Similarly, two extra $\U1$'s arise from the hidden sector
(which means with origin in the second $\E8$).
 The generators $t_i, i=1,\dots,5,$ 
 of the five $\U1$ factors \index{U1@$\U1$!generators}
  have to be orthogonal to the simple roots (\ref{eq:srootsbulk}). We choose
 a basis as specified in Table \ref{tab:t}, which also shows a number of other Abelian factors that
 will become important later.
 The charge of a state $\ket{{\bf p}_{\rm sh}}$ under one of these $\U1$'s with generator $t_i$
  is then calculated by 
 \begin{align}
 \label{eq:getQ}
	Q_i \ket{{\bf p}_{\rm sh}}&=t_i \cdot H \ket{{\bf p}_{\rm sh}} = t_i \cdot {\bf p}_{\rm sh} \ket{{\bf p}_{\rm sh}}\,, 
\end{align}
where $H$ denotes the vector of the 16 Cartan generators of $\E8 \times \E8$.

Recalling that the ${\bf q}$ vectors in (\ref{eq:pVgf'}) imply a transformation as a vector multiplet in
six dimensions, we have found the ingredients of a Super-Yang-Mills theory in the bulk, 
with gauge group (\ref{eq:bulkgaugegroup}).
Note that the group factor $\SU6$ is large enough to inherit all standard model gauge fields.

\subsubsection*{Gauge vector projections}

All gauge symmetry of the orbifold model  has its origin in the vector multiplet
of $\E8 \times \E8$, which arises from the untwisted sector. However,
at the orbifold fixed point the symmetry is generically reduced and some of the bulk states are projected out.
This is required since at fixed points, supersymmetry is reduced and its breaking is related to the
orbifold twist, in complete analogy to the local breaking of the gauge group.

For the fixed points $z_{g_f}^i$ of the effective GUT in $T^2_{\rm large}$, this follows from the
additional local projection conditions (\ref{eq:pceff3}). Let the  elements $g_{f'}$  which lead
to a localization at $z_{g_f}^i$ be
characterized by labels $\tilde n_2=n_2, \tilde n_2'=n_2'$, and
\begin{align}
  	&g_{f'}\,: & \tilde k&=1,3,5\,, & {\bf \tilde n}&\,:\;
	\left\{\begin{array}{l l l}
	\tilde n_6=0, & \tilde n_3=0,1,2,    & \text{if}\;\;\tilde  k=1,5 \\
	\tilde n_6=0,1, & \tilde n_3=0,    & \text{if}\;\;\tilde  k=3.
	\end{array}\right.
\end{align}
Then the additional conditions (\ref{eq:2ndstep}) become
\begin{align}
	 \tilde k \left({\bf p} \cdot  \left( {\bf V}_6+n_2{\bf W}_{(2)}\right)
	- {\bf q}\cdot{\bf v}_6\right) &=\jmod 01\,, & {\bf p}\cdot {\bf W}_{(3)}&=\jmod 01\,.
\end{align}
Here we are only interested in states which after the projection form vector multiplets at the fixed points,
the remaining states will be discussed later.
Local gauge vector states \index{Multiplet!vector!four dimensions}
 fulfill the additional constraint ${\bf q}\cdot {\bf v}_6=\jmod 01$, with solutions
\begin{align}
	{\bf q}&=(0,0,0,\pm1)\,, \hspace{1cm} \text{or}  \hspace{1cm} 
	 {\bf q}=\pm \big(\frac12,\frac12,\frac12,\frac12\big)\,.
\end{align}
The left-moving momenta are then constrained by
\begin{align}
 	 {\bf p} \cdot {\bf V}_{g_f}  &=\jmod 01\,, & {\bf p}\cdot {\bf W}_{(3)}&=\jmod 01\,,
\end{align}
which has 50 solutions for $n_2=0$ and 28 for $n_2=1$. They again can be interpreted as simple roots
\index{Simple roots} 
of non-Abelian gauge groups, given in Table \ref{tab:srloc}.
\begin{table}
\centering
\footnotesize
\begin{tabular}{c|c|c}
 \jvsb $n_2$ & non-Abelian gauge groups & Simple roots \\
\hhline{===}
 \jvsb$0$ & $SU(5) \times \big[ SU(3) \times SO(8) \big]$ & 
 $\big\{ \big( 0^3,1,-1,0^3 \big),\big( 0^4,1,-1,0^2 \big),\big( 0^5,1,-1,0 \big),\big( 0^6,1,-1 \big) \big\}$ \\ \jvsb
 & & $\otimes 
 \big\{ \big( 0,1,-1,0^5 \big),\big( 0^2,1,-1,0^4 \big); \big.$ \\ \jvsb
 & & $\big. \big( 0^4,1,-1,0^2 \big),\big( 0^5,1,-1,0 \big),
\big( 0^6,1,-1 \big),\big( 0^6,1,1 \big) \big\}$ \\
\hline
 \jvsb1 & $SU(2) \times SU(4) \times \big[ SU(2)' \times SU(4)' \big]$ &
$\big\{ \big( 0^3,1,-1,0^3 \big); \big( 1,0^4,-1,0^2 \big),\big( 0^5,1,-1,0 \big),\big( 0^6,1,-1 \big) \big\}$ \\ \jvsb
& & $\otimes 
\big\{  \big( 0,1,-1,0^5 \big); \big( 0^4,1,-1,0^2 \big),\big( 0^5,1,-1,0 \big),\big( 0^6,1,1 \big) \big\}$ \\
\hline
\hline
 \jvsb$\cap$ & $SU(2) \times SU(3) \times \big[ SU(2)' \times SU(4)' \big]$
& $\big\{ \big( 0^3,1,-1,0^3 \big);\big( 0^5,1,-1,0 \big),\big( 0^6,1,-1 \big) \big\}$ \\ \jvsb
&& $\otimes \big\{\big( 0,1,-1,0^5 \big);  \big( 0^4,1,-1,0^2 \big),\big( 0^5,1,-1,0 \big),\big( 0^6,1,1\big)  \big\}$ \\
\end{tabular}
\caption{The local gauge groups \index{Gauge symmetry!local} at the fixed points   $n_2=0$ and $n_2=1$, and
their intersection. The latter is given by the product of the electroweak $\SU2$ and the color $\SU3$ of the standard
model in the visible sector. The knowledge of the explicit embedding of the associated simple roots
allows the calculation of  the representations of matter states under the given gauge groups.}
\label{tab:srloc}
\end{table}
The corresponding gauge groups  \index{Gauge symmetry!local}
are
\begin{align}
\label{eq:ggn20}
	n_2&=0\,: & \SU5 \times \U1^4 \times &\left[
	\SU3 \times \SO8 \times \U1^2 \right]\,, \\
	\label{eq:ggn21}
	n_2&=1\,: & \SU2\times \SU4 \times \U1^4  \times &\left[
	\SU2' \times \SU4'  \times \U1^4 \right]\,,
\end{align}
and the intersection in the visible sector contains the standard model gauge group\index{Gauge symmetry!four dimensions},
 \begin{align}
	  \SU3 \times \SU2 \times \U1^5 \times &\left[
	\SU2' \times \SU4' \times \U1^4 \right]\,.
\end{align}
 The Abelian factors have generators $t_i$ which are defined in Table \ref{tab:t}.
 Note that the inequivalent fixed points $n_2=0$ and $n_2=1$ share
 the five $\U1$ factors from the bulk, but extend it by one more factor, 
 with generator
 denoted by  $t_6^0$ or $t_6^1$, respectively. The latter are neither collinear nor
 orthogonal in $\E8 \times \E8$ weight space. 
 
 The spontaneous breakdown of
 all Abelian groups, except for hypercharge, under the constraint of a quasi-realistic
 phenomenology is the topic of the next chapter. Here we are  interested in the
 spectrum of the model and the transformation of physical states under the full gauge
 groups given above.
 
 In particular, we shall be interested in the $\SU5$ GUT which arises at the fixed points
 with $n_2=0$. Even though the underlying orbifold model \cite{bhx06} was designed
 to contain two generations of the standard model as ${\bf 16}$-plets
 of  $\SO{10}$, as will be apparent later, 
 the  anisotropic limit under consideration results in a local $\SU5$ model. 
 The breaking to the standard model gauge group is due to the non-trivial Wilson line
 ${\bf W}_{(2)}$, or in other words, due to the existence of a second pair of 
 inequivalent fixed points,
 corresponding to $n_2=1$.
 Since that is the only purpose of this second pair, and since  it turns out that 
 all matter which is localized at these fixed
 points is exotic and has to be decoupled, we do not expect that models with different
 Wilson lines ${\bf W}_{(2)}$ are fundamentally different in this class of compactifications.

\subsubsection{Matter states}
\label{sec:matter}

Matter states $\ket{{\bf p}_{\rm sh},{\bf R},\gamma',\gamma}$ are in general charged under 
all or some of the generators associated with the relevant gauge group. 
We now describe how these states can be combined  into multiplets. 

As summarized in Appendix \ref{app:lie},
a representation is fully specified by its Dynkin labels $l=(l_1,\dots,l_r)$, where
$r$ is the total rank of the gauge group $G$. The vector $l$ arises from multiplication of the 
weight ${\bf p}_{\rm sh}$
  with the simple roots ${\bf p}_i$,   taken from (\ref{eq:srootsbulk}) in the case of bulk states or
  from Table \ref{tab:srloc} for localized states. We evaluate the Cartan--Weyl
  labels \index{Cartan--Weyl labels} $l_i$ for arbitrary weights, defined by
\begin{align}
\label{eq:getli}
    l_i&\equiv2\frac{{\bf p}_i \cdot {\bf p}_{\rm sh}}{|{\bf p_i}|^2}\,, & &{\bf p}_i \,:\;\text{simple roots of $G$}\,, 
    & i&=1,\dots,r={\rm rank}(G)\,.
\end{align}
The Dynkin labels \index{Dynkin labels} of a representation are then given by the vector
$l$ of the highest weight state. Our approach here is to calculate $l$ for all present states, and then identify the
ones which belong to the same representation of the relevant gauge group.
\begin{figure}
\begin{center}
\includegraphics[height=3.6cm]{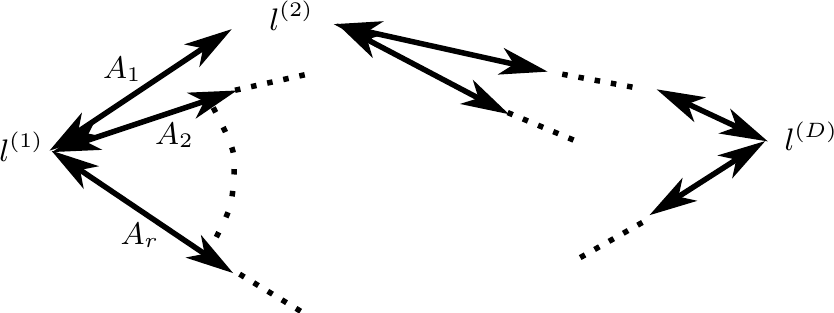}
\caption{Sketch of a $D$-dimensional representation of a Lie group $G$. Shown are some of the Cartan--Weyl
labels $l^{(n)}, n=1,\dots,D$, defined in (\ref{eq:getli}). They transform under the raising and lowering operators,
represented by columns $A_i$ of the Cartan matrix,
$i=1,\dots,r={\rm rank}(G)$. They arise from the subgroups $\SU2_i$, which are not orthogonal.
Note that $A_i$ ($-A_i$) can only be applied if the $i-th$ entry of $l$ is negative (positive).
Details can be found in Appendix \ref{app:lie}.}
\label{fig:dynkin}
\end{center}
\end{figure}

The $l_i$ are integer numbers. Recall that every simple root ${\bf p}_i$ is associated with a specific $\SU2_i$ subgroup
of the gauge group, which can be associated with raising and lowering operators. The fact that
all these $\SU2$ factors are interrelated is encoded in the off-diagonal entries of the Cartan matrix 
\index{Cartan matrix} $A_{ij}$.
The  labels $l_i$ can then be understood as the number of times the   lowering operator 
(or raising operator for $l_i<0$) 
which corresponds to $\SU2_i$ can be
applied to the state, before it gives zero. The action of these operators on a state $l$ is   given by  subtracting (or adding)
column number $i$ of the
Cartan matrix, which we denote as $A_i$. 
In summary, a representation of dimension $D$ consists of $D$ states\footnote{The adjoint
representation is an exception, since the Cartan generators are not covered by the presented
method. In that case, $D$ has to be replaced by the dimension of the group minus its rank.} with   labels
$l^{(n)}, n=1,\dots,D$, which are related by the action of $A_{i}, i=1,\dots,r$. A sketch of this relation is given in
Figure \ref{fig:dynkin}. A simple example is the ${\bf 3}$ of $\SU3$,
\begin{align}
&l^{(1)}=\left(\begin{array}{r}1\\0\end{array}\right)
\stackrel{-A_1}{\longmapsto}l^{(2)}=
\left(\begin{array}{r}-1\\1\end{array}\right)
\stackrel{-A_2}{\longmapsto}l^{(3)}=
\left(\begin{array}{r}0\\-1\end{array}\right)\,, &
A_{ij}&=\left( \begin{array}{rr}
2 & -1\\
-1&2
\end{array}\right)\,. 
\end{align}

Similarly, we now calculate the representations that make up the sets $\{{\bf p}_{\rm sh}\}$,
 which we obtain from solving the
projection conditions for bulk matter and localized states.
     
\subsubsection*{Bulk matter from the untwisted sector}

Not all states from the untwisted sector \index{Sector!untwisted}
are contained in vector multiplets. The projection conditions $\Phi_{g_{f'}}=1$ for
states with ${\bf q}\cdot {\bf v}_3 \neq \jmod 01$ then give rise to hypermultiplets in six dimensions.
With elements $g_{f'}$ determined by the quantum numbers $\tilde k=2,4$, $\tilde n_6=0,2$, $\tilde n_3=0,1,2$ 
and $\tilde n_2=\tilde n_2'=0$ the conditions (\ref{eq:2ndstep}) become
\begin{align}
\label{eq:pc1}
	{\bf p} \cdot {\bf V}_3-{\bf q}\cdot {\bf v}_3&=\jmod 01\,, & {\bf p}\cdot {\bf W}_{(3)}&=\jmod 01\,,
\end{align}
where $l_3=j=n_2=0$ and ${\bf \tilde N}^*={\bf \tilde N}=0$ for untwisted states was used.
The conditions (\ref{eq:pc1}) have 384 solutions $({\bf p},{\bf q})$, which form 48 hypermultiplets
with right-moving momenta ${\bf q}$ as in (\ref{eq:qSO4b}),
\begin{align}
\label{eq:qu2}
    	{\bf q} &=  \left(\underline{\pm1,0},0,0\right)
	\quad \text{or} \quad 
	{\bf q}= 	 \pm\Big( \frac 12,-\frac 12,\underline{\frac 12,-\frac 12}
	\Big) \,.
\end{align}

As described above, one can now calculate the labels $l_i$ by evaluating Equation (\ref{eq:getli}) for
each of the appearing left-moving momenta ${\bf p}_{\rm sh}={\bf p}$. For example, 
consider ${\bf q}=(-\frac12,\frac12,\frac12,-\frac12)$ and
\begin{align}
\label{eq:astate1}
	&\left.\begin{array}{l}
	{\bf p}_i,: \; \text{simple roots of $\SU6$ from (\ref{eq:srootsbulk}), } \\
	{\bf p}_{\rm sh}  =(\frac12,-\frac12, \frac12, \frac12,\frac12, -\frac12, -\frac12, -\frac12)(0^8)
	\end{array}\right\} & l&=(0,0,1,0,0)\,.
\end{align}
The resulting $l$ is the Dynkin label  of the ${\bf 20}$ of $\SU6$. In fact, there are exactly 19 other non-trivial $l$ vectors
with respect to the simple roots of $\SU6$ among the 48 solutions, 
which transform as expected and complete the representation. We can now also calculate the five
$\U1$ charges \index{U1@$\U1$!charges} of the multiplet. According to (\ref{eq:getQ}) this is done by multiplying the
vector ${\bf p}_{\rm sh}$ from (\ref{eq:astate1}) with the generators $t_i, i=1,\dots,5$, from
Table \ref{tab:t}, with result
\begin{align}
{\bf Q} \equiv \left(t_1\cdot   {\bf p}_{\rm sh}, \dots,t_5\cdot   {\bf p}_{\rm sh}\right) &= \left(
-\frac 12,\frac12,0,0,0\right)\,.
\end{align}
We have thus found a hypermultiplet $({\bf 20};\mathbbm 1,\mathbbm1)$ of $\SU6 \times [\SU3\times
\SO8]$, with charge vector ${\bf Q}$. The charge conjugated states have momenta $(-{\bf p},-{\bf q})$, and therefore
  inverted $\U1$ charges, $-{\bf Q}$.

The 28 remaining ${\bf p}$ vectors can be treated similarly. In summary,
the untwisted matter states $\ket{{\bf p},{\bf q}}$ are given by combinations of
\begin{align}
\label{eq:mults0}
    	&{\bf q}\,:\; \text{8 states of a hypermultiplet, given in (\ref{eq:qu2})}\,, \\ \label{eq:multis1}
	&{\bf p}\,: \; ({\bf 20};\mathbbm 1,\mathbbm1)+(\mathbbm 1,\mathbbm1,{\bf 8})+
	(\mathbbm 1,\mathbbm1,{\bf 8}_s)+(\mathbbm 1,\mathbbm1,{\bf 8}_c)+4\cdot
	(\mathbbm 1,\mathbbm1,\mathbbm 1)\,.
\end{align}
 Here the vector representation ${\bf 8}$ of $\SO8$
with Dynkin labels $(1,0,0,0)$ appears, as well as the two spinor representations ${\bf 8}_s,{\bf 8}_c$,
 with Dynkin labels
$(0,0,0,1)$ and $(0,0,1,0)$, respectively. The corresponding $\U1$ charges of the states can be inferred
from Table \ref{tab:bulkU}.
\begin{table}[t]
\centering
\footnotesize
\begin{tabular}{c|c||c|c|c|c|c}
\multicolumn{2}{c||}{Multiplet} & $t_1$ & $t_2$ & $t_3$ & $t_4$ & $t_5$\jvsb  \\
\hhline{=======}
\jvsb$({\bf 20}, \mathbbm 1,\mathbbm 1)$ && $-\frac{1}{2}$ & $\frac{1}{2}$ & $0$ & $0$ & $0$ \\
\jvsb$(\mathbbm 1, \mathbbm 1,{\bf 8})$& & $0$ & $0$ & $0$ & $-1$ & $0$ \\
\jvsb$(\mathbbm 1, \mathbbm 1,{\bf 8}_s)$ && $0$ & $0$ & $0$ & $\frac{1}{2}$ & $\frac{3}{2}$ \\
\jvsb$(\mathbbm 1, \mathbbm 1,{\bf 8}_c)$ && $0$ & $0$ & $0$ & $\frac{1}{2}$ & $-\frac{3}{2}$ \\
\jvsb$(\mathbbm 1, \mathbbm 1,\mathbbm 1)$ &$U_1$& $\frac{1}{2}$ & $\frac{1}{2}$ & $3$ & $0$ & $0$ \\
\jvsb$(\mathbbm 1, \mathbbm 1,\mathbbm 1)$ &$U_2$& $\frac{1}{2}$ & $\frac{1}{2}$ & $-3$ & $0$ & $0$ \\
\jvsb$(\mathbbm 1, \mathbbm 1,\mathbbm 1)$ &$U_3$& $1$ & $-1$ & $0$ & $0$ & $0$ \\
\jvsb$(\mathbbm 1, \mathbbm 1,\mathbbm 1)$ &$U_4$& $-1$ & $-1$ & $0$ & $0$ & $0$ \\
\end{tabular}
\caption{Bulk matter multiplets \index{Spectrum!bulk}
 from the untwisted sector. Shown are states with ${\bf q}=(-1,0,0,0),(-\frac12,\frac12,
\frac12,-\frac12);(0,1,0,0)(-\frac12,\frac12,-\frac12,\frac12)$, corresponding to $H=(H_L,H_R)$. Additionally  the charge
conjugated states are present, such that each of the gauge multiplets appearing in the table corresponds to
a full hypermultiplet. Singlets from the untwisted sector are called $U_i$.}
\label{tab:bulkU}
\end{table}

\subsubsection*{Bulk matter from the twisted sectors}

We now consider bulk states $\ket{{\bf p}_{\rm sh},{\bf R};\gamma',\gamma}_{[g]}$
of $T^2_{\rm large}$ which are localized in $T^4_{\rm small}$.
They correspond to twisted sectors\index{Sector!twisted}, 
which are generated by space group elements $g=(\theta^k,m_a{\bf e}_a)$
with $k=2,4$, cf.~Table \ref{tab:fppos}. 
The localization of the states on the orbifold is specified by the labels 
 \begin{align}
 \label{eq:gu2}
	g\,: \quad \quad k=2,4\,, \quad \quad  n_6=0,2\,, \quad \quad  n_3=0,1,2\,.
\end{align}
Additionally, states at ${\bf z}_g$ have the geometrical quantum numbers $\gamma'$ and
$\gamma$, specified by integers $l_3=0,1,2$
and $j=0,1,2$, respectively.

 The projection condition for bulk states (\ref{eq:pceff2}) requires $\Phi_{g_{f'}}=1$
 for all fixed point elements $g_{f'}$ which leave $T^2_{\rm large}$ invariant.
 Here we follow the two-step projection procedure presented in section
 \ref{sec:genpc}. We first solve the condition (\ref{eq:1ststep}) 
 for $g_{f'}=g$,
\begin{align}
\label{eq:redcond2}
{\bf p}_{\rm sh} \cdot {\bf V}_{g}
   -{\bf R} \cdot {\bf v}_{g}&=\jmod01\,.
\end{align}
For all $g$ from (\ref{eq:gu2}), this in total gives  6480 massless solutions. 
They form 810 hypermultiplets, which transform under the non-Abelian gauge group
$\SU6\times[\SU3\times \SO8]$ as
\begin{align}
\label{eq:psol24a}
	 n_6&=0\,
	 :
	  &\hspace{-.1cm} 3 \cdot 3 \cdot&\left[({\bf 6},\mathbbm 1,\mathbbm 1)+({\bf \bar 6},\mathbbm 1,\mathbbm 1)
	 +(\mathbbm 1,{\bf 3},\mathbbm 1) +(\mathbbm 1,{\bf \bar 3},\mathbbm 1)\right. \nonumber \\
	 &  & &\left.
	\hspace{3.5cm} +(\mathbbm 1,\mathbbm 1,{\bf 8}^*)+4 \times (\mathbbm 1,\mathbbm 1,\mathbbm 1)\right],\\
	\label{eq:psol24b}
	 n_6&=2\,
	 :
	  &\hspace{-.1cm} 2 \cdot 3\cdot 3 \cdot &\left[({\bf 6},\mathbbm 1,\mathbbm 1)+({\bf \bar 6},\mathbbm 1,\mathbbm 1)
	 +(\mathbbm 1,{\bf 3},\mathbbm 1)+(\mathbbm 1,{\bf \bar 3},\mathbbm 1)\right. \nonumber \\
	 &  & &\left.
	 \hspace{3.5cm}+(\mathbbm 1,\mathbbm 1,{\bf 8}^*)+4 \times (\mathbbm 1,\mathbbm 1,\mathbbm 1)\right],
\end{align}
where ${\bf 8}^*={\bf 8}_v,{\bf 8}_s,{\bf 8}_c$, for $n_3=0,1,2$, respectively. Here the overall multiplicities
arise from $n_3\in\{0,1,2\}$,  $l_3\in\{0,1,2\}$ and $j=0$ for $n_6=0$, or $j\in\{1,2\}$
 otherwise.

However, according to (\ref{eq:pceff2})
consistent bulk states are additionally constrained by $\Phi_{g_{f'}}=1$, for other fixed point elements
$g_{f'}\neq g$. They also arise from the twisted sectors $T_2$ and $T_4$, and have $\tilde n_3=0,1,2$,
$\tilde n_2=n_2=0$.
This leads to the additional
projection condition (\ref{eq:2ndstep}),
\begin{align}
\label{eq:gammapc24}
	 \frac{l_3k}{3} &=\jmod{ {\bf p}_{\rm sh}\cdot {\bf W}_{(3)}}{1}\,.
\end{align}
For a given vector ${\bf p}_{\rm sh}$ that is a solution of (\ref{eq:redcond2}), 
this condition  specifies the superposition in the $\SU3$ sub-lattice which is compatible with the orbifold symmetries,
cf.~(\ref{eq:setchphaseeffect}). Thus the multiplicity of three in (\ref{eq:psol24a}) and
(\ref{eq:psol24b}) which is related to $l_3$  is reduced to one.
For each multiplet, one value of $l_3\in\{0,1,2\}$ is selected by (\ref{eq:gammapc24}).

Note that there is no similar restriction as (\ref{eq:gammapc24}) on the rotational
superposition quantum number $j=0,1$ in this sector,
neither do new constraints for the $R$-charges arise. 

In total, we have found 270 hypermultiplets
from the $T_2$ and $T_4$ twisted sectors, which are bulk states of the six-dimensional effective
theory. Their quantum numbers and gauge charges are summarized in Table \ref{tab:bulkT2T4}.
\begin{table}[t!!hp]
\centering
\footnotesize
\begin{tabular}{c||c|c|c||c|c|c|c|c||c|c}
\jvsb Multiplet & $n_3$ & $n_6/\gamma$ & $\gamma_3'$& $t_1$ & $t_2$ & $t_3$ & $t_4$ & $t_5$ 
& ${\bf \tilde N}$ & ${\bf \tilde N}^*$\\
\hhline{===========}
\jvsb$({\bf  6},\mathbbm 1, \mathbbm 1)$ & $0$ & $0,\frac 12,1$ & $\frac{1}{3}$ & $0$ & $-\frac{1}{3}$ & $1$ & $\frac{2}{3}$ & $0$ \\
\jvsb$({\bf \bar 6},\mathbbm 1, \mathbbm 1)$ & $0$ & $0,\frac 12,1$ & $0$ & $0$ & $-\frac{1}{3}$ & $-1$ & $\frac{2}{3}$ & $0$ \\
\jvsb$(\mathbbm 1, {\bf 3},\mathbbm 1)$ & $0$ & $0,\frac 12,1$ & $\frac{1}{3}$ & $0$ & $\frac{2}{3}$ & $0$ & $-\frac{1}{3}$ & $1$ \\
\jvsb$(\mathbbm 1, {\bf \bar 3},\mathbbm 1)$ & $0$ & $0,\frac 12,1$ & $0$ & $0$ & $\frac{2}{3}$ & $0$ & $-\frac{1}{3}$ & $-1$ \\
\jvsb$(\mathbbm 1, \mathbbm 1,{\bf 8})$ & $0$ & $0,\frac 12,1$ & $\frac{2}{3}$ & $0$ & $\frac{2}{3}$ & $0$ & $-\frac{1}{3}$ & $0$ \\
\jvsb$(\mathbbm 1, \mathbbm 1,\mathbbm 1)$ & $0$ & $0,\frac 12,1$ & $\frac{2}{3}$ & $1$ & $-\frac{1}{3}$ & $0$ & $\frac{2}{3}$ & $0$ \\
\jvsb$(\mathbbm 1, \mathbbm 1,\mathbbm 1)$ & $0$ & $0,\frac 12,1$ & $\frac{2}{3}$ & $-1$ & $-\frac{1}{3}$ & $0$ & $\frac{2}{3}$ & $0$ \\
\hline
\jvsb$(\mathbbm 1, \mathbbm 1,\mathbbm 1)$ & $0$ & $0,\frac 12,1$ & $\frac{2}{3}$ & $0$ & $\frac{2}{3}$ & $0$ & $\frac{2}{3}$ & $0$ &$(0,1,0,0)$ & $(0,0,0,0)$\\
\jvsb$(\mathbbm 1, \mathbbm 1,\mathbbm 1)$ & $0$ & $0,\frac 12,1$ & $\frac{2}{3}$ & $0$ & $\frac{2}{3}$ & $0$ & $\frac{2}{3}$ & $0$ &$(0,0,0,0)$ & $(1,0,0,0)$\\
\hline
\jvsb$({\bf  6},\mathbbm 1, \mathbbm 1)$ & $1$ & $0,\frac 12,1$ & $\frac{1}{3}$ & $0$ & $-\frac{1}{3}$ & $-1$ & $-\frac{1}{3}$ & $-1$ \\
\jvsb$({\bf \bar 6},\mathbbm 1, \mathbbm 1)$ & $1$ & $0,\frac 12,1$ & $0$ & $\frac{1}{2}$ & $\frac{1}{6}$ & $0$ & $-\frac{1}{3}$ & $-1$ \\
\jvsb$(\mathbbm 1, {\bf 3},\mathbbm 1)$ & $1$ & $0,\frac 12,1$ & $\frac{1}{3}$ & $-\frac{1}{2}$ & $\frac{1}{6}$ & $1$ & $\frac{2}{3}$ & $0$ \\
\jvsb$(\mathbbm 1, {\bf \bar 3},\mathbbm 1)$ & $1$ & $0,\frac 12,1$ & $0$ & $-\frac{1}{2}$ & $\frac{1}{6}$ & $1$ & $-\frac{1}{3}$ & $1$ \\
\jvsb$(\mathbbm 1, \mathbbm 1,{\bf 8}_s)$ & $1$ & $0,\frac 12,1$ & $\frac{2}{3}$ & $-\frac{1}{2}$ & $\frac{1}{6}$ & $1$ & $\frac{1}{6}$ & $\frac{1}{2}$ \\
\jvsb$(\mathbbm 1, \mathbbm 1,\mathbbm 1)$ & $1$ & $0,\frac 12,1$ & $\frac{2}{3}$ & $\frac{1}{2}$ & $-\frac{5}{6}$ & $1$ & $-\frac{1}{3}$ & $-1$ \\
\jvsb$(\mathbbm 1, \mathbbm 1,\mathbbm 1)$ & $1$ & $0,\frac 12,1$ & $\frac{2}{3}$ & $0$ & $\frac{2}{3}$ & $-2$ & $-\frac{1}{3}$ & $-1$ \\
\hline
\jvsb$(\mathbbm 1, \mathbbm 1,\mathbbm 1)$ & $1$ & $0,\frac 12,1$ & $\frac{2}{3}$ & $-\frac{1}{2}$ & $\frac{1}{6}$ & $1$ & $-\frac{1}{3}$ & $-1$&$(0,1,0,0)$ & $(0,0,0,0)$ \\
\jvsb$(\mathbbm 1, \mathbbm 1,\mathbbm 1)$ & $1$ & $0,\frac 12,1$ & $\frac{2}{3}$ & $-\frac{1}{2}$ & $\frac{1}{6}$ & $1$ & $-\frac{1}{3}$ & $-1$ &$(0,0,0,0)$ & $(1,0,0,0)$\\
\hline
\jvsb$({\bf  6},\mathbbm 1, \mathbbm 1)$ & $2$ & $0,\frac 12,1$ & $\frac{1}{3}$ & $\frac{1}{2}$ & $\frac{1}{6}$ & $0$ & $-\frac{1}{3}$ & $1$ \\
\jvsb$({\bf \bar 6},\mathbbm 1, \mathbbm 1)$ & $2$ & $0,\frac 12,1$ & $0$ & $0$ & $-\frac{1}{3}$ & $1$ & $-\frac{1}{3}$ & $1$ \\
\jvsb$(\mathbbm 1, {\bf 3},\mathbbm 1)$ & $2$ & $0,\frac 12,1$ & $\frac{1}{3}$ & $-\frac{1}{2}$ & $\frac{1}{6}$ & $-1$ & $-\frac{1}{3}$ & $-1$ \\
\jvsb$(\mathbbm 1, {\bf \bar 3},\mathbbm 1)$ & $2$ & $0,\frac 12,1$ & $0$ & $-\frac{1}{2}$ & $\frac{1}{6}$ & $-1$ & $\frac{2}{3}$ & $0$ \\
\jvsb$(\mathbbm 1, \mathbbm 1,{\bf 8}_c)$ & $2$ & $0,\frac 12,1$ & $\frac{2}{3}$ & $-\frac{1}{2}$ & $\frac{1}{6}$ & $-1$ & $\frac{1}{6}$ & $-\frac{1}{2}$ \\
\jvsb$(\mathbbm 1, \mathbbm 1,\mathbbm 1)$ & $2$ & $0,\frac 12,1$ & $\frac{2}{3}$ & $\frac{1}{2}$ & $-\frac{5}{6}$ & $-1$ & $-\frac{1}{3}$ & $1$ \\
\jvsb$(\mathbbm 1, \mathbbm 1,\mathbbm 1)$ & $2$ & $0,\frac 12,1$ & $\frac{2}{3}$ & $0$ & $\frac{2}{3}$ & $2$ & $-\frac{1}{3}$ & $1$ \\
\hline
\jvsb$(\mathbbm 1, \mathbbm 1,\mathbbm 1)$ & $2$ & $0,\frac 12,1$ & $\frac{2}{3}$ & $-\frac{1}{2}$ & $\frac{1}{6}$ & $-1$ & $-\frac{1}{3}$ & $1$&$(0,1,0,0)$ & $(0,0,0,0)$ \\
\jvsb$(\mathbbm 1, \mathbbm 1,\mathbbm 1)$ & $2$ & $0,\frac 12,1$ & $\frac{2}{3}$ & $-\frac{1}{2}$ & $\frac{1}{6}$ & $-1$ & $-\frac{1}{3}$ & $1$ &$(0,0,0,0)$ & $(1,0,0,0)$\\
\end{tabular}
\caption{Bulk hypermultiplets \index{Spectrum!bulk} from twisted sectors. All shown states have $k=2$ and 
${\bf q}_{\rm sh}=(-\frac{1}{3},-\frac{2}{3},0,0),(\frac{1}{6},-\frac{1}{6},\frac12,-\frac12);
(\frac{2}{3},\frac{1}{3},0,0),(\frac{1}{6},-\frac{1}{6},-\frac12,\frac12)$, corresponding to
 $H=(H_L,H_R)$. The charge conjugated states arise from the $k=4$
sector, so each gauge multiplet corresponds to a complete hypermultiplet.
The case of a zero entry in the $n_6/\gamma$ column represents the case $n_6=\gamma=0$.
non-vanishing entries specify the value of $\gamma\neq 0$ and imply
$n_6=2$.
Note that there are 18 bulk multiplets with non-vanishing oscillator numbers in
internal directions.
}
\label{tab:bulkT2T4}
\end{table}

\subsubsection*{Matter states at $n_2=0$}

We now consider matter states from a sector $\mathcal H_{[g_f]}$
which are localized at the fixed point $n_2=0$ in the large plane
$T^2_{\rm large}$. These are massless states (\ref{eq:masslessL}), (\ref{eq:masslessR}),
which fulfill the projection conditions (\ref{eq:pceff1}),
\begin{align}
\label{eq:pceffn20}
&\begin{array}{ll}
    \Phi_{g_{f'}}=e^{2 \pi i  \left( {\bf p}_{\rm sh} \cdot {\bf V}_{g_{f'}}
   -{\bf R} \cdot {\bf v}_{g_{f'}}
    \right)} \tilde \Phi_{g_{f'}}
    =1\,, & \tilde \Phi_{g_{f'}}=e^{2 \pi i  \gamma_a' ( \tilde k m_a- k \tilde m_a)}e^{2 \pi i \gamma \tilde k}\,,
    \end{array}\\ \nonumber
    &\hspace{2cm}\text{for all fixed point elements}\; g_{f'}=(\theta^{\tilde k},\tilde m_a{\bf e}_a) \in S\,.
\end{align}
These states are also zero modes in four dimensions.
The representative space group element $g_f$, which characterizes the localized states, can be chosen as
in Table \ref{tab:fppos}. It is defined by the labels $k, {\bf n}=(n_6,n_3,n_2,n_2')$, with $n_2=0$ and
\begin{align}
	&g_f\,: & k&=1,3,5\,, & {\bf n}&\,:\;
	\left\{\begin{array}{l l l l}
	n_6=0, & n_3=0,1,2,   & n_2'=0,1,& \text{if}\;\; k=1,5 \\
	n_6=0,1, & n_3=0,   & n_2'=0,1,& \text{if}\;\; k=3.
	\end{array}\right.
\end{align}
The corresponding states  $\ket{{\bf p}_{\rm sh},{\bf R};\gamma',\gamma}_{[g_f]}$ 
involve superposition quantum numbers $\gamma',\gamma$ which are specified by
 integers $l_3,l_5 ,j$,
\begin{align}
\label{eq:mts3}
	l_3&=\left\{\begin{array}{ll}
		1,2,3, & \text{if}\;\;k=1,5,\\
		0, & \text{if}\;\;k=3,
	\end{array}\right. & l_5&=1,2\,,&
	j&=\left\{\begin{array}{ll}
		0, & \text{if}\;\;k=1,5,\\
		0,1,2,3, & \text{if}\;\;k=3,
	\end{array}\right. &
\end{align}
such that $ \gamma_a'm_a = \frac{l_3}{3} k n_3, \gamma=\frac  j3$.

The above conditions can be solved by the two-step procedure of Section \ref{sec:genpc}.
First, one solves for $g_{f'}=g$,
 \begin{align}
\label{eq:redcond3}
{\bf p}_{\rm sh} \cdot {\bf V}_{g_{f}}
   -{\bf R} \cdot {\bf v}_{g_f}&=\jmod01\,,
\end{align}
and finds 3872 solutions, forming 968 chiral multiplets. These  enormous numbers include multiplicities
of three and four for $k=1,5$ and $k=3$, respectively,  related to the constants $l_3,l_5,j$ from
 (\ref{eq:mts3}). We expect that
this multiplicity is reduced   by the additional constraints  (\ref{eq:2ndstepS})
 due to $g_{f'}\neq g_f$. These conditions are
\begin{align}
\label{eq:pcn204}
	   {\bf p}_{\rm sh}\cdot\left( {\bf V}_6+n_3 {\bf W}_{(3)}\right)-{\bf R}\cdot {\bf v}_6+\frac j3&=\jmod 01\,,
\end{align}
\begin{align}
	\frac{l_3 k}{3}&=\jmod{{\bf p}_{\rm sh}\cdot{\bf W}_{(3)}}{1} \,, &
	 \frac{l_5k }{2} &=\jmod{{\bf p}_{\rm sh}\cdot {\bf W}_{(2)}}{1}\,, 
\end{align}
and indeed, for each solution they select specific values for the superposition quantum numbers $l_3,l_5,j$.
However, for $k=3$   the value of $l_3$  is fixed to zero by (\ref{eq:mts3}), and a projection
condition for ${\bf p}_{\rm sh}$ arises from the above:
\begin{align}
	k=3\,:\;\hspace{2cm}  {\bf p}_{\rm sh}\cdot{\bf W}_{(3)}&=\jmod 01\,.
\end{align}
Note that for $k=1,5$ the value of $j=0$ is also fixed, but in that case (\ref{eq:pcn204})
is identical to the projection condition (\ref{eq:redcond3}), and  not a new constraint.

With the full set of consistency conditions, 88 chiral multiplets survive at fixed points with $n_2=0$.
Recall from Figure \ref{fig:so4} that there are two equivalent fixed points with $n_2=0$ in the geometry.
Since there is no Wilson line in the direction ${\bf e}_6$, their spectra are identical and
thus consist of 44 chiral multiplets each. Up to the value for $l_5$, they form complete gauge representations
with respect to the local gauge group\footnote{$l_5$ is a quantum number of the four-dimensional
theory and not of the local $\SU5$ GUT, 
since the simple root $(0^4,1,-1,0^2)$ from Table \ref{tab:srloc} is not orthogonal to ${\bf W}_{(2)}$,
cf.~the discussion below (\ref{eq:pceff1b}).},
as listed in Table \ref{tab:n2T}.
All listed states also appear in the four-dimensional zero mode spectrum. The latter
was already studied in \cite{bhx06}. Appendix \ref{app:4d} repeats the findings in our notation, including the
  translational superposition quantum numbers $\gamma'$.
\begin{table}[t]
\centering
\footnotesize
\begin{tabular}{c|c||c|c|c|c||c|c|c|c|c|c||c|c}
 \multicolumn{2}{c||}{Multiplet}&  $k$ & $n_3$ & $\gamma$ & $\gamma_3'$ & $t_1$ & $t_2$ & $t_3$ & $t_4$ & $t_5$ & $t_6^0$ &${\bf \tilde N}$ & ${\bf \tilde N}^*$ \jvsb \\
\hline
\hline
\jvsb$({\bf 10}, \mathbbm 1,\mathbbm 1)$ &${\bf 10}_{(1)}$ & $1$ & $0$ & $0$ & $1$ & $0$ & $-\frac{1}{6}$ & $-\frac{1}{2}$ & $\frac{1}{3}$ & $0$ & $\frac{1}{2}$   \\
\jvsb$({\bf \bar 5},\mathbbm 1, \mathbbm 1)$ & ${\bf \bar 5}_{(1)}$& $1$ & $0$ & $0$ & $\frac{2}{3}$ & $0$ & $-\frac{1}{6}$ & $\frac{3}{2}$ & $\frac{1}{3}$ & $0$ & $-\frac{3}{2}$   \\
\jvsb$(\mathbbm 1, \mathbbm 1,\mathbbm 1)$ & $N^c_{(1)}$&  $1$ & $0$ & $0$ & $\frac{1}{3}$ & $0$ & $-\frac{1}{6}$ & $-\frac{5}{2}$ & $\frac{1}{3}$ & $0$ & $\frac{5}{2}$   \\
\cline{1-12}
\jvsb$(\mathbbm 1, \mathbbm 1,{\bf 8}_c)$ & &  $1$ & $1$ & $0$ & $\frac{1}{3}$ & $0$ & $-\frac{1}{6}$ & $-\frac{1}{2}$ & $-\frac{1}{6}$ & $-\frac{1}{2}$ & $\frac{5}{2}$   \\
\jvsb$(\mathbbm 1, {\bf 3},\mathbbm 1)$ &  & $1$ & $2$ & $0$ & $\frac{2}{3}$ & $0$ & $-\frac{1}{6}$ & $\frac{3}{2}$ & $\frac{1}{3}$ & $0$ & $\frac{5}{2}$   \\
\jvsb$(\mathbbm 1, \mathbbm 1,\mathbbm 1)$ & $S_8$ & $1$ & $2$ & $0$ & $\frac{2}{3}$ & $0$ & $-\frac{1}{6}$ & $\frac{3}{2}$ & $-\frac{2}{3}$ & $-1$ & $\frac{5}{2}$   \\
\hline
\jvsb$(\mathbbm 1, \mathbbm 1,\mathbbm 1)$ & $S_1$  & $1$ & $0$ & $0$ & $\frac{1}{3}$ & $-\frac{1}{2}$ & $-\frac{2}{3}$ & $\frac{1}{2}$ & $\frac{1}{3}$ & $0$ & $\frac{5}{2}$ & $(0,0,0,0)$ & $(1,0,0,0)$ \\
\jvsb$(\mathbbm 1, \mathbbm 1,\mathbbm 1)$ & $S_2$  &  $1$ & $0$ & $0$ & $1$ & $\frac{1}{2}$ & $-\frac{2}{3}$ & $-\frac{1}{2}$ & $\frac{1}{3}$ & $0$ & $-\frac{5}{2}$ & $(0,0,0,0)$ & $(1,0,0,0)$ \\
\jvsb$(\mathbbm 1, \mathbbm 1,\mathbbm 1)$ &$S_3$    & $1$ & $0$ & $0$ & $\frac{1}{3}$ & $\frac{1}{2}$ & $\frac{1}{3}$ & $\frac{1}{2}$ & $\frac{1}{3}$ & $0$ & $\frac{5}{2}$ & $(0,0,0,0)$ & $(0,1,0,0)$ \\
\jvsb$(\mathbbm 1, \mathbbm 1,\mathbbm 1)$ & $S_4$  & $1$ & $0$ & $0$ & $\frac{1}{3}$ & $\frac{1}{2}$ & $\frac{1}{3}$ & $\frac{1}{2}$ & $\frac{1}{3}$ & $0$ & $\frac{5}{2}$ & $(0,0,0,0)$ & $(2,0,0,0)$ \\
\jvsb$(\mathbbm 1, \mathbbm 1,\mathbbm 1)$ & $S_5$ &  $1$ & $0$ & $0$ & $1$ & $-\frac{1}{2}$ & $\frac{1}{3}$ & $-\frac{1}{2}$ & $\frac{1}{3}$ & $0$ & $-\frac{5}{2}$ & $(0,0,0,0)$ & $(0,1,0,0)$ \\
\jvsb$(\mathbbm 1, \mathbbm 1,\mathbbm 1)$ & $S_6$  & $1$ & $0$ & $0$ & $1$ & $-\frac{1}{2}$ & $\frac{1}{3}$ & $-\frac{1}{2}$ & $\frac{1}{3}$ & $0$ & $-\frac{5}{2}$ & $(0,0,0,0)$ & $(2,0,0,0)$ \\
\jvsb$(\mathbbm 1, {\bf \bar 3},\mathbbm 1)$ &  & $1$ & $1$ & $0$ & $1$ & $0$ & $-\frac{1}{6}$ & $-\frac{1}{2}$ & $\frac{1}{3}$ & $0$ & $\frac{5}{2}$ & $(0,0,0,0)$ & $(1,0,0,0)$ \\
\jvsb$(\mathbbm 1, \mathbbm 1,\mathbbm 1)$ & $S_7$  & $1$ & $1$ & $0$ & $\frac{1}{3}$ & $0$ & $-\frac{1}{6}$ & $-\frac{1}{2}$ & $-\frac{2}{3}$ & $1$ & $\frac{5}{2}$ & $(0,0,0,0)$ & $(1,0,0,0)$ \\
\hline
\jvsb$(\mathbbm 1, {\bf 3},\mathbbm 1)$ &  & $3$ & $0$ & $\frac{1}{3}$ & $0$ & $-\frac{1}{2}$ & $0$ & $\frac{1}{2}$ & $0$ & $1$ & $\frac{5}{2}$   \\
\jvsb$(\mathbbm 1, {\bf \bar 3},\mathbbm 1)$ &  & $3$ & $0$ & $\frac{1}{3}$ & $0$ & $\frac{1}{2}$ & $0$ & $-\frac{1}{2}$ & $0$ & $-1$ & $-\frac{5}{2}$   \\
\end{tabular}
\caption{Chiral multiplets at $(n_2,n_2')=(0,0)$ from the twisted sectors.
A copy of this spectrum \index{Spectrum!GUT fixed points} arises at $(n_2,n_2')=(0,1)$.
Shown are states with ${\bf q}_{\rm sh}=(-\frac16,-\frac13,-\frac12,0),
(\frac13,\frac16,0,-\frac12)$ for the case $k=1$, and  ${\bf q}_{\rm sh}=(\frac12,0,-\frac12,0),
(0,\frac12,0,-\frac12)$ for $k=3$. The corresponding charge conjugate states have twist $6-k$
and are not listed.
Two families of standard model chiral
multiplets arise from $k=1,n_3=0$, with family index $(i)=(1),(2)$, for $n_2'=0,1$, respectively.
Furthermore,we introduce  the nomenclature $S_i, S_i'$  for localized 
singlets at $(n_2, n_2')=(0,0),(0,1)$, respectively. Note that seven out of eight gauge
singlet chiral multiplets have oscillator numbers in internal directions. 
The value $\gamma=0$ implies $n_6=0$,  states with $k=3$ have $n_6=1$.}
\label{tab:n2T}
\end{table}

The most striking feature of the local spectrum is the appearance of one complete standard model
quark-lepton generation \index{Standard model families}
 at each of the two fixed points with $n_2=0$. Recall from (\ref{eq:ggn20}) that the
local non-Abelian gauge group at these points is the unifying group $\SU5$ in the visible sector.
The only contribution with $k=1,n_3=0, {\bf \tilde N}={\bf N}=0$ to the spectrum with respect to that group is
\begin{subequations}
\label{eq:fam12}
\begin{eqnarray}
	  (n_2,n_2') = (0,0)\,:&& \hspace{.5cm}\T_{(1)}+\Fb_{(1)}+N^c_{(1)}\,,\\
	  (n_2,n_2') = (0,1)\,:&& \hspace{.5cm}\T_{(2)}+\Fb_{(2)}+N^c_{(2)}\,,
\end{eqnarray}
\end{subequations}
where $N^c_{(i)}$ denotes a right-handed neutrino and $(i)=(1),(2)$ is a family index.
The third generation is combined from bulk states, as discussed in the next chapter.
 
\subsubsection*{Matter states at $n_2=1$}

The fixed points with localization label $n_2=1$ play the role of  `hidden branes'\index{Hidden brane}
 in our model. This is meant 
in the sense that it is physically separated from the GUT fixed point, and
all degrees of freedom which are localized there are exotics and have to be decoupled
from the low-energy spectrum. The states at these fixed points are determined by the Wilson line
${\bf W}_{(2)}$, whose main purpose is to break the gauge group to the standard model gauge group
in the low-energy limit. It has no direct effect on the spectrum at the $\SU5$ GUT fixed point.
However, it does influence the zero modes of bulk fields and hence cannot be chosen arbitrarily
for a valid phenomenology. 

Here we take the viewpoint that there is much less to learn from the hidden brane than from the
physical one. Once the details there are understood satisfactorily, we expect that a further adjustment of
${\bf W}_{(2)}$ can lead to further improvements. 

In the following we shall thus restrict our attention mainly to the bulk and the GUT fixed points,
characterized by $n_2=0$. However, for the sake of completeness the spectrum at the hidden brane
 for the chosen gauge embedding (\ref{eq:gaugeemb}) is given in Appendix \ref{app:n21}. Here
 we just state the presence of exotics \index{Exotics} with non-vanishing charges under the electroweak
 $\SU2_L$ and the hypercharge $\U1_Y$: \index{Spectrum!exotic fixed points}
  \begin{align}
	&\text{Exotics from $n_2=1$:} &
	8\cdot{\bf 2}_0+16\cdot\left(
	 \mathbbm 1_{1/2}+ \mathbbm1_{-1/2}\right)\,.
\end{align} 
It is known that these states can be decoupled from the effective theory in four
dimensions~\cite{bhx06}. We assume that this is also possible in the situation
discussed  in this work, but do not study this issue any further.

\subsubsection{The effective orbifold GUT field theory}
\label{sec:locmodel}

In the previous sections, bulk and brane states on the geometry $\mathcal M_{\rm eff}$ were calculated
by solving the string mass equations and the relevant orbifold projection conditions. 
The assumption behind this construction was that the $\SO4$-plane is much larger than the $\G 2$- and
$\SU3$-planes, whose scale is given by the string scale in our setup.
 Below that scale, one then has the effective description in terms of an orbifold field theory on $\mathbbm C/
 (\Gamma_{\SO4}\times\Z2)$. \index{Effective orbifold!field theory}
 Note that the introduction of six-dimensional fields 
 goes beyond the zero mode description and includes Kaluza-Klein modes of the two large compact dimensions.
  \begin{table}[t]
\begin{center}
	\footnotesize
	\begin{tabular}{c||c|c|c|c||c|c|c|c|c|c}
	\jvsb	Bulk & $n_2=0$  & V & $\phi$ & ${t_6}^0$& $n_2=1$  & V & $\phi$ & ${t_6}^1$ & $t_7$ & $t_8$  \\
		\hline
		\hline
	\jvsb	$({\bf 35},\mathbbm 1,\mathbbm 1)$ & $({\bf 24},\mathbbm 1,\mathbbm 1)$  & $+$ & $-$ & 0 
		  & $({\bf 3},\mathbbm 1,\mathbbm 1,\mathbbm 1)$  & $+$ & $-$ & 0 & 0 & 0  \\
	\jvsb	  & $({\bf 5},\mathbbm 1,\mathbbm 1)$  & $-$ & $+$ &   $-6$
		           	 & $(\mathbbm 1,{\bf 15},\mathbbm 1,\mathbbm 1)$  & $+$ & $-$ & 0 & 0 & 0  \\
	\jvsb			 & $({\bf \bar 5},\mathbbm 1,\mathbbm 1)$  & $-$ & $+$ & $6$ 
				 & $({\bf 2},{\bf 4},\mathbbm 1,\mathbbm 1)$  & $-$ & $+$ & $15$ & 0 & 0  \\ 
		\jvsb		 & $(\mathbbm 1,\mathbbm 1,\mathbbm 1)$  & $+$ & $-$ & 0
				 & $({\bf 2},{\bf \bar{4}},\mathbbm 1,\mathbbm 1)$  & $-$ & $+$ & $-  15$& 0 & 0  \\ 
		\jvsb		 &    &   &   &   
				 & $(\mathbbm 1,\mathbbm 1,\mathbbm 1,\mathbbm 1)$  & $+$ & $-$ & 0& 0 & 0  \\ 
		\hline		 
		\jvsb$(\mathbbm 1,{\bf 8},\mathbbm 1)$ & $(\mathbbm 1,{\bf 8},\mathbbm 1)$  & $+$ & $-$ & 0
		& $(\mathbbm 1,\mathbbm 1,{\bf 3},\mathbbm 1)$  & $+$ & $-$ & 0 & 0 & 0  \\
		\jvsb		  &    &   &   & 
				  & $(\mathbbm 1,\mathbbm 1,{\bf 2},\mathbbm 1)$  & $-$ & $+$ & 0 & 3 & 0 \\
		\jvsb		  &     &   &   & 
				  & $(\mathbbm 1,\mathbbm 1,{\bf 2},\mathbbm 1)$  & $-$ & $+$ & 0 & $-3$ & 0  \\
		\jvsb		  &      &   &   & 
				  & $(\mathbbm 1,\mathbbm 1,\mathbbm 1,\mathbbm 1)$  & $+$ & $-$ & 0 & 0 & 0  \\ 
		\hline
	\jvsb	$(\mathbbm 1,\mathbbm 1,{\bf 28})$ & $(\mathbbm 1,\mathbbm 1,{\bf 28})$  & $+$ & $-$ & 0
		& $(\mathbbm 1,\mathbbm 1,\mathbbm 1,{\bf   15})$  & $+$ & $-$ & 0 & 0 & 0 \\
		\jvsb		   &  &   &   & 
				   & $(\mathbbm 1,\mathbbm 1,\mathbbm 1,{\bf 6})$  & $-$ & $+$ & 0 & 0 & 2  \\
		\jvsb		   &  &   &   & 
				   & $(\mathbbm 1,\mathbbm 1,\mathbbm 1,{\bf 6})$  & $-$ & $+$ & 0 & 0 & $-2$  \\
		\jvsb		   &  &   &   & 
				   & $(\mathbbm 1,\mathbbm 1,\mathbbm 1,\mathbbm 1)$  & $+$ & $-$ & 0 & 0 & 0  \\
		\end{tabular}
\end{center}
\caption{Local decomposition of vector multiplets at $n_2=0$ and $n_2=1$. 
\index{Spectrum!GUT fixed points}\index{Spectrum!exotic fixed points} The representations
refer to $\SU6\times[\SU3\times\SO8]$ in the bulk and $\SU5\times[\SU3\times\SO8]$ or
$\SU2\times\SU4\times[\SU2'\times\SU4']$ at $n_2=0,1$, respectively. Shown are the charges under the 
strictly local $\U1$
factors and the parities under the local $\Z2$ twist, for the 4d $\mathcal N=1$
 components of the six-dimensional vector, $A=(V,\phi)$.}
\label{tab:locdec}
\end{table}
 \begin{table}[t]
\centering
\footnotesize
\begin{tabular}{c|c||c||c|c||c|c|c|c|c|c}
\multicolumn{2}{c||}{Bulk} &  $n_2=0$ & $H_L$ & $H_R$ & $t_1$ & $t_2$ & $t_3$ & $t_4$ & $t_5$ & $t_6^0$\jvsb  \\
\hline
\hline
\jvsb $({\bf 20}, \mathbbm 1,\mathbbm 1)$& & $({\bf 10}, \mathbbm 1,\mathbbm 1)$   
& $+$ & $-$ & $-\frac{1}{2}$ & $\frac{1}{2}$ & $0$ & $0$ & $0$ & $3$ \\
\jvsb  & & $({\bf \bar {10}}, \mathbbm 1,\mathbbm 1)$  
 & $-$ & $+$ & $-\frac{1}{2}$ & $\frac{1}{2}$ & $0$ & $0$ & $0$ & $-3$ \\
\hline
\jvsb  $(\mathbbm 1, \mathbbm 1,{\bf 8})$& &$(\mathbbm 1, \mathbbm 1,{\bf 8})$  & $-$ & $+$ & $0$ & $0$ & $0$ & $-1$ & $0$ & $0$ \\
\hline
\jvsb  $(\mathbbm 1, \mathbbm 1,{\bf 8}_s)$& &$(\mathbbm 1, \mathbbm 1,{\bf 8}_s)$   & $+$ & $-$ & $0$ & $0$ & $0$ & $\frac{1}{2}$ & $\frac{3}{2}$ & $0$ \\
\hline
\jvsb  $(\mathbbm 1, \mathbbm 1,{\bf 8}_c)$& &$(\mathbbm 1, \mathbbm 1,{\bf 8}_c)$  & $+$ & $-$ & $0$ & $0$ & $0$ & $\frac{1}{2}$ & $-\frac{3}{2}$ & $0$ \\
\hline
\jvsb  $(\mathbbm 1, \mathbbm 1,\mathbbm 1)$ & $U_1$& $(\mathbbm 1, \mathbbm 1,\mathbbm 1)$  & $-$ & $+$ & $\frac{1}{2}$ & $\frac{1}{2}$ & $3$ & $0$ & $0$ & $0$ \\
\hline
\jvsb  $(\mathbbm 1, \mathbbm 1,\mathbbm 1)$ & $U_2$& $(\mathbbm 1, \mathbbm 1,\mathbbm 1)$ & $+$ & $-$ & $\frac{1}{2}$ & $\frac{1}{2}$ & $-3$ & $0$ & $0$ & $0$ \\
\hline
\jvsb  $(\mathbbm 1, \mathbbm 1,\mathbbm 1)$ &$U_3$ & $(\mathbbm 1, \mathbbm 1,\mathbbm 1)$ & $+$ & $-$ & $1$ & $-1$ & $0$ & $0$ & $0$ & $0$ \\
\hline
\jvsb  $(\mathbbm 1, \mathbbm 1,\mathbbm 1)$& $U_4$& $(\mathbbm 1, \mathbbm 1,\mathbbm 1)$  & $+$ & $-$ & $-1$ & $-1$ & $0$ & $0$ & $0$ & $0$ \\
\end{tabular}
\caption{Chiral projection of  bulk hypermultiplets from the untwisted sector to the fixed points $n_2=0$.
\index{Spectrum!GUT fixed points}
The bulk and fixed point representations correspond to $\SU6 \times [\SU3 \times \SO8]$
and $\SU5 \times [\SU3 \times \SO8]$, respectively. Shown are the charges under the six local $\U1$
factors and the parities under the local $\Z2$ twist, for the hypermultiplet components $H=(H_L,H_R)$.}
\label{tab:n20U}
\end{table}
\begin{table}[t]
\centering
\footnotesize
\begin{tabular}{c||c|c||c|c||c|c||c|c|c|c|c|c}
\jvsb Bulk & \multicolumn{2}{c||}{$n_2=0$} & $n_3$ & $\gamma_3'$ &$H_L$ & $H_R$ & 
 $t_1$ & $t_2$ & $t_3$ & $t_4$ & $t_5$ & $t_6^0$  \\
\hline
\hline
\jvsb $(\mathbbm 1, \mathbbm 1,\mathbbm 1)$ & $(\mathbbm 1, \mathbbm 1,\mathbbm 1)$ &  $Y_0^*$ & $0$ & $\frac{2}{3}$ & $-,+,-$ & $+,-,+$ & $0$ & $\frac{2}{3}$ & $0$ & $\frac{2}{3}$ & $0$ & $0$ \\
\hline
\jvsb  $(\mathbbm 1, \mathbbm 1,\mathbbm 1)$& $(\mathbbm 1, \mathbbm 1,\mathbbm 1)$ & $\bar Y_0^*$  & $0$ & $\frac{2}{3}$ & $+,-+$ & $-,+,-$ & $0$ & $\frac{2}{3}$ & $0$ & $\frac{2}{3}$ & $0$ & $0$ \\
\hline
\hline
\jvsb $(\mathbbm 1, \mathbbm 1,\mathbbm 1)$ & $(\mathbbm 1, \mathbbm 1,\mathbbm 1)$ &  $Y_1^*$  & $1$ & $\frac{2}{3}$ & $-,+,-$ & $+,-,+$ & $-\frac{1}{2}$ & $\frac{1}{6}$ & $1$ & $-\frac{1}{3}$ & $-1$ & $0$ \\
\hline
\jvsb $(\mathbbm 1, \mathbbm 1,\mathbbm 1)$ & $(\mathbbm 1, \mathbbm 1,\mathbbm 1)$ &   $\bar Y_1^*$ & $1$ & $\frac{2}{3}$ & $+,-+$ & $-,+,-$ & $-\frac{1}{2}$ & $\frac{1}{6}$ & $1$ & $-\frac{1}{3}$ & $-1$ & $0$ \\
\hline
\hline
\jvsb  $(\mathbbm 1, \mathbbm 1,\mathbbm 1)$ & $(\mathbbm 1, \mathbbm 1,\mathbbm 1)$ &  $Y_2^*$ & $2$ & $\frac{2}{3}$ & $-,+,-$ & $+,-,+$ & $-\frac{1}{2}$ & $\frac{1}{6}$ & $-1$ & $-\frac{1}{3}$ & $1$ & $0$ \\
\hline
\jvsb $(\mathbbm 1, \mathbbm 1,\mathbbm 1)$ & $(\mathbbm 1, \mathbbm 1,\mathbbm 1)$ & $\bar Y_2^*$  & $2$ & $\frac{2}{3}$ & $+,-+$ & $-,+,-$ & $-\frac{1}{2}$ & $\frac{1}{6}$ & $-1$ & $-\frac{1}{3}$ & $1$ & $0$ \\
\end{tabular}
\caption{Projection of  bulk hypermultiplets with oscillator numbers from twisted sectors $T_2$ and $T_4$
 to the fixed points $n_2=0$. 
 \index{Spectrum!GUT fixed points}
 The singlets $Y_{n_3}^*$ have ${\bf \tilde N}=(0,1,0,0)$, ${\bf \tilde N}^*=(0,0,0,0)$,
 the singlets $\bar Y_{n_3}^*$ have ${\bf \tilde N}=(0,0,0,0)$, ${\bf \tilde N}^*=(1,0,0,0)$. The three parities for the chiral multiplet components
 $H_L, H_R$ correspond to $\gamma=0,\frac12,1$. The localization in the $\G2$-plane
 is given by $n_6=0$ for $\gamma=0$, and $n_6=2$ otherwise. }
\label{tab:n20T2T4*}
\end{table}

 All states which were found before are now interpreted as states which arise from corresponding
 fields on the six-dimensional space $T^2_{\rm large}/\Z2 \times M_4$. For the 
 $\mathcal N=2$ vector and hypermultiplets
  of the six-dimensional bulk
 we shall find it convenient to use a decomposition into
  $\mathcal N=1$ multiplets \cite{agw02}
  \begin{align}
	&\text{6d vector multiplet $A$:} & A&=(V,\phi)\,, & &\left\{ \begin{array}{ll}
	V:& \text{4d vector multiplet},\\ \phi: & \text{4d  left-handed chiral multiplet},
	\end{array}\right. \label{eq:ah1}\\ \label{eq:ah2}
	&\text{6d hypermultiplet $H$:} & H&=(H_L,H_R)\,, & &\left\{ \begin{array}{ll}
	H_L:& \text{4d left-handed chiral multiplet},\\ H_R: & \text{4d right-handed chiral multiplet},
	\end{array}\right.
\end{align}
where $A$ transforms in the adjoint representation of the bulk gauge group (\ref{eq:bulkgaugegroup}),
and $H$ as a multiplet from Tables \ref{tab:bulkU}, \ref{tab:bulkT2T4}. The gauge interaction 
Lagrangian in the six-dimensional bulk can then be written as
\begin{align}
  \begin{split}
    {\cal L}_H &= \int d^4\theta\left(H_L^\dagger e^{2gV} H_L 
      + H_R^{c\dagger} e^{-2gV} H_R^c\right)\\
    &\quad + \int d^2\theta\ H_R^c\left(\partial +\sqrt{2}g\phi\right)H_L + \text{h.c.}\,,
  \end{split}\label{eq:tyuk1}
\end{align}
 where $H^c_L,H^c_R$ are the charge conjugated fields.
 
 The $\Z2$ orbifold projection conditions \index{Orbifold!projection conditions}
 at the fixed points $z_{g_f}^3$ in the $\SO4$-plane 
 can now be formulated for the above bulk fields. For $g_f=(\theta^k,m_a{\bf e}_a), k=1,3,5,$ and a
 coordinate $z^3$ in $T^2_{\rm large}$ one has
 \begin{align}
	 g_f ( z^3_{g_f}+z^3)&=z_{g_f}^3-z^3\,,
\end{align}
The local projection condition at  $z_{g_f}^3$ for a generic bulk field $\phi(z^3)$ is then given by
  \begin{align}
	P_f\,: \hspace{1cm} \phi(z^3_{g_f}+z^3)&= \eta_{f}(\phi) \phi(z^3_{g_f}-z^3)\,,\\
	\label{eq:etabulk}
	\eta_{f}(\phi)&=
	e^{2 \pi i \left( {\bf p}_{\rm sh} \cdot {\bf V}_6-{\bf R}\cdot {\bf v}_6+\frac j2\right)}\,,
\end{align}
where $j=0$ for untwisted   fields is understood.
 The above phase $\eta_{f}(\phi)$ has to be evaluated for 
the quantum numbers of the states associated with the field $\phi$. It follows from evaluating
the condition (\ref{eq:pceff3}) for the bulk states with
 $k=0,2,4; n_3=0,1,2; \gamma'_3 k={\bf p}_{\rm sh}\cdot {\bf W}_{(3)}$, and $j=0,1,2$, as found before.
 The corresponding projected matter states at the fixed point with $n_2=0$ are given in Tables \ref{tab:n20U}, 
 \ref{tab:n20T2T4*} and \ref{tab:n20T2T4}. A similar projection to the fixed points with $n_2=1$ is done
 in Appendix \ref{app:n21proj}. 
\begin{table}[p]
\centering
\footnotesize
\begin{tabular}{c||c|c||c|c||c|c||c|c|c|c|c|c}
\jvsb Bulk & \multicolumn{2}{c||}{$n_2=0$} & $n_3$ & $\gamma_3'$ &$H_L$ & $H_R$ & 
 $t_1$ & $t_2$ & $t_3$ & $t_4$ & $t_5$ & $t_6^0$  \\
\hline
\hline
\jvsb  $({\bf  6},\mathbbm 1, \mathbbm 1)$& $({\bf  5},\mathbbm 1, \mathbbm 1)$ &   & $0$ & $\frac{1}{3}$ & $-,+,-$ & $+,-,+$ & $0$ & $-\frac{1}{3}$ & $1$ & $\frac{2}{3}$ & $0$ & $-1$ \\
\jvsb  & $(\mathbbm 1, \mathbbm 1,\mathbbm 1)$ &  $X_0$ & $0$ & $\frac{1}{3}$ & $+,-+$ & $-,+,-$ & $0$ & $-\frac{1}{3}$ & $1$ & $\frac{2}{3}$ & $0$ & $5$ \\
\hline
\jvsb  $({\bf \bar 6},\mathbbm 1, \mathbbm 1)$ & $({\bf \bar 5},\mathbbm 1, \mathbbm 1)$ &   & $0$ & $1$ & $-,+,-$ & $+,-,+$ & $0$ & $-\frac{1}{3}$ & $-1$ & $\frac{2}{3}$ & $0$ & $1$ \\
\jvsb  & $(\mathbbm 1, \mathbbm 1,\mathbbm 1)$ & $\bar X_0$  & $0$ & $1$ & $+,-+$ & $-,+,-$ & $0$ & $-\frac{1}{3}$ & $-1$ & $\frac{2}{3}$ & $0$ & $-5$ \\
\hline
\jvsb  $(\mathbbm 1, \mathbbm 1,\mathbbm 1)$ & $(\mathbbm 1, \mathbbm 1,\mathbbm 1)$ & $Y_0$  & $0$ & $\frac{2}{3}$ & $+,-+$ & $-,+,-$ & $1$ & $-\frac{1}{3}$ & $0$ & $\frac{2}{3}$ & $0$ & $0$ \\
\hline
\jvsb   $(\mathbbm 1, \mathbbm 1,\mathbbm 1)$& $(\mathbbm 1, \mathbbm 1,\mathbbm 1)$ & $\bar Y_0$  & $0$ & $\frac{2}{3}$ & $+,-+$ & $-,+,-$ & $-1$ & $-\frac{1}{3}$ & $0$ & $\frac{2}{3}$ & $0$ & $0$ \\
\hline
\hline
\jvsb  $(\mathbbm 1, {\bf 3},\mathbbm 1)$& $(\mathbbm 1, {\bf 3},\mathbbm 1)$ &   & $0$ & $\frac{1}{3}$ & $-,+,-$ & $+,-,+$ & $0$ & $\frac{2}{3}$ & $0$ & $-\frac{1}{3}$ & $1$ & $0$ \\
\hline
\jvsb $(\mathbbm 1, {\bf \bar 3},\mathbbm 1)$ & $(\mathbbm 1, {\bf \bar 3},\mathbbm 1)$ &   & $0$ & $1$ & $-,+,-$ & $+,-,+$ & $0$ & $\frac{2}{3}$ & $0$ & $-\frac{1}{3}$ & $-1$ & $0$ \\
\hline
\jvsb $(\mathbbm 1, \mathbbm 1,{\bf 8})$ & $(\mathbbm 1, \mathbbm 1,{\bf 8})$ &   & $0$ & $\frac{2}{3}$ & $-,+,-$ & $+,-,+$ & $0$ & $\frac{2}{3}$ & $0$ & $-\frac{1}{3}$ & $0$ & $0$ \\
\hline
\hline
\jvsb $({\bf  6},\mathbbm 1, \mathbbm 1)$ & $({\bf  5},\mathbbm 1, \mathbbm 1)$ &   & $1$ & $\frac{1}{3}$ & $+,-+$ & $-,+,-$ & $0$ & $-\frac{1}{3}$ & $-1$ & $-\frac{1}{3}$ & $-1$ & $-1$ \\
\jvsb  & $(\mathbbm 1, \mathbbm 1,\mathbbm 1)$ &  $X_1$  & $1$ & $\frac{1}{3}$ & $-,+,-$ & $+,-,+$ & $0$ & $-\frac{1}{3}$ & $-1$ & $-\frac{1}{3}$ & $-1$ & $5$ \\
\hline
\jvsb $({\bf \bar 6},\mathbbm 1, \mathbbm 1)$ & $({\bf \bar 5},\mathbbm 1, \mathbbm 1)$ &  & $1$ & $1$ & $+,-+$ & $-,+,-$ & $\frac{1}{2}$ & $\frac{1}{6}$ & $0$ & $-\frac{1}{3}$ & $-1$ & $1$ \\
\jvsb  & $(\mathbbm 1, \mathbbm 1,\mathbbm 1)$ & $\bar X_1$  & $1$ & $1$ & $-,+,-$ & $+,-,+$ & $\frac{1}{2}$ & $\frac{1}{6}$ & $0$ & $-\frac{1}{3}$ & $-1$ & $-5$ \\
\hline
\jvsb $(\mathbbm 1, \mathbbm 1,\mathbbm 1)$ & $(\mathbbm 1, \mathbbm 1,\mathbbm 1)$ & $Y_1$  & $1$ & $\frac{2}{3}$ & $+,-+$ & $-,+,-$ & $0$ & $\frac{2}{3}$ & $-2$ & $-\frac{1}{3}$ & $-1$ & $0$ \\
\hline
\jvsb  $(\mathbbm 1, \mathbbm 1,\mathbbm 1)$& $(\mathbbm 1, \mathbbm 1,\mathbbm 1)$ &  $\bar Y_1$ & $1$ & $\frac{2}{3}$ & $+,-+$ & $-,+,-$ & $\frac{1}{2}$ & $-\frac{5}{6}$ & $1$ & $-\frac{1}{3}$ & $-1$ & $0$ \\
\hline
\hline
\jvsb  $(\mathbbm 1, {\bf 3},\mathbbm 1)$& $(\mathbbm 1, {\bf 3},\mathbbm 1)$ &   & $1$ & $\frac{1}{3}$ & $-,+,-$ & $+,-,+$ & $-\frac{1}{2}$ & $\frac{1}{6}$ & $1$ & $\frac{2}{3}$ & $0$ & $0$ \\
\hline
\jvsb  $(\mathbbm 1, {\bf \bar 3},\mathbbm 1)$& $(\mathbbm 1, {\bf \bar 3},\mathbbm 1)$ &   & $1$ & $1$ & $-,+,-$ & $+,-,+$ & $-\frac{1}{2}$ & $\frac{1}{6}$ & $1$ & $-\frac{1}{3}$ & $1$ & $0$ \\
\hline
\jvsb $(\mathbbm 1, \mathbbm 1,{\bf 8}_s)$ & $(\mathbbm 1, \mathbbm 1,{\bf 8}_s)$ &   & $1$ & $\frac{2}{3}$ & $+,-+$ & $-,+,-$ & $-\frac{1}{2}$ & $\frac{1}{6}$ & $1$ & $\frac{1}{6}$ & $\frac{1}{2}$ & $0$ \\
\hline
\hline
\jvsb  $({\bf  6},\mathbbm 1, \mathbbm 1)$ & $({\bf  5},\mathbbm 1, \mathbbm 1)$ &   & $2$ & $\frac{1}{3}$ & $-,+,-$ & $+,-,+$ & $\frac{1}{2}$ & $\frac{1}{6}$ & $0$ & $-\frac{1}{3}$ & $1$ & $-1$ \\
\jvsb  & $(\mathbbm 1, \mathbbm 1,\mathbbm 1)$ & $X_2$  & $2$ & $\frac{1}{3}$ & $+,-+$ & $-,+,-$ & $\frac{1}{2}$ & $\frac{1}{6}$ & $0$ & $-\frac{1}{3}$ & $1$ & $5$ \\
\hline
\jvsb $({\bf \bar 6},\mathbbm 1, \mathbbm 1)$  & $({\bf \bar 5},\mathbbm 1, \mathbbm 1)$ &   & $2$ & $1$ & $+,-+$ & $-,+,-$ & $0$ & $-\frac{1}{3}$ & $1$ & $-\frac{1}{3}$ & $1$ & $1$ \\
\jvsb  & $(\mathbbm 1, \mathbbm 1,\mathbbm 1)$ & $\bar X_2$  & $2$ & $1$ & $-,+,-$ & $+,-,+$ & $0$ & $-\frac{1}{3}$ & $1$ & $-\frac{1}{3}$ & $1$ & $-5$ \\
\hline
\jvsb  $(\mathbbm 1, \mathbbm 1,\mathbbm 1)$& $(\mathbbm 1, \mathbbm 1,\mathbbm 1)$ & $Y_2$  & $2$ & $\frac{2}{3}$ & $+,-+$ & $-,+,-$ & $0$ & $\frac{2}{3}$ & $2$ & $-\frac{1}{3}$ & $1$ & $0$ \\
\hline
\jvsb  $(\mathbbm 1, \mathbbm 1,\mathbbm 1)$& $(\mathbbm 1, \mathbbm 1,\mathbbm 1)$ & $\bar Y_2$  & $2$ & $\frac{2}{3}$ & $-,+,-$ & $+,-,+$ & $\frac{1}{2}$ & $-\frac{5}{6}$ & $-1$ & $-\frac{1}{3}$ & $1$ & $0$ \\
\hline
\hline
\jvsb $(\mathbbm 1, {\bf 3},\mathbbm 1)$ & $(\mathbbm 1, {\bf 3},\mathbbm 1)$ &   & $2$ & $\frac{1}{3}$ & $+,-+$ & $-,+,-$ & $-\frac{1}{2}$ & $\frac{1}{6}$ & $-1$ & $-\frac{1}{3}$ & $-1$ & $0$ \\
\hline
\jvsb $(\mathbbm 1, {\bf \bar 3},\mathbbm 1)$ & $(\mathbbm 1, {\bf \bar 3},\mathbbm 1)$ &   & $2$ & $1$ & $+,-+$ & $-,+,-$ & $-\frac{1}{2}$ & $\frac{1}{6}$ & $-1$ & $\frac{2}{3}$ & $0$ & $0$ \\
\hline
\jvsb  $(\mathbbm 1, \mathbbm 1,{\bf 8}_c)$& $(\mathbbm 1, \mathbbm 1,{\bf 8}_c)$ &   & $2$ & $\frac{2}{3}$ & $-,+,-$ & $+,-,+$ & $-\frac{1}{2}$ & $\frac{1}{6}$ & $-1$ & $\frac{1}{6}$ & $-\frac{1}{2}$ & $0$ \\
\end{tabular}
\caption{Projection of  bulk hypermultiplets from twisted sectors $T_2$ and $T_4$ with vanishing oscillator numbers
 to the fixed points $n_2=0$. \index{Spectrum!GUT fixed points}
 For excited states see Table \ref{tab:n20T2T4*}.
 The three parities for the chiral multiplet components
 $H_L, H_R$ correspond to $\gamma=0,\frac12,1$. 
 $H_L$ is associated with ${\bf q}=(-\frac13,-\frac23,0,0),(\frac16,-\frac16,\frac12,-\frac12)$ and
 $H_R$ with ${\bf q}=(\frac23,\frac13,0,0),(\frac16,-\frac16,-\frac12,\frac12)$.
 The localization in the $\G2$-plane
 is given by $n_6=0$ for $\gamma=0$, and $n_6=2$ otherwise.}
\label{tab:n20T2T4}
\end{table}

The local parities \index{Local parities} of the vector fields with respect to the local $\Z2$ twists are
summarized in table  \ref{tab:locdec}. This demonstrates how gauge symmetry breaking
and supersymmetry breaking are linked. Consider for example the ${\bf 35}$ of $\SU6$,
the adjoint representation. In the six-dimensional bulk, the corresponding degrees of freedom
can be expressed by a four-dimensional $\mathcal N=1$ vector multiplet $V$ and a
chiral multiplet $\phi$, cf.~(\ref{eq:ah1}). The parities in Table \ref{tab:locdec} now
state that either the vector multiplet or the chiral multiplet survive the projection to one of the fixed points.
Since at $n_2=0$ the 
gauge symmetry is broken to $\SU5\times \U1$, with adjoints ${\bf 24}+\mathbbm 1$,
 this implies the presence of two additional
chiral multiplets whose origin lies in the six-dimensional gauge sector,
\begin{align}
	&\text{Chiral multiplets at $n_2=0$ from the 6d gauge vector:} &
	\F,\;\Fb\hspace{.5cm}\text{of $\SU5$}\,.
\end{align}
These fields will play an important role in the discussion of  the phenomenology of the model 
in the next chapter.
The term `gauge-Higgs unification'\index{Gauge-Higgs unification} describes a situation where these two fields are
interpreted as the two  Higgs multiplets of a supersymmetric $\SU5$   GUT at $n_2=0$.
Note that they have four-dimensional zero modes which transform as weak doublets, so
they may be suitable.
If only one of them is associated with a Higgs multiplet, one refers to that as `partial gauge-Higgs unification'.
\index{Partial gauge-Higgs unification}
As we shall see, the interpretation of multiplets as Higgs or matter fields
is a question of the choice of vacuum of the scalar potential.

\subsection{Anomaly cancelation}
\label{sec:anomalies}

The  effective orbifold GUT field theory which was developed in the last sections is a complicated interacting theory.
Additionally to the supergravity multiplet, the dilaton multiplet and the 76 vector multiplets, there are
320 hypermultiplets in the six-dimensional bulk, and 136 localized chiral multiplets. All fields participate
in the gravitational interaction. Furthermore, the theory has local gauge symmetry which
contains non-Abelian and Abelian
factors, and the matter fields are in general  charged under both the visible and the hidden sector  $\U1$'s. 

The statement that the associated quantum theory is consistent up to its cutoff at the string scale
is highly non-trivial. Here we explicitly check that all anomalies vanish or can be canceled by
the Green--Schwarz mechanism \cite{gs84}. This shows that the spectrum of the effective GUT model as
derived from the heterotic string leads to a consistent gauge theory at the quantum level.
Furthermore, an important outcome of the calculation is the presence of two distinct anomalous 
$\U1$ factors at the inequivalent fixed points. They exactly add to a single anomalous $\U1$
in the effective four-dimensional model, as it was found in \cite{bhx06}.
 \begin{figure}[t]
\begin{center}
\includegraphics[height=4cm]{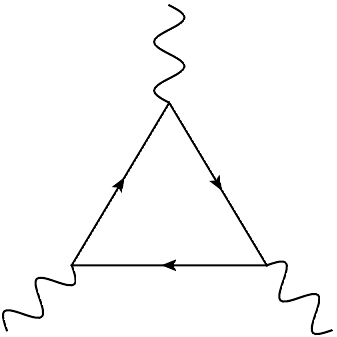}
\hspace{3cm}
\includegraphics[height=4cm]{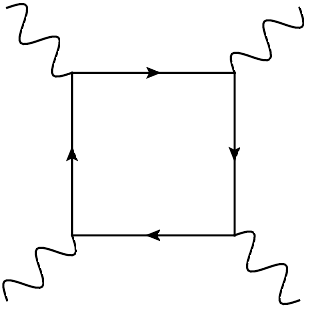}
\caption{The one-loop diagrams which induce anomalies in four (left) and six (right)
dimensions. Lines with arrows represent fermions. Wiggly lines represent gauge bosons
or gravitons.}
\label{fig:tribox}
\end{center}
\end{figure}

Here we do not give a broad review of the topic of anomalies, but 
restrict to a brief introduction. The interested reader  is  referred to 
 \cite{s04,a06} for more details and further references.
 
A quantum field theory with classical action $S$ and a symmetry transformation $\delta_\epsilon$
with parameter $\epsilon$ is anomalous \index{Anomaly} if
its one-loop effective action $\Gamma$ is not invariant under the symmetry,
\begin{align}
\label{eq:anomaly}
	\text{Anomaly}: \hspace{1cm} \delta_\epsilon S = 0 \,, \hspace{.5cm} \text{but}
	\hspace{.5cm}   \mathcal I(\epsilon) \equiv \delta_\epsilon \Gamma \neq 0\,. 
\end{align}
In gauge theories, the loss of gauge invariance due to an anomaly is accompanied
by the loss of unitarity and renormalizability, and therefore the available description as
a quantum field theory breaks down. It is thus a standard requirement for field theoretical
model building that the spectrum has to be anomaly-free. This requirement also applies
to string theory in ten dimensions. String theories are  as well anomaly free by construction, 
and since our effective model
is derived from one of them we shall benefit from this fact also in six dimensions.

Anomalies are known and studied  since the end 1960's, and they have had important impact
on the understanding of the standard model. Pion to photon decays could only be consistently
described after the discovery of the chiral anomaly \cite{a69}, also other hadronic and
semileptonic decays as well as meson masses \cite{wz71}
and the QCD $\U1$ problem \cite{h76} are related to anomalies. 
Note that these   refer to global symmetries, such as the axial symmetry in QCD, and
do not harm the quantum consistency of the standard model. With respect to its local
symmetries, the latter is anomaly-free.

In general, anomalies can arise in any model with even dimensions. They are a consequence of the
presence of Weyl  fermions, which have
a non-invariant measure of integration in the path integral  \cite{f79}. 
Additionally, in $d=4n+2$ dimensions gravitational anomalies may appear,
related to local coordinate transformations. Their complete calculation   was
done in \cite{aw83}. In six dimensions, we thus have to deal with both types
of anomalies, and also combinations of gauge and gravitational anomalies appear,
called mixed anomalies.

Anomalies in $d=2n$ dimensions can be related to divergent one-loop Feynman diagrams
with $n+1$ external gauge boson legs and chiral fermions running in the loop.
In four dimensions the relevant diagram is the triangle diagram,
in six dimensions it is the box diagram, as shown in Figure \ref{fig:tribox}.
These diagrams are divergent and the anomaly is related to the finite part of
the corresponding expressions.

\subsubsection{The Stora--Zumino descent relations}

For local symmetries, the transformation parameter $\epsilon$ in (\ref{eq:anomaly}) is space-time dependent. 
Furthermore,
it  takes values in the Lie-algebra of the symmetry group.
Thus two successive transformations have to be compatible with the corresponding non-Abelian group structure:
\begin{align}
	\left[ \delta_{\epsilon_1},\delta_{\epsilon_2} \right] &= \delta_{\left[ \epsilon_1, \epsilon_2\right]}\,.
\end{align}
This basic requirement implies the Wess--Zumino consistency relations \cite{wz71},
\begin{align}
\label{wq:wz}
	\delta_{\epsilon_1} \mathcal I(\epsilon_2)-\delta_{\epsilon_2} \mathcal I(\epsilon_1)
	&=\mathcal I([\epsilon_1,\epsilon_2])\,,
\end{align}
and an anomaly is called consistent \index{Anomaly!consistent}
 if this relation holds. The general solution in  even dimensions $d=2n$
for an arbitrary local symmetry can be deduced from the Stora--Zumino descent relations 
\index{Stora--Zumino descent relations} \cite{s83}, which will also 
be the main tool for the calculation of the anomalies in the six-dimensional orbifold model under investigation.
They read\footnote{The following results are obtained in Euclidean space.
 It is expected that they can be Wick-rotated to Minkowski space.}
\begin{align}
\label{eq:sz}
    	\mathcal I(\epsilon) &= 2 \pi i \int_{M_{2n}} \Omega_{2n}^{(1)}(\epsilon) \,, &
	\rmd \Omega_{2n}^{(1)}&=\delta_{\epsilon} \Omega_{2n+1}^{(0)} \,, &
	\rmd \Omega_{2n+1}^{(0)} &= \Omega_{2n+2}\,,
\end{align}
and will be explained in the following.

The main observation is that they relate the anomaly $\mathcal I$ in $d=2n$ dimensions to a
$(2n+2)$-form $\Omega_{2n+2}$.
This $(2n+2)$-form is a gauge invariant function of the field strength $F$ and the curvature $R$. In fact,
it describes the chiral anomaly in $2n+2$ dimensions \cite{gg84} and can be calculated 
by topological methods. This means that the quantity $\Omega_{2n+2}(F,R)$ is known,
and the appearing invariants can be evaluated for the given spectrum.

Chiral anomalies \index{Anomaly!chiral} in $d=2n+2=2m$ dimensions
arise from Dirac fermions $\psi$, in contrast to the local gauge and gravitational anomalies,
that we are originally interested in. The latter rely on the presence of Weyl fermions. The chiral anomaly
corresponds to global chiral transformations of the Dirac spinors, generated by the chirality matrix
$\gamma_{2m+1}$. It has the classically conserved current
\begin{align}
	J^{M}_{2m +1} = \bar \psi_A \gamma_{2m+1} \gamma^M \psi^A\,,
\end{align}
where $M$ is an euclidean space-time index and $A$ labels the states of the gauge representation $\mathcal R$ of $\psi$.
This current is not conserved in the quantum theory,
\begin{align}
	\langle \partial_M J^M_{2m+1} \rangle &= 2 i \int_{M_{2m}} \Omega_{2m}\,,
\end{align}
which is identical to the statement that the chiral symmetry is anomalous. 
The quantity on the right hand side can then be related to topological quantities of the gauge bundle
and the tangent bundle, which are associated with gauge and gravitational interactions \cite{aw83}.
They are constructed from the Yang-Mills field strength \index{Field-strength two-form} and the Riemann curvature 
\index{Riemann curvature two-form} two-form,
\begin{align}
	F&=\frac 12 F_{MN}\, \rmd x^M \wedge \rmd x^N=\rmd A+A \wedge A\,, \\
	R&=\frac 12 R_{MN}\, \rmd x^M \wedge \rmd x^N=\rmd \omega+\omega \wedge \omega\,, 
\end{align}
where $A$ and $\omega$ denote the gauge and spin connection, respectively. Note that matrix indices 
were suppressed, 
${F_{MN\,i}}^j$ and ${R_{MN\,a}}^b$, where $i,j$
label states of the gauge representation $\mathcal R$,  and $a,b$ correspond to the
vector representation of the Lorentz group in $2m$ dimensions. 
 For
  the case of spin $1/2$ the chiral anomaly is expressed by 
  the product of the Chern character ${\rm ch}_{\mathcal R}(F)$ \index{Chern character}
 and the Dirac genus $\hat A(R)$\index{Dirac genus}. One can then obtain
the result for $\Omega_{2m}$ from 
a power series expansion:
\begin{subequations}
\label{eq:om2m}
\begin{eqnarray}
\label{eq:om2m1}
	\Omega_{2m}&=& \left. {\rm ch}_{\mathcal R}(F) \hat A(R) \right|_{\text{$2m$-form component}}\,,\\
	{\rm ch}_{\mathcal R}(F)&=&r+\frac{i}{2 \pi}{\rm tr}_{\mathcal R}\,F 
	 -\frac{1}{2(2\pi)^2} {\rm tr}_{\mathcal R}
	\,F^2\nonumber \\ \label{eq:om2m2}
	&&-\frac{i}{6(2\pi)^3} {\rm tr}_{\mathcal R}\,F^3
	+\frac{1}{24(2\pi)^4} {\rm tr}_{\mathcal R}\,F^4+\cdots\,,\\ \label{eq:om2m3}
	\hat A(R)&=&1+\frac{1}{48(2 \pi)^2}{\rm tr}\,R^2+
	\frac{1}{4!(2\pi)^4}\left[ \frac{1}{192}({\rm tr}\,R^2)^2
	+\frac{1}{240} {\rm tr}\,R^4\right]+\cdots\,.
\end{eqnarray}
\end{subequations}
Here ${\rm tr}_{\mathcal R}$ denotes the trace evaluated in the representation $\mathcal R$ of dimension $r$,
  and wedge products are understood. Similar expressions arise for fermions of higher spins.
By the use of the Stora--Zumino relations (\ref{eq:sz}) one can now calculate the gauge and
gravitational anomalies in $d=2n$ dimensions from the above result for the chiral anomaly in
$2m=2n+2$ dimensions. However, the analogy between the two anomalies holds up to a minus sign,
which is related to the chirality of the fermions on the side of the gauge and gravitational anomaly.
Furthermore, the total $(2n+2)$-form is the sum of the contributions of all fermions in the spectrum.
We will now evaluate the corresponding anomalies for the six-dimensional model of interest.

\subsubsection{The bulk anomaly polynomial}

In six dimensions, the gauge and gravity anomalies (cf.~\cite{abc03,gq03})
 can be deduced from an eight-form, studied in \cite{gsw84,e94,hm02}. 
It is 
convenient to introduce the anomaly polynomial $I_8$\index{Anomaly!polynomial!bulk}, 
\begin{align}
	I_8 &\equiv -16 (2 \pi)^{3} (2 \pi   \Omega_8)\,,
\end{align}
since the factor $1/16(2\pi)^{3}$ appears throughout the calculation. The result (\ref{eq:om2m}) corresponds
to the contribution of a single fermion of spin $1/2$. 
In total, the anomaly polynomial arises from the sum of all fermionic
contributions,
\begin{align}
	I_8&=\underbrace{I_{(3/2)}(R)}_{\text{gravitino}}-\underbrace{I_{(1/2)}(R)}_{\text{dilatino}}
	+\underbrace{I_{(1/2)}(F,R)}_{\text{gauginos}}-\sum_i s^i \underbrace{I^i_{(1/2)}(F,R)}_{\text{hyperinos}}\,,
\end{align}
where $s^i$ denotes the multiplicity of the representation $\mathcal R_i$ of the bulk hypermultiplets.
Note that the effect from the self-dual antisymmetric two-form $B_{MN}^+$ is canceled by
the terms due to its anti-self dual partner, $B_{MN}^-$, so they do not appear in the anomaly polynomial.
The different signs for the contributions of the dilatino and the hyperinos as compared to the gravitino and the
gauginos stem from the fact that they have opposite 6d chirality, cf.~(\ref{eq:qSO4}), (\ref{eq:qvec2}), 
\index{Weyl fermion!six dimensions}\index{Chirality|see{Weyl fermion}}
(\ref{eq:qu2}).
\begin{table}[t]
  \centering 
  \footnotesize
  \begin{tabular}{l|| l| l| c}
	\jvsb ${\rm tr}_{\mathcal R} F^n$ & \multicolumn{2}{c|}{$\SU N$} & 
	\multicolumn{1}{c}{$\SO N$} \\
\hline
\hline
\jvsb ${\rm tr}_{\mathcal A} F^2$ & \multicolumn{2}{c|}{$2 N \,{\rm tr}\, F^2$} & $(N-2) \,{\rm tr}\, F^2$ \\
\hline
\jvsb ${\rm tr}_{\mathcal A} F^4$ & $N=2,3$ & $\frac 12({\rm tr}\, F^2)^2 $& 
$(N-8) \,{\rm tr}\, F^4+3({\rm tr}\, F^2)^2$ \\
\jvsb   &  $N>3$ & $2 N \,{\rm tr}\, F^4+6({\rm tr}\, F^2)^2 $ & \\
\hline
\jvsb ${\rm tr}_{\mathcal S} F^2$ & \multicolumn{2}{l|}{ }& 
$2^{(N-8)/2} \,{\rm tr}\, F^2$ \\
\cline{1-1}
\cline{4-4}
\jvsb ${\rm tr}_{\mathcal S} F^4$ & \multicolumn{2}{l|}{ }& 
$-2^{(N-10)/2} \,{\rm tr}\, F^4+3(2^{(N-14)/2})({\rm tr}\, F^2)^2$ \\
\hline
\jvsb ${\rm tr}_{\text{\cal a}^{ij}} F^3$ & $N>2$ & $(N-4) {\rm tr}\, F^3$ &\\
\cline{1-3}
\jvsb ${\rm tr}_{\text{\cal a}^{ijk}} F^2$ & \multicolumn{2}{c|}{$\frac12(N^2-5N+6){\rm tr}\,F^2$}\\
\cline{1-3}
\jvsb ${\rm tr}_{\text{\cal a}^{ijk}} F^3$ & \multicolumn{2}{c|}{$\frac12(N^2-9N+18){\rm tr}\,F^3$} \\
\cline{1-3}
\jvsb ${\rm tr}_{\text{\cal a}^{ijk}} F^4$ & \multicolumn{2}{c|}{$\frac12(N^2-17N+54){\rm tr}\,F^4
+(3N-12)({\rm tr}\, F^2)^2$}
\end{tabular}
  \caption{A collection of trace identities \index{Trace identities}
   for representations $\mathcal R$ of $\SU N$ and $\SO N$.
Here $\mathcal A$ and $\mathcal S$ denote adjoint and spinor representations
and $ \text{\cal a}^{ijk}$ the third rank antisymmetric tensor representation, respectively.
The trace in the fundamental representation is denoted as ${\rm tr}$.
}
\label{tab:traces}
\end{table}

For the spin $3/2$ gravitino\index{Gravitino}, the contribution to the anomaly polynomial is \cite{aw83}
\begin{align}
	I_{(3/2)}(R) &=-\frac{16}{4!} \left[ \frac{245}{240}{\rm tr}\, R^4 -\frac{43}{192}({\rm tr}\, R^2)^2 \right]\,,
\end{align}
and therefore the sum of all  purely curvature dependent  terms is
\begin{align}
\label{eq:redan1}
I_{\rm grav}(R)&=-\frac{16}{4!240} \underbrace{\left( 244+y-s\right)}_0{\rm tr}\, R^4
-\frac{16}{4!192} \left( -44+y-s\right)({\rm tr}\, R^2)^2 \,,
\end{align}
where $y=76$ counts the total number of gauginos, and $s=320$ the number of hyperinos in the
six-dimensional bulk. This demonstrates that the gravitational $R^4$ anomaly cancels in the model.
This is a necessary requirement, since there are no alternative mechanisms to the direct
cancelation, for example by additional counter terms. Anomalies with that property
are called `irreducible'\index{Anomaly!irreducible}, 
in contrast to reducible\index{Anomaly!reducible} ones which can for example be canceled by the
Green--Schwarz mechanism, which will be discussed later.

All remaining terms in the anomaly polynomial involve traces 
over the representation indices of powers of the field strength two-forms. Recall that the bulk gauge group
is given by (\ref{eq:bulkgaugegroup}). Let $A$ and $u$ label the non-Abelian and the Abelian gauge factors, 
respectively,
\begin{align}
	 A&=
	 \SU6, \SU3, \SO8
	 \,, \hspace{1cm}
	 u=1,\dots,5\,.
\end{align}
One can then convert all traces ${\rm tr}_{\mathcal R}F^n$ into expressions proportional to the
trace in the fundamental representation, denoted as ${\rm tr}$.
This is summarized in Table \ref{tab:traces} for $\SU N$ and $\SO N$, which collects trace identities
as given in \cite{e94,hm02}.

According to (\ref{eq:om2m1}), the anomaly polynomial for the spin 1/2 fermions is given by all
combinations of terms from  (\ref{eq:om2m2}) and  (\ref{eq:om2m3}) which lead to an eight form.
We can collect them into expressions $I_{(1/2)}(F^n,R^m)$, where $n+m=4$:
\begin{align}
	I_{(1/2)}(F^4,R^0)=&-\frac{2}{3}\underbrace{ \left( 12+6\, s_{\bf 20} -s_{\bf 6}
	-s_{\bf \bar 6} \right)}_0 {\rm tr}\, F_{\SU 6}^4 
	-\frac{2}{3}\underbrace{ \left( \frac 12\, s_{{\bf 8}_s}+ \frac 12\, s_{{\bf 8}_c} -s_{\bf 8}
	 \right)}_0 {\rm tr}\, F_{\SO 8}^4\nonumber \\
	 &-\frac 23  \underbrace{\left(6-6 s_{\bf 20}\right)}_0({\rm tr}\,F^2_{\SU6})^2
	 -\frac 23  \underbrace{\left(3-\frac 38  s_{{\bf 8}_s}-\frac 38  s_{{\bf 8}_c}
	 \right)}_0({\rm tr}\,F^2_{\SO8})^2
	   \nonumber \\
	   &+I_{(1/2)}^{\rm Abelian}(F^4,R^0) \\
	 I_{(1/2)}(F^2,R^2)=&\frac 16 \big[ \underbrace{\left(12-6\, s_{\bf 20} -s_{\bf 6}
	-s_{\bf \bar 6} \right)}_{-12} {\rm tr}\, F_{\SU 6}^2+
	\underbrace{\left(6-s_{\bf 3}
	-s_{\bf \bar 3} \right)}_{-12} {\rm tr}\, F_{\SU 3}^2 \big. \nonumber \\
	&\big.\hspace{.8cm}+
	\underbrace{\left(6-s_{\bf 8}-s_{{\bf 8}_s}-s_{{\bf 8}_c} \right)
	}_{-6} {\rm tr}\, F_{\SO8}^2
	\big]
	  {\rm tr}\, R^2\nonumber \\
	   &+I_{(1/2)}^{\rm Abelian}(F^2,R^2)\,.\label{eq:redan2}
\end{align}
The number of hyperinos in the representation $\mathcal R$ is denoted as $s_{\mathcal R}$. From Tables
\ref{tab:bulkU} and \ref{tab:bulkT2T4} one can read off
\begin{align}
	s_{\bf 20}&=1\,, & s_{\bf 6}&=s_{\bf \bar 6}=9\,,&s_{\bf 3}&=s_{\bf \bar 3}=9\,, &
	s_{\bf 8}&=s_{{\bf 8}_s}=s_{{\bf 8}_c}=4\,,
\end{align}
which implies the numerous cancelations denoted above. They are necessary for a consistent model,
since the corresponding anomalies are irreducible.
 
What remains to be studied are the contributions to the anomaly polynomial which involve
$\U1$ factors, denoted as $I_{(1/2)}^{\rm Abelian}(F^4,R^0)$ and $I_{(1/2)}^{\rm Abelian}(F^2,R^2)$
above. They can be parametrized by constants $a^u_A,b_A^{uv},c^{uvwx},d^{uv}$,
\begin{subequations}
\label{eq:iabs}
\begin{eqnarray}
	I_{(1/2)}^{\rm Abelian}(F^4,R^0)&=&\frac 83 \sum_A \sum_{u=1}^5 a_A^{u} \big({\rm tr}\,F^3_A\big) \hat F_u
	+4 \sum_A \sum_{u,v=1}^5 b_A^{uv} \big({\rm tr}\,F^2_A\big)
	 \hat F_u \hat F_v 
	 \nonumber \\
	 &&+ \frac 23 \sum_{u,v,w,x=1}^5 c^{uvwz} \hat F_u  \hat F_v  \hat F_w  \hat F_x  \,,\\
	 I_{(1/2)}^{\rm Abelian}(F^2,R^2)&=&-\frac 16 ({\rm tr}\, R^2)  \sum_{u,v=1}^5 d^{uv}  \hat F_u  \hat F_v\,,
\end{eqnarray}
\end{subequations}
where $\hat F_u=\rmd \hat A_u+\hat A^2_u$ is the field strength two-form
which is associated with the canonically normalized generator \index{U1@$\U1$!canonical normalization}
 $\hat t_u$,
\begin{align}
\label{eq:hattu}
    \hat F_u, \hat A_u\,: \hspace{1cm} \hat t_u \equiv \frac{t_u}{\sqrt 2 |t_u|}\,.
\end{align}

In the literature, to our knowledge only cases with a single Abelian factor were studied (for example in \cite{e94}).
However, the bulk spectrum of the orbifold model derived before has five $\U1$ factors,
which leads to a complication of the analysis. Especially the cancelation of the related anomalies
by the Green--Schwarz mechanism is a non-trivial check of the consistency of the model, as will be discussed
later. Due to its technical nature we evaluate the (partly) Abelian traces 
in Appendix \ref{app:traces}, and directly state the 
result:
\begin{subequations}
\label{eq:abcd}
\begin{eqnarray}
	a_{\SU6}^{u}&=&a_{\SU3}^{u}=a_{\SO8}^{u}=0\,, \\
	b_{\SU6}^{uv}&=&b_{\SU3}^{uv}=2 b_{\SO8}^{uv}=\frac12 \beta^{uv}\,,\\
	c^{uvwx}&=&\frac{3}{2|\sigma(u,v,w,x)|} \left( \delta^{uv}\beta^{wx}+\text{permutations} \right)\,,\\
	d^{uv}&=&6\left(\beta^{uv}+\delta^{uv}\right)\,.
\end{eqnarray}
\end{subequations}
Here $|\sigma(u,v,w,x)|$ is a symmetry factor, which is one if all its variables are distinct,
and equal to the number of possibilities to assign the identical labels to the four 
$F$'s otherwise. Furthermore, $\beta^{uv}$ is a symmetric
$5 \times 5$ matrix,
\begin{align}
	(\beta^{uv})&=\left(\begin{array}{ccccc}
	3 & -1 & 0 & -1 & 0 \\
	-1 & 3 & 0 & -1 & 0 \\
	0 & 0 & 2 & 0 & \sqrt 2 \\
	-1 & -1 & 0 & 4 & 0 \\
	0 & -0 & \sqrt 2 & 0 & 4
	\end{array}\right)\,,
\end{align}
demonstrating that the model has non-trivial (partly) Abelian anomalies in the bulk. However, as we will
demonstrate in the next section, all anomalies can be canceled by a classical term in the action.

Summarizing the results (\ref{eq:redan1}), (\ref{eq:redan2}) and (\ref{eq:iabs}) with (\ref{eq:abcd}), 
the total remaining anomaly polynomial in the six-dimensional bulk is \index{Anomaly!polynomial!bulk}
\begin{align}
\label{eq:redanomaly}
    	I_8=&({\rm tr}\,R^2)^2-\left(2\,{\rm tr}\,F^2_{\SU6}+2\,{\rm tr}\,F^2_{\SU3}+\,{\rm tr}\,F^2_{\SO8}
	+{\bf  \hat F} \cdot \beta \cdot{\bf \hat F}+{\bf \hat F}^2\right){\rm tr}\, R^2 \nonumber \\
	&+   \left(2 {\rm tr}\,F^2_{\SU6}+2{\rm tr}\,F^2_{\SU3}
	+{\rm tr}\,F^2_{\SO8}+{\bf \hat F}^2\right)
	{\bf  \hat F} \cdot \beta \cdot{\bf \hat F}\,,
\end{align}
with vector notation ${\bf \hat F}=(\hat F_1,\dots,\hat F_5)$ for the Abelian field strength two-forms.

\subsubsection{Anomaly polynomials at fixed points}
\label{sec:anpolFP}

Similarly to the bulk case, we can now evaluate the anomaly polynomials $I_{n_2,6}$ which are
related to the spectra at the fixed points, labeled by $n_2=0,1$.
In that case two dimensions are fixed.
Hence the localized fields only propagate  in four dimensions and the anomaly polynomial is
 a six-form, \index{Anomaly!polynomial!local}
\begin{align}
	I_{n_2,6}\equiv -48 (2\pi)^3i \Omega_6\,,
\end{align}
where $ \Omega_6$ follows from (\ref{eq:om2m}).

Following the same program as before, it can be parametrized as
\begin{align}
\label{eq:I61}
	I_6=& {\rm tr}\, R^2\, \sum_v \tilde a^v \hat F_v
	-8\, \sum_{B} \tilde b_B \,{\rm tr} F^3_B
	-24\, \sum_{B,v} \tilde c^v_B \,{\rm tr} F^2_B\,\hat F_v
	-8\, \sum_{v,w,x} \tilde d^{vwx}_B \,\hat F_v\hat F_w\hat F_x\,,
\end{align}
where $\tilde a^v, \tilde b_B, \tilde c^v_B,\tilde d^{vwx}$ are coefficients, and
$B$ and $v,w,x$ label the non-Abelian and the Abelian factors at the fixed point, respectively,
\begin{align}
	B&=\left\{\begin{array}{ll}
	\SU5,\SU3,\SO8, & \text{if $n_2=0$},\\
	\SU2,\SU4,\SU2',\SU4',& \text{if $n_2=1$},
	\end{array} \right.&
	v&=\left\{\begin{array}{ll}
	1,\dots,6, & \text{if $n_2=0$},\\
	1,\dots,8,& \text{if $n_2=1$}.
	\end{array}\right. \label{eq:Bv}
\end{align}
The coefficients in (\ref{eq:I61}) are determined in Appendix \ref{app:loctraces}, with results
\begin{align}
\label{eq:I6n20}
	n_2&=0\,:&
	I_6 &= \left( {\rm tr}\, R^2-2\, {\rm tr}\, F_{\SU5} -2 \,{\rm tr}\, F_{\SU3} - {\rm tr}\, F_{\SO8}
	-{\bf \hat F}^2 \right) 2 \sqrt{37} \hat F_{\rm an}^0\,,\\
	\label{eq:I6n21}
	n_2&=1\,:&
	I_6 &= \Big( {\rm tr}\, R^2-2\, {\rm tr}\, F_{\SU2} -2\, {\rm tr}\, F_{\SU4} \Big. \nonumber \\
	&&&\hspace{1.7cm} \left.- \,2 \,{\rm tr}\, F_{\SU2'}
	 -2\,{\rm tr}\, F_{\SU4'}-{\bf \hat F}^2 \right) 2 \sqrt{10} \hat F_{\rm an}^1\,,	
\end{align}
where $\hat F_{\rm an}^0$ and $\hat F_{\rm an}^0$ denote field strength two-forms associated
with the canonically normalized generators $\hat t_{\rm an}^0$ and $\hat t_{\rm an}^1$, whose 
non-canonically normalized versions can be found in Table \ref{tab:t}. They are linear
combinations of the local $\U1$ factors,
\begin{align}
	t_{\rm an}^0 &\equiv -4 t_2+5t_4-t_5+t_6^0\,, & \sqrt 2 | t_{\rm an}^0|&=2 \sqrt{37}\,,\\
	t_{\rm an}^1 &\equiv 3 t_1-t_2+t_3-4t_4+4t_5 \,, &  \sqrt 2 |t_{\rm an}^1|&=4\sqrt{10}\,.
\end{align}
All generators of the Abelian groups in (\ref{eq:I6n20}) and (\ref{eq:I6n21})
are chosen orthogonal to each other, furthermore the generators $\hat t_{\rm an}^0$
and $\hat t_{\rm an}^1$ correspond to one of the indices $v$ and the corresponding
contributions also appear inside the bracket.

The above vectors $t_{\rm an}^0 $ and $t_{\rm an}^1$ are the generators of the local
anomalous $\U1$'s \index{U1@$\U1$!anomalous} at the inequivalent fixed points $n_2=0$ and $n_2=1$.
We stress that they are neither collinear nor orthogonal. It is an interesting observation
that the model provides two different anomalous $\U1$ factors, localized at different 
fixed points, since it is a well-known fact that there can be at most one anomalous $\U1$
in the four-dimensional low-energy theory, after compactification of six dimensions of
 heterotic string theory. However, this is not contradicted by our findings. 
 Summing over the zero modes, one finds that there is only one anomalous $\U1$ in
 the effective four-dimensional
 theory, with generator \cite{bhx06}
 \begin{align}
	t_{\rm an}^{\rm 4d}&=\frac 13 \left(  t_{\rm an}^0+\frac 12 t_{\rm an}^1\right) \,.
\end{align}
It is given by the sum of the localized anomalous generators, weighted by the 
size of the anomalous traces.

\subsubsection{Green--Schwarz anomaly cancelation in six dimensions}

The Green--Schwarz mechanism \index{Green--Schwarz mechanism}
 originally refers to string derived supergravities in ten dimensions \cite{gs84}.
It states that models with an anomalous spectrum can be anomaly free, due to the variation of counter terms
in the action which are related to an antisymmetric tensor field. This  is a remarkable statement, since it means
that a one-loop quantum effect is balanced by a classical term in the action. The six-dimensional version
of the mechanism is studied in \cite{e94,lnz04,a06}. Here we  closely follow \cite{s04}, for the special
case of six dimensions.

In our model, the antisymmetric two-form \index{Antisymmetric tensor field}
is called $B_2$ and arises from the gravitational sector,
cf.~(\ref{eq:qgrav}). Its kinetic term is determined by the three-form field strength
\begin{align}
\label{eq:H3}
    	H_3&=\rmd B_2+\sqrt{2 \pi} \xi X_3^{(0)}\,,
\end{align}
where $\xi$ is a dimensionless parameter.
The introduction of the three-form $ X_3^{(0)}$ is necessary if $B_2$ transforms inhomogeneously
under gauge transformations\footnote{Here and everywhere else in this work,
 $\delta_\epsilon$ unifies gauge transformations and diffeomorphisms. In the following we refer to the combined
 transformations
 as gauge transformations.},
\begin{align}
\label{eq:deltaB}
	\delta_\epsilon B_2&=-\sqrt{2\pi} \xi   X_2^{(1)}\,, \hspace{1cm}
	\rmd X_2^{(1)}=\delta_\epsilon X_3^{(0)}\,, \hspace{1cm}
	\rmd X_3^{(0)}=X_4\,.
\end{align}
Here $X_4$ was introduced since it will appear in calculations soon.
Thus the extra term in the definition of the field strength $H_3$ in (\ref{eq:H3}) 
is responsible for invariance of the latter under local symmetry transformations.

The Green--Schwarz action in six dimensions is
\begin{align}
\label{eq:SGS}
    	S=&\int_{M_6} \left[ \frac 12 \left| H_3 \right|^2+i \frac{ \sqrt{2 \pi}}{\xi} B_2 Y_4
	-2\pi i \left(\frac12+\alpha\right)  X_3^{(0)} Y_3^{(0)} \right]\,,
\end{align}
where $\alpha,\xi$ are parameters and $Y_4$ is a four-form, related to $Y_3^{(0)}$ 
and another two-form $Y_2^{(1)}$ by
\begin{align}
	\rmd Y_3^{(0)}&=Y_4\,, \hspace{1cm} \delta_\epsilon Y_3^{(0)}=\rmd Y_2^{(1)}\,.
\end{align}
The above action is not invariant under gauge transformations. One can calculate the variation as
\begin{align}
	\delta_\epsilon S=&-2 \pi i \int_{M_6} \left[ X_2^{(1)} Y_4 +\left(\frac 12+\alpha\right) \rmd X_2^{(1)}Y_3^{(0)}
	+\left(\frac 12+\alpha\right) X_3^{(0)} \rmd Y_2^{(1)} \right] \nonumber \\
	=&-2 \pi i \int_{M_6} \left[ X_2^{(1)} Y_4 -\left(\frac 12+\alpha\right)  X_2^{(1)}Y_4
	+\left(\frac 12+\alpha\right) X_4  Y_2^{(1)} \right] \nonumber \\ \label{eq:deltaS1}
	=&-2 \pi i \int_{M_6} \left[ \left(\frac 12-\alpha\right)  X_2^{(1)}Y_4
	+\left(\frac 12+\alpha\right) X_4  Y_2^{(1)} \right]\equiv-2\pi i \int_{M_6} \Omega_6^{(1)}\,,
\end{align}
where partial integration was used.
The action (\ref{eq:SGS}) can thus cancel the anomalous variation of the quantum contribution to the
effective action $\Gamma$, if the anomaly is described by $\mathcal I=2 \pi i \int \Omega_6^{(1)}$,
where $\Omega_6^{(1)}$ takes the form which appears in the bracket.

This is the case if it descents from an eight form which factorizes into $X_4$ and $Y_4$,
\begin{align}
	\Omega_8&=X_4 Y_4 & &\Rightarrow&
	 \Omega_6^{(1)} &=\left(\frac 12-\alpha\right)  X_2^{(1)}Y_4
	+\left(\frac 12+\alpha\right) X_4  Y_2^{(1)}\,,
\end{align}
where $\alpha$ is a free parameter and all field strengths are treated symmetrically. 

Now reconsider the anomaly polynomial (\ref{eq:redanomaly}), which we found for the bulk theory in our model.
It can be factorized as \index{Anomaly!polynomial!bulk}
\begin{align}
\label{eq:I8fact}
    	I_8=&\left( {\rm tr}\,R^2-2\,{\rm tr}\,F^2_{\SU6}-2\,{\rm tr}\,F^2_{\SU3}-\,{\rm tr}\,F^2_{\SO8}
	-{\bf \hat F}^2   \right)\left( {\rm tr}\,R^2-{\bf \hat F}\cdot \beta \cdot {\bf \hat F}  \right)\,,
\end{align}
and thus we have shown that all anomalies either cancel among themselves, or are canceled
by the variation of the Green--Schwarz action (\ref{eq:SGS}). \index{Anomaly!reducible}

In fact, the coefficients $\alpha_A$ in the first bracket are generic and related to the dual Coxeter number
\index{Dual Coxeter number} of the
gauge group \cite{e94},
\begin{align}
	\alpha_{\SU N}&=2\,, \hspace{1cm} \alpha_{\SO N}=1\,, \hspace{1cm}
	\alpha_{\U1}=1\,,
\end{align}
where the latter corresponds to canonical normalization of the $\U1$ generator, as in (\ref{eq:hattu}).
The second factor in (\ref{eq:I8fact}) carries model dependent information, condensed into the matrix $\beta$.
Note that the identification of the factors $X_4,Y_4$ also determines the classical action (\ref{eq:SGS}) of the
antisymmetric tensor field $B_2$, as well as its gauge transformation (\ref{eq:deltaB}), up to two free
constants $\alpha,\xi$.

\subsubsection{Green--Schwarz anomaly cancelation on the orbifold}

So far we have demonstrated that the bulk theory of the effective orbifold GUT field theory is
a consistent gauge theory with respect to quantum fluctuations. However, at the fixed points
additional matter is generated, originating from twisted sectors, and  the local consistency
of the spectrum remains to be checked.

The Green--Schwarz \index{Green--Schwarz mechanism}
 action (\ref{eq:SGS}) can be extended to include fixed points $z^3_f$,
associated with delta-functions $\delta^f_2\equiv\delta(x_5-x_5^f)
\delta(x_6-x_6^f)\rmd x^5 \wedge \rmd x^6$ and local two-forms $Y_2^f$ \cite{s04}:
\begin{align}
\label{eq:SGS2}
    	S=\int_{M_6} &\left[ \frac 12 \left| H_3 \right|^2+i \frac{ \sqrt{2 \pi}}{\xi}
	 B_2 \left(Y_4 +\sum_f \delta^f_2 Y^f_2 \right)
	\right.\nonumber\\
	&\left.
	-2\pi i \left(\frac12+\alpha\right)  X_3^{(0)} 
	\left(Y_3^{(0)}+\sum_f \delta^f_2 Y^{f(0)}_1 \right)
	 \right]\,.
\end{align}
Here $\xi$ and $\alpha$ are two  dimensionless parameters which may be chosen arbitrarily
without affecting the anomaly cancelation mechanism.
The variation of this action contains local contributions,
\begin{align}
	\delta_\epsilon S=&-2 \pi i \int_{M_6} \left[ \Omega_6^{(1)}
	+\sum_f \delta^f_2 \Omega_4^{f  (1)} \right]\,,
\end{align}
where $\Omega_6^{(1)}$ was defined before in (\ref{eq:deltaS1}), and
\begin{align}
	\Omega_4^{f  (1)}&=
	 \left(\frac 12-\alpha\right)  X_2^{(1)}Y_2^f
	+\left(\frac 12+\alpha\right) X_4  Y_0^{f (1)}\,.
\end{align}

The action (\ref{eq:SGS2}) is thus suitable for the cancelation of reducible anomalies \index{Anomaly!reducible}
with an anomaly polynomial of the form
\begin{align}
	\label{eq:I8ct}
	I_8&=X_4 Y_4 + \sum_f \delta^f_2 X_4 Y_2^f\,.
\end{align}

The factorization of the bulk anomaly polynomial was discussed in the previous section, with
\begin{align}
	 X_4&= {\rm tr}\,R^2-2\,{\rm tr}\,F^2_{\SU6}-2\,{\rm tr}\,F^2_{\SU3}-\,{\rm tr}\,F^2_{\SO8}
	-{\bf \hat F}^2  \,.
\end{align}
At the fixed points some of the field strength forms vanish due to the boundary conditions
of the fields.
The localized anomalies from (\ref{eq:I6n20}) and (\ref{eq:I6n21}) thus only
contain a projection of this factor, \index{Anomaly!polynomial!local}
\begin{align}
	n_2&=0\,:&
	I_6 &= \left( {\rm tr}\, R^2-2\, {\rm tr}\, F_{\SU5} -2 \,{\rm tr}\, F_{\SU3} - {\rm tr}\, F_{\SO8}
	-{\bf \hat F}^2 \right) 2 \sqrt{37} \hat F_{\rm an}^0\,,\\
	n_2&=1\,:&
	I_6 &= \Big( {\rm tr}\, R^2-2\, {\rm tr}\, F_{\SU2} -2\, {\rm tr}\, F_{\SU4} \Big. \nonumber \\
	&&&\hspace{1.7cm} \left.- \,2 \,{\rm tr}\, F_{\SU2'}
	 -2\,{\rm tr}\, F_{\SU4'}-{\bf \hat F}^2 \right) 2 \sqrt{10} \hat F_{\rm an}^1\,.
\end{align}
Note that the definitions of the $\U1$ factors are different here, as discussed below (\ref{eq:I6n21}).

We have thus shown that all anomalies of the model cancel. Either among the fields of the spectrum,
in the case of irreducible anomalies, or by classical counter terms, in the case of reducible anomalies.
With respect to quantum corrections, the model is a consistent gauge field theory on an orbifold in six
dimensions.

\section{Vacua and phenomenology}
\label{cpt:vacuum}

The six-dimensional GUT model, as discussed so far, does not describe the visible world.
It is obvious that it contains many exotic particles, which are inconsistent with observations.
Furthermore, even though all anomalies are canceled by a bulk field, the non-vanishing
traces of anomalous $\U1$ factors imply that supersymmetry is broken radiatively 
by induced Fayet--Iliopoulos $D$-terms at a
large scale \cite{ads87}, if the vacuum expectation values of all fields are zero.
 What was described in the last chapter thus cannot be a stable vacuum
solution of a compactification of perturbative string theory, which should be supersymmetric.
One rather expects the presence of
a tachyonic direction in the effective potential which drives the system towards its
physical vacuum.

The main point of this chapter is to establish a relation between the unbroken symmetries
in this new vacuum, which are mainly discrete symmetries, and the interactions
in the effective GUT theory. The latter arise from higher-dimensional operators, 
after integrating out singlet fields. The symmetries will guide us to an algorithm which determines 
vacuum configurations with the property that phenomenologically disfavored couplings
are forbidden to all orders in the singlet fields, as long as supersymmetry
is unbroken. This result and its application to the $\mu$-term  \cite{bs08}
is reviewed in Sections \ref{sec:symms} and \ref{sec:kernel}.
Furthermore, Section \ref{sec:allorders} discusses
the reverse argument, namely the generation of the superpotential terms to all orders in the singlets,
in a given vacuum. 

The abovementioned problems, the presence of exotics and supersymmetry
breaking Fayet--Iliopoulos terms $\xi$,  are by no means specific to our model. 
They generically arise in orbifold
compactifications of the heterotic string. The usual approach  to overcoming them is
to assume expectation values of singlet fields, thereby guessing the location of the
supersymmetric vacuum in the potential, \index{False vacuum} \index{Physical vacuum}
\begin{align}
\label{eq:falsevac00}
	\text{False vacuum:} \hspace{1cm} \langle s_i \rangle &= 0 \hspace{.5cm}\text{for all singlets $s_i$},
	\hspace{1cm} D \sim \xi \neq 0 \,,\\ \label{eq:physvac00}
	\text{Physical vacuum:} \hspace{1cm} \langle s_i \rangle &\neq 0 \hspace{.5cm}\text{for some singlets $s_i$},
	\hspace{1cm} D = 0 \,,
\end{align}
 However, due to the size of the
supersymmetry breaking $D$-term this new vacuum is not necessarily nearby the origin,
and it is not clear that the findings from the false vacuum can be extrapolated.
This worry is linked to the observation that vacuum expectation values of localized
fields imply  a resolution of the orbifold singularities (cf.~\cite{chx85,fix88,a94,ht06,ngx07,lrx08,nhx09}). The orbifold result,
shifted to a vacuum remote from the origin, may then 
correspond to
 an effective description of a compactification on a smooth manifold. 
 Here we shall take the optimistic viewpoint that orbifolds, even after the transition to
 the supersymmetric vacuum, are a first-order approximation to the results one would
 obtain in a fully dynamical calculation, if the latter was doable. 
 
 We therefore now turn to the study
 of possibly interesting physical vacua of the model at hand, and return to the issue of
 contact with smooth geometries in the following chapter. Here our analysis is guided 
 by phenomenological requirements. We thus hope to reveal some of its mechanisms which determine the role
 of the various fields in the spectrum. We furthermore hope to understand
  the presence or absence of couplings
 among the fields of the local GUT theory, and consequently also of the low-energy zero modes.

\subsection[The $\SU5$ GUT superpotential]{\boldmath The $\SU5$ GUT superpotential}
\label{sec:SU5}

The effective orbifold GUT in six dimensions which was derived in the previous chapter contains
fixed points with $\SU5$ gauge symmetry\index{SU5 GUT@$\SU5$ GUT}, 
localized and bulk-projected matter. What we are aiming
at in this chapter is the construction of a viable local $\SU5$ GUT theory at these fixed points. For
that it is instructive to recall the ingredients and interactions of standard $\SU5$ GUTs.

The group $\SU5$ was the first simple group which was suggested for grand unification\footnote{The
first proposal for extended gauge symmetry by
Pati and Salam was based on a non-simple Lie group \cite{ps73}, corresponding to
three independent gauge couplings.}, by Georgi and
Glashow in 1974 \cite{gg74}, and it
is somehow the simplest choice. In the following we will need its supersymmetric version, therefore
we associate every gauge multiplet with a chiral multiplet of $\mathcal N=1$ supersymmetry in
four dimensions.

One matter generation consists of one antisymmetric tensor representation of rank two 
 and one 
anti-fundamental representation, \index{Standard model families}
\begin{subequations}
\label{eq:TFbSU5}
\begin{align}
\label{eq:TSU5}
	\T&=\underbrace{({\bf 3},{\bf 2})_{1/6}}_{q} + 
	\underbrace{({\bf \bar 3},\mathbbm 1)_{-2/3}}_{u^c} + 
	\underbrace{(\mathbbm 1,\mathbbm 1)_1}_{e^c} \,,\\
	\label{eq:FbSU5}
	\Fb &= \underbrace{({\bf \bar 3},\mathbbm 1)_{1/3}}_{d^c} + 
	\underbrace{(\mathbbm 1,{\bf 2})_{-1/2}}_{l}\,,
\end{align}
\end{subequations}
where the decomposition is given with respect to the standard model
gauge group $\SU3\times\SU2\times\U1_Y$. Note that the right-handed neutrino 
is not automatically included and has to be added as an extra singlet. This
is in contrast to the GUT group $\SO{10}$ \cite{g75}, where a single ${\bf 16}$
contains the above matter as well as the right-handed neutrino, ${\bf 16}={\bf 10}
+\Fb+\mathbbm 1$.

In addition to the matter fields two Higgs multiplets are required for the generation of
masses for the standard model particles, 
\begin{subequations}
\label{eq:HuHdSU5}
\begin{align}
	H_u&=\F=\underbrace{({\bf 3},\mathbbm 1)}_{\text{exotic}}+\underbrace{(\mathbbm 1,{\bf 2})_{1/2}}_{h_u}\,,\\
	H_d&=\Fb=\underbrace{({\bf \bar 3},\mathbbm 1)}_{\text{exotic}}+\underbrace{(\mathbbm 1,{\bf 2})_{-1/2}}_{h_d}\,,
\end{align}
\end{subequations}
and the fact that the exotic triplet and anti-triplet have to be decoupled from the low-energy theory
is the doublet-triplet splitting problem. In orbifold models it is solved by the fact that
the exotics have odd boundary
conditions, while the doublets correspond to zero modes in four dimensions.  

In the language of $\mathcal N=1$ supersymmetry, interactions are then expressed
by the superpotential
\begin{align}
\label{eq:WSU5}
W =& \mu H_uH_d + \mu_i H_u\Fb_{(i)} + C_{ij}^{(u)}\T_{(i)}\T_{(j)} H_u 
+ C_{ij}^{(d)}\Fb_{(i)}\T_{(j)}H_d \nonumber\\
& + C_{ijk}^{(R)}\Fb_{(i)}\T_{(j)}\Fb_{(k)} 
+ C_{ij}^{(L)}\Fb_{(i)}H_u\Fb_{(k)}H_u 
+  C_{ijkl}^{(B)}\T_{(i)}\T_{(j)}\T_{(k)}\Fb_{(l)}\;, 
\end{align}
where we included dimension-five operators but no higher non-renormalizable couplings.
$\mu_i$ and $C^{(R)}$ yield the well 
known renormalizable baryon (B) and lepton (L) number violating interactions, 
and the coefficients $C^{(L)}$ and $C^{(B)}$ of the dimension-five operators 
are usually obtained by integrating out states with masses 
$\mathcal{O}(M_\mathrm{GUT})$.

In supergravity theories
also the expectation value of the superpotential is important since it 
determines the gravitino mass. One expects
\begin{align}
\label{eq:muEW}
\langle W \rangle \sim \mu \sim M_{\mathrm{EW}}\,,
\end{align}
if the scale $M_{\mathrm{EW}}$ of electroweak symmetry breaking is related to 
supersymmetry breaking.

Experimental bounds on the proton lifetime and lepton number violating 
processes imply 
\begin{align}
\mu_i &\ll \mu\,, & C^{(R)} &\ll 1\,, &
C^{(B)} &\ll 1/M_\mathrm{GUT}\,.
\end{align} 
 Furthermore, one has to accommodate the
hierarchy between the electroweak scale and the GUT scale,
$M_{\mathrm{EW}}/M_\mathrm{GUT} = \mathcal{O}(10^{-14})$. On the other hand, 
lepton number violation should not be too much suppressed, since 
$C^{(L)} \sim 1/M_\mathrm{GUT}$ yields the right order of magnitude for 
neutrino masses.

Convincing realizations of the above bounds can be found from symmetry arguments.
Exact symmetries completely forbid the presence of couplings, while
`mildly broken' symmetries induce `approximate symmetries' \cite{kx08} and
may lead to sufficient suppression.

Dimension-four proton decay operators \index{Proton decay!dimension-four}
can be forbidden by `matter parity' \index{Matter parity} \cite{drw82}, 
which arises as an
unbroken $\Z2$ subgroup of $\U1_{B-L}$ \index{U1@$\U1$!B-L@$B-L$} in the standard model. This mechanism can be 
lifted to the $\SU5$ GUT description by the extension of the symmetry by the additional
Abelian factor $\U1_X$, \index{U1@$\U1$!X@$X$}
\begin{align}
\label{eq:U1XSU5}
	\SU5 \times \U1_X \subset \SO{10}\,,
\end{align}
with charge assignments
\begin{align}
\label{eq:PXSU5}
t_X(\T) = \frac{1}{5}\;, \quad t_X(\Fb) = -\frac{3}{5}\;, \quad
t_X(H_u) = -\frac{2}{5}\;, \quad t_X(H_d) = \frac{2}{5}\,.
\end{align}
This symmetry has the consequence $\mu_i = C^{(R)} = C^{(L)} = 0$.
Together with the hypercharge in $\SU5$ it yields $\U1_{B-L}$,
\begin{align}
\label{eq:tXSU5}
	\SU5 \times \U1_X &\supset G_{\rm SM} \times \U1_{B-L} \,, & 
	t_{B-L} &= t_X + \frac{4}{5}\ t_Y \,.
\end{align}
The wanted result, $\mu_i = C^{(R)} = 0$, $C^{(L)} \neq 0$, can be obtained
with a $\Z2^X$ subgroup of $\U1_X$, which contains the matter parity
$P_X$,
\begin{align}
P_X(\T) = P_X(\Fb) = -1\;, \quad  P_X(H_u) = P_X(H_d) = 1 .
\end{align}
Matter parity, however, does not solve the problem $C^{(B)} \neq 0$, and also 
the hierarchy $M_\mathrm{EW}/M_\mathrm{GUT} \ll 1$ remains unexplained.

Note that the effective model, as derived from the heterotic string, does not
contain a continuous $\U1_R$ symmetry which forbids the $\mu$-term or
 dimension-five proton decay
operators. Instead, three discrete $R$-symmetries are present, and in principle
they can be sufficient. However, as we shall see, the phenomenology
of a model strongly depends on the chosen vacuum and it is difficult
to fulfill all low-energy constraints simultaneously.
It is a main motivation for the present work to get a better understanding of the relation
between orbifold vacua and phenomenology, and to explore new tools for
the selection of candidates for physical vacua.

\subsection[Standard model families and a unique $\U1_{B-L}$]{\boldmath Standard model families and a unique $\U1_{B-L}$}
\label{sec:SM}

In the previous chapter,  the spectrum of an effective orbifold GUT in six dimensions was calculated,
which followed from a specific compactification of the heterotic string. The latter was an orbifold
which was designed to produce the content of a ${\bf 16}$ of the grand unifying group $\SO{10}$,
localized at two fixed points \cite{bhx06}. The effective local GUT at the fixed points $n_2=0,1$
is a $\SU5$ theory, and therefore these two families \index{Standard model families}
 take the form (cf.~(\ref{eq:fam12}), (\ref{eq:TFbSU5}))
\begin{subequations}
\label{eq:locfams}
\begin{eqnarray}
	  (n_2,n_2') = (0,0)\,:&& \hspace{.5cm}\T_{(1)}+\Fb_{(1)}+N^c_{(1)}\,,\\
	  (n_2,n_2') = (0,1)\,:&& \hspace{.5cm}\T_{(2)}+\Fb_{(2)}+N^c_{(2)}\,.
\end{eqnarray}
\end{subequations}
They arise from the first twisted sector and are localized at the origin of the compact $\SU3$-plane,
 described by $n_3=0$, cf.~Figure \ref{fig:z6II}. The charges of the chiral multiplets
 under the six local $\U1$ factors
 at the fixed points are listed in Table \ref{tab:n2T}.

The fields which combine to the third standard model family in four dimensions have to come
from the bulk. This is interesting, because also the Higgs multiplets have to be bulk fields, since
they should couple directly to both localized families. One can then expect that the bulk family
is naturally heavier than the two localized ones, which matches with the structure of the low-energy
standard model. Table \ref{tab:n20U} shows that at $n_2=0$, there are in fact two ten-plets with origin
in the bulk, both coming from the untwisted sector,
\begin{align}
\label{eq:ten34}
	\T_{(3)} &\equiv \T \,, \hspace{1cm} \T_{(4)}\equiv \Tb^c\,.
\end{align}
Note that $\T$ is the left-handed chiral multiplet component $H_L$ of the bulk hypermultiplet $H=(H_L,H_R)$,
while $\Tb^c$ denotes the charge conjugated right-handed component $H_R^c$. This then contains
a left-handed fermion and transforms as a $\T$ of $\SU5$. Also the Abelian charges of this field
are the inverse of those given for the hypermultiplet $H$ in Table \ref{tab:n20U}.

The interesting observation now is that both of these ten-plets are split multiplets and
only the sum of their zero modes combines to the expected content of a $\T$ of $\SU5$,
given in (\ref{eq:TSU5}):
\begin{align}
	&\text{Zero modes of $\T_{(3)}$}: \hspace{.5cm} ({\bf \bar 3},\mathbbm 1) = u^c\,, \hspace{.5cm}
	(\mathbbm 1,\mathbbm 1)=e^c\,,\\
	&\text{Zero modes of $\T_{(4)}$}: \hspace{.5cm} ({\bf  3},{\bf 2}) = q\,.
\end{align}
This is explicitly shown in Table \ref{tab:4d} in Appendix \ref{app:4d}.
The representations are here given with respect to the non-Abelian part of the standard model
gauge group, $\SU3\times \SU2$. One also observes that all $\Fb$-plets of $\SU5$
in the projected bulk spectrum are split multiplets. Thus we conclude that
the model requires not three but four families in the
six-dimensional description. This avoids the mostly incorrect mass relations of standard four-dimensional
$\SU5$ GUTs and gives room for an interesting flavor structure of the Yukawa couplings.

The identification of quark-lepton $\Fb$-plets is more complicated. For the left-handed chiral
bulk multiplets  $H_L, H_R^c$, associated with the states
from Tables  \ref{tab:n20U}, \ref{tab:n20T2T4} and their charge conjugates, respectively, we introduce the notation
\begin{subequations}
\begin{eqnarray}
	k=0: \hspace{1cm} && \F,\; \Fb,\\
	k=2: \hspace{1cm} && \F_{n_3, \gamma},\;\Fb_{n_3, \gamma},\\
	k=4: \hspace{1cm} && \F^c_{n_3, \gamma},\;\Fb^c_{n_3, \gamma}\,.
\end{eqnarray}
\end{subequations}
The local $\SU5$ GUT at $n_2=0$ provides a hypercharge generator $t_Y$, under which the
standard model lepton $l,e^c$, the quarks $q,d^c,u^c$ and the Higgs doublets $H_u,H_d$ have the
following charges:
\begin{align}
	Q_Y(l)=-2Q_Y(e^c)=Q_Y(H_d)=-Q_Y(H_u)&=-\frac 12\,, \\
	Q_Y(u^c)=-2 Q_Y(d^c)= -4Q_Y(q)&=-\frac 23\,.
\end{align}
Inspection of the zero modes, listed
in Table \ref{tab:4d} in Appendix \ref{app:4d}, then suggests the following classification for the $\F, \Fb$-plets at $n_2=0$,
cf.~(\ref{eq:FbSU5}), (\ref{eq:HuHdSU5}):
\begin{subequations}
\label{eq:5s}
\begin{align}
\label{eq:5s1}
	&\text{Quark-like zero modes, $d^c=({\bf \bar 3},\mathbbm 1)$}: &
	& \Fb_{0, \frac12} ,\; \F^c_{0, 0} ,\; \F^c_{0, 1} ,\; \Fb_{2, 0} ,\; \Fb_{2, 1}\,,\\
	&\text{Lepton- or $H_d$ -like zero modes, $(\mathbbm 1, {\bf 2})$}: &
	& \Fb_{1, 0} ,\;\Fb_{1, 1} ,\; \F^c_{1, \frac12} ,\; \F^c_{2, 0} ,\;\F^c_{2, 1},\;\Fb\,,\\
	&\text{$H_u$ -like zero modes, $(\mathbbm 1, {\bf 2})$}: &
	& \F_{1, 0} ,\;\F_{1, 1} ,\;\Fb^c_{1,\frac12} ,\; \F_{2,\frac12},\; \F\,,\\
	&\text{Exotic zero modes}: &
	& \F_{0, \frac12} ,\;\Fb^c_{0,0} ,\;\Fb^c_{0,1} ,\; \Fb^c_{2, \frac12} \,.\label{eq:5s4}
\end{align}
\end{subequations}
So far the spectrum still seems to be rather ambiguous. 
%
However, we can now benefit from the local $\SU5$ GUT structure at $n_2=0$.  As it was described in 
Section \ref{sec:SU5},
dangerous dimension-four proton decay operators can be forbidden in the $\SU5$ language by an additional Abelian
factor, called $\U1_X$. This symmetry plays the role
of $\U1_{B-L}$  and in fact only differs from it by hypercharge.
Such a symmetry should also be present at $n_2=0$. We can make an ansatz
for the generator  (\ref{eq:tXSU5}),
\begin{subequations}
\begin{align}
	 G_{\rm SM}\times\U1_{B-L}
	 &\subset\SU5 \times \U1_X \subset \SU6 \times \U1^5\,, \\
	 t_X&=\sum_{i=1}^6 a_i t_i \,, \hspace{1cm} t_6 \equiv t_6^0\,, \\
	 t_{B-L}&=t_X+\frac 45 \, t_Y\,,
\end{align}
\end{subequations}
with coefficients $a_1,\dots,a_6$.
We have already identified two localized families in (\ref{eq:locfams}), and two $\T$-plets of 
two bulk families in (\ref{eq:ten34}). This fixes four of the coefficients, 
$a_1=a_2=2a_4, a_3=-1/3, a_6=1/15$, and leads to
\begin{align}
	Q_X(\F)&=Q_X(\Fb_{0,0}^c)=Q_X(\Fb_{0,1}^c)=-\frac 25\,,\\
	 Q_X(\Fb)&=Q_X(\Fb_{0,\frac12})=Q_X(\F_{0,0}^c)=Q_X(\F_{0,1}^c)=\frac 25\,.
\end{align}
From the list of candidates for $d^c$-quarks in (\ref{eq:5s1}), only $\Fb_{2,0}$
and $\Fb_{2,1}$ can be consistent with such a symmetry. 
These fields have identical charges, and the requirement that both
the charge under $\U1_X$ and $\U1_{B-L}$ should take the physical values fixes
the remaining two coefficients, $a_1=1, a_5=1/6$.
We have thus found a unique $\U1_X$  \index{U1@$\U1$!X@$X$}
 at $n_2=0$, implying a unique $\U1_{B-L}$ 
\index{U1@$\U1$!B-L@$B-L$} in four dimensions:
\begin{align}
\label{eq:tX}
	\U1_X\,&: & t_X&=t_1+t_2-\frac 13 t_3+\frac12 t_4+\frac 16 t_5+\frac{1}{15}t_6^0\,, & &\phantom{AAA}\\
	\label{eq:tBL}
	\U1_{B-L}&: & t_{B-L}&=t_X+\frac 45 t_Y\,.
\end{align} 
The embedding of these generators into $\E8\times\E8$ can be found in Table \ref{tab:t}.

Consistency with $\U1_X$ and $\U1_{B-L}$ reshuffles the list of physical and unphysical \index{Exotics}
\index{Standard model families}
$\F$- and $\Fb$-plets,
\begin{subequations}
\label{eq:5sb}
\begin{align}
\label{eq:5s1b}
	&\text{Quark-like zero modes, $d^c=({\bf \bar 3},\mathbbm 1)$}: &
	&  \Fb_{2, 0} ,\; \Fb_{2, 1}\,,\\
	&\text{Lepton-like zero modes, $(\mathbbm 1, {\bf 2})$}: &
	&  \F^c_{2, 0} ,\;\F^c_{2, 1}\,,\\
	&\text{$H_d$ -like zero modes, $(\mathbbm 1, {\bf 2})$}: &
	& \Fb_{1, 0} ,\;\Fb_{1, 1} ,\; \F^c_{1, \frac12} ,\; \Fb\,,\\
	&\text{$H_u$ -like zero modes, $(\mathbbm 1, {\bf 2})$}: &
	& \F_{1, 0} ,\;\F_{1, 1} ,\;\Fb^c_{1,\frac12} ,\; \F\,,\\
	&\text{Exotic zero modes}: &
	& \F_{0, \frac12} ,\;\Fb^c_{0,0} ,\;\Fb^c_{0,1} ,\; \Fb^c_{2, \frac12},\;
	 \Fb_{0, \frac12} ,\; \F^c_{0, 0} ,\; \F^c_{0, 1}\; \F_{2,\frac12}.
	\label{eq:5s4b}
\end{align}
\end{subequations}
Up to the doubling due to $\gamma=0,1$, we have uniquely identified the missing matter $\Fb$-plets
of the two bulk families. 


\subsection[$F$-terms and $D$-terms]{\boldmath $F$-terms and $D$-terms}
\label{sec:FD}
 
The driving force for non-zero vacuum expectation values of  fields present in the model
are the conditions for unbroken supersymmetry (cf.~\cite{ccx98}). 
As stated before, the false vacuum (\ref{eq:falsevac00}),
  which was obtained  compactifying the heterotic string on an orbifold,
  does not maintain supersymmetry since the corresponding symmetry transformations 
  do not close in the presence of non-vanishing Fayet--Iliopoulos $D$-terms. The latter
  correspond to auxiliary fields of the vector multiplets that are associated with the 
    $\U1$ factors which locally  have non-zero traces. The corresponding generators
    and the traces were calculated in Section \ref{sec:anpolFP}.

The $\mathcal N=2$ vector multiplet has in fact three
auxiliary fields $D_1,D_2,D_3$ which form a triplet
under $\SU2_R$ and must all vanish
in the bulk. However, at
the fixed points half of the supersymmetry is broken and
the local $\mathcal N=1$ vector multiplet
has an effective $D$-term $D\equiv- D_3+F_{56}$,
where $F_{56}$ is the associated field strength in the $z^5, z^6$
direction. Thus the local $D$-term \index{D-terms@$D$-terms} cancelation condition 
at $n_2=0$ (cf.~\cite{ads87,lnz04,gno02}),
\begin{align}
\label{eq:Dterm}
	&D_3^a = F^a_{56}=\frac{g M_P^2}{384 \pi^2}\frac{\tr\, t_a}{|t_a|^2} 
		+ \sum_i q^a_i | s_i |^2 , 
\end{align}
where $q^a_i$ is the $\U1_a$ charge of the singlet $s_i$,
has always a solution, even for non-vanishing right-hand-side.
This means that in principle localized
FI-terms
do not necessarily induce singlet vevs and
the corresponding $\U1$ can remain unbroken. However,
since our model has distinct anomalous $\U1$ factors
at the inequivalent fixed points $n_2=0,1$ and a
non-vanishing net anomalous $\U1$ in four dimensions,
its global $D$-flat solution 
cannot be of that kind. We rather expect a mixture of
singlet vevs and a nontrivial gauge background $\langle F_{56}^{\rm an} \rangle$.

Note that the Fayet--Iliopoulos term \index{Fayet--Iliopoulos term} 
in (\ref{eq:Dterm}) is of GUT scale size,
\begin{align}
\label{eq:FIterm}
	\xi \sim \frac{M_P}{\sqrt{384 \pi^2}} \sim M_{\rm GUT}\,,
\end{align}
and appears as an explicit scale in the effective model.
In fact, the abovementioned expectation value of a bulk hypermultiplet
can in principle be used to transfer this scale to the effective potential of the volume modulus.
After supersymmetry breaking, this contribution in combination with 
the Casimir energy can indeed stabilize the size of the effective
orbifold at the GUT scale \cite{bcs08,bms09}.

Further conditions arise from the $F$-terms\index{F-terms@$F$-terms}. Using the formulation 
 in terms of $\mathcal N=1$
superfields, which was introduced in Section \ref{sec:locmodel}, the 
$F$-term conditions are\footnote{These are $F$-terms with respect to global supersymmetry 
transformations. All vacua that we will study in the following have the property $\langle W \rangle=0$,
such that they coincide with the $F$-terms of a supergravity description.}
\begin{align}
\label{eq:Fterm}
	F_\phi&=\frac{\partial W}{\partial \phi} = 0\,, \hspace{1cm}\text{for all fields $\phi$}. 
\end{align}
In the following we shall assume that the equations (\ref{eq:Dterm}) and (\ref{eq:Fterm})
can be solved by giving expectation values to a selection of singlets in the model.
The collection of the latter is denoted by the symbol $\mathcal S$ and called a vacuum of the model.
However, the proof that this can be done for the suggested sets will not be given in this work.
Due to the large number of contributing fields this is difficult to study, especially in the six-dimensional setup. One may
hope that phenomenological requirements are a good guide towards solutions, since
the physical vacuum, if existent among all possibilities within the model,
has to be  supersymmetric\footnote{Note that the
scale of the Fayet--Iliopoulos term in (\ref{eq:Dterm}) is too large for a phenomenologically interesting
supersymmetry breaking scenario via the above $D$-terms.}.
We will only briefly return to the explicit calculation of  the $F$-terms at the end of this chapter.

\subsection{Decoupling of exotics}
\label{sec:decoupling}

The most severe phenomenological requirement at this point is the decoupling \index{Decoupling} of the exotic fields
at the GUT fixed points $n_2=0$. Namely, these are 16 $\F$- or $\Fb$-plets of $\SU5$ from
the list (\ref{eq:5s}) which have unwanted zero modes,
additionally to  the required third matter generation.

The exotics \index{Exotics} in the model are vector-like. One thus has the possibility to decouple pairs of $\F$'s and $\Fb$'s
by generating supersymmetric mass terms for these fields,
\begin{align}
	W&=m\,\F\Fb=\langle s_1\cdots s_M\rangle \F \Fb\,, & s_i\,:&\;\;\text{Singlets}.
\end{align}
This requires that the fields $s_i$ are singlets of the gauge group that should remain unbroken in this
process, which contains the local $\SU5$ in the case of interest. Furthermore, one assumes
that the scalar potential has a supersymmetric minimum which corresponds to non-vanishing vacuum
expectation values (vevs) for the $s_i$, as discussed in the previous section. 

The analysis of the local decoupling at the $\SU5$ fixed points is much easier than its analogue
in the effective four-dimensional field theory. The reason is that many exotics can be decoupled
in one step, by the use of only three singlets. From Table \ref{tab:n20T2T4} we read off that
for each left-handed chiral multiplet from the bulk, there is also at least one right-handed
chiral multiplet with exactly the same charges. They only differ by the superposition
quantum number $\gamma$. One can thus build the complete gauge invariants
\begin{align}
	&\F_{n_3} \F_{n_3}^c\,, \hspace{.5cm}\Fb_{n_3} \Fb_{n_3}^c\,, &
	&\Rightarrow & {\bf Q}&=0\,, & {\bf R}&=\left(-1,-1,0,0\right)\,,
\end{align}
where we dropped the sub-label $\gamma$ and gave the $R$-charges of the
bosonic components. Allowed superpotential terms have to fulfill the rule (\ref{eq:ruleR}),
$\sum {\bf R}=(-1,-1,-1,0)$ for bosonic components, up to the order of the twist in the three planes. 
One can thus generate a universal mass term \index{Universal mass term} for all above pairs by one monomial
of singlets, of order three:
\begin{align}
	\label{eq:Ms}
	M_*&=\langle \bar X_0^c S_2 S_5 \rangle & &\Rightarrow & {\bf Q}&=0\,, &
	{\bf R}&=(0,0,-1,0)\,.
\end{align}
This combination is allowed by all selection rules (\ref{eq:srules}).
Note that this is not the only possible choice, the complete list at order three is given in
Table \ref{tab:Msmon}.
\begin{table}
  \centering 
  \begin{tabular}{|lll|}
	\hline
	\jvsb $S_3 S_7 X_2^c$, & $S_3 S_6 Y_0^{*c}$, & $S_2 S_3 Y_0^c$, \\
	\jvsb $S_1 S_3 X_0^c$, & $S_4 S_5 Y_0^{*c}$, & $S_1 S_5 \bar Y_0^c$,\\
	\jvsb $S_2 S_5 \bar X_0^c$. & &\\
	\hline
\end{tabular}
  \caption{Monomials to order three with the properties ${\bf Q}=0$, ${\bf R}=(0,0,-1,0)$,
  up to the order of the twist in the three compact planes.}
  \label{tab:Msmon}
\end{table}
For each choice, a specific combination of the six local $\U1$ factors is 
broken\footnote{In \cite{bls07} we chose the combination $S_1S_5\bar Y_0^c$.
As it turned out, this does not allow the definition of a $\U1_{B-L}$ which
is spontaneously broken to matter parity.}.

The mass (\ref{eq:Ms}) leads to a universal decoupling of six pairs of vector-like exotics:
 \begin{align}
 \label{eq:dec1}
	W \supset M_* \left( \F_0 \F_0^c+\Fb_0 \Fb_0^c+ \F_1 \F_1^c+\Fb_1 \Fb_1^c+ \F_2 \F_2^c+\Fb_2 \Fb_2^c\right)\,.
\end{align}
Here we suppressed the $\gamma$ values of the decoupled fields, and from now on we shall also drop them
for the remaining combinations. In Section~\ref{sec:symmsfalse}
it will be shown that this is justified, since
 they do not influence any of the relevant couplings in $W$.
We are now in the position to summarize the four quark-lepton families
\index{Standard model families} which are present in the model:
\begin{subequations}
\label{eq:4families}
\begin{eqnarray}
	\text{Family $(1)$:} \hspace{1cm}  &\T_{(1)},\,\Fb_{(1)}\,, &\hspace{1cm}(n_2,n_2')=(0,0),\\
	\text{Family $(2)$:} \hspace{1cm}  &\T_{(2)},\,\Fb_{(2)}\,, &\hspace{1cm}(n_2,n_2')=(0,1),\\
	\text{Family $(3)$:} \hspace{1cm}  &\T_{(3)}=\T,\;\Fb_{(3)}=\F_2^c\,, &\hspace{1cm}\text{(bulk)},\\\
	\text{Family $(4)$:} \hspace{1cm}  &\T_{(4)}=\Tb^c,\,\Fb_{(4)}=\Fb_2\,, &\hspace{1cm}\text{(bulk)}.
\end{eqnarray}
\end{subequations}
The right-handed neutrinos $N^c_{(i)}$ are omitted in the discussion, due to the large number of singlets
in the model. Recall from (\ref{eq:5sb}) or from Table \ref{tab:4d} in Appendix \ref{app:4d}
that the two bulk families give rise to only one standard model family in four dimensions.
For the Higgs multiplets $H_u, H_d$ \index{Higgs candidates} an ambiguity remains,
\begin{align}
\label{eq:HuHdcand}
\underbrace{\F,\; \F_1}_{\text{$H_u$ -like}}\,,
    	\hspace{1cm}\underbrace{\Fb,\; \Fb_1}_{\text{$H_d$ -like}}\,.
\end{align}
Furthermore, one is left with one single pair of vector-like exotics 
\index{Decoupling}\index{Exotics} $\F_0^c, \Fb_0^c$,
which can again be decoupled with a set of three singlets,
\begin{align}
\label{eq:dec2}
	W&\supset M_*' \, \Fb_0^c \F_0^c\,, & M_*'&=\langle X_0 S_2 S_5 \rangle\,.
\end{align}

In summary, the local $\SU5\times \U1_X$ 
GUT structure at the fixed points $n_2=0$ leads to a significant simplification of the discussion.
In (\ref{eq:tX}) and (\ref{eq:tBL}) we identified uniquely the generators of $\U1_X$ and $\U1_{B-L}$.
After a first step of universal decoupling (\ref{eq:dec1}) using only three singlet fields, these symmetries
 fixed the definition of four 
standard model families in the $\SU5$ language, as given in (\ref{eq:4families}). 
One remaining pair of exotics could again be decoupled by a singlet monomial of order three,
extending the list of singlets needed for the decoupling to
\begin{align}
	X_0,\; \bar X_0^c,\;S_2,\;S_5\,.
\end{align}
So far unsolved is the question which combination of fields from (\ref{eq:HuHdcand})
plays the role of the two Higgs multiplets $H_u,H_d$. The  remaining multiplets will then 
constitute another pair of exotics, and consequently require the generation of another 
  mass term.

\subsubsection{A minimal vacuum}
\label{sec:S0}

The transition from the false vacuum, $\langle s_i\rangle=0$ for all $s_i$, to the physical vacuum $\mathcal S$,
\index{Physical vacuum}
  \begin{align}
	\mathcal S &= \left\{ \left. s \right| \langle s \rangle \neq 0 \right\}\,,
\end{align}
is required for three reasons: Supersymmetry has to be maintained in the perturbative vacuum solution,
all exotics have to be decoupled, and the additional gauge symmetry  has to be broken, under which the
standard model fields transform in the effective theory. 

The latter point states that a Higgs mechanism is needed in order to make the gauge fields heavy, which are
associated with the nine extra $\U1$ factors
in four dimensions, including the six at $n_2=0$.
This implies the spontaneous breaking of $\U1_X$ and $\U1_{B-L}$. 
However, the dimension-four proton decay operators
only remain absent if a global matter parity survives the breaking. 
In the present model, this can be  achieved by
assuming vevs of singlets with charges $\pm2$ under $\U1_X$. 
This leads to the breaking of the latter factor to a parity (or a group of even order), 
denoted as $P_X$. From Table \ref{tab:chs} in Appendix \ref{app:discrete} 
one finds that there are only two such singlet fields
 in the model, $U_2$ and $U_4$ from the untwisted sector,
 \begin{align}
 \label{eq:U2U4}
	&\langle U_2 \rangle\neq 0 \; \text{or}\; \langle U_4 \rangle \neq 0\,: \hspace{1cm}
	\U1_X \rightarrow \Z M^X \supset P_X\,,
\end{align}
for some even integer $M$.
Note that these fields do not have zero modes in four dimensions, but are present as bulk fields
of the six-dimensional theory, which have positive parity at the GUT fixed points $n_2=0$.
Their corresponding mass term is then expected to be of the size of the compactification scale.
However, if this is of the order of the GUT scale it coincides with the size of the Fayet--Iliopoulos scale,
and vevs of the fields can contribute to vacua.

With the above requirement, we can formulate a minimal vacuum
\index{Minimal vacuum} $\mathcal S_0$ which 
leads to the desired decoupling of exotics (\ref{eq:dec1}), (\ref{eq:dec2}), and breaks $\U1_X$ to the
parity $P_X$, which yields matter parity in four dimensions. We shall thus demand that the vacuum
\begin{align}
\label{eq:vacS0}
    \mathcal S_0=\left\{ X_0, \bar X_0^c, U_2, U_4, S_2, S_5 \right\}
\end{align}
is contained in all sets of singlets which are studied as candidates for physical vacua in the 
following\footnote{It is not minimal in the strict sense: Only the presence of one of
the fields $U_2, U_4$ is required to fulfill the constraints. However, we will not
study complete scans over all possibilities in the following, but only
focus on some interesting examples. }.

\subsection{Unbroken symmetries}
\label{sec:symms}

The phenomenology of a string model is determined by the spectrum and the interactions of the
low-energy effective theory. 
The knowledge of the unbroken symmetries in a given vacuum $\mathcal S$ is essential for an understanding
of the latter. In the present setup, couplings among fields of
 the effective theory \index{Couplings} arise from higher-dimensional operators, after the
transition to the physical vacuum,
\begin{align}
	W&=\langle s_1 \cdots s_N \rangle \phi_1 \cdots \phi_M \,, &
	s_i&\in\mathcal S\,, & \phi_i &\notin \mathcal S\,.
\end{align}
  Here the fields $\phi_i$ can play the role of matter multiplets, for which the effective
  Lagrangian is desired. Furthermore, the mass terms (\ref{eq:dec1}), (\ref{eq:dec2}) are
  examples of couplings which have to be generated for a viable phenomenology.
  
  In the following we describe how the unbroken symmetries in a physical vacuum can be determined,
  and study the minimal vacuum $\mathcal S_0$ from (\ref{eq:vacS0}) as an example.
  The methods introduced here enable us to identify it as a vacuum which does not
  suffer from dimension-four proton decay operators\footnote{The absence of dimension-five proton
  decay operators in the vacuum $\mathcal S_0$  is shown in Section~\ref{sec:kernel}.}, but 
  contains two light Higgs pairs in the spectrum. 

\subsubsection{Symmetries of the false vacuum}
\label{sec:symmsfalse}

As discussed in Sections \ref{sec:selectionrules} and \ref{sec:sgrulesZ6II},
the terms which appear in the superpotential are subject to a set of constraints, the string selection rules.
They arise from string theory calculations, but can be interpreted in terms of symmetries of the effective 
field theory. 

In the present model, the rules give rise to an effective symmetry
\begin{align}
\label{eq:Gfalse}
	G&=G_{\rm gauge} \times G_{\rm discrete}\,, 
\end{align}
where, at $n_2=0$, the gauge part is
\begin{align}
	 G_{\rm gauge} &=\SU5\times\U1^4\times\left[\SU3\times\SO8\times\U1^2\right]\,,
\end{align}
and the discrete symmetry can be identified as (cf.~(\ref{eq:ruleR}), (\ref{eq:rulek}), (\ref{eq:srules}))
\begin{align}
	\label{eq:Gdisc}
	G_{\rm discrete}& = \left[ \tZ6^{R^1} \times  \tZ3^{R^2} \times  \tZ2^{R^3} \times 
	\Z6^k \right]_R \times \Z3^{n_3k} \times \Z2^{n_2k} \times \Z2^{n_2'k}\,.
\end{align}
Here we use the notation that the group $\Z M^q$ acts on a state with charge $q$ 
as
\begin{align}
	&\Z M^q\,:& \ket{q} &\mapsto g_M^a(q) \ket{q}\,, & g_M^a(q)&=e^{2 \pi i \frac{aq}{M}}\,,&a&=0,\dots,(M-1)\,.
\end{align} 
The tilde in $ \tZ M^q$ indicates that the $M$-fold application of the generating element
does not reproduce the identity, but some other transformation which is generated by the remaining group
factors. For example, the symmetries $ \tZ{l_i}^{R^i}, i=1,2,3$, with elements $\tilde g_{l_i}^a(R^i)$,
 give for the choice $a=\jmod{0}{l_i}$
 an element of  $\Z 6^k$. This follows from 
the observation 
\begin{align}
R^i=\jmod{-\frac{k}{l_i}}{1}\,,
\end{align}
and allows us to identify
\begin{align}
\label{eq:identifytZ}
	 \tilde g_{l_i}^{l_i}(R^i)&\equiv g_6^{-6/l_i}(k)=g_{l_i}^{-1}(k)\, 
	&l_i&={6,3,2}\,, & i&=1,2,3\,.
\end{align}
This is a consequence of   the statement that the selection rule $\sum k=\jmod 06$ can
be understood as part of the discrete $R$-symmetries, cf.~the discussion before  (\ref{eq:rulek}).

The last three factors of (\ref{eq:Gdisc}) resemble the order of the sub-lattices 
 in the planes $\SU3$ and
$\SO4=\SU2\times\SU2$, respectively. A peculiarity arises from the $\G2$-plane.  
There, the twists $k=2$ and $k=3$ correspond to two different sub-lattices $\Lambda_k$,
whose sum is equivalent to the sub-lattice of full $\Z6$ rotations, $\Lambda_1$.
This leads to an additional symmetry factor, which only applies to couplings with
the property that all its contributing fields come from equivalent twisted sectors,
\begin{align}
\label{eq:GG2}
	G_{\rm discrete}^{G_2} &= \left\{ \begin{array}{ll}
		\Z6^{n_6k}, & \text{if all fields have the same $k$, modulo 6 and up to a sign},\\
		\mathbbm 1, & \text{otherwise.}
	\end{array}\right.
\end{align}
The additional restriction does not apply to all couplings in the superpotential. Therefore, it cannot
be interpreted as an additional symmetry of the full system. However, physical interactions
have to fulfill this constraint, and thus we are forced to write the total superpotential $W_{\rm tot}$ as
the sum of two terms,
\begin{align}
\label{eq:splitW}
    	W_{\rm tot}&= W_0+W\,, & &\begin{array}{ll}
	W_0: &    \text{Symmetry $G_{\rm gauge} \times G_{\rm discrete}\times\Z6^{n_6k}$},\\
	W\;: &    \text{Symmetry $G_{\rm gauge} \times G_{\rm discrete}$}.
	\end{array}
\end{align}
All couplings which only involve fields with equivalent 
twists, in the sense of (\ref{eq:GG2}), are collected in $W_0$.
The rest, which is the majority of the couplings of interest, is contained in $W$.
Note that the form (\ref{eq:splitW}) is the outcome of a string calculation for the
effective low-energy field theory. The split structure is a direct consequence
of the fact that the $\Z{\rm 6-II}$ twist has two sub-twists, which both affect
the torus $G_2$, and  therefore seems to be unavoidable.

However, in the following analysis of the symmetries
we shall ignore the contributions from $W_0$. This is justified, since
order-by-order scans show that the latter play practically no role. For example,
direct Yukawa-couplings among fields of one generation only arise through 
couplings which involve singlets from other twisted sectors.
We will comment more on the terms in $W_0$  in Section \ref{sec:allorders},
when we discuss the superpotential to all orders.

\subsubsection{Symmetries of physical vacua}

We now study the question which of the symmetries (\ref{eq:Gfalse}) remains
unbroken after the transition form the false vacuum to the physical vacuum, \index{Unbroken symmetries}
\index{Physical vacuum}
\begin{align}
	G_{\rm tot} &\rightarrow G_{\rm vac}(\mathcal S)\,, &
	 \mathcal S &= \left\{ \left. s \right| \langle s \rangle \neq 0 \right\}\,.
\end{align}
For simplicity, we shall assume that the non-Abelian gauge symmetry in $G_{\rm gauge}$ 
is not broken during this process, and focus on the Abelian factors and the discrete
symmetries (\ref{eq:Gdisc}) in the following.

Consider the transformation of a singlet $s_i\in\mathcal S$ under the symmetries of the false
vacuum. We can collect the relevant charges of the continuous and the discrete symmetries
in two vectors ${\bf Q}$ and $\mathcal K$, respectively,
\index{Continous symmetry!charge vector Q@charge vector ${\bf Q}$}
\index{Discrete symmetry!charge vector K@charge vector $\mathcal K$}
\begin{align}
\label{eq:QK}
   	{\bf Q}&=\left( Q_1,\dots,Q_6\right)\,, & \mathcal K&=\left( R^1,R^2,R^3,k,n_3 k,n_2 k,n_2' k \right)\,, 
\end{align}
where $Q_u$ denotes the charge with respect to the $\U1$ which is generated by $t_u$.
The transformation of $s_i$ can then be written as
\begin{align}
	s_i &\rightarrow e^{2 \pi i \left( \alpha \cdot {\bf Q}+{\bf r}\cdot  \mathcal K \right)} s_i \,,
\end{align}
with two parameter vectors 
\begin{align}
\label{eq:alphar}
    	\alpha&=\left(\alpha_1,\dots,\alpha_6\right) \,, & \alpha_u &\in \mathbbm R\,, &  
	{\bf r}&=\left(\frac{r_1}{6},\frac{r_2}{3},\frac{r_3}{2},\frac{r_4}{6},\frac{r_5}{3},\frac{r_6}{2},\frac{r_7}{2}\right)\,, &
	r_i &\in \Z{ }\,.
\end{align}
Note that this way of writing the transformation ignores the identifications (\ref{eq:identifytZ})
and thus double-counts the common factors of the $R$-symmetries and $\Z 6^k$ for $r_i\geqslant l_i$,
$i=1,2,3$.

The symmetry group of the vacuum $\mathcal S$ is then given by the set of transformations which
act as the identity on each of the individual vevs $\langle s_i \rangle$,
\begin{align}
\label{eq:Gvacdef}
	&G_{\rm vac}(\mathcal S)\,: & \langle s_i\rangle &= 
	e^{2 \pi i \left( \alpha' \cdot {\bf Q}+{\bf r}' \cdot  \mathcal K \right)} \langle s_i \rangle\,, &
	&\text{for all $s_i\in \mathcal S$}. 
\end{align}
Here we characterized the unbroken subgroup of the symmetries of the false vacuum by $\alpha',{\bf r}'$.

\subsubsection*{Example: The minimal vacuum $\mathcal S_0$}

As an example, consider the vacuum $\mathcal S_0$, defined in (\ref{eq:vacS0}). The physical vacuum
is an extension of this set and its symmetry will be given by a subgroup of $G_{\rm vac}(\mathcal S_0)$.
\begin{table}
  \centering 
  \footnotesize
  \begin{tabular}{c| c| c}
	 \jvsb Symmetry & $\alpha'$ & ${\bf r}'$ \\
	\hline
	\hline
	\jvsb $[\tZ{24}]_R$ & $\left(-\frac{1}{6},0,-\frac{1}{12},-\frac{1}{24},0,\frac{1}{30}\right)$&
	$\left(\frac{1}{6},0,0,0,0,0,0\right)$\\
	\jvsb $[\Z{6}]_R$ & $\left(\frac{5}{6},-\frac{5}{6},0,0,0,\frac{1}{6}\right)$&
	$\left(0,\frac{5}{3},0,0,0,0,0\right)$\\
	\hline
	\jvsb $\Z{120}$ &$\left(\frac{1}{2},\frac{1}{2},-\frac{1}{6},\frac{1}{4},0,\frac{1}{30}\right)$&
	$\left(0,0,0,0,0,0,0\right)$\\
\end{tabular}
  \caption{The generators of the discrete symmetries in the vacuum $\mathcal S_0$.
  The notation $\tZ M$ stands for a discrete rotation of order $M$, up to elements of the other symmetry factors.
 This could alternatively be expressed as $\Z A/\Z B$, where $A/B=M$ and $\Z B$ is a subgroup of the remaining
 discrete symmetry.
 }\label{tab:vacS0}
\end{table}

The generating elements of the unbroken symmetry directly follow from (\ref{eq:Gvacdef}). From Tables
\ref{tab:n2T}, \ref{tab:n20U} and \ref{tab:n20T2T4} one reads off that none of the singlets in $\mathcal S_0$
is charged under the $\U1_5$ with generator $t_5$, thus this will remain a continuous symmetry of the new vacuum.
However, all other Abelian factors are broken to discrete subgroups. For the total symmetry, acting on all
fields present at $(n_2,n_2')=(0,0)$, we find
\begin{align}
\label{eq:GvacS0}
    	G_{\rm vac}(\mathcal S_0) &=
	\left[ \tZ{24} \times \Z 6 \right]_R \times
	 \U1_5 \times \Z{120} \,.
\end{align}
The enlargement of the orders of the appearing $R$-symmetries is due to the mixing of the latter with
generators of the six $\U1$ factors. The list of generators of $G_{\rm vac}(\mathcal S_0)$ is given
in Table \ref{tab:vacS0}. It shows the complicated embedding of the present discrete symmetries into
the symmetries of the false vacuum. The tables in Appendix \ref{app:discrete} 
explicitly confirm that the given symmetry is unbroken in the vacuum, furthermore they list
the corresponding transformation phases for all fields present at $(n_2,n_2')=(0,0)$.

Note the appearance of the large factor $\Z{120}$. Up to the entry  corresponding
to the unbroken $\U1_5$, its generator  is proportional to the generator $t_X$
of $\U1_X$, 
defined in (\ref{eq:tX}). We will see that this factor is indeed responsible for
the absence of dimension-four proton decay operators,
as argued in (\ref{eq:U2U4}).
In fact, allowed couplings should be invariant under all possible symmetry transformations,
including combinations of the discrete generators and arbitrary choices for
the continuous parameter $\alpha_5$, for example $\alpha_5=1/12$.
Together with the generator of the $\Z{120}$, the latter choice adds to $\frac 12 t_X$, associated with
 a $\Z{60}$.
The naive guess for the matter parity generator would then be $15 t_X$. However,
as it turns out this $\Z2$ acts trivially on all singlets of the hidden $\SO8$
gauge group, including
the standard model families. 
The symmetry which corresponds to matter parity for the singlets of the hidden sector
is in fact a $\Z4$ symmetry with respect to the full spectrum,
\begin{align}
\label{eq:PX}
    \text{Matter parity}:\hspace{1cm}P_X=e^{2 \pi i \left(\frac{15}{2} t_X\right)}\,.  
\end{align}
This definition yields the correct parity assignments for the matter generations (\ref{eq:4families})
and the Higgs candidates (\ref{eq:HuHdcand}),
as shown in Table \ref{tab:chns} in Appendix \ref{app:discrete}. 
Together with the other unbroken symmetries in (\ref{eq:GvacS0}),
it determines whether a coupling may be allowed in the effective theory in
the vacuum $\mathcal S_0$, or not. 
As will become  apparent,  large unbroken symmetries can be very restrictive and forbid
phenomenologically favored interactions. 

\subsubsection{Symmetries and interactions}

The origin of the symmetries that we study in this chapter are the string selection rules (\ref{eq:ruleR}), (\ref{eq:srules}) for couplings in the superpotential $W$.
 Expressed in terms of the $R$-charges of the bosonic components of chiral multiplets, 
a coupling \index{Couplings} which consists of two fermions and bosons otherwise, has to fulfill
\begin{align}
R^i(W)&=\jmod{-1}{l_i}\,, & l_i&={6,3,2}\,, & i&=1,2,3\,.
\end{align}
In the notation of (\ref{eq:QK}), the full set of constraints can be expressed as
\begin{align}
\label{eq:Wvalid}
e^{2 \pi i \left( \alpha' \cdot {\bf Q}(W)+{\bf r}' \cdot \mathcal K(W)
\right)}&=e^{2 \pi i \,{\bf r}' \cdot \mathcal K_{\rm vac}}\,, 
\end{align}
\begin{align}
\label{eq:Kvac}
\mathcal K_{\rm vac}&=\left( \jmod{-1}{6},\jmod{-1}{3},\jmod{-1}{2},\jmod 06,\jmod 03,\jmod 02,\jmod 02 \right)\,,
\end{align}
for all possible choices of  the parameters $\alpha'$ and ${\bf r}'$ which describe the symmetry of the vacuum. 
Once the explicit form of these parameter sets is known, the above formulation of the selection
rules allows for a fast and easy check if a given coupling is present in the superpotential,
or forbidden by an exact symmetry of the vacuum. In the latter case it will be absent to all
orders in the involved singlet fields. 

\subsubsection*{Example: The minimal vacuum $\mathcal S_0$}

We illustrate this with the simple six-field vacuum $\mathcal S_0$ from (\ref{eq:vacS0}), whose symmetries were
discussed in the previous section. Consider the $\mu$-terms \index{mu-term@$\mu$-term} of the Higgs candidates
\index{Higgs candidates} from (\ref{eq:HuHdcand}),
which now is a $2\times 2$ matrix,
\begin{align}
	W  &= \mu_{ij} H_u^i H_d^j \,, & (H_u^i)&=(\F, \F_1)\,,& (H_d^i)&=(\Fb, \Fb_1)\,,
\end{align}
where each entry of $\mu_{ij}$ consists of expectation values of monomials of singlets,
\begin{align}
	\mu_{ij}&=\langle s_{ij}^{(1)} \cdots s_{ij}^{(N_{ij})} \rangle \,, & s_{ij}^{(k)} &\in \mathcal S_0\,, & i,j&=1,2\,.
\end{align}
The generators of the unbroken symmetries $\alpha', {\bf r}'$ can be found in Table \ref{tab:vacS0}.
Be definition of the latter, $\mu_{ij}$ is invariant under all symmetry transformations.
The condition (\ref{eq:Wvalid}) thus only depends on the transformation of the Higgs candidates
$H_u^i, H_d^i$. For the generator of the $R$-symmetry $[\tZ{24}]_R$,
\begin{align}
	\alpha'&=\left(-\frac{1}{6},0,-\frac{1}{12},-\frac{1}{24},0,\frac{1}{30}\right)\,,&
	{\bf r}'&=\left(\frac{1}{6},0,0,0,0,0,0\right)\,,
\end{align}
the condition (\ref{eq:Wvalid}) reads
\begin{align}
	 \underbrace{
	 \alpha' \cdot {\bf Q}\left(H_u^iH_d^j\right)+\frac 16 \cdot  R^1\left(H_u^iH_d^j\right)
	 }_{{\displaystyle \frac{1}{24}}\left(\begin{array}{ll}
	 0 & 17 \\ 5& 11
	 \end{array}\right)}
	&=\jmod{-\frac 16}{1}\,.
\end{align}
From this we conclude that all $\mu$-terms are forbidden by a discrete $R$-symmetry,
\index{R-symmetry@$R$-symmetry}
\begin{align}
\label{eq:nomuS0}
	\big[ \tZ{24}\big]_R\,: \hspace{1cm}\mu_{ij} &= 0\,.
\end{align}
This states that all four Higgs candidate fields are massless in the vacuum $\mathcal S_0$.
This is a two-Higgs model, which may be phenomenologically acceptable, but is inconsistent with
gauge coupling unification.

Similarly, one can study the dimension-four and dimension-five proton decay operators
\index{Proton decay!dimension-four}\index{Proton decay!dimension-five}
\begin{align}
\label{eq:proton45b}
	W&=C^{(R)}_{ijk} \Fb_{(i)} \T_{(j)} \Fb_{(k)}+C^{(B)}_{ijkl}  \T_{(i)} \T_{(j)} \T_{(k)} \Fb_{(l)}\,,
\end{align}
where $(i)$ labels the four families, defined in (\ref{eq:4families}).
As expected, the generator of the $\Z{120}$, which is related to $\U1_X$, \index{U1@$\U1$!X@$X$}
 forbids the presence of the dimension-four term,
\begin{align}
	\Z{120}\,: \hspace{1cm} C^{(R)}_{ijk} =0\,.
\end{align}
The dimension-five operators are also forbidden in the present vacuum,
	 $C^{(B)}_{ijkl} =0$,
but this does not fully follow from the unbroken symmetries. This is discussed in Appendix~\ref{app:discrete}
and in the following section.

Furthermore, we can study Yukawa interactions\index{Yukawa couplings!six dimensions},
 which now depend on the choice of Higgs fields
$H_u,H_d$,
\begin{align}
	W&=C^{(u)}_{ij} \T_{(i)} \T_{(j)} H_u+C^{(d)}_{ij}  \Fb_{(i)} \T_{(j)} H_d\,.
\end{align}
The appearing matrices are related to the Yukawa couplings \index{Yukawa couplings!four dimensions}
of the three standard
model families of the effective four-dimensional theory,
 \begin{align}
  W_\mathrm{Yuk} = Y_{ij}^{(u)} u_i^c q_j H_u + Y_{ij}^{(d)} d_i^c q_j H_d +  
  Y_{ij}^{(l)} l_i e_j^c H_d \,,
 \end{align}
 by
 \begin{subequations}
 \label{eq:Yuk4to3}
 \begin{align}
	Y^{(u)} &= \left( \begin{array}{c c c}
	C^{(u)}_{11} & C^{(u)}_{12} & C^{(u)}_{14} \\
	C^{(u)}_{21} & C^{(u)}_{22} & C^{(u)}_{24} \\
	C^{(u)}_{31} & C^{(u)}_{32} & C^{(u)}_{34} \\
	\end{array} \right) \,, \\
	Y^{(d)} &= \left( \begin{array}{c c c}
	C^{(d)}_{11} & C^{(d)}_{12} & C^{(d)}_{14} \\
	C^{(d)}_{21} & C^{(d)}_{22} & C^{(d)}_{24} \\
	C^{(d)}_{41} & C^{(d)}_{42} & C^{(d)}_{44} \\
	\end{array} \right)  \,, &
	Y^{(l)} &= \left( \begin{array}{c c c}
	C^{(d)}_{11} & C^{(d)}_{12} & C^{(d)}_{13} \\
	C^{(d)}_{21} & C^{(d)}_{22} & C^{(d)}_{23} \\
	C^{(d)}_{31} & C^{(d)}_{32} & C^{(d)}_{33} \\
	\end{array} \right) \,.	
\end{align}
\end{subequations}
In the vacuum $\mathcal S_0$, the couplings are forbidden for the choice $H_u=\F_1$ and $H_d=\Fb$,
\begin{align}
	&\U1_5\,:  \hspace{.5cm}C^{(u)}_{H_u=\F_1}=0\,, &
	&[\tZ{24}]_R\,:  \hspace{.5cm}C^{(d)}_{H_d=\Fb}=0\,.
\end{align}
On the other hand, for $H_u=\F$ and $H_d=\Fb_1$ they are (partly) allowed,
\begin{align}
\label{eq:YukuS0}
	C^{(u)}_{H_u=\F}&=\left(\begin{array}{cccc}
		S_2 S_5 (\bar X_0^c)^2 & S_2' S_5 (\bar X_0^c)^2 &(S_2)^2S_5(\bar X_0^c)^2 & S_2 (S_5)^2 (\bar X_0^c)^2\\
		S_2' S_5 (\bar X_0^c)^2 & S_2' S_5' (\bar X_0^c)^2 &(S_2)^2S_5'(\bar X_0^c)^2 & S_2' (S_5)^2 (\bar X_0^c)^2\\
		(S_2)^2S_5(\bar X_0^c)^2 & (S_2)^2S_5'(\bar X_0^c)^2 & (S_2)^3 S_5 (\bar X_0^c)^2 &g\\
		S_2 (S_5)^2 (\bar X_0^c)^2 & S_2' (S_5)^2 (\bar X_0^c)^2 & g & S_2 (S_5)^3 (\bar X_0^c)^2
	\end{array}\right)\,,
\end{align}
\begin{align}
\label{eq:YukdS0}
	C^{(d)}_{H_d=\Fb_1}&=\left(\begin{array}{cccc}
	0 & 0 &0&0\\
	0&0&0&0\\
	0&0&0&0\\
	S_5 & S_5' & S_2 S_5 & (S_5)^2
	\end{array}\right)\,,
\end{align}
where primes denote localization at $n_2'=1$ and the matrices contain examples for lowest order monomials.
Note that all zeros are present up to infinite order in the singlets, since they are protected by symmetries,
cf.~Appendix \ref{app:discrete}.
The up-type Yukawa coupling between the two bulk families is proportional to the gauge coupling. This
is an important consequence of the choice $H_u=\F$,  due to the origin of that multiplet in the
six-dimensional gauge vector. The unification of gauge interactions and Yukawa couplings
in six dimensions will imply a large mass for the top quark of the four-dimensional effective theory, and is thus
phenomenologically desired. However, the down-type Yukawa couplings, which are
also responsible for the lepton masses in four dimensions, are obviously not satisfactory.
It seems that the generation of the latter with acceptable entries also for the light families,
 and the simultaneous vanishing of the $\mu$-term 
is difficult to realize. This appears to be a generic problem of $\Z{\rm 6-II}$ orbifold
models. If the symmetries which protect the zeros in $Y^{(d)}$ are completely broken, we expect that also the $\mu$-term
will be generated. 

The lessons that we learn from the vacuum $\mathcal S_0$ are twofold. First,
new $R$-symmetries with a complicated embedding can be responsible
for vanishing $\mu$-terms and dimension-five proton decay operators.
Concerning the Yukawa couplings, the Higgs choice $H_u=\F$ (and $H_d=\Fb_1$) seems to be preferred,
where $\F$ is part of the gauge vector multiplet in six dimensions.

Second, more generally, one finds that the restriction to a small
set of singlets implies a large unbroken symmetry, which forbids many of the interactions.
This can have positive effects for the phenomenology, like stability of the proton, but also
negative, like vanishing down-type Yukawa couplings. We expect that
adding more singlets to the list breaks more symmetry and thus allows the generation of
more terms in the superpotential. However, also dangerous or disfavored terms, like the
$\mu$-term, may then be generated at a high scale. The art of orbifold model building is thus to
find the best compromise between desired and problematic terms, or in other words, 
too much or too little unbroken symmetry.
In the following, we will study this issue more and develop an algorithm which
enables one to find the largest vacua for which a choice of properties of the superpotential
remains intact, thereby exploring the phenomenological limits of a given orbifold compactification.

\subsection{Vacuum selection}
\label{sec:kernel}

Pure guesswork is an unsatisfactory approach for the definition of a physical vacuum. At the GUT fixed points
 there are 67 non-Abelian singlets (and 72 in four dimensions), leading 
 to an enormous number
 of different possibilities.
 %
 On the other hand, a dynamical approach starting from the explicit
 scalar potential is excluded for the same reason, there are too many degrees of freedom in order
 to handle the full problem. Furthermore, the numerical coefficients of the various terms
 in the superpotential have to be calculated from string theory for the explicit approach,
 which is a non-trivial task.
 
 One therefore usually assumes that there is a supersymmetric
 vacuum which can describe the MSSM to some accuracy. If that is the case, the list
 of singlets contributing to it should be
 constrained by phenomenological requirements: \index{Phenomenological requirements}
 \begin{itemize}
  \item The low-energy spectrum contains the supersymmetric standard model, and no exotics.
  \item Supersymmetry is unbroken: $D_a=0, F_i=0$ for all $a$ and $i$.
  \item The model has non-vanishing Yukawa interactions.
  \item The proton is stable (or has a sufficiently long lifetime).
  \item The $\mu$ term vanishes (or is strongly suppressed).
  \item The gravitino is massless, $\langle W \rangle=0$.
\end{itemize}
The list could be enlarged by many more details, but, as we will see, the above constraints
are already quite restrictive. Typically, the vacuum expectation values of singlets are
of the order of the localized Fayet--Iliopoulos terms, which means they are comparable
to $M_{\rm GUT}\sim 10^{16} {\rm GeV}$. 
For that reason it is difficult to generate a $\mu$-term of the weak scale by higher dimensional
operators. Also the breaking of supersymmetry is expected to arise from some 
additional mechanism, which we do not study in this work. If the latter is linked to
the electroweak scale, one expects the relation (\ref{eq:muEW}),
\begin{align}
\label{eq:muW}
    	\langle W \rangle \sim \mu \sim M_{\rm EW}\,,
\end{align}
after supersymmetry breaking, and $\langle W \rangle = \mu =0$ before.

An important guideline for the definition of the vacuum is the decoupling of exotics.
In Section \ref{sec:decoupling} we described how the definitions of a unique $\U1_X$
and a unique $\U1_{B-L}$ selected between physical multiplets and exotics, up to an
additional doubling due to the orbifold breaking of the $\mathcal N=2$ supersymmetry in the
six-dimensional bulk. The embedding of the corresponding generators into $\E8\times \E8$
is expected to depend on the choice of the Wilson line in the $\SU3$-plane, which
is also linked to the presence of complete standard model families at the fixed points.
There is also only little choice in the breaking of $\U1_X$ to a matter parity, since
only two singlets in the model have the required even charge under this symmetry.
These arguments lead to the definition of the minimal vacuum $\mathcal S_0$ in
(\ref{eq:vacS0}), which we take as a starting point for further steps.

The ambiguity in the model up to this point consists mainly in the identification of the
Higgs multiplets $H_u$ and $H_d$. Here the crucial issue is the
vanishing $\mu$-term in the effective superpotential. We shall now describe how 
a maximal vacuum can be defined, which contains the maximally possible number
of singlets, such that $\mu=0$ holds up to infinite order.
However, the arguments are general and can be used to forbid also any other
disfavored interaction.

\subsubsection{Maximal vacua for vanishing couplings}
\label{sec:maxvac}

Consider a vacuum $\mathcal S$ and a superpotential term
\begin{align}
\label{eq:W22}
   	W&=\lambda \Phi\,, & \lambda&=s_1^{n_1} \cdots s_N^{n_N}\,, & \Phi&=\phi_1 \cdots \phi_M\,,&
	 \;n_i,N,M &\in \mathbbm N\,,
\end{align}
where $s_i \in \mathcal S$ and $\phi_i \notin \mathcal S$.
This coupling is allowed in the false vacuum by the string selection rules, provided \index{Discrete symmetry}
\index{Continous symmetry}
\begin{align}
\label{eq:condWQ}
    	\text{Continuous symmetries:} \hspace{1cm}
	 {\bf Q}\left( \lambda \Phi \right) &=0\,, \\\label{eq:condWK}
	\text{Discrete symmetries:} \hspace{1cm}
	  \mathcal K\left( \lambda \Phi \right) &=\mathcal K_{\rm vac}\,,
\end{align}
where  the notation from (\ref{eq:QK}) and (\ref{eq:Kvac}) was used.
\index{Continous symmetry!charge vector Q@charge vector ${\bf Q}$}
\index{Discrete symmetry!charge vector K@charge vector $\mathcal K$}
These two conditions can now be solved separately. Equation (\ref{eq:condWQ}) has
multiple solutions, which can be written as
\begin{align}
\label{eq:lambdas}
    	\lambda&=\lambda_0\,\lambda_s\,, \hspace{1cm}
	\left\{\begin{array}{r}
	{\bf Q}\left(\lambda_s \Phi \right)=0\,, \\
	{\bf Q}\left(\lambda_0 \right)=0\,.
	\end{array}\right.
\end{align}
Here $\lambda_s$ corresponds to a particular solution, and the infinite multiplicity
is contained in the homogeneous solutions $\lambda_0$. The latter form the `charge kernel'
\index{Charge kernel}
\begin{align}
\label{eq:kerQ}
    {\rm ker}\,Q\left( \mathcal S \right)&= \left\{ \left. \lambda_0=s_1^{n_1}\cdots s_N^{n_N} \right|
    s_i \in \mathcal S, n_i \in \Z{ }, N \in \mathbbm N, {\bf Q}\left( \lambda_0\right)=0\right\}\,,
\end{align}
which is spanned by a set $\mathcal B$ of basis monomials.
Here by convention $s^{-1}$ denotes a singlet with negative charge vectors ${\bf Q}, \mathcal K$, as
compared to the singlet $s$.
Note that the ${\rm ker}\,Q$ is a module, not a vector space, due
to the requirement of integer exponents $n_i\in \Z{ }$.
In the end, a projection onto physical monomials $\lambda=\lambda_0\lambda_s$ with integer
exponents will be necessary,
\begin{align}
\label{eq:PN}
	\lambda_{\rm phys} = \mathcal P_{\mathbbm N} \lambda
	\equiv\left\{ \begin{array}{ll}
	\lambda, &\text{if all $n_i\geqslant0$}\,,\\
	0,&\text{otherwise}\,.
	\end{array}\right.
\end{align}
 Keeping this in mind, we can also allow negative exponents
for the particular solution $\lambda_s$ in (\ref{eq:lambdas}).
In fact, the solution of systems of linear integer equations with positive integer solutions are
the subject of a branch of applied mathematics, called linear programming.
The details of this are  beyond the scope of this work. In the following, all results will be formulated
with projectors $\mathcal P_{\mathbbm N}$ and elements of 
the above basis $\mathcal B$.

Once the particular solution $\lambda_s$ is fixed, covariance of the superpotential under
the discrete symmetries (\ref{eq:condWK}) selects a subset of the elements in the charge kernel,
\begin{align}
\label{eq:condK2}
	\mathcal K\left( \lambda_0 \right) &=\mathcal K_{\rm vac}- \mathcal K\left( \lambda_s \Phi \right)\,.
\end{align}
This condition is easily evaluated. Since ${\rm ker}\,Q(\mathcal S)$ only depends on the
vacuum $\mathcal S$, and not on the interaction $\Phi$, it has to be calculated only once.
Given its basis $\mathcal B$, one can then consider solutions of the condition (\ref{eq:condK2})
term by term, for each coupling $\Phi$.

In the following, we will turn the argument around and study unwanted couplings, such as the $\mu$-term.
It is then possible to use  (\ref{eq:condK2}) as a tool for the definition of maximal vacua, 
\index{Maximal vacuum} in which
the disfavored interaction is forbidden to infinite order in the singlets. For this one starts with a small
enough set of singlets, which either does not contain a particular solution, or 
 whose charge kernel is too small to fulfill (\ref{eq:condK2}).
  These sets can then be subsequently enlarged by more singlets, until it
is impossible to circumvent the fulfillment of the constraints. This will now be illustrated by two phenomenologically
interesting examples.

\subsubsection{Gauge-Higgs unification}
\label{sec:GHU}

Recall from (\ref{eq:HuHdcand}) that after the first step of decoupling, the local GUT theory
at the fixed points $n_2=0$ contained two pairs of candidate multiplets for the Higgs fields.
We now consider the case of gauge-Higgs unification, \index{Gauge-Higgs unification}
\begin{align}
	H_u&=\F \,, \hspace{.5cm} H_d=\Fb\,, &
	\text{exotics:} \hspace{.5cm}\F_1, \Fb_1\,,
\end{align}
corresponding to a vacuum $\mathcal S_1$
 that extends the minimal vacuum from (\ref{eq:vacS0}), $\mathcal S_1\supset \mathcal S_0$. 
The vacuum $\mathcal S_1$ is now constructed as a maximal vacuum with respect to the
constraints
\begin{itemize}
  \item Matter parity $P_X$ is unbroken,
  \item The $\mu$-term vanishes to all orders,
\end{itemize}
since we want to keep the feature that the dimension-four proton decay operators remain forbidden.
The $\mu$-term \index{mu-term@$\mu$-term} can be written as
\begin{align}
\label{eq:muGHU}
	W&=\mu \Phi\,, & \Phi&=H_uH_d\,, & {\bf Q}\left( \Phi \right)&=0\,, &\mathcal K\left(\Phi \right)&=0\,,
\end{align}
leading to $\lambda_s=1,$ and 
\begin{align}
\label{eq:condK3}
\mu=\lambda_0 &\in {\rm ker}\, Q\,,\hspace{1cm}
	\mathcal K\left(\lambda_0\right)=\mathcal K_{\rm vac}\,.
\end{align}
\begin{table}[t]
\begin{center}
  \begin{tabular}{c|c|r r r|c|c}
	  \jvsb Name & Monomial & $R_1$ & $R_2$ & $R_3$ & $k$ & $k  n_3$ \\
	  \hline
	  \jvsb   $\Omega_1$ & $\bar X_0^c S_2 S_5$ & 0 & 0 & $-1$ & 6&0\\
	  \jvsb   $\Omega_2$ & $X_1 Y_2 S_2 S_5$ & 0 & $-1$ & $-1$ & 6&6 \\
	  \jvsb   $\Omega_3$ & $X_0 \bar X_1 S_5 S_7$ & 0 & $-1$ & $-1$ & 6&3 \\
	  \jvsb   $\Omega_4$ & $X_0 \bar X_1 Y_2 U_2 U_4$ & $-3$ & $-2$ & 0 & 6&6\\
  \end{tabular}
  \caption{Basis monomials of ${\rm ker} Q(\mathcal{S}_1)$ and the 
  corresponding discrete charges. All monomials have $k n_2 = k n_2'=0$.}
\label{tab:basisS1}
\end{center}
\end{table}
\begin{table}[t]
\begin{center}
  \begin{tabular}{c|c|c||c|c}
   \jvsb   Add & Mass term for $\F\Fb$ & Order& Mass term for $\F_1 \Fb_1$ & Order\\
	  \hline
	  \jvs   $\bar Y_2$ & 
	  $(X_0 \bar X_0^c \bar X_1 \bar Y_2 (S_5)^2 )^2 \Omega_1 \Omega_4$  & 20
	  &$ (X_0)^2 X_1 \bar X_1 (Y_2)^2 (\bar Y_2)^2 (S_5)^4 \Omega_2 \Omega_4$ & 21\\
	  \jvsb   
	  $\bar Y_2^c$ & 
	  $(\bar Y_2^c S_2 S_7)^2 \Omega_1 \Omega_4$  & 14
	  &$X_0 Y_2 (\bar Y_2^c)^2 (S_2)^3 (S_5)^2 (S_7)^3 \Omega_2 \Omega_4$&21\\
  \jvsb 	  
	  $U_1^c$ & $(X_0 \bar X_1 Y_2 U_1^c) \Omega_2$  & 8 
	  &$X_0 (Y_2)^2 U_1^c S_2 S_5$&6\\
	  \jvsb   
	  $U_3$ &
	  $(\bar X_0^c U_3 (S_5)^2)^2 \Omega_2 \Omega_4$  & 17
	  &$X_0 \bar X_0^c (Y_2)^2 U_2 (U_3)^2 U_4 (S_5)^4 \Omega_1$&15\\
  \jvsb 	  
	  $S_6$ &
	  $(X_1 Y_2 S_2 S_6) \Omega_4$&9 
	  &$X_0 (Y_2)^2 U_2 U_4 S_2 S_6$&7\\
  \end{tabular}
  \caption{Addition of any further field to $\mathcal{S}_1$ generates monomials
  which induce mass terms for $\F \Fb$ and $\F_1 \Fb_1$.
  Shown are lowest order examples. The monomials $\Omega_i$ are defined in Table \ref{tab:basisS1}.
  Singlets which complete pairs of the form $A^c A$ are not listed, since they always allow for 
  mass terms proportional to $\Omega_1 A^c A$.
  We do only consider singlets which conserve matter parity.
  }
\label{tab:proof-S1}
\end{center}
\end{table}

From (\ref{eq:dec1}) one infers that the combination $\Omega_1=\bar X_0^c S_2 S_5$ is a basis monomial of the charge kernel
of the minimal vacuum $\mathcal S_0$. In fact, the addition of $X_0, U_2, U_4$ does not give rise
to new uncharged combinations, and the dimension of ${\rm ker}\, Q(\mathcal S_0)$ is one. 
Since 
\begin{align}
\mathcal K(\Omega_1)=(0,0,-1,0,0,0,0)\neq\mathcal K_{\rm vac}\,,
\end{align}
the condition (\ref{eq:condK3}) is violated and the $\mu$-term is absent in the vacuum $\mathcal S_0$,
in agreement with (\ref{eq:nomuS0}). As a second example, consider the dimension-five proton
decay operator \index{Proton decay!dimension-five}
\begin{align}
\label{eq:lsprot5}
\Phi&=\T_{(1)}\T_{(1)}\T_{(1)}\Fb_{(1)} \,, \hspace{1cm} \Rightarrow \hspace{1cm}
\lambda_s=(X_0)^{-1} \bar X_0^c\,.
\end{align} 
Here $\lambda_s$ involves a negative exponent which cannot be canceled by the singlets in  $\Omega_1$.
Thus the projection (\ref{eq:PN}) yields zero for any possible term, and the coupling vanishes. This cannot
be explained  by the unbroken symmetries, since invariance of $X_0$ automatically
 implies invariance of $X_0^{-1}$.

We now aim to extend the vacuum $\mathcal S_0$, without generating a $\mu$-term or breaking matter parity.
The latter requires to consider only such singlets for the enlargement which are uncharged
under $\U1_X$, or have even charges. Table \ref{tab:chs} from Appendix \ref{app:discrete} 
shows that these are the singlets
\begin{align}
\label{eq:singsX0}
Q_X=0\,:\hspace{1cm}	
U_1^c, U_3, S_6, S_7, X_0^c, \bar X_0, X_1, X_1^c, \bar X_1, \bar X_1^c, Y_2, Y_2^c, 
	\bar Y_2,\bar Y_2^c\,.
\end{align}
We find that a maximal vacuum \index{Maximal vacuum} is given by
\begin{align}
\label{eq:S1}
    	\mathcal S_1&=\mathcal S_0 \cup \left\{
		X_1, \bar X_1, Y_2, S_7
	\right\}\,,
\end{align}
with a charge kernel \index{Charge kernel} of dimension four. The corresponding basis monomials are listed in table
\ref{tab:basisS1}. Note that from the basis monomials it is clear that the condition $R^1=\jmod{-1}{6}$ is
violated for all monomials in ${\rm ker}\,Q(\mathcal S_1)$.
 Table \ref{tab:proof-S1} shows that further addition of singlets 
indeed generates mass terms for the Higgs fields. On the other hand, we checked to order
500 in the singlets that the $\mu$-term is absent in the vacuum $\mathcal S_1$.

For the case of gauge-Higgs unification, any monomial which appears as a contribution to $\mu$
also contributes to $\langle W \rangle$ \cite{lx07c}, whence
\begin{align}
	\langle W \rangle = \mu = 0\, \hspace{1cm}\text{to all orders}. 
\end{align}
 This follows from the observation
(\ref{eq:muGHU}), which states that $H_u H_d$ is invariant under all symmetry transformations.

We can furthermore study the possibilities to generate a mass term for the pair of
exotics \index{Exotics} $\F_1, \Fb_1$,
\begin{align}
	W&=m \Phi\,, \hspace{1cm} \Phi=\F_1\Fb_1\,, \hspace{1cm} m=\lambda_s \lambda_0\,.
\end{align}
With the choice $\lambda_s=(X_1\bar X_1)^{-1}$ one obtains
\begin{align}
	{\bf Q}\left(\lambda_s \F_1\Fb_1 \right)&=0\,,& \mathcal K \left(\lambda_s \F_1\Fb_1 \right)&=0\,, &
	\mathcal K\left(\lambda_0\right)&=\mathcal K_{\rm vac}\,.
\end{align}
The latter condition is by definition not fulfillable by singlets from the vacuum $\mathcal S_1$.
Hence the two Higgs pairs \index{Two Higgs pairs} property of the vacuum $\mathcal S_0$ is also present here.
We conclude that gauge-Higgs unification is impossible in the present model.
The absence or presence of 
mass terms for $\F, \Fb$  is closely linked, even when going beyond the vacuum $\mathcal S_1$,
as demonstrated in Table \ref{tab:proof-S1}. This table also shows that the generation of $\mu$-terms
at high order seems possible, when starting from a maximal vacuum\footnote{Any
symmetry breakdown corresponds to an approximate symmetry, valid up to
some finite order \cite{kx08}. Here the latter may be large enough for an interesting phenomenology.}.

\subsubsection*{Unbroken symmetries and interactions}

The vacuum $\mathcal S_0$ suggested that $\F$ is preferred over $\F_1$ as the
up-Higgs from the viewpoint of the Yukawa couplings, which vanish for the latter
in that vacuum, cf.~(\ref{eq:YukuS0}).
Here we repeat the same analysis for the vacuum $\mathcal S_1$, and 
find\footnote{The differences to \cite{bs08} are due to
the inclusion of hidden sector multiplets in the analysis. We recover the generators
given there, but here they are associated with larger symmetries. Furthermore,
additional symmetry factors arise.} \index{Unbroken symmetries}
\begin{align}
G_{\rm vac}\left(\mathcal S_1\right)&=\left[\Z6\times\tZ2 \right]_R
\times\Z{60} \times \Z2\,,
\end{align}
with generators as in Table \ref{tab:vacS1}\,.
\begin{table}
  \centering 
  \footnotesize
  \begin{tabular}{c|c|c}
	\jvsb Symmetry & $\alpha'$ & ${\bf r}'$ \\
	\hline
	\hline
\jvsb	$[\Z6]_R$ & $\left(\frac{5}{6},\frac{5}{2},0,\frac{5}{6},-\frac{5}{6},\frac{1}{6}\right)$&
	$\left(\frac{5}{3},0,0,0,0,0,0\right)$ \\
\jvsb	$[\tZ2]_R$ & 	$\left(\frac{1}{2},-\frac{1}{2},0,-\frac{1}{4},\frac{1}{4},0\right)$&
	$(0,0,1,0,0,0,0)$\\
	\hline
\jvsb	$\Z{60}$ & $\left(\frac{1}{2},\frac{1}{2},-\frac{1}{6},\frac{1}{4},\frac{1}{12},\frac{1}{30}\right)$&
	$(0,0,0,0,0,0,0)$\\
\jvsb	$\Z2$ & 	$\left(\frac{75}{2},-\frac{15}{2},5,\frac{15}{2},-\frac{5}{2},\frac{7}{2}\right)$&
	$(0,0,0,0,0,0,0)$ \\	
\end{tabular}
  \caption{The generators of the unbroken symmetry in the vacuum $\mathcal S_1$.}\label{tab:vacS1}
\end{table}
Note that the symmetry $\Z{60}$ is generated by $\frac 12 t_X$, where $t_X$ is the generator of
$\U1_X$ \index{U1@$\U1$!X@$X$} 
from (\ref{eq:tX}). The symmetry thus contains the matter parity \index{Matter parity} 
(\ref{eq:PX}), as expected.

The dimension-four proton decay operators from (\ref{eq:proton45b}) are therefore absent.
However, dimension-five operators \index{Proton decay!dimension-five}
which lead to rapid proton decay are generated, for example
\begin{align}
	C^{(B)}_{1111}&=(S_5)^3 (S_7)^3 (X_0)^2 \bar X_0^c (\bar X_1)^3\,.
\end{align}
This follows from the basis monomial $\Omega_3$ from Table \ref{tab:basisS1},
which was not present in the vacuum $\mathcal S_0$ and 
allows for the cancelation of the negative exponent in (\ref{eq:lsprot5}).

A further difference of the vacuum $\mathcal S_1$ compared to $\mathcal S_0$
arises for the  Yukawa couplings. \index{Yukawa couplings!six dimensions} In the vacuum $\mathcal S_0$
all couplings for the choice $H_u=\F_1$ and $H_d=\Fb$ were forbidden by  symmetry, while the
ones for $H_u=\F$ and some for $H_d=\Fb_1$ were allowed and of low order, as shown in (\ref{eq:YukuS0}),
(\ref{eq:YukdS0}).
In the present vacuum, this statement is weakened. In terms of the four $\SU5$ families from (\ref{eq:4families})
the couplings for the choice $H_u=\F_1$ and $H_d=\Fb$ are
\begin{align}
	C^{(u)}_{H_u=\F_1}&=\left(\begin{array}{llll}
	s^9 & s^9 & s^{10} & s^{10} \\
	s^9 & s^9 & s^{10} & s^{10 }\\
	s^{10} & s^{10} & s^{11} & s^{11}\\
	s^{10} & s^{10} & s^{11} & s^{11}
	\end{array}\right)\,, &
	C^{(d)}_{H_d=\Fb}&=\left(\begin{array}{llll}
	0&0&0&0\\
	0&0&0&0\\
	0&0&0&0\\
	s^5&s^5&s^6&s^6
	\end{array}\right)\,,
\end{align}
where $s^n$ denotes an monomial of singlets of order $n$. The explicit expressions can be found in 
Table \ref{tab:Yuku2S1}.
\begin{table}
  \centering 
  \footnotesize
  \begin{tabular}{c|c|c||c|c|c}
\jvsb  $i$ & $j$ & $Y^{(u)}_{ij}$&$i$ & $j$ & $Y^{(u)}_{ij}$\\
  \hline
 \jvsb  $1$ & $1$ & $(S_5)^2(S_7)^2(X_0)^2(\bar X_1)^2 Y_2$ &  $2$ & $3$ & $S_2' (S_5')^2(S_7')^2(X_0)^2(\bar X_1)^2Y_2$ \\
\jvsb   $1$ & $2$ & $S_5S_5'(S_7)^2(X_0)^2(\bar X_1)^2 Y_2$&$2$ & $4$ & $(S_5')^3(S_7')^2(X_0)^2(\bar X_1)^2 Y_2$\\
 \jvsb  $1$ & $3$ & $S_2 (S_5)^2(S_7)^2(X_0)^2(\bar X_1)^2Y_2$ & $3$ & $3$ & $(S_2)^2(S_5)^2(S_7)^2(X_0)^2(\bar X_1)^2Y_2$\\
\jvsb   $1$ & $4$ & $(S_5)^3(S_7)^2(X_0)^2(\bar X_1)^2 Y_2$&$3$ & $4$ & $S_2(S_5)^3(S_7)^2(X_0)^2(\bar X_1)^2Y_2$\\
\jvsb   $2$ & $2$ & $(S_5')^2(S_7')^2(X_0)^2(\bar X_1)^2 Y_2$ &$4$ & $4$ & $(S_5)^4(S_7)^2(X_0)^2(\bar X_1)^2Y_2$\\
\hline
\hline
\jvs  $i$ & $j$ & $Y^{(d)}_{ij}$&$i$ & $j$ & $Y^{(d)}_{ij}$\\
  \hline
 \jvsb $4$ & $1$ & $S_2 (S_5)^2 \bar X_0^c \bar X_1$ & $4$ & $3$ & $(S_2)^2(S_5)^2\bar X_0^c\bar X_1$ \\
   \jvsb $4$ & $2$ & $S_2' (S_5')^2 \bar X_0^c \bar X_1$ & $4$ & $4$ & $S_2(S_5)^3\bar X_0^c\bar X_1$ 
\end{tabular}
  \caption{Examples of leading order monomials for the Yukawa couplings $Y^{(u)}_{ij}=Y^{(u)}_{ji}$ 
  and $Y^{(d)}_{ij}$ for
  the choice $H_u=\F_1$ and $H_d=\Fb$ in the vacuum $\mathcal S_1$.}\label{tab:Yuku2S1}
\end{table}
The vacuum $\mathcal S_1$ thus indeed is a two Higgs pairs vacuum, and for both pairs a similar 
pattern  occurs in the Yukawa couplings. 
 However, assuming that each of the singlets contributes
a suppression factor in the appropriate units, one observes a 
 significant hierarchy in the coupling to matter for the two Higgs pair choices. 

We do not follow the phenomenology of this vacuum any further, since we do not expect that gauge
coupling unification is compatible with the two Higgs scenario. 

\subsubsection{Partial gauge-Higgs unification}
\label{sec:PGHU}

The main reason for considering gauge-Higgs unification in the context of this model is the 
large top-quark mass which follows from identifying the up-type Yukawa coupling with 
the gauge interaction in the bulk. However, this result does only rely on the choice $H_u=\F$
and is independent of the down-Higgs. We are thus free to study
the assignment
\begin{align}
	H_u&=\F \,, \hspace{.5cm} H_d=\Fb_1\,, &
	\text{exotics:} \hspace{.5cm}\F_1, \Fb\,,
\end{align}
as suggested by the Yukawa matrices (\ref{eq:YukuS0})  and (\ref{eq:YukdS0}) for the vacuum $\mathcal S_0$.
We call this scenario `partial gauge-Higgs unification',
\index{Partial gauge-Higgs unification} since only the $\F$ arises from
 bulk gauge interactions, while $\Fb_1$ is a field from the second twisted sector.

The $\mu$-term \index{mu-term@$\mu$-term} can then be written as
\begin{align}
	W&=\mu \Phi\,, & \Phi&=\F\Fb_1\,, & \mu&=\lambda_0\lambda_s\,,
\end{align}
and the choice $\lambda_s=(\bar X_1)^{-1}$ leads to
\begin{align}
	{\bf Q}\left(\lambda_s \F \Fb_1\right)&= 0\,, & \mathcal K\left(\lambda_s \F \Fb_1\right)&=(0,0,-1,0,0,0,0)\,.
\end{align}
Up to the projection (\ref{eq:PN}), the conditions for the monomials $\lambda_0$ then read
\begin{align}
	{\bf Q}\left(\lambda_0\right)&=0\,, \\
	\mathcal K\left(\lambda_0\right)&=(\jmod{-1}{6},\jmod{-1}{3},\jmod 02,\jmod 06,\jmod 03,\jmod 02,\jmod02)\,.
\end{align}
By subsequently adding singlets from the list (\ref{eq:singsX0}), we find the maximal vacuum
\index{Maximal vacuum}
\begin{align}
\label{eq:S2}
    	\mathcal S_2&=\mathcal S_0 \cup \left\{ X_1^c,\bar X_1,Y_2^c, \bar Y_2, U_1^c,U_3,S_6,S_7\right\}\,.
\end{align}
This vacuum contains 14 singlets. Its charge kernel \index{Charge kernel}
 has dimension eight, with basis monomials as
given in Table \ref{tab:basisS2}.
\begin{table}
\centering
\footnotesize
\begin{center}
  \begin{tabular}{c|c|r r r|c|c}
\jvsb	  Name&Monomial & $R_1$ & $R_2$ & $R_3$ & $k$ & $k n_3$\\
	  \hline
\jvsb		  $\Omega_1'$ & $\bar X_0^c S_2 S_5$ & 0 & 0 & $-1$ & 6&0\\
\jvsb		  $\Omega_2'$ & $\bar X_0^c X_1^c Y_2^c$ & $-2$ & $-1$ & 0 & 12&12 \\
\jvsb		  $\Omega_3'$ & $\bar X_0^c (S_5)^2 U_3$ & $-2$ & 1 & $-1$ & 6&0 \\
\jvsb		  $\Omega_4'$ & $X_0 \bar X_1 S_5 S_7$ & 0 & $-1$ & $-1$ & 6&3 \\
\jvsb		  $\Omega_5'$ & $X_0 \bar X_0^c X_1^c \bar X_1 U_1^c$ & $-2$ & $-3$ & 0 & 12&6 \\
\jvsb		  $\Omega_6'$ & $X_0 \bar X_0^c \bar X_1 \bar Y_2 (S_5)^2$ & $-2$ & $-1$ & $-1$ & 12&6 \\
\jvsb		  $\Omega_7'$ & $X_0 \bar X_0^c \bar X_1 \bar Y_2 (S_6)^2$ & $2$ & $-3$ & $-1$ & 12&6 \\
\jvsb		  $\Omega_8'$ & $X_0 \bar X_0^c X_1^c \bar X_1 U_2 U_4$ & $-4$ & $-2$ & 0 & 12&6 \\
  \end{tabular}
  \caption{Basis monomials of ${\rm ker}\, Q(\mathcal{S}_2)$ and their discrete charges.
  All monomials have $kn_2=k n_2'=0$.}
\label{tab:basisS2}
\end{center}
\end{table}
Again, the condition $R^1=\jmod{-1}{6}$ is obviously violated by all monomials in ${\rm ker}\, Q(\mathcal S_2)$.
The vacuum is maximal, since the only possibility to 
enlarge it without breaking matter parity is to add singlets $A$ ($A^c$) 
whose $\mathcal{N} = 2$ superpartners $A^c$ ($A$) already belong to 
$\mathcal{S}_2$. One then obtains the $\mu$-term
\begin{align}
\mu = \lambda_0\lambda_s, \hspace{1cm} \lambda_0 = A A^c (\Omega_5')^3,
\end{align}
which is of order 16 in the singlets. Similar to the case of $\mathcal S_1$, going beyond
the maximal vacuum leads to a generation of the $\mu$-term with large suppression.
Order 16 may even be enough for the fulfillment of the phenomenological requirements.

The vacuum expectation value of the superpotential follows from monomials $\lambda_0$
with $\mathcal K(\lambda_0)=\mathcal K_{\rm vac}$. Again, this condition cannot be
fulfilled by the monomials from Table \ref{tab:basisS2}, since $\sum R^1$
is always even. Also for the case of partial gauge-Higgs unification one therefore has
the property
\begin{align}
	\langle W \rangle = \mu = 0\hspace{1cm}\text{to all orders,}
\end{align}
as it is the case for full gauge-Higgs unification.

We now consider the mass term of the exotic \index{Exotics} pair of Higgs field candidates,
\begin{align}
	W&=m \Phi\,, \hspace{1cm} \Phi=\F_1\Fb\,, \hspace{1cm} m=\lambda_s \lambda_0\,,
\end{align}
and choose $\lambda_s=(X_1)^{-1}$\,. One then obtains
\begin{align}
	  \mathcal K\left(\lambda_s \F \Fb_1\right)&=(0,0,-1,0,0,0,0) \hspace{1cm}
	  \Rightarrow \hspace{1cm} m=0
\end{align}
in the vacuum $\mathcal S_2$.
This looks as if the situation is the same as for $\mathcal S_1$, where we found a
two Higgs pair vacuum. However, there the full $2\times 2$ mass matrix was zero, whereas
here one finds
\begin{align}
	W &= \langle X_1^c \rangle \, \F_1 (\Fb + \epsilon \, \Fb_1) 
	\;, \quad 
	\epsilon = \langle X_0 \bar X_0^c X_1^c Y_2^c S_6 S_7 \rangle\;.
\end{align}
This shows that $\F_1$ decouples
together with a linear combination of $\Fb$ and $\Fb_1$.
The orthogonal linear combination is the down-type Higgs,
\begin{align}
	H_d &= \Fb_1 - \epsilon \; \Fb \;.
\end{align}
The vacuum $\mathcal S_2$ thus leads to a down-type Higgs with a dominant component
from the second twisted sector, as it was already suggested by the Yukawa matrices (\ref{eq:YukdS0})
of the vacuum $\mathcal S_0$.
In contrast, the up-type Higgs remains a pure gauge field in six dimensions, $H_u=\F$, 
giving an explanation of  the large top-quark mass within our orbifold GUT model.

\subsubsection*{Unbroken symmetries and interactions}

The unbroken symmetry \index{Unbroken symmetries} of the vacuum $\mathcal S_2$ is
\begin{align}
\label{eq:GvacS2}
    G_{\rm vac}\left(\mathcal S_2\right)&=\left[ \tZ4 \times \Z2 \right]_R\times \Z{60}\,, 
\end{align}
with generators as shown in Table \ref{tab:vacS2}.
\begin{table}
  \centering 
  \footnotesize
  \begin{tabular}{c|c|c}
	\jvsb Symmetry & $\alpha'$ & ${\bf r}'$ \\
	\hline
	\hline
\jvsb	$[\tZ4]_R$ & $\left(\frac{1}{2},0,-\frac{1}{12},\frac{5}{8},\frac{1}{24},-\frac{1}{30}\right)$&
	$\left(\frac{1}{2},0,0,0,0,0,0\right)$ \\
\jvsb	$[\Z2]_R$ & $\left(-\frac{5}{2},\frac{5}{2},0,-5,0,\frac{1}{2}\right)$&
	$\left(0,0,0,\frac{5}{6},0,0,0\right)$	 \\
	\hline
\jvsb	$\Z{60}$ & $\left(\frac{1}{2},\frac{1}{2},-\frac{1}{6},\frac{1}{4},\frac{1}{12},\frac{1}{30}\right)$&
	$(0,0,0,0,0,0,0)$\\
\end{tabular}
  \caption{The generators of the unbroken symmetry in the vacuum $\mathcal S_2$.}\label{tab:vacS2}
\end{table}
Again, one obtains the factor $\Z{60}$ which was also present in the vacuum $\mathcal S_1$. It
has the generator $\frac 12 t_X$, where $t_X$ is the generates  $\U1_X$ \index{U1@$\U1$!X@$X$}
 in the false vacuum. This
implies the presence of the matter parity \index{Matter parity} 
from (\ref{eq:PX}) in both studied  vacua which extend $\mathcal S_0$.
Dimension-four proton decay operators are therefore absent in both cases, however, dimension-five
proton decay \index{Proton decay!dimension-five} occurs also in the vacuum $\mathcal S_2$,
\begin{align}
	C^{(B)}_{1111}&=S_6 S_7 X_1^c (\bar X_0^c)^2 \bar X_1 Y_2^c\,.
\end{align}
So far, there is no known model with vanishing $\mu$-term, interesting Yukawa couplings, preserved supersymmetry,
no exotic particles, and a stable proton. In our setup, full proton stability seems to be linked to the existence of
two Higgs pairs, as in the vacuum $\mathcal S_0$. We interpret the fact that we are not able to solve this problem
as further need to understand the structure of the model, and maybe relate the question
of decoupling to underlying smooth compactifications. We will comment more on this in the following chapter.

The $\mu$-term vanishes in the vacuum $\mathcal S_2$ by construction, we checked this
explicitly by order-by-order scans up to order 500. The up-type Yukawa couplings are
at lowest order given by the result (\ref{eq:YukuS0}) for the vacuum $\mathcal S_0$.
Hence we again benefit from partial gauge-Higgs unification by the generation of a large 
top-quark mass through the gauge interaction in the bulk.
The down-type Yukawa couplings \index{Yukawa couplings!six dimensions} are given by
\begin{align}
	C^{(d)}_{H_d=\Fb_1}&=\left(\begin{array}{llll}
	0&0&0&0\\
	0&0&0&0\\
	s^{10}&s^{10}&s^{6}&s^6\\
	s^1&s^1&s^2&s^2
	\end{array}\right)\,,
\end{align}
with explicit lowest order monomials as in Table \ref{tab:YukdS2}. Note
that all zeros are protected by symmetries and thus present at arbitrary order in the singlets.
This can be approved by close inspection of Table \ref{tab:chns} in Appendix \ref{app:discrete}.
\begin{table}
  \centering 
  \footnotesize
  \begin{tabular}{c|c|c||c|c|c}
\jvs  $i$ & $j$ & $Y^{(d)}_{ij}$&$i$ & $j$ & $Y^{(d)}_{ij}$\\
  \hline
 \jvsb $3$ & $1$ & $X_0$$($ $\bar{X}_0^c$ $)^2$  $($ $X_1^c$ $)^2$  $\bar{X}_1$  $\bar{Y}_2$  $U_2$  $U_4$  $S_5$  & $4$ & $1$ &  $S_5$ \\
   \jvsb $3$ & $2$ & $X_0$$($ $\bar{X}_0^c$ $)^2$  $($ $X_1^c$ $)^2$  $\bar{X}_1$  $\bar{Y}_2$  $U_2$  $U_4$  $S_5'$   & $4$ & $2$ & $S_5'$\\
    \jvsb $3$ & $3$ & $X_0$ $X_1^c$  $\bar{X}_1$  $\bar{Y}_2$  $S_6$  $S_7$   & $4$ & $3$ &  $S_2S_5$ \\
   \jvsb $3$ & $4$ & $\bar{X}_0^c$ $($ $X_1^c$ $)^2$  $Y_2^c$  $S_6$  $S_7$  & $4$ & $4$ &   $(S_5)^2$
\end{tabular}
  \caption{Examples of leading order monomials for the Yukawa couplings  $Y^{(d)}_{ij}$
   in the vacuum $\mathcal S_2$.}\label{tab:YukdS2}
\end{table}
From the above matrix and the up-type analogue (\ref{eq:YukuS0}) one deduces the Yukawa couplings 
\index{Yukawa couplings!four dimensions} of the
four-dimensional effective theory, as described in (\ref{eq:Yuk4to3}),
 \begin{align}
  W_\mathrm{Yuk} = Y_{ij}^{(u)} u_i^c q_j H_u + Y_{ij}^{(d)} d_i^c q_j H_d +  
  Y_{ij}^{(l)} l_i e_j^c H_d ,
 \end{align}
 with
 \begin{align}
	Y^{(u)} &=
	\left( \begin{array}{c c c}
	 s^4 &   s^4  &  s^5  \\
	  s^4  & s^4  & s^5 \\
	  s^5 &  s^5 & g \\
	\end{array} \right) \;, &
	Y^{(d)} &=
	\left( \begin{array}{c c c}
	0 & 0  & 0  \\
	0  & 0  & 0 \\
	  s^1 &   s^1 &   s^2 \\
	\end{array} \right) \;, &
	Y^{(l)} &=	\left( \begin{array}{c c c}
	0 & 0  & 0  \\
	0  & 0  & 0 \\
	  s^{10} &   s^{10} &   s^6 \\
	\end{array} \right) \;.	
\end{align}
This explicitly shows that the top-quark mass is proportional to the gauge coupling $g$ in six dimensions.
The lepton and down-quark  mass matrices do not look convincing. In fact, they predict 
$m_e=m_\mu=m_d=m_s=0$, which  excludes the studied model. One can now
either understand this as a problem of the specific choice of gauge embedding that
we chose and study other possibilities, as claimed in \cite{lx07}.
An alternative approach may be to further explore correlations between the decoupling of exotics
and the identification of matter and Higgs fields, and the generated interactions.
Possibly the details of the physical vacuum are closely linked to the dynamics of the underlying geometry,
once the restriction to the orbifold point is relaxed.

\subsection{The all-order superpotential}
\label{sec:allorders}

So far symmetry arguments were given for the vanishing of couplings to all orders in the singlets.
This was then used as a tool for the construction of partly-realistic vacua. The non-vanishing couplings
were calculated by order-by-order scans over all monomials within a given vacuum $\mathcal S$.
Here we argue that the methods introduced in Section \ref{sec:kernel} can be extended
to an algorithm which allows for the calculation of the superpotential to all orders, without relying on the naive
order-by-order approach.

In (\ref{eq:lambdas}) we separated the continuous and the discrete transformation
behavior of a superpotential term $\lambda \Phi$, where the monomial
 $\lambda=s_1\cdots s_N$, $s_i\in\mathcal S$, determines the
coupling \index{Couplings} of the interaction $\Phi=\phi_1\cdots \phi_M$, $\phi_i \notin \mathcal S$,
\begin{align}
	W&\supset\lambda_0\lambda_s^\Phi\Phi\,,& {\bf Q}\left(\lambda_s^\Phi\Phi\right)={\bf Q}\left(\lambda_0\right)&=0\,,&
	\mathcal K\left(\lambda_0\right)&=\mathcal K_{\rm vac}-\mathcal K\left(\lambda_s\Phi\right)\,.
\end{align}
In fact, the total superpotential is a sum of such terms, which we did not write out explicitly. 
It will have an overall factor which is completely invariant under all symmetries. The corresponding
terms, written as $\omega_0\in {\rm ker}\, Q\left(\mathcal S\right)$, are then universal contributions to $\lambda_0$,
\begin{align}
	\lambda_0&=\omega_0\lambda_0^\Phi\,, & 
	\mathcal K\left(\lambda_0^\Phi\right)&=\mathcal K_{\rm vac}-\mathcal K\left(\lambda_s^\Phi\Phi\right)\,, &
	\mathcal K\left(\omega_0\right)&=\mathcal K_0\,,
\end{align}
where
\begin{align}
\label{eq:K0}
	\mathcal K_0&=(\jmod 06,\jmod 03,\jmod 02,\jmod 06,\jmod 03,\jmod 02,\jmod 02)\,.
\end{align}
We denote the defining properties of $\omega_0$ as $\omega_0\in{\rm ker}\, Q \cap {\rm ker}\, \mathcal K$.
The latter set of monomials can now be calculated for a given vacuum configuration $\mathcal S$.
Since the $\omega_0$ are universal, this has to be done only once for the full superpotential.

However, what remains to be found for each interaction $\Phi$ are the monomials $\lambda_s^\Phi \in \mathcal S$
and $\lambda_0^\Phi \in {\rm ker}\, Q(\mathcal S)$. If one can find a basis of ${\rm ker}\, Q \cap {\rm ker}\,
\mathcal K$, it is sufficient to find only one solution.
In the end, the projection (\ref{eq:PN}) to physical monomials with positive exponents has to be evaluated. The
total superpotential to all orders \index{All-order superpotential} can then be written as
\begin{align}
\label{eq:Wallorders}
	W&=\mathcal P_{\mathbbm N}
	\left(\sum_{\omega_0} \omega_0 \right)\left(\sum_\Phi \lambda_0^\Phi\lambda_s^\Phi\Phi\right)\,,
\end{align}
where the omnipresent and unknown coefficients are not shown.

Recall from Section \ref{sec:sgrulesZ6II} that the discrete symmetries discussed so far are
not sufficient for the description of all interactions of the effective theory. There we found
that is was necessary to discuss the terms which contain only fields from equivalent
twisted sectors separately.  They can be collected into a part $W_0$ of the full superpotential,
\begin{align}
	W_{\rm tot}&=W+W_0\,,
\end{align}
which is then subject to additional $\Z6^{n_6k}$ symmetry constraints.
If such terms appear in the potential (\ref{eq:Wallorders}) the corresponding
condition
\begin{align}
	\sum n_6k&=\jmod 06
\end{align}
has to be checked a posteriori. 

\subsubsection*{Example: The vacuum $\mathcal S_1$}

As an example for the algorithm suggested above by (\ref{eq:Wallorders}), 
we calculate the superpotential for the vacuum $\mathcal S_1$ from (\ref{eq:vacS0}).
The corresponding basis monomials of the charge kernel were given in Table \ref{tab:basisS1}.
The basis of ${\rm ker}\, Q \cap {\rm ker}\, \mathcal K$ can then be found by calculating
\begin{align}
	{\rm ker}\, \left(
\begin{array}{rrrrrrrrr}
 0 & 0 & 0 & -3 & 6 & 0 & 0 & 0 & 0 \\
 0 & -1 & -1 & -2 & 0 & 3 & 0 & 0 & 0 \\
 -1 & -1 & -1 & 0 & 0 & 0 & 2 & 0 & 0 \\
 6 & 6 & 6 & 6 & 0 & 0 & 0 & 6 & 0 \\
 0 & 6 & 3 & 6 & 0 & 0 & 0 & 0 & 3
\end{array}
\right)\,,
\end{align}
\begin{table}
  \centering 
  \footnotesize
  \begin{tabular}{l|l}
	\jvsb $\tilde \Omega_1$ & $(S_2)^{-1} X_0  (X_1)^{-1} \bar X_1 (Y_2)^{-1} S_7$ \\
	\jvsb $\tilde \Omega_2$ & $(S_5)^2 (S_2)^2 (\bar X_0^c)^2$\\
	\jvsb $\tilde \Omega_3$ &
	$(U_4)^{-2} (U_2)^{-2} (S_5)^2 (S_2)^{-2} (X_0)^2 (\bar X_0^c)^{-2}  
 	(\bar X_1)^2 (Y_2)^{-2} (S_7)^4$\\
	\jvsb $\tilde \Omega_4$ & $(S_2)^{-6} (X_0)^6 (\bar X_0^c)^{-3} (X_1)^{-3}
	 (\bar X_1)^6 (Y_2)^{-3} (S_7)^6$
\end{tabular}
  \caption{Basis monomials of ${\rm ker}\, Q \cap {\rm ker}\,\mathcal K$ for the
  vacuum $\mathcal S_1$.}\label{tab:tOm1}
\end{table}
where the first four columns contain the quantum numbers 
$R^1,R^2,R^3,k,n_3k$ of the four basis monomials $\Omega_i$ of ${\rm ker}\,Q(\mathcal S_1)$, and
the last five columns are added in order to find solutions modulo the corresponding order, as
given in (\ref{eq:K0}). This yields a basis $\mathcal B_0$ for ${\rm ker}\, Q \cap {\rm ker}\, \mathcal K$,
\begin{align}
	\mathcal B_0&=\big\{\underbrace{( \Omega_2 )^{-1} \Omega_3}_{\tilde \Omega_1},
	\underbrace{ (\Omega_1)^{2}}_{\tilde \Omega_2},
	\underbrace{ (\Omega_1)^{-2}(\Omega_3)^4
	(\Omega_4)^{-2}}_{\tilde \Omega_3},
	\underbrace{(\Omega_1)^{-3} (\Omega_2)^{-3}(\Omega_3)^6}_{\tilde \Omega_4}\big\}\,.
\end{align}
The monomials $\tilde \Omega_i$ are then completely invariant under the continuous and the discrete
symmetries. Their explicit form is given in Table \ref{tab:tOm1}.
Note the appearance of inverse exponents, which will eventually be subject to the projection $\mathcal P_{\mathbbm N}$.
Consider for example the down-type Yukawa coupling
\begin{align}
	\Phi&=\Fb_{(4)} \T_{(4)} \Fb_1\,,
\end{align}
for which one finds
\begin{align}
	\lambda_s^\Phi&= S_5 (S_2)^{-1} (\bar X_0^c)^{-1}\,, & 
	\lambda_0^\Phi&=(\Omega_1)^{-1}=(S_2)^{-1}(S_5)^{-1}(\bar X_0^c)^{-1}\,.
\end{align}
Table \ref{tab:tOm1} shows that the smallest term with only positive exponents
arises for $\omega_0=\tilde \Omega_2$ and the corresponding coupling is
\begin{align}
	C^{(d)}_{44} &= \tilde \Omega_2 \lambda_0^\Phi \lambda_s^\Phi = (S_5)^2\,,
\end{align}
in agreement with (\ref{eq:YukdS0}). The next terms then follow from
\begin{align}
	\omega_0&= (\tilde \Omega_2)^2,(\tilde \Omega_1)^{-3}(\tilde \Omega_2)^2 \tilde \Omega_4,
	(\tilde \Omega_1)^{-4}(\tilde \Omega_2)^2 \tilde \Omega_4,
	(\tilde \Omega_1)^{-5}(\tilde \Omega_2)^2 \tilde \Omega_4,
	(\tilde \Omega_1)^{-6}(\tilde \Omega_2)^2 \tilde \Omega_4,
\end{align}
where the last four choices all produce  couplings of order 13.

The evaluation of the projection to physical monomials corresponds to solving Diophantine equations. 
We find that for the
ten-field vacuum $\mathcal S_1$, the 
algorithm (\ref{eq:Wallorders}) can be solved to very high orders without any restrictions due to limits of computing
power (order 50 in less than a minute). We expect that the efficiency changes if one  studies much larger vacua.  
Here, the method correctly reproduces the Yukawa matrices (\ref{eq:YukuS0}), (\ref{eq:YukdS0}) and higher
order terms, which we do not explicitly list.

\subsubsection[The calculation of $F$-terms]{\boldmath The calculation of $F$-terms}

The method (\ref{eq:Wallorders}) can also be used for the calculation of $F$-terms 
\index{F-terms@$F$-terms} $F_u=\partial W/\partial u$, by
choosing $\Phi=u \notin \mathcal S$. 
The expectation value of $F_u$ only gets non-zero contributions from terms in which $u$ is the only
field in the monomial which is not contained in $\mathcal S$,
\begin{align}
	W&=\lambda u\,, & \lambda=&s_1\cdots s_N\,,&s_i \in \mathcal S\,.
\end{align}
Now consider $u$ from the second or the fourth twisted sector and
 the case that the vacuum $\mathcal S$ contains the charge conjugate field $u^c$.
For the maximal vacuum $\mathcal S_2$, this is the case for all $u$ with non-trivial $F$-terms.
Then one can choose
\begin{align}
	\lambda_s^u&=u^c\,, & \lambda_0^u&=\Omega_0\in{\rm ker}\,Q\,, & \mathcal K
	\left(\Omega_0 u^c u \right)&=\mathcal K_{\rm vac}\,,
\end{align}
with a universal monomial $\Omega_0$. For the vacua $\mathcal S_1, \mathcal S_2$
this is the monomial which was used for the universal decoupling, $\Omega_0=S_2 S_5 \bar X_0^c$.
With (\ref{eq:Wallorders}) the $F$-terms take the form
\begin{align}
	F_u&=\mathcal P_{\mathbbm N}  \left( \sum_ {\omega_0} \omega_0 \right) \Omega_0u^c \,,
\end{align}
up to coefficients. Due to $u^c\in\mathcal S$ the conditions $F_u=0$, required for unbroken supersymmetry, are
not universal. The projection to physical monomials depends on $u^c$ and has to be evaluated for each case
separately. We do not perform this calculation here, and simply
 assume the existence of solutions in the studied
vacua,
which are needed for consistency of the model.

\section{Connection with smooth geometries}
\label{cpt:K3}

In the previous chapters, we have described in detail an effective local orbifold GUT model in six dimensions,
obtained from an anisotropic compactification of the heterotic string.
It was found by first compactifying four internal dimensions on an orbifold $T^4/\Z3$, whose volume was
assumed to be small, the other two on a torus of GUT scale size, and then dividing
out another $\Z2$ symmetry. This short chapter is meant to be an outlook on work in progress,
concerned with the possibility that the
four small internal dimensions are not compactified on an orbifold, but on the smooth manifold $K3$.
The orbifold $T^4/\Z3$ is a singular limit of the latter, and thus a comparison of the two compactifications,
\begin{align}
\label{eq:compare}
	&\frac{T^4/\Z3 \times T^2}{\Z2} \hspace{.5cm}\text{versus} \hspace{.5cm}
	\frac{K3 \times T^2}{\Z2}\,,
\end{align}
may yield insights on the role of the fields that were obtained on the orbifold side. 
Note that this idea is somehow diametrical to the standard blow-up approach (cf.~\cite{chx85,fix88,a94,ht06,ngx07,lrx08,nhx09}).
We hope that the above comparison may 
help to improve the understanding of the decoupling procedure and the structure
of physical vacua. Here we present first results in this direction\footnote{We
are grateful to Taizan Watari for related discussions.}.

\subsection[Compactification on the $K3$ surface]{Compactification on the {\boldmath $K3$} surface}

The $K3$ \index{K3 surface} 
surface is a standard example for string compactifications on a manifold (see for example \cite{gsw84,a96}). 
It is the only 
non-trivial Calabi--Yau in two complex dimensions. Its topology has 
  Euler characteristic \index{Euler characteristic} $\chi=24$, corresponding to the result
 \begin{align}
	\frac{1}{16 \pi^2} \int_{K3} \tr\, R^2&=24\,.
\end{align}
This has to be understood together with the `tadpole cancelation condition',
\index{Tadpole cancelation on K3}
\begin{align}
	\int_{K3} \left(  \tr\, R^2-\tr\,F^2\right)&=0\,,
\end{align}
which is required for the absence of anomalies, \index{Anomaly} cf.~Section \ref{sec:anomalies}. 
The above formulas imply that consistent compactifications on $K3$ are tied to the existence of a non-vanishing
gauge background. This necessarily breaks the $\E8\times\E8$ gauge group, and
the possible effective theories that can be obtained in six dimensions 
correspond to the different choices for the embedding of the broken 
generators into the full gauge group \cite{gsw84,bix96,ht06}.
The non-vanishing gauge background of the compactified theory is then often referred to as the
`gauge bundle' \index{Gauge bundle} in the literature. 

The topological invariant on the
gauge theory side is called the `instanton number', \index{Instanton number}
and as a consequence of the tadpole
cancelation condition it also has to sum to 24. The breaking of the gauge group can then be characterized
by the distribution of the instantons among subgroups of $\E8\times\E8$. If they all sit in a subgroup
$H$, its commutant $G$ gives the unbroken gauge symmetry, 
defined such that $G\times H$ is a maximal subgroup
of the original symmetry.

Both the geometry and the gauge symmetry background are associated with moduli fields, which
comprise the degrees of freedom that describe fluctuations around that background. Including
fluctuations of the antisymmetric tensor field and the volume, the geometrical part has a moduli space 
of dimension 80 and is essentially given by \cite{a96} \index{Moduli space!geometrical}
\begin{align}
	\frac{O(4,20)}{O(4)\times O(20)} \hspace{.5cm} \Rightarrow
	\hspace{.5cm} \text{80 degrees of freedom $\sim$  $20$ hypermultiplets.}
\end{align}

For the gauge bundles, the moduli space depends on the model. 
We refer the reader to table
3 of \cite{bix96} 
for a classification of various possibilities to break $\E8\times\E8$ to a subgroup,
and focus only on one particular case of interest for the comparison with the orbifold model
discussed in this work. In Chapter \ref{cpt:lGUT} the
gauge group and the matter content of the six-dimensional bulk theory were found to be
\index{Gauge symmetry!bulk}
\index{Spectrum!bulk}
\begin{align}
	T^4/\Z3\,:\hspace{1cm}&\SU6 \times \U1^3 \times \left[ \SU3 \times \SO8 \times \U1^2 \right]\,,\\
	& ({\bf 20},\mathbbm 1,\mathbbm 1) + 9 \cdot \left[({\bf 6},\mathbbm 1,\mathbbm 1)
	+({\bf \bar 6},\mathbbm 1,\mathbbm 1)\right]
	+9 \cdot \left[(\mathbbm 1,{\bf 3},\mathbbm 1)+(\mathbbm 1,{\bf \bar 3},\mathbbm 1)\right] \nonumber \\
	&+4\cdot\left[
	(\mathbbm 1,\mathbbm 1,{\bf 8})+ (\mathbbm 1,\mathbbm 1,{\bf 8}_s)
	+(\mathbbm 1,\mathbbm 1,{\bf 8}_c) \right] +42\cdot(\mathbbm 1,\mathbbm 1,\mathbbm 1)\,. \label{eq:specorb}
\end{align}
In the following we shall compare this model to a compactification
 on $K3$ which yields the following gauge symmetry: \index{Gauge symmetry!K3}
\begin{equation}
	K3 :\hspace{.8cm}\SU6 \times \left[ \SO8 \right]\,. \hspace{6.7cm}\phantom{A}
\end{equation}
We thus assume that all $\U1$ factors as well as the hidden sector $\SU3$ gauge group are remnants of
the singular orbifold limit in the geometrical moduli space. 

The corresponding instanton distribution is \cite{bix96,r08}
\begin{align}
	\text{Visible sector:}& &  \E8 &\supset \SU6 \times \underbrace{ \langle \SU2 \rangle}_{6\, \text{instantons}}
	\times \underbrace{\langle \SU3 \rangle}_{6 \,\text{instantons}}\,,& &\phantom{AA} \\
	\text{Hidden sector:}& &  \E8 &\supset \SO8 \times \underbrace{\langle \SO8 \rangle}_{12\, \text{instantons}}\,.
	& &\phantom{AA}
\end{align}
From these numbers one can also infer the number of massless matter multiplets in the effective
theory in six dimensions. The result is \index{Spectrum!K3}
\begin{align}
\label{eq:matterK3}
	K3:\hspace{1cm} ({\bf 20},\mathbbm 1) + 9 \cdot \left[ ({\bf 6},\mathbbm 1)
	+({\bf \bar 6},\mathbbm 1)\right]+4\cdot\left[
	(\mathbbm 1,{\bf 8})+ (\mathbbm 1,{\bf 8}_s)+(\mathbbm 1,{\bf 8}_c) \right]\,,
\end{align}
and perfectly matches the spectrum of the orbifold result (\ref{eq:specorb}), as
far as the common gauge group $\SU6 \times [\SO8]$ is concerned.
The discrepancies will be discussed shortly.

The instanton numbers also specify the gauge bundle moduli space. The
corresponding degrees of freedom can be understood in terms of
 hypermultiplets of the six-dimensional effective theory: \index{Moduli space!gauge}
\begin{subequations}
\begin{align}
	&\phantom{AA}& \text{$\SU2$ gauge bundle:}& & \text{9 hypermultiplets}, & & \phantom{AA}\\
	&&\text{$\SU3$ gauge bundle:}& & \text{10 hypermultiplets},\\
	&&\text{$\SO8$ gauge bundle:}& & \text{44 hypermultiplets}.
\end{align}
\end{subequations}

In summary, the $K3$ compactification with the above gauge bundles,
 to which we want to compare the orbifold results,
has a smaller unbroken gauge group and matter spectrum, but predicts 83 moduli hypermultiplets,
\begin{align}
\label{eq:moduliK3}
	K3 :\hspace{1cm} \underbrace{19}_{\text{vis. $\E8$}} + 
	\underbrace{44}_{\text{hid. $\E8$}}+\underbrace{20}_{\text{geom.}} =83 \; \;\text{moduli}\,.
\end{align}
On the other hand, the orbifold provides a larger gauge group and additional matter
transforming under the latter, as well as 40 singlets and two 
geometrical moduli \index{Geometrical moduli} hypermultiplets, denoted as $C_1,C_2$ in (\ref{eq:C1C2}).
However, additionally there are non-zero Fayet--Iliopoulos terms (\ref{eq:FIterm}), and
 parts of the gauge symmetry will be broken.
As will be described in the following, 
this breaking is the key for a successful
matching of the two approaches.

\subsection{Identifying moduli}

The 320 hypermultiplets (\ref{eq:specorb}) of the $T^4/\Z3$ compactification arise from different
sectors of the orbifold,
\begin{subequations}
\label{eq:l}
\begin{align}
\label{eq:l1}
&\left.\begin{array}{ccccl}
\multicolumn{1}{l}{3}&
\multicolumn{4}{r}{\hspace{-.6cm} \,\hspace{-.3cm}
\left[(\mathbbm 1,\mathbbm 1,{\bf 8})+ (\mathbbm 1,\mathbbm 1,{\bf 8}_s)
	+(\mathbbm 1,\mathbbm 1,{\bf 8}_c)\right]}\\
	3\hspace{.5cm}&  \hspace{.15cm}\cdot&\hspace{.2cm} 3&\hspace{.2cm} \cdot &\left[({\bf 6},\mathbbm 1,\mathbbm 1)
	+({\bf \bar 6},\mathbbm 1,\mathbbm 1)\right]\\
	3& \hspace{.15cm} \cdot& \hspace{.2cm} 3&\hspace{.2cm} \cdot &
	\left[(\mathbbm 1,{\bf 3},\mathbbm 1)+(\mathbbm 1,{\bf \bar 3},\mathbbm 1)\right]\\
	3&  \hspace{.15cm}\cdot& \hspace{.2cm} 3& \hspace{.2cm}\cdot &
	\left[(\mathbbm 1, \mathbbm 1,\mathbbm 1)+(\mathbbm 1, \mathbbm 1,\mathbbm 1)\right]
\end{array}\right\} &
&\begin{array}{l}
k=2,4, \\
\text{no oscillators},
\end{array}\\
&\hspace{-.2cm}\underbrace{3}_{\gamma=0,\frac12,1}\cdot
	\underbrace{3}_{n_3=0,1,2\vphantom{\frac12}}\cdot \hspace{.25cm}
	\left[ (\mathbbm 1, \mathbbm 1,\mathbbm 1)+(\mathbbm 1, \mathbbm 1,\mathbbm 1)\right] :&
&\begin{array}{l}
k=2,4, \\
\text{non-zero oscillator numbers},
\end{array}	\label{eq:lS}	\\\label{eq:lU}
&\hspace{.7cm}\left.\begin{array}{r}
(\mathbbm 1,\mathbbm 1,{\bf 8})+ (\mathbbm 1,\mathbbm 1,{\bf 8}_s)
	+(\mathbbm 1,\mathbbm 1,{\bf 8}_c) \\
	({\bf 20},\mathbbm 1,\mathbbm 1)+4 \cdot (\mathbbm 1, \mathbbm 1,\mathbbm 1) 
\end{array}\right\}
& &\hspace{.2cm}\text{untwisted sector, charged},\\
\label{eq:lC}
&\hspace{4.6cm}2 \cdot (\mathbbm 1, \mathbbm 1,\mathbbm 1) : & &\hspace{.2cm}\text{untwisted sector, uncharged}.
\end{align} 
\end{subequations}
The singlets in (\ref{eq:lS}) carry
non-vanishing oscillator numbers ${\bf N}, {\bf N}^*, {\bf \tilde N}, {\bf \tilde N}^*$, defined in (\ref{eq:N}).
The singlets in (\ref{eq:lU}) and (\ref{eq:lC}) were called $U_1,\dots,U_4$ and $C_1,C_2$
in the previous chapters, respectively. The two latter hypermultiplets
 play the role of geometrical moduli of the orbifold geometry.

We can only make contact with the assumed $K3$ configuration, if we also break the  
additional gauge group factors $\U1^3 \times [\SU3 \times \U1^2]$ on the orbifold side
by a Higgs mechanism. This means that some of the degrees of freedom in the
above list are traded for masses of the corresponding gauge bosons and gauginos,
\begin{subequations}
\begin{align}
	\U1^3\,:\hspace{1cm}&\text{3 hypermultiplets},\\  
	\SU3 \times \U1^2\,:\hspace{1cm}&\text{$8+2=10$ hypermultiplets.}
\end{align}
\end{subequations}
Let us now count the degrees of freedom that remain on the orbifold side. We are left with $\SU6\times[\SO8]$
gauge symmetry after the Higgs mechanism. From the representations of the hidden sector $\SU3$
one obtains $9 \cdot (3+3)=54$ hypermultiplets. 
For the visible sector, one can show that for each localization quantum number $n_3=0,1,2$, the
multiplets from (\ref{eq:l1}) arise from different embeddings of the group $\SO{14}$
into $\E8\times\E8$. This fact was observed in \cite{bhx06} and not stressed much 
in the previous chapters. For each value of $n_3$ and $\gamma$ one has the decomposition
${\bf 14}={\bf 6}+{\bf \bar 6}+2\cdot\mathbbm 1$, and we can associate $3\cdot3\cdot2=18$
singlets with vanishing oscillator numbers with the visible sector.
Amending the latter by the four charged untwisted singlets, one obtains after the Higgs mechanism
\begin{align}
	&\text{Visible sector}: & 18+4-3 &=19 & &\text{hypermultiplets},\\
	&\text{Hidden sector}: & 54-10&=44 & &\text{hypermultiplets}\,.
\end{align}
The degrees of freedom from the twisted sectors of the orbifold 
match the gauge bundle moduli in (\ref{eq:moduliK3}) of the $K3$ compactification,
after breaking the additional gauge symmetry.

Furthermore, 18 twisted singlets  from oscillator excitations 
in the internal directions are present
on the orbifold, as well as two  from the untwisted sector with the same property.
Together, they may correspond to the geometrical moduli of $K3$:
\begin{align}
 &\text{Oscillator excitations}: & 18+2&=20 & &\text{hypermultiplets}\,.
\end{align}
We have thus found a correspondence between the gauge bundle moduli and the geometrical moduli of a smooth
$K3$ compactification and the  hypermultiplets of the orbifold, after spontaneous breakdown of
the additional symmetries.

It is a remarkable fact that the hypermultiplets which play the role of geometrical moduli
are charged under the $\U1$'s of the orbifold. This may be related to the fact that we chose a
gauge background on $K3$ without Abelian factors and therefore they are specific
to the orbifold point. This issue and the interpretation of the above is still under debate.

\subsection[Outlook: A  local GUT based on $K3$]{Outlook: A  local GUT based on {\boldmath $K3$}}

The next step would be to complete the comparison (\ref{eq:compare}) by defining
the action of $\Z2$ on the manifold $K3$ and work out the local spectra at the fixed points
of $T^2/\Z2$. The involutions on $K3$ were classified in \cite{b97} and their application
to our model is work in progress. We hope that the
comparison with the results obtained for the local orbifold GUT may lead
to significant progress in understanding the validity of the orbifold description
and its interpretation in terms of their smooth counterparts.

\section{Conclusions}
\label{cpt:conclusions}

Field theoretical orbifold GUTs can be derived from heterotic string theory by the consideration
of anisotropic compactifications. We explicitly constructed an effective $T^2/\Z2$ model
in six dimensions with some interesting properties. For vanishing vacuum expectation values,
it has $\SU6$ gauge symmetry in the bulk, which is broken to the GUT group $\SU5$ at
two fixed points of the geometry. There, two complete standard model families are localized
and it was possible to uniquely determine a $\U1_X$ gauge symmetry, which combines
with the hypercharge to a unique $\U1_{B-L}$ in the effective four-dimensional theory.
This was then used as a tool to interpret the  bulk fields of the model, with the result that two additional
complete quark-lepton generations in the $\SU5$ language could be identified. They are split multiplets, yielding
the missing third standard model family in four dimensions as the sum of their zero modes.
This as such is an interesting setup for a higher dimensional GUT model, since it resembles
the well-known and successful orbifold solution of the doublet-triplet splitting problem of the GUT Higgs
multiplets in the matter sector. Phenomenologically, this mechanism avoids the problematic
mass relations obtained in standard $\SU5$ GUTs in four dimensions. However, also the
successful predictions are lost, and the effective Yukawa matrices are to a high degree 
vacuum dependent. 

In fact, this statement generalizes to all superpotential
interactions.
A main point that we learn from the explicit calculation of the spectrum 
is therefore that a better
  understanding of the vacuum structure of orbifold models 
is required in order to 
reveal the mechanisms of string theory behind phenomenological observations, if existent. 

Closely linked to that issue is the problem of the presence of exotic fields in the model, which are
unavoidable in the false vacuum. Since they are vector-like, it is possible to generate
mass terms for them which imply their decoupling from the effective low-energy theory.
However, this is inevitably linked to the simultaneous generation of interactions, since
the total superpotential
originates from the sum of all consistent  higher-dimensional operators. 
The latter are restricted by continuous  gauge symmetries and 
a set of discrete symmetries. We explicitly calculated these symmetries for example vacua and found that
they explain the absence or presence of interactions in the effective superpotential. This provides
a tool for reversing the order of the construction of phenomenologically interesting vacua:
Instead of throwing dice and calculating the results, it is possible to construct vacua such that
they allow or forbid specific terms of phenomenological interest. Furthermore, we found an algorithm
which allows the calculation of the superpotential to arbitrary order in the singlets with better efficiency than
the naive order-by-order approach. 

All exotics at the $\SU5$ GUT fixed points are bulk fields in the studied model, almost
all arise from the second or the fourth twisted sectors of the $\Z{\rm 6-II}$ orbifold. 
Due to the existence of
inequivalent superpositions of states from equivalent Hilbert spaces, the latter
hypermultiplets have an overall multiplicity of three in the bulk. 
The intertwining of  gauge  and supersymmetry breaking
at the orbifold fixed points then allows for the universal decoupling of many exotics,
since uncharged pairs of fields with different superposition quantum numbers can be formed.
This mechanism is independent of the specific gauge embedding of the model. 

This universal decoupling required only four singlet fields with a non-zero vacuum expectation value.
We extended this set to two different maximal vacua, which cannot be enlarged further without either
violating matter parity or generating a $\mu$-term. This showed that in the studied model, the unification
of both the up- and the down-type Higgs with gauge fields of the six-dimensional bulk is impossible. 
Instead, we found that either a model with two Higgs pairs can be realized, which
is incompatible with gauge coupling unification, or partial gauge Higgs unification. The latter
corresponds to the situation that only the up-type Higgs is contained in the gauge vector,
with the desirable prediction of a large top-quark mass. In fact, the case of full gauge-Higgs
unification is phenomenologically not preferred over the partial gauge-Higgs unification scenario.
One may even argue that the symmetry between the fundamental and the anti-fundamental
representation of $\SU5$ is already broken by the identification of matter with the latter.
However, both vacua predict massless electrons and down-quarks and are thus excluded
as physical vacua.

This observation is not surprising. The whole analysis was based on one particular model,
contained in a whole landscape of possibilities. The motivation for the work was
not to find the standard model vacuum by pure luck, but to consider one model in
detail and gain a better understanding of the problems that form the typical obstacles 
on the way towards fully realistic vacua.

The main part of this paper was concerned with the issue of vacuum selection and 
the question of relating the latter to phenomenology.
Another source of problems may arise from the misinterpretation of remnants of the
orbifold construction. It is apparent from the  calculation of anomalies that the 
effective orbifold GUT model has localized Fayet--Iliopoulos terms of GUT scale at the fixed points.
Thus even though we could show explicitly that all anomalies of the effective orbifold field
theory vanish or can be canceled, the  trivial vacuum is inconsistent,
since it is not supersymmetric. This implies that singlet fields acquire large
vacuum expectation values during the transition to the supersymmetric solution, thereby 
 blowing up  singularities. The orbifold description can thus only be effective, and its
 validity as an approximate description is not clear a priori. For the bulk theory of the
 effective orbifold GUT in six dimensions it was possible to identify the fields of the orbifold
 with the moduli of a compactification of four dimensions of heterotic string theory on the $K3$ manifold.
This assigns new interpretations to the singlets of the orbifold model. One may hope
that the analogue orbifold GUT can be constructed in terms of an anisotropic compactification,
where the small dimensions  are compactified on $K3$, modulo an involution.
If the resulting model can also be matched with the $\Z{\rm 6-II}$ orbifold case, new
insights on the validity and interpretation of orbifolds may be gained, including
 new guiding principles for the search for physical vacua.

\section*{Acknowledgements}

I would like to thank Wilfried Buchm\"uller, Rolf Kappl, Christoph L\"udeling,  Jan M\"oller, Sa\'ul Ramos--S\'anchez, Kai Schmidt--Hoberg and Hagen Triendl for many valuable discussions and comments.

\appendix

\section{Orbifold details}
\label{app:orbifolds}

\subsection{Solving the fixed point equation}
\label{app:solveZ}

For any element $g=(\theta^k,m_a{\bf e}_a)\in S$ one can write down the fixed point condition
\begin{align}
	(\mathbbm 1 -g){\bf Z}_g &= 0\,. 
\end{align}
Thus the corresponding fixed point or fixed plane coordinates form the kernel of $(\mathbbm 1-g)$.
They can be expressed as
\begin{align}
	\left(1-\vartheta_{(i)}^k\right) Z^i_g &=\pi m_ae_a^i\,.
\end{align}
This equation has no solution if $g$ describes a pure translation in the plane $i$, 
which means $\vartheta_{(i)}^k=1, m_ae_a^i \neq 0$. Otherwise the solutions 
\index{Fixed point/plane/torus!solutions}
are given by
\begin{align}
\label{eq:Zisol}
Z^i_g &=\left\{\begin{array}{ll}
(1-\vartheta_{(i)}^k)^{-1} \pi m_a e_a^i\,, & \text{if $\vartheta_{(i)}^k \neq 1$}\,,\\
\text{unconstrained} \,, & \text{if $\vartheta_{(i)}^k = 1, m_a e_a^i=0$}\,.
\end{array}\right.
\end{align}
Here the first line describes a localization in the plane $i$, while for non-trivial $g$ 
the latter corresponds to a fixed plane solution.

In summary, the space group $S$ splits into two regimes. One generates winding modes
\index{Winding modes/numbers}
in at least one of the planes
 ($Z^i \rightarrow Z^i+m_ae_a^i\neq Z^i$).
All other  elements fulfill a fixed pointed equation. These regimes can now be decomposed
into disjoint conjugacy classes, yielding a non-redundant description of the space group.

\subsection[Space group selection rules for $\Z{\rm 6-II}$]{\boldmath Space group selection rules for $\Z{\rm 6-II}$}
\label{app:sgselrules}

A coupling
\begin{align}
    W &= \alpha \phi_1 \cdots \phi_M\,
\end{align}
 of $M$ multiplets at fixed points with space group elements
$g_{(l)}=(\theta^{k_{(l)}},m_a^{(l)})$, $l=1,\dots,M$, can
be present if the space group selection rule \index{String selection rules!space group}
(\ref{eq:summ0L}) is fulfilled,
\begin{align}
\label{eq:summ3}
    \sum_l  m_a^{(l)}{\bf e}_a &= \jmod{0}{\sum_l  \Lambda_{k_{(l)}}} \,.
\end{align}
The sub-lattices $\Lambda_k$ are given in Table \ref{tab:sublas}.
We count the number of multiplets $\phi_{l}$ with quantum numbers $k,n_i$ by
$N_{k,n_i}$, where $n_i\equiv n_6,n_3$ and $n_i\equiv n_2,n_2'$ in the $G_2,\SU3$- and the $\SO4$-plane, respectively.
 With Table \ref{tab:fppos} the left hand side of (\ref{eq:summ3}) then gives for the three planes
 \begin{subequations}
  \begin{align}
 \label{eq:summG2}
	&G_2 \,:  & \sum_l m_a^{(l)} {\bf e}_a =&
	 \left(N_{3,1}+N_{2,2}\right){\bf e}_1+
	 \left( N_{3,1}+N_{2,2}-N_{4,2} \right){\bf e}_2\,, \\
	\label{eq:summSU3}
	&\SU3 \,:  & \sum_l m_a^{(l)} {\bf e}_a =& 
	\left(N_{1,2}+N_{2,1}+N_{2,2}+N_{4,2} +N_{5,1}+N_{5,2}\right) {\bf e}_3 \nonumber \\
	& & &+
	 \left(N_{1,1}+N_{1,2}+N_{2,1}+N_{4,1}+N_{4,2} +N_{5,1}\right) {\bf e}_4 \,, \\
	&\SO4 \,:  & \sum_l m_a^{(l)} {\bf e}_a =&
	\left( N_{1,1,0}+N_{1,1,1}+N_{3,1,0}+N_{3,1,1}+N_{5,1,0}+N_{5,1,1}\right) {\bf e}_5 \nonumber \\
	&&&+\left( N_{1,0,1}+N_{1,1,1}+N_{3,0,1}+N_{3,1,1}+N_{5,0,1}+N_{5,1,1}\right) {\bf e}_6\,.	
\end{align}
\end{subequations}
The $\G2$-plane was discussed in Section \ref{sec:sgrulesZ6II} of the main text. Recall from table 
\ref{tab:sublas} that the sub-lattices
relevant for the $\SU3$- and the $\SO4$-plane are 
\begin{subequations}
\begin{align}
&\SU3 \,: &\{a{\bf e}_1+3b{\bf e}_2|a,b\in \Z{ }\}&\subset\Lambda_2\,, & &\phantom{AA}\\
&\SO4\,: &\{2 a{\bf e}_5+2{\bf e}_6 |a,b\in \Z{ }\}&\subset\Lambda_3\,.
\end{align}
\end{subequations}
The condition (\ref{eq:summ3}) can thus be evaluated as
\begin{subequations}
\begin{align}
&\SU3 \,: & 2\left( N_{1,2}+N_{2,1}+N_{4,2} +N_{5,1}\right) + N_{1,1}
+N_{2,2}+N_{4,1} +N_{5,2}&=\jmod 03\,,  \\
&\SO4 \,: & N_{1,1,0}+N_{1,1,1}+N_{3,1,0}+N_{3,1,1}+N_{5,1,0}+N_{5,1,1} &=\jmod 02 \,, \\
&& N_{1,0,1}+N_{1,1,1}+N_{3,0,1}+N_{3,1,1}+N_{5,0,1}+N_{5,1,1}& =\jmod 02 \,.
\end{align}
\end{subequations}
These conditions are equivalent to the rules
\begin{subequations}
\begin{align}
&\SU3 \,: & \sum_l k^{(l)} n^{(l)}_3 &=\jmod 03\,, \\
&\SO4 \,: & \sum_l k^{(l)} n^{(l)}_2 &=\jmod 02\,, \\
&& \sum_l k^{(l)} {n_2'}^{(l)} &=\jmod 02\,, & &\phantom{AAA}
\end{align}
\end{subequations}
as stated in (\ref{eq:srules}).

\subsection{The gamma phases}
\label{app:gammas}

\subsubsection{The general case}

Consider a sector $\mathcal H_{[g]}$, with representative space group element $g=(\theta^k,m_a{\bf e}_a)$.
As explained in Section \ref{sec:quantization}, physical states are superpositions 
of states which are localized at the various fixed points ${\bf z}_l \in \mathbbm C^3$, 
corresponding to the elements $l \in [g]$, 
 \begin{align}
\label{eq:chig3}
    \ket{\chi}_{[g]} &= \sum_{l \in [g]} \chi_l \ket{{\bf z}_l}\,,
\end{align}
with coefficients  $\chi_l$. The gamma phases are then defined by the eigenvalues of the action of
other elements $h\in S$ on $\mathcal H_{[g]}$. 

Any element of the conjugacy class $l\in[g]$ arises from the representative $g$ by conjugation with
a twist $(\theta^n,0)$ and a translation $(\mathbbm 1,m_a'{\bf e}_a)$,
\begin{align}
l&=\big(\mathbbm 1,m_a'{\bf e}_a\big)\big(\theta^n,0\big)g\big(\theta^{-n},0\big)
\big(\mathbbm 1,-m_a'{\bf e}_a\big)\,.
\end{align}
For the associated fixed point ${\bf z}_l$ this means
\begin{align}
	\ket{{\bf z}_l} &= \big(\mathbbm 1,m_a'{\bf e}_a\big)\big(\theta^n,0\big) \ket{{\bf z}_g}\,,
\end{align}
and a natural ansatz for the superposition (\ref{eq:chig3}) is
\begin{align}
\label{eq:chig4}
 \ket{\chi}_{[g]} &=  \underbrace{\left( 
 \sum_{m_a'{\bf e}_a \in \Lambda}
      \chi_{m'_a{\bf e}_a}' \left(\mathbbm 1,m_a'{\bf e}_a\right) \right)}_{X}
    \underbrace{\left( \sum_{n=0}^{N-1} 
      \chi_{n} \big(\theta^n,0\big) \right)
      \vphantom{
      \left( 
 \sum_{m_a'{\bf e}_a \in \Lambda}
      \chi_{m'_a{\bf e}_a}' \left(\mathbbm 1,m_a'{\bf e}_a\right) \right)
      }}_{Y} \ket{{\bf z}_{g}} \,.
\end{align}
The first bracket $X$ sums over infinitely many terms which correspond to infinitely many
lattice translations within $\Lambda$.
The second bracket $Y$ represents $N$ rotations of the 
translational part of  $g\in [g]$, where $N$ is the order of the orbifold.

The coefficients $\chi_{m'_a{\bf e}_a}'$ and $\chi_n$ in (\ref{eq:chig4}) are not fully arbitrary. 
First, we require compatibility with the group multiplication rule,
\begin{align}
\label{eq:chipchi1}
	\chi_{(m_a+m'_a){\bf e}_a}'&=\chi_{m_a{\bf e}_a}'\chi_{m'_a{\bf e}_a}'  \,, &
	\chi_{n+n'}&=\chi_{n}\chi_{n'}&
	 \chi_{\theta^k m_a{\bf e}_a}'&=\chi_{m_a{\bf e}_a}'\,,
\end{align}
where the latter relation follows from the explicit form of the inverse element (\ref{eq:inverse})
and should hold for any $k$.
The rotation can be expressed on the coefficients $m_a'$ 
by use of the Coxeter element \index{Coxeter element} $C^T$ of the geometry, which is the rotational matrix defined as
 \begin{align}
 \label{eq:C0}
	\theta  {\bf e}_a &\equiv C^T_{ab}  {\bf e}_b\,.
\end{align}
The requirements (\ref{eq:chipchi1}) are then fulfilled by the exponential ansatz
\begin{align}
\label{eq:chipchi2}
	\chi_{m'_a{\bf e}_a}'&= e^{2\pi i x_a m'_a}\,,  
	\hspace{1cm}
     \chi_n =    e^{2\pi i y n}\,,
\end{align}
where $x_1,\dots,x_6$ and $y$ are some coefficients, constrained by
\begin{align}
\label{eq:cnstr0}
	x_a\big(\delta_{ab}-C^{-1}_{ab}\big) &=\jmod 01\,, \hspace{1cm}
	N y = \jmod 01\,.
\end{align}

Second, elements
$h'=(\theta^n,m_a'{\bf e}_a)$ which do commute with the representative
$g=(\theta^k,m_a{\bf e}_a)$
do not affect the state $\ket{{\bf z}_g}$, and hence should
not appear with different coefficients. The property $[g,h']=0$ can be expressed as
\begin{align}
\label{eq:hggh0}
	\big(\mathbbm 1-\theta^n\big)m_a{\bf e}_a&=\big(\mathbbm 1-\theta^k\big)m'_a{\bf e}_a \,.
\end{align}
For the coefficients $m'_a$ this reads
\begin{align}
\label{eq:hggh1}
	\big(\delta_{ab}-C_{ab}^n\big)m_b&=\big(\delta_{ab}-C_{ab}^k\big)m'_b \,.
\end{align}
We now aim to associate the right-hand side of this equation with  $x_a m_a'$ in (\ref{eq:chipchi2}),
and the left-hand side with $y n$, with a relative minus sign.
However, this requires a linear expression in $n$, up to integers.
With a new set of constrained constants $\eta_a$ the linearization can be realized:
\begin{align}
	 &\eta_a \big(\delta_{ab}-C_{ab}^n\big) = \jmod{n \,\eta_a \big(\delta_{ab}-C_{ab}\big)}{1}\,,
	\hspace{1cm}\text{for all $n$}\,.
\end{align}
This leads to the improved, but still preliminary ansatz
\begin{align}
\label{eq:chipchi3}
    \chi_{m'_a{\bf e}_a}'&= e^{2\pi i  k \gamma_a' m'_a}\,, &
     \chi_n &= e^{-2\pi i  n \gamma_a' m_a} \,, &
      \gamma_a'&=  \eta_b \big(\delta_{ba} -C_{ba}\big)\,,
\end{align}
for which the constraints (\ref{eq:cnstr0}) become
\begin{align}
\label{eq:cnstr1}
	k\, \eta_a \big(\delta_{ab} -C_{ab}\big)\big(\delta_{bc} -C_{bc}^{-1}\big) &=\jmod 01\,, &
	N  \gamma_a' m_a&=\jmod 01\,.
\end{align}
Note that the two brackets combine to a diagonal matrix, whose entries are given by $2(1-{\rm Re}\,\vartheta_{(i)})$,
where $\vartheta_{(i)}$ is the twist in the corresponding plane.

So far we have realized the group transformation laws on the coefficients, and we have ensured that 
elements which commute with the representative element have unit coefficients. They are thus 
equivalent to the trivial element $(\mathbbm 1,0)$ within the product of sums $XY$. However,
one can add an additional phase to the terms in $Y$ which is also trivial for commuting elements, without
cancelation from the phase in $X$. 
Let $N_{(g)}$ be the smallest positive integer for which 
\begin{align}
\label{eq:Ng0}
   	\big(\mathbbm 1-\theta^{N_{(g)}}\big) m_a{\bf e}_a &
	\in \Lambda_k\,, \hspace{1cm}
	1 \leqslant N_{(g)} \leqslant k\,.
\end{align}
Note that $N_{(g)}$ is a divisor of $N$.
By comparison with (\ref{eq:hggh0}) one finds that for commuting elements $(\theta^n,m_a'{\bf e}_a)$,
$n$ must be a multiple of $N_{(g)}$. This means that the phase 
\begin{align}
	e^{-2 \pi i \gamma n}\,, \hspace{1cm} N_{(g)} \gamma = \jmod 01\,,
\end{align}
is one for commuting elements, and we have introduced a new constant $\gamma$.
Our final ansatz is then
\begin{align}
\label{eq:chipchi3}
    \chi_{m'_a{\bf e}_a}'&= e^{2\pi i  k \gamma_a' m'_a}\,, \hspace{1cm}
     \chi_n = e^{-2\pi i  n \gamma_a' m_a} e^{-2\pi i  \gamma n}\,,
\end{align}
and one can now study 
the transformation behavior of the superposition (\ref{eq:chig4}), with
the suggested
coefficients.

 As we will show, they fulfill an eigenvalue equation with respect to arbitrary operators $h=(\theta^{\tilde k},
\tilde m_a{\bf e}_a)\in  S$, of the form
\begin{align}
	 h \ket{\chi}_{[g]}= \big(\theta^{\tilde k},\tilde m_a{\bf e}_a\big) X Y \ket{{\bf z}_g}&=
	 \big(\mathbbm 1,\tilde m_a{\bf e}_a\big) \big(\theta^{\tilde k},0\big) XY  \ket{{\bf z}_g}\nonumber \\
	 &=
	  \Big( \big(\mathbbm 1,\tilde m_a{\bf e}_a\big) X \Big)
	 \left(  \big(\theta^{\tilde k},0\big) Y \right) \ket{{\bf z}_g} \nonumber \\
	 &=
	 \tilde \lambda_{\tilde m } \lambda_{\tilde k} \ket{\chi}_{[g]}\,,
\end{align}
where 
 $\tilde \lambda_{\tilde m }$ and $ \lambda_{\tilde k}$ are  eigenvalues of $X$ and $Y$
 under translations and rotations, respectively.
This follows the spirit of the separation of space group elements into a
translational and a rotational part, which is the crucial feature of the decomposition (\ref{eq:chig4}).

On $X$, rotations $(\theta ,0)$ act  as
\begin{align}
	\big(\theta,0\big) X &=
	 \sum_{m_a'{\bf e}_a \in \Lambda}
      e^{2\pi i  k \gamma_a' m'_a}\big( \mathbbm 1, 
      \underbrace{\theta   m_a'{\bf e}_a}_{m_a''{\bf e}_a} \big)\big(\theta,0\big)
       \nonumber\\
      &=\sum_{m_a''{\bf e}_a \in \Lambda}
      e^{2\pi i k \gamma_a' C_{ab}^{-1} m_b''}\big( \mathbbm 1, m_a''{\bf e}_a \big)\big(\theta ,0\big)
      =X\big(\theta,0\big)
      \,,\label{eq:Xtrans0}
\end{align}
where the rotational invariance of $k \gamma_a' m_a'$ from (\ref{eq:cnstr0}) was used.
Thus $X$ commutes with rotations, as a direct consequence of (\ref{eq:chipchi1}).

Furthermore, $X$ fulfills an eigenvalue equation with respect to translations:
 \begin{align}
 	\big(\mathbbm 1,\tilde m_a{\bf e}_a\big)X     &=
	 \sum_{m_a'{\bf e}_a \in \Lambda}
      e^{2 \pi i k \gamma_a'    m_a' }\big( \mathbbm 1, 
      \underbrace{  m_a'{\bf e}_a+\tilde m_a{\bf e}_a}_{m_a''{\bf e}_a} \big) 
       \nonumber\\ 
      &=e^{-2 \pi i  k \gamma_a'  \tilde m_a  }
      \sum_{m_a''{\bf e}_a \in \Lambda}
      e^{2 \pi i k \gamma_a'  m_a'' }\big( \mathbbm 1, m_a''{\bf e}_a \big) 
      =e^{-2 \pi i k \gamma_a'  \tilde m_a  } X\,.\label{eq:Xtrans1}
\end{align}
 
Finally we consider the action of rotations $\theta$ on $Y$:
 \begin{align}
	\big(\theta ,0\big) Y&= \sum_{n=0}^{N -1} 
        e^{-2 \pi i n \gamma_a' m_a} e^{-2\pi i  \gamma n}\big(\underbrace{\theta^{n+1}}_{\theta^{n'}},0\big) \nonumber \\
       &= e^{2 \pi i \gamma_a' m_a }e^{2\pi i  \gamma}\sum_{n'=0}^{N }  e^{-2 \pi i n' \gamma_a' m_a }
       e^{-2\pi i  \gamma n'} \big(\theta^{n'},0\big)
       =e^{2 \pi i \gamma_a' m_a } e^{2\pi i  \gamma}Y\,.
        \label{eq:Ytrans0}
\end{align}

In summary, the eigenstates \index{Space group!eigenstate}
of $h=(\theta^{\tilde k},\tilde m_a{\bf e}_a)$ in the sector $\mathcal H_{[g]}$,
$g=(\theta^{k},m_a{\bf e}_a)$, are \index{Space group!transformation!phase}
\begin{align}
\label{eq:eigenstate0}
\ket{ \gamma',\gamma }_{[g]}  &=   \left( 
 \sum_{m_a'{\bf e}_a \in \Lambda}
      e^{2 \pi i k \gamma_a' m_a' }\left(\mathbbm 1,m_a'{\bf e}_a\right) \right) 
     \left( \sum_{n=0}^{N-1} 
     e^{-2 \pi i n \gamma_a' m_a  }e^{-2\pi i  \gamma n}\big(\theta^n,0\big) \right)  \ket{{\bf z}_{g}}\,,	\\
    \ket{\gamma',\gamma  }_{[g]}
     & \stackrel{h}{\mapsto} 
    e^{2 \pi i  \gamma_a' ( \tilde  k m_a-k \tilde m_a)}e^{2 \pi i \gamma \tilde k}
 \ket{ \gamma',\gamma  }_{[g]}\,,
\end{align}
where the constants in $\gamma'=(\gamma_1',\dots,\gamma_6')$ and $\gamma$ 
\index{Gamma-phases} fulfill the constraints
\begin{subequations}
\label{eq:condglist0}
\begin{eqnarray}
\gamma_a' &=&\eta_b \big(\delta_{ba} -C_{ba}\big)\,, \label{eq:cgl1} \\
N_{(g)} \gamma &=& \jmod 01\,, \label{eq:cgl2} \\
 \eta_a \big(\delta_{au}-C_{au}^n\big) &=& \jmod{n\, \eta_a \big(\delta_{au}-C_{au}\big)}{1}\,,
	\hspace{1cm}\text{for all $n,u$}\,, \label{eq:cgl3} \\
	2 k \left(1-{\rm Re}\,\vartheta_{(i_u)}\right) \eta_u &=&\jmod 01\,, \hspace{3.55cm}\text{for all $u$}\,,
	 \label{eq:cgl4} 
\end{eqnarray}
\end{subequations}
where $(i_1,\dots,i_6)=(1,1,2,2,3,3)$ are the labels of the planes associated with the lattice vectors ${\bf e}_u$,
and $N_{(g)}$ was defined in (\ref{eq:Ng0}).
Note that the  condition 
\begin{align}
	N \gamma_a' m_a &=\jmod 01
\end{align}
is automatically satisfied for solutions of the above constraints. This immediately follows from (\ref{eq:cgl2})
with the choice $n=N$.

\subsubsection{\boldmath The $\Z{\rm 6-II}$ example}

For the $\Z{\rm 6-II}$ geometry, the transpose $C$ of the Coxeter element
 \index{Coxeter element} from (\ref{eq:C0}) reads
\begin{align}
C&=\left( \begin{array}{cccccc}
-1 &1 & \\
-3 & 2 & \\
 & & -1 & 1\\
 & & -1 & 0\\
 & & & & -1 &0\\
 & & & & 0& -1
 \end{array}\right)\,,
\end{align}
where all entries which are not shown are zero.
The conditions (\ref{eq:condglist0}) for the space group eigenstates
(\ref{eq:eigenstate0}) for the $\Z{\rm 6-II}$ geometry
are then solved by \index{Z6-II orbifold@$\Z{\rm 6-II}$ orbifold!quantum numbers $\gamma'$}
\begin{align}
	 \gamma'=\left(0,0,\frac{l_3}{3},\frac{l_3}{3},\frac{l_5}{2},\frac{l_6}{2}\right)\,, 
\end{align}
in terms of integers $l_3,l_5,l_6$.
Thus the order of the $a$-th entry of $\gamma'$ corresponds to the order of the lattice vector ${\bf e}_a$,
and the two entries for the $\SU3$-plane are identical since ${\bf e}_3$ and ${\bf e}_4$ are
connected by a twist.

 \section{Lie Group Basics}
\label{app:lie}

This section is a very brief review of selected well known facts about Lie groups\index{Lie group/algebra}. Some of them are needed in the main text, 
especially for the identification of the representations of states, which arise in the effective theories derived from
the heterotic string.

\subsubsection*{Weights}
A Lie group $G$ has an associated Lie algebra $\mathcal{G}$. Its commuting generators form
 the Cartan sub-algebra \index{Cartan generators}
  $\mathcal{H}= \lbrace H_i \rbrace $ with dimension $r$, dubbed the rank of $G$.
 
For any representation of dimension $D$, the action of the elements of
the Cartan sub-algebra $H_i\in \mathcal{H}$ can be simultaneously diagonalized. The states then obey a relation
\begin{equation}
H_i \ket{\mu ,  D} = \mu_i \ket{\mu ,  D}
\end{equation}
where the weight vector $\mu=(\mu_1,\dots,\mu_r)$ collects the weights $\mu_i$ \index{Weights} 
of the representation $D$.
In the main text, the weight vectors are given by the shifted left-moving momenta
${\bf p}_{\rm sh}$ of the generated states.

\subsubsection*{Roots}
Roots $\alpha_i$ \index{Roots} are the weights of the adjoint representation.
They obey
\begin{equation}
[H_i,E_\alpha]= \alpha_i E_\alpha , \hspace{1cm} 
[E_\alpha,E_{-\alpha}]= \alpha \cdot H
\end{equation}
and the corresponding states $E_\alpha$ act as raising and lowering operators:
\begin{equation}
[E_\alpha,E_\beta]\sim E_{\alpha + \beta} , \hspace{1cm} 
H_i E_{\pm \alpha} \ket{\mu, D} = (\mu_i \pm \alpha_i) E_{\pm \alpha} \ket{\mu, D}.
\end{equation}
For each root eigenstate $E_\alpha$, there is a natural $SU(2)$ subgroup generated by
\begin{equation}
E^\pm \equiv |\alpha|^{-1} E_{\pm \alpha} , \hspace{1cm} 
E_3 \equiv |\alpha|^{-2} \alpha \cdot H ,
\end{equation}
such that the $E^\pm$ raise and lower the eigenvalue of $E_3$ of a state, respectively.
In general,
\begin{equation}
E_3 \ket{\mu,D} = \frac {\alpha \cdot \mu}{\alpha^2} \ket{\mu,D}\,,
\end{equation}
and $ \alpha \cdot \mu/\alpha^2$ in any representation is an integer or half integer.
As always for representations of $SU(2)$ there are a non-negative integers $p,q$ such that
\begin{subequations}
\label{eq:raislow}
\begin{eqnarray}
(E^+)^p \ket{\mu,D} \neq 0 , \hspace{1cm}  & &  (E^+)^{p+1} \ket{\mu,D} = 0  \\
(E^-)^q \ket{\mu,D} \neq 0 , \hspace{1cm}  & & (E^-)^{q+1} \ket{\mu,D} = 0 
\end{eqnarray}
\end{subequations}
In that case, the highest and lowest weights are $(\mu + p \alpha)$ and $(\mu - q \alpha)$, respectively, and
\begin{subequations}
\begin{eqnarray}
E_3 \ket{\mu+ p \alpha,D} & = & \left( \frac {\alpha \cdot \mu}{\alpha^2} + p \right) \ket{\mu+ p \alpha,D}
 \equiv j \ket{\mu+ p \alpha,D} , \\
E_3 \ket{\mu - q \alpha,D} & = & \left( \frac {\alpha \cdot \mu}{\alpha^2} -q \right) \ket{\mu - q \alpha,D}
 \equiv- j \ket{\mu - q \alpha,D} , 
\end{eqnarray}
\end{subequations}
where $j$ is the spin of the representation.
One then finds the formula
\begin{equation}
\label{eq:mastereqB}
\frac{\alpha \cdot \mu}{\alpha^2} = -\frac 12 ( p-q )\,.
\end{equation}
The right hand side gives a label to the state with weight $\mu$, compare e.g. the $j=3/2$ representation, where $(p-q)/2 \in \lbrace -3/2,-1/2,1/2,3/2 \rbrace$. The left hand side, when applied
to the adjoint representation, gives the cosine of the angle between roots. The integer nature of $p,q$
then restricts the possible angles to 90,120,135 or 150 degrees. These are the angles appearing  in Dynkin- and root diagrams.

\subsubsection*{Simple Roots}

A root is said to be positive, if the first non-zero entry in some fixed basis is positive. It is
negative, if the first non-zero entry is negative. With this definition in the adjoint representation 
the positive roots correspond to raising, the negative to lowering operators.
For roots $\mu,\nu$ we define
\begin{center}
$\mu > \nu$ \quad if $(\mu - \nu)$ is positive.
\end{center}

Simple roots \index{Simple roots} are positive roots that cannot be written as a sum of other positive roots.
The number of simple roots equals the rank $r$ of the group. All other roots and thus the whole algebra
can be deduced from linear combinations of simple roots.

\subsubsection*{Cartan Matrix}
The Cartan matrix \index{Cartan matrix} is defined as
\begin{equation}
A_{ij} \equiv \frac{2}{\alpha_i^2} \left( \alpha_i \cdot \alpha_j \right)\,.
\end{equation}
For fixed $j$, its entries give  
the $(p-q)$ values of the $SU(2)_i$'s associated with the simple roots
$\alpha_i$, when acting on a the state $E_{\alpha_j}$. 
The Cartan matrix is characteristic for a group, it contains the same information as
the associated Dynkin diagram. Non-zero off-diagonal entries, or equivalently non-zero angles between roots, describe
a non-trivial overlap of the contained $\SU2_i$ subgroups, generated by the simple roots $E_{\alpha_i}$.

\subsubsection*{Dynkin Labels and Fundamental Weights}
In the spirit of (\ref{eq:mastereqB}) one can define the `Cartan--Weyl labels' $l_i$ 
\index{Cartan--Weyl labels} for
 any state $\ket{\mu,D}$ of a irreducible representation $D$
(here no summation over $i$)
\begin{align}
\label{eq:myli}
l_i\equiv \frac{2}{\alpha_i^2} \left( \alpha_i \cdot \mu \right) & = -\frac12(p_i-q_i)\,, & i&=1,\dots,r\,.
\end{align}
The integers $p_i,q_i$ specify how often the simple root $E_{\alpha_i}$ can be applied 
before the state reaches zero, cf. (\ref{eq:raislow}).
Now consider the application of $E_{\pm\alpha_j}$ to the state $\ket{\mu,D}$. It results in
$\mu \mapsto \mu \pm \alpha_j$, and
\begin{align}
	l_i \stackrel{E_{\pm\alpha_j}}{\longmapsto} l_i\pm\frac{2}{\alpha_i^2} \left( \alpha_i \cdot \alpha_j \right)
	=l_i\pm A_{ij} \,.
\end{align}
The difference of the vectors $l=(l_1,\dots, l_r)$ for two different states of a representation
is thus always given by one of the columns of the Cartan matrix. This is the property
which is used in the main text in order to identify the weights $\mu={\bf p}_{\rm sh}$
that correspond to the same representation $D$.

For any irreducible representation $D$
 there is a maximal weight $\mu$ such that
\begin{equation}
E_{\alpha_i} \ket{\mu, D} =0 \hspace{1cm} \forall \, i.
\end{equation}
Thus for this weight one has
$p_i=0$ for all $i$, and the set of integers 
\begin{equation}
(l_1, \dots , l_r)
\end{equation} 
fully determines the representation. These labels $l_i$ are called Dynkin labels 
\index{Dynkin labels} of the representation $D$.

Furthermore, one defines fundamental weights $\mu^i$,
\begin{equation}
 \frac{2}{\alpha_i^2} \left( \alpha_i \dots \mu^j \right) = \delta_i^j\,,
\end{equation}
which imply the decomposition of general weights as
\begin{equation}
\label{eq:decommu}
\mu = \sum_{i=1}^r l_i \mu^i .
\end{equation}
This gives another interpretation for the quantities $l_i$ from (\ref{eq:myli}).
In (\ref{eq:decommu}) they
play the role of coefficients of the weight vector with respect
to a basis in weight space, spanned by the fundamental weights $\mu^j$.

 \section{Anomalies with Abelian factors}
 \label{app:anomalies}

 \subsection{Bulk anomalies}
 \label{app:traces}

The contributions to the bulk anomaly polynomial \index{Anomaly!polynomial!bulk}
 $I_8$ which involve Abelian field strength
two-forms \index{Field strength two-form} 
$\hat F_u, u=1,\dots,5$, canonically normalized as in (\ref{eq:hattu}), take the form (\ref{eq:iabs}),
\begin{subequations}
\label{eq:iabs0}
\begin{eqnarray}
	I_{(1/2)}^{\rm Abelian}(F^4,R^0)&=&\frac 83 \sum_A \sum_{u=1}^5 a_A^{u} \big({\rm tr}\,F^3_A\big) \hat F_u
	+4 \sum_A \sum_{u,v=1}^5 b_A^{uv} \big({\rm tr}\,F^2_A\big)
	 \hat F_u \hat F_v 
	 \nonumber \\
	 &&+ \frac 23 \sum_{u,v,w,x=1}^5 c^{uvwz} \hat F_u  \hat F_v  \hat F_w  \hat F_x  \,,\\
	 I_{(1/2)}^{\rm Abelian}(F^2,R^2)&=&-\frac 16 ({\rm tr}\, R^2)  \sum_{u,v=1}^5 d^{uv}  \hat F_u  \hat F_v\,,
\end{eqnarray}
\end{subequations}
where $A=\SU6,\SU3,\SO8$ labels the non-Abelian gauge groups in the bulk.
The different relative multiplicities arise from the various choices of associating the non-Abelian
groups with some of the field strength two-forms, and the Abelian group factors with the others.

The coefficients $a_A^u,b_A^{uv},c^{uvwx},d^{uv}$ follow from summations over all hyperinos $\tilde h$
in the bulk. Each of them transforms in a representation of the gauge group and has charges under the $\U1$'s,
\begin{align}
\label{eq:hyperinos}
	\tilde h\,: \hspace{1cm} \left\{\begin{array}{l}
	\text{representation $\mathcal R_A(\tilde h)$ w.r.t. group $A$},\\ 
	\text{charge $Q_u(\tilde h)$ w.r.t. $ U(1)_u$}.
	\end{array}\right.
\end{align}
For the conversion of this information into quantities of relevance for (\ref{eq:iabs0})
we use the identities from Table \ref{tab:traces} and the definition (\ref{eq:hattu}),
\begin{align}
\label{eq:vw}
	{\rm tr}_{\mathcal R} F^2_A&\equiv v_A^{\mathcal R} \,{\rm tr}\, F^2_A\,,&
	{\rm tr}_{\mathcal R} F^3_A&\equiv w_A^{\mathcal R} \,{\rm tr}\, F^3_A\,, &
	Q_u&= \sqrt 2 |t_u|\, \hat Q_u\,.
\end{align}
The hyperino charge sums then read
\begin{subequations}
\begin{eqnarray}
	a_A^u&=&\sum_{\tilde h} w_A^{\mathcal R(\tilde h)} \hat Q_u(\tilde h)\,,\\
	b_A^{uv}&=&\sum_{\tilde h} v_A^{\mathcal R(\tilde h)} \hat Q_u(\tilde h) \hat Q_v(\tilde h)\,,\\
	c^{uvwx}&=&\sum_{\tilde h}  \hat Q_u(\tilde h) \hat Q_v(\tilde h) \hat Q_w(\tilde h) \hat Q_x(\tilde h)\,,\\
	d^{uv}&=&\sum_{\tilde h}   \hat Q_u(\tilde h) \hat Q_v(\tilde h)\,.
\end{eqnarray}
\end{subequations}

We now evaluate these sums for the bulk hypermultiplets listed in the Tables \ref{tab:bulkU} and
\ref{tab:bulkT2T4}. With $w_A^{\mathcal R}$ from Table \ref{tab:vw}
\begin{table}
  \centering 
  \footnotesize
  \begin{tabular}{l| l|| l| l|| l| l|| l| l|| l}
	\multicolumn{2}{c||}{$\SU3$} &\multicolumn{2}{c||}{$\SU4$} &\multicolumn{2}{c||}{$\SU5$}&
	\multicolumn{2}{c||}{$\SU6$} & \multicolumn{1}{c}{$\SO8$} \jvsb \\
	\hline
	\jvsb $v^{\bf 3}=1$ & $w^{\bf 3}=1$ &
	 $v^{\bf 4}=1$ & $w^{\bf 4}=1$ &
	 $v^{\bf 5}=1$ & $w^{\bf 5}=1$ &
	 $v^{\bf 6}=1$ & $w^{\bf 6}=1$
	&$v^{\bf 8}=1$ \\
	\jvsb $v^{\bf \bar 3}=1$ & $w^{\bf \bar 3}=-1$ &
	 $v^{\bf \bar 4}=1$ & $w^{\bf \bar 4}=-1$ &
	  $v^{\bf \bar 5}=1$ & $w^{\bf \bar 5}=-1$ &
	 $v^{\bf \bar 6}=1$ & $w^{\bf \bar 6}=-1$
	&$v^{{\bf 8}_s}=1$   \\
	\jvsb   &   &
	   $v^{\bf 6}=2$ & $w^{\bf 6}=0$ &
	   $v^{\bf 10}=3$ & $w^{\bf 10}=1$&
	 $v^{\bf 20}=6$ & $w^{\bf 20}=0$
	&$v^{{\bf 8}_c}=1$  \\
	\jvsb   &   &
	   &   &
	   $v^{\bf \bar {10}}=3$ & $w^{\bf \bar {10}}=-1$&
	   &  
	&  
\end{tabular}
  \caption{The constants $v_A^{\mathcal R}$ and $w_A^{\mathcal R}$ for a choice
  of gauge groups $A$ and representations $\mathcal R$.
  Note that $\SO8$ has no third order invariant.}\label{tab:vw}
\end{table}
and the given charges follows immediately that
\begin{align}
	a_{\SU6}^u&=a_{\SU3}^u=a_{\SO8}^u=0 \,, \hspace{1cm} \text{for all $u=1,\dots,5$.}
\end{align}
\begin{table}[t]
\centering
\footnotesize
\begin{tabular}{c|c|c|c|c||c|c|c|c|c||c|c|c|c|c||c|c|c|c|c}
\jvsb $u$ & $v$ & $w$ & $x$ & $h$ & $u$ & $v$ & $w$ & $x$ & $h$ &$u$ & $v$ & $w$ & $x$ & $h$ &$u$ & $v$ & $w$ & $x$ & $h$  \\
\hline
\hline
\jvsb$1$ & $1$ & $1$ & $1$ & $\frac{9}{2}$ & $1$ & $2$ & $2$ & $5$ & $0$ & $2$ & $2$ & $2$ & $3$ & $0$ & $2$ & $5$ & $5$ & $5$ & $0$ \\
\jvsb$1$ & $1$ & $1$ & $2$ & $-\frac{3}{4}$ & $1$ & $2$ & $3$ & $3$ & $-\frac{1}{4}$ & $2$ & $2$ & $2$ & $4$ & $-\frac{3}{4}$ & $3$ & $3$ & $3$ & $3$ & $3$ \\
\jvsb$1$ & $1$ & $1$ & $3$ & $0$ & $1$ & $2$ & $3$ & $4$ & $0$ & $2$ & $2$ & $2$ & $5$ & $0$ & $3$ & $3$ & $3$ & $4$ & $0$ \\
\jvsb$1$ & $1$ & $1$ & $4$ & $-\frac{3}{4}$ & $1$ & $2$ & $3$ & $5$ & $0$ & $2$ & $2$ & $3$ & $3$ & $\frac{5}{4}$ & $3$ & $3$ & $3$ & $5$ & $\frac{3}{2 \sqrt{2}}$ \\
\jvsb$1$ & $1$ & $1$ & $5$ & $0$ & $1$ & $2$ & $4$ & $4$ & $-\frac{1}{4}$ & $2$ & $2$ & $3$ & $4$ & $0$ & $3$ & $3$ & $4$ & $4$ & $\frac{3}{2}$ \\
\jvsb$1$ & $1$ & $2$ & $2$ & $\frac{3}{2}$ & $1$ & $2$ & $4$ & $5$ & $0$ & $2$ & $2$ & $3$ & $5$ & $\frac{1}{2 \sqrt{2}}$ & $3$ & $3$ & $4$ & $5$ & $0$ \\
\jvsb$1$ & $1$ & $2$ & $3$ & $0$ & $1$ & $2$ & $5$ & $5$ & $-\frac{1}{4}$ & $2$ & $2$ & $4$ & $4$ & $\frac{7}{4}$ & $3$ & $3$ & $5$ & $5$ & $\frac{3}{2}$ \\
\jvsb$1$ & $1$ & $2$ & $4$ & $-\frac{1}{4}$ & $1$ & $3$ & $3$ & $3$ & $0$ & $2$ & $2$ & $4$ & $5$ & $0$ & $3$ & $4$ & $4$ & $4$ & $0$ \\
\jvsb$1$ & $1$ & $2$ & $5$ & $0$ & $1$ & $3$ & $3$ & $4$ & $-\frac{1}{4}$ & $2$ & $2$ & $5$ & $5$ & $\frac{7}{4}$ & $3$ & $4$ & $4$ & $5$ & $\frac{1}{2 \sqrt{2}}$ \\
\jvsb$1$ & $1$ & $3$ & $3$ & $\frac{5}{4}$ & $1$ & $3$ & $3$ & $5$ & $0$ & $2$ & $3$ & $3$ & $3$ & $0$ & $3$ & $4$ & $5$ & $5$ & $0$ \\
\jvsb$1$ & $1$ & $3$ & $4$ & $0$ & $1$ & $3$ & $4$ & $4$ & $0$ & $2$ & $3$ & $3$ & $4$ & $-\frac{1}{4}$ & $3$ & $5$ & $5$ & $5$ & $\frac{3}{2 \sqrt{2}}$ \\
\jvsb$1$ & $1$ & $3$ & $5$ & $\frac{1}{2 \sqrt{2}}$ & $1$ & $3$ & $4$ & $5$ & $0$ & $2$ & $3$ & $3$ & $5$ & $0$ & $4$ & $4$ & $4$ & $4$ & $6$ \\
\jvsb$1$ & $1$ & $4$ & $4$ & $\frac{7}{4}$ & $1$ & $3$ & $5$ & $5$ & $0$ & $2$ & $3$ & $4$ & $4$ & $0$ & $4$ & $4$ & $4$ & $5$ & $0$ \\
\jvsb$1$ & $1$ & $4$ & $5$ & $0$ & $1$ & $4$ & $4$ & $4$ & $-\frac{3}{4}$ & $2$ & $3$ & $4$ & $5$ & $0$ & $4$ & $4$ & $5$ & $5$ & $2$ \\
\jvsb$1$ & $1$ & $5$ & $5$ & $\frac{7}{4}$ & $1$ & $4$ & $4$ & $5$ & $0$ & $2$ & $3$ & $5$ & $5$ & $0$ & $4$ & $5$ & $5$ & $5$ & $0$ \\
\jvsb$1$ & $2$ & $2$ & $2$ & $-\frac{3}{4}$ & $1$ & $4$ & $5$ & $5$ & $-\frac{1}{4}$ & $2$ & $4$ & $4$ & $4$ & $-\frac{3}{4}$ & $5$ & $5$ & $5$ & $5$ & $6$ \\
\jvsb$1$ & $2$ & $2$ & $3$ & $0$ & $1$ & $5$ & $5$ & $5$ & $0$ & $2$ & $4$ & $4$ & $5$ & $0$ &  &  &  &  &  \\
\jvsb$1$ & $2$ & $2$ & $4$ & $-\frac{1}{4}$ & $2$ & $2$ & $2$ & $2$ & $\frac{9}{2}$ & $2$ & $4$ & $5$ & $5$ & $-\frac{1}{4}$ &  &  &  &  &  \\
\end{tabular}
\caption{The totally symmetric coefficients $c^{uvwx}$.}
\label{tab:h}
\end{table}
Furthermore, we find
\begin{align}
	b_{\SU6}^{uv}&=b_{\SU3}^{uv}=2 b_{\SO8}^{uv}=\frac12 \beta^{uv}\,,&
	d^{uv}&=6\left(\beta^{uv}+\delta^{uv}\right)\,,
\end{align}
with a  symmetric matrix $\beta$,
\begin{align}
	(\beta^{uv})&=\left(\begin{array}{ccccc}
	3 & -1 & 0 & -1 & 0 \\
	-1 & 3 & 0 & -1 & 0 \\
	0 & 0 & 2 & 0 & \sqrt 2 \\
	-1 & -1 & 0 & 4 & 0 \\
	0 & 0 & \sqrt 2 & 0 & 4
	\end{array}\right)\,.  
\end{align}

The coefficients $c^{uvwx}$ are listed in Table \ref{tab:h}. Close inspection shows that they can be written as
\begin{align}
c^{uvwx}&=\frac{3}{2|\sigma(u,v,w,x)|} \left( \delta^{uv}\beta^{wx}+\text{permutations} \right)\,.
\end{align}
Here $|\sigma(u,v,w,x)|$ is a symmetry factor, which is only different from one if two or more of its
variables are equal. In that case it counts the possibilities to associate these equal labels
with equally many of the four $F$-factors. For example,
\begin{align}
	|\sigma(3,4,4,5)|&=\left(\begin{array}{l}4\\2\end{array}\right)=6 \hspace{1cm}
	\Rightarrow \hspace{1cm} c^{3445}=\frac{3}{2\cdot 6}(\delta^{44}\beta^{35}+0)=\frac{1}{2\sqrt2}\,.
\end{align}

 \subsection{Fixed point anomalies}
 \label{app:loctraces}

The anomaly polynomial at the fixed points \index{Anomaly!polynomial!local}
characterized by $n_2=0,1$ is a six-form of the form
\begin{align}
\label{eq:I62}
	I_6=& {\rm tr}\, R^2\, \sum_v \tilde a^v \hat F_v
	-8\, \sum_{B} \tilde b_B \,{\rm tr} F^3_B
	-24\, \sum_{B,v} \tilde c^v_B \,{\rm tr} F^2_B\,\hat F_v
	-8\, \sum_{v,w,x} \tilde d^{vwx}_B \,\hat F_v\hat F_w\hat F_x\,,
\end{align}
where $\tilde a^v, \tilde b_B, \tilde c^v_B,d^{vwx}$ are coefficients, and $B$ and $v,w,x$ label the local
non-Abelian and Abelian gauge group factors, respectively, cf. (\ref{eq:Bv}). The canonically
normalized $\U1$ factors $\hat F_v$ were defined in (\ref{eq:hattu}).

The coefficients are calculated by summing over the chiral fermions 
which are present at the fixed point. They can either arise from twisted sectors, in which
case we denote them as $\psi_l$, or they are contained in a bulk hypermultiplet, denoted as $\psi_b$.
As detailed calculations show, see for example \cite{lnz04}, the contributions of the latter are weighted
with a relative factor of $\frac 14$ compared to the localized fermions. This is interpreted as a
democratic distribution of their effect among the four fixed points of the geometry.

The coefficients in (\ref{eq:I62}) are given by the sums
\begin{subequations}
\label{eq:coefI6}
\begin{eqnarray}
	\tilde a^v&=&\sum_{\psi_l} \hat Q_v(\psi_l)+\frac 14 \sum_{\psi_b} \hat Q_v(\psi_b)\,,\\
	\tilde b_B&=&\sum_{\psi_l} w_B^{\mathcal R(\psi_l)}+\frac 14 \sum_{\psi_b} w_B^{\mathcal R(\psi_b)}\,,\\
	\tilde c^v_B&=&\sum_{\psi_l} v_B^{\mathcal R(\psi_l)}\hat Q_v(\psi_l)+
	\frac 14 \sum_{\psi_b} v_B^{\mathcal R(\psi_b)} \hat Q_v(\psi_b)\,,\\
	\tilde d^{vwx}&=&\sum_{\psi_l} \hat Q_v(\psi_l)\hat Q_w(\psi_l)\hat Q_x(\psi_l)
	+\frac 14 \sum_{\psi_b} \hat Q_v(\psi_b)\hat Q_w(\psi_b)\hat Q_x(\psi_b)\,,
\end{eqnarray}
\end{subequations}
where $v_B^{\mathcal R}$ and $w_B^{\mathcal R}$ were defined in (\ref{eq:vw}).

With Tables \ref{tab:n2T}, \ref{tab:n20U}, \ref{tab:n20T2T4}, \ref{tab:n20T2T4*}
and \ref{tab:n21T}, \ref{tab:n21U}, \ref{tab:n21T2T4*}, \ref{tab:n21T2T4}, \ref{tab:vw}
we evaluate these sums and find
\begin{align}
n_2&=0\,: & \tilde a&=\left(0,-4 \sqrt{2},0,5 \sqrt{2},-\sqrt{6},2 \sqrt{15}\right)\,,\hspace{1cm}
 \tilde b=(0,0,0)\,,\\
&& \tilde c_{\SU5}&=\tilde c_{\SU3}=2 \tilde c_{\SO8}=
\left(0,-\frac{\sqrt{2}}{3},0,\frac{5}{6 \sqrt{2}},-\frac{1}{2 \sqrt{6}},\frac 12 \sqrt{\frac{5}{3}}\right)\,,\\
n_2&=1\,: & \tilde a&=
\left(\frac{19}{8 \sqrt{2}},-\frac{19}{24 \sqrt{2}},\frac{19}{8 \sqrt{3}},-\frac{19}{6 \sqrt{2}},\frac{19}{2 \sqrt{6}},0,0,0\right)
\,,\hspace{.5cm}
 \tilde b=(0,0,0,0)\,, \\
 && \tilde c_{\SU2}&=\tilde c_{\SU4}= \tilde c_{\SU2'}=\tilde c_{\SU4'}\nonumber \\
 &&&=\left(\frac{1}{4 \sqrt{2}},-\frac{1}{12 \sqrt{2}},\frac{1}{4 \sqrt{3}},-\frac{1}{3 \sqrt{2}},\frac{1}{\sqrt{6}},0,0,0\right)\,.
\end{align}
We can bring this into a much simpler form by changing the basis of the Abelian generators. 
With \index{U1@$\U1$!anomalous}
\begin{align}
	t_{\rm an}^0 &\equiv -4 t_2+5t_4-t_5+t_6^0\,, & \hat t_{\rm an}^0&=\frac{t_{\rm an}^0}{\sqrt 2 |t_{\rm an}^0|}\,,\\
	t_{\rm an}^1 &\equiv 3 t_1-t_2+t_3-4t_4+4t_5 \,, & \hat t_{\rm an}^1&=\frac{t_{\rm an}^1}{\sqrt 2 |t_{\rm an}^1|}\,,
\end{align}
and orthogonal sets of generators which extend $\hat t_{\rm an}^0$ and $\hat t_{\rm an}^1$
 to a basis at $n_2=0$ or $n_2=1$, respectively,
the above results become
\begin{align}
n_2&=0\,: & \sum_v \tilde a^v \hat F_v&= 2\sqrt{37}  \,\hat F_{\rm an}^0\,,\hspace{1cm}
 \tilde b=(0,0,0)\,,\\
 && \sum_v \tilde c_{\SU5}^v \hat F_v&=\sum_v \tilde c_{\SU3}^v \hat F_v=2\sum_v \tilde c_{\SO8}^v \hat F_v=
 \frac 16 \sqrt{37} \,\hat F_{\rm an}^0\,,\\
 n_2&=1\,: & \sum_v \tilde a^v \hat F_v&= 2 \sqrt{10} \,\hat F_{\rm an}^1\,,\hspace{1cm}
 \tilde b=(0,0,0,0)\,,\\
 && \sum_v \tilde c_{\SU2}^v \hat F_v&=\sum_v \tilde c_{\SU4}^v \hat F_v=\sum_v \tilde c_{\SU2'}^v \hat F_v
 =\sum_v \tilde c_{\SU4'}^v \hat F_v\nonumber \\
&&&= \frac 16 \sqrt{10}  \,\hat F_{\rm an}^1\,.
 \end{align}
 Note that the appearing square roots are a consequence of
 \begin{align}
 \sqrt 2 |t_{\rm an}^0|&=2 \sqrt{37}\,, \hspace{1cm} \sqrt 2 |t_{\rm an}^1|=4 \sqrt{10}\,.
 \end{align}
 Finally, we calculate the sums $d^{vwx}$:
 \begin{align}
	n_2&=0\,: \hspace{1cm} \sum_{vwx} d^{vwx}  \hat F_v \hat F_w \hat F_x
	=\frac 18 \left( \sum_{uv} \hat F_u \frac{\hat t_u\cdot \hat t_v}{|\hat t_u| \,|\hat t_v|} \hat F_v \right)
	2 \sqrt{37}\, \hat F_{\rm an}^0\,,\\
	n_2&=1\,: \hspace{1cm} \sum_{vwx} d^{vwx}  \hat F_v \hat F_w \hat F_x
	=\frac 18 \left( \sum_{uv} \hat F_u \frac{\hat t_u\cdot \hat t_v}{|\hat t_u| \,|\hat t_v|} \hat F_v \right)
	2 \sqrt{10}\, \hat F_{\rm an}^1\,.	
\end{align}
 For an orthogonal basis, the brackets sum to $\sum_u \hat F_u \hat F_u$.

 \section{The hidden brane}
\label{app:n21}

The effective orbifold GUT in six dimensions which is described in Chapter \ref{cpt:lGUT} has two pairs
of inequivalent fixed points. They differ in the localization label $n_2$ and their role in the compactification:
\index{Hidden brane}
\begin{align}
	n_2&=0\,: \hspace{1cm}\text{Physical brane, with local $\SU5$ GUT}\,,\\
	n_2&=1\,: \hspace{1cm}\text{Hidden brane with exotics, breaks GUT group.}
\end{align}
The hidden brane fixed point is responsible for the breaking of the GUT gauge group to the
standard model gauge group in the effective theory in four dimensions.

 \section*
 {Localized states at {\boldmath $n_2=1$}}

We now calculate the matter spectrum at  $n_2=0$.
This requires to solve the mass equations 
 (\ref{eq:masslessL}), (\ref{eq:masslessR}), and simultaneously the
  projection conditions (\ref{eq:pceff1}).
Consider a sector $\mathcal H_{[g_f]}$ which gives rise to localized states at $(n_2,n_2')=(1,0)$. The
case   $(n_2,n_2')=(1,1)$ will be completely analogous.
The elements $g_f$ are then further characterized by 
\begin{align}
	&g_f\,: & k&=1,3,5\,, & {\bf n}&\,:\;
	\left\{\begin{array}{l l l}
	n_6=0, & n_3=0,1,2,   & \text{if}\;\; k=1,5 \\
	n_6=0,1, & n_3=0,    & \text{if}\;\; k=3.
	\end{array}\right. \\
	& &  && &\hspace{1cm}\gamma_a'm_a^{(g)}= \frac{l_3}{3}n_3 \,, 
	 \hspace{1cm} \gamma=\frac  j3\,,
\end{align}
where
\begin{align}
\label{eq:mts4}
	l_3&=\left\{\begin{array}{ll}
		1,2,3, & \text{if}\;\;k=1,5,\\
		0, & \text{if}\;\;k=3,
	\end{array}\right.\,, &
	j&=\left\{\begin{array}{ll}
		0, & \text{if}\;\;k=1,5,\\
		0,1,2,3, & \text{if}\;\;k=3.
	\end{array}\right. &
\end{align}
By the  arguments of Section \ref{sec:genpc} 
we find projection conditions
\begin{subequations}
 \begin{eqnarray}
 {\bf p}_{\rm sh} \cdot {\bf V}_{g_{f}}
   -{\bf R} \cdot {\bf v}_{g_f}&=&\jmod01\,, \\
   {\bf p}_{\rm sh}\cdot\left( {\bf V}_6+n_3 {\bf W}_{(3)}
	   + {\bf W}_{(2)}\right)-{\bf R}\cdot {\bf v}_6+\frac j3&=&\jmod 01\,,\\
	{\bf p}_{\rm sh}\cdot{\bf W}_{(3)}-\frac{l_3 k}{3}&=&\jmod{0}{1}\,.
\end{eqnarray}
\end{subequations}
These equations have 96 solutions which can be casted into multiplets of the local
gauge symmetry $\SU2\times\SU4\times[\SU2'\times\SU4']$. The result is given in Table \ref{tab:n21T}.
All states are exotics and have to be decoupled from the low-energy spectrum. 
An example for this is given in \cite{bhx06}.

The chiral projection of bulk hypermultiplets to the fixed points $n_2=1$ is summarized in Tables \ref{tab:n21U},
\ref{tab:n21T2T4}.
\begin{table}[t]
\centering
\footnotesize
\begin{tabular}{c||c|c|c|c||c|c|c|c|c|c|c|c||c}
\jvsb Multiplet & $k$ & $n_3$ & $\gamma$ & $\gamma_3'$ & $t_1$ & $t_2$ & $t_3$ & $t_4$ & $t_5$ & $t_6$& $t_7$ & $t_8$ & $t_Y$ \\
\hline
\hline
\jvsb$({\bf 2}, \mathbbm 1,\mathbbm 1,\mathbbm 1)$ & $1$ & $0$ & $0$ & $1$ & $-\frac{1}{2}$ & $-\frac{1}{6}$ & $0$ & $-\frac{5}{12}$ & $\frac{1}{4}$ & $0$ & $1$ & $-1$ & $0$ \\
\jvsb$(\mathbbm 1, \mathbbm 1,\mathbbm 1,\mathbbm 1)$ & $1$ & $0$ & $0$ & $\frac{2}{3}$ & $\frac{1}{2}$ & $-\frac{1}{6}$ & $-1$ & $-\frac{5}{12}$ & $\frac{1}{4}$ & $10$ & $1$ & $-1$ & $-\frac{1}{2}$ \\
\jvsb$(\mathbbm 1, \mathbbm 1,\mathbbm 1,\mathbbm 1)$ & $1$ & $0$ & $0$ & $\frac{1}{3}$ & $\frac{1}{2}$ & $-\frac{1}{6}$ & $1$ & $-\frac{5}{12}$ & $\frac{1}{4}$ & $-10$ & $1$ & $-1$ & $\frac{1}{2}$ \\
\jvsb$({\bf 2}, \mathbbm 1,\mathbbm 1,\mathbbm 1)$ & $1$ & $1$ & $0$ & $\frac{1}{3}$ & $0$ & $\frac{1}{3}$ & $-1$ & $\frac{1}{12}$ & $\frac{3}{4}$ & $0$ & $-1$ & $1$ & $0$ \\
\jvsb$(\mathbbm 1, \mathbbm 1,\mathbbm 1,\mathbbm 1)$ & $1$ & $1$ & $0$ & $\frac{2}{3}$ & $0$ & $-\frac{2}{3}$ & $0$ & $\frac{1}{12}$ & $\frac{3}{4}$ & $-10$ & $-1$ & $1$ & $\frac{1}{2}$ \\
\jvsb$(\mathbbm 1, \mathbbm 1,\mathbbm 1,\mathbbm 1)$ & $1$ & $1$ & $0$ & $1$ & $\frac{1}{2}$ & $-\frac{1}{6}$ & $1$ & $\frac{1}{12}$ & $\frac{3}{4}$ & $10$ & $-1$ & $1$ & $-\frac{1}{2}$ \\
\jvsb$({\bf 2}, \mathbbm 1,\mathbbm 1,\mathbbm 1)$ & $1$ & $2$ & $0$ & $\frac{1}{3}$ & $0$ & $\frac{1}{3}$ & $1$ & $-\frac{5}{12}$ & $\frac{1}{4}$ & $0$ & $-1$ & $-1$ & $0$ \\
\jvsb$({\bf 2}, \mathbbm 1,\mathbbm 1,\mathbbm 1)$ & $1$ & $2$ & $0$ & $1$ & $0$ & $\frac{1}{3}$ & $1$ & $\frac{1}{12}$ & $\frac{3}{4}$ & $0$ & $1$ & $1$ & $0$ \\
\jvsb$(\mathbbm 1, \mathbbm 1,\mathbbm 1,\mathbbm 1)$ & $1$ & $2$ & $0$ & $\frac{2}{3}$ & $\frac{1}{2}$ & $-\frac{1}{6}$ & $-1$ & $-\frac{5}{12}$ & $\frac{1}{4}$ & $-10$ & $-1$ & $-1$ & $\frac{1}{2}$ \\
\jvsb$(\mathbbm 1, \mathbbm 1,\mathbbm 1,\mathbbm 1)$ & $1$ & $2$ & $0$ & $\frac{2}{3}$ & $0$ & $-\frac{2}{3}$ & $0$ & $\frac{1}{12}$ & $\frac{3}{4}$ & $10$ & $1$ & $1$ & $-\frac{1}{2}$ \\
\jvsb$(\mathbbm 1, \mathbbm 1,\mathbbm 1,\mathbbm 1)$ & $1$ & $2$ & $0$ & $\frac{1}{3}$ & $\frac{1}{2}$ & $-\frac{1}{6}$ & $-1$ & $\frac{1}{12}$ & $\frac{3}{4}$ & $-10$ & $1$ & $1$ & $\frac{1}{2}$ \\
\jvsb$(\mathbbm 1, \mathbbm 1,\mathbbm 1,\mathbbm 1)$ & $1$ & $2$ & $0$ & $1$ & $0$ & $-\frac{2}{3}$ & $0$ & $-\frac{5}{12}$ & $\frac{1}{4}$ & $10$ & $-1$ & $-1$ & $-\frac{1}{2}$ \\
\hline
\jvsb$(\mathbbm 1, \mathbbm 1,\mathbbm 1,\mathbbm 1)$ & $3$ & $0$ & $0$ & $0$ & $\frac{1}{2}$ & $\frac{1}{2}$ & $1$ & $-\frac{1}{4}$ & $-\frac{1}{4}$ & $-10$ & $-1$ & $1$ & $\frac{1}{2}$ \\
\jvsb$(\mathbbm 1, \mathbbm 1,\mathbbm 1,\mathbbm 1)$ & $3$ & $0$ & $0$ & $0$ & $0$ & $0$ & $2$ & $\frac{1}{4}$ & $\frac{1}{4}$ & $10$ & $1$ & $-1$ & $-\frac{1}{2}$ \\
\jvsb$(\mathbbm 1, \mathbbm 1,\mathbbm 1,\mathbbm 1)$ & $3$ & $0$ & $\frac{1}{3}$ & $0$ & $\frac{1}{2}$ & $\frac{1}{2}$ & $-1$ & $\frac{1}{4}$ & $\frac{1}{4}$ & $10$ & $1$ & $-1$ & $-\frac{1}{2}$ \\
\jvsb$(\mathbbm 1, \mathbbm 1,\mathbbm 1,\mathbbm 1)$ & $3$ & $0$ & $\frac{1}{3}$ & $0$ & $-\frac{1}{2}$ & $-\frac{1}{2}$ & $1$ & $-\frac{1}{4}$ & $-\frac{1}{4}$ & $-10$ & $-1$ & $1$ & $\frac{1}{2}$ \\
\jvsb$(\mathbbm 1, \mathbbm 1,\mathbbm 1,\mathbbm 1)$ & $3$ & $0$ & $\frac{2}{3}$ & $0$ & $0$ & $0$ & $-2$ & $-\frac{1}{4}$ & $-\frac{1}{4}$ & $-10$ & $-1$ & $1$ & $\frac{1}{2}$ \\
\jvsb$(\mathbbm 1, \mathbbm 1,\mathbbm 1,\mathbbm 1)$ & $3$ & $0$ & $\frac{2}{3}$ & $0$ & $-\frac{1}{2}$ & $-\frac{1}{2}$ & $-1$ & $\frac{1}{4}$ & $\frac{1}{4}$ & $10$ & $1$ & $-1$ & $-\frac{1}{2}$ \\
\jvsb$(\mathbbm 1, \mathbbm 1,\mathbbm 1,\mathbbm 1)$ & $3$ & $0$ & $1$ & $0$ & $\frac{1}{2}$ & $\frac{1}{2}$ & $1$ & $-\frac{1}{4}$ & $-\frac{1}{4}$ & $-10$ & $-1$ & $1$ & $\frac{1}{2}$ \\
\jvsb$(\mathbbm 1, \mathbbm 1,\mathbbm 1,\mathbbm 1)$ & $3$ & $0$ & $1$ & $0$ & $0$ & $0$ & $2$ & $\frac{1}{4}$ & $\frac{1}{4}$ & $10$ & $1$ & $-1$ & $-\frac{1}{2}$ \\
\end{tabular}
\caption{Multiplets of $\SU2_L \times \SU4 \times[\SU2'\times \SU4']$
 at $(n_2,n_2')=(1,0)$ from the twisted sectors. 
 \index{Spectrum!exotic fixed points}The spectrum at $(n_2,n_2')=(1,1)$ 
is an exact copy. All states are exotics, they are either weak doublets or are charged under hypercharge
with generator $t_Y$. }
 \label{tab:n21T}
\end{table}

 \subsection*{Projection of bulk states}
 \label{app:n21proj}

The local projection condition for bulk fields to fixed points $z^3_{g_f}$ with $n_2=1$ is
  \begin{align}
	P_f\,: \hspace{1cm} \phi(z^3_{g_f}+z^3)&= \eta_{f}(\phi) \phi(z^3_{g_f}-z^3)\,,\\
	\eta_{f}(\phi)&=
	e^{2 \pi i \left( {\bf p}_{\rm sh} \cdot \left({\bf V}_6+{\bf W}_{(2)}\right)-{\bf R}\cdot {\bf v}_6+\frac j2\right)}\,,
\end{align}
where $j=0$ is understood for fields from the untwisted sector. The results are listed in Tables \ref{tab:n21U}, 
 \ref{tab:n21T2T4*} and \ref{tab:n21T2T4}.
\begin{table}[t]
\centering
\footnotesize
\begin{tabular}{c|c||c||c|c||c|c|c|c|c|c|c|c}
\multicolumn{2}{c||}{Bulk} & $n_2=1$& $H_L$ & $H_R$ & $t_1$ & $t_2$ & $t_3$ & $t_4$ & $t_5$ & $t_6^1$ & $t_7$ & $t_8$ \jvsb \\
\hline
\hline
\jvsb $({\bf 20},\mathbbm 1,\mathbbm 1)$& & $({\bf 2},{\bf 6},\mathbbm 1,\mathbbm 1)$ & $-$ & $+$ & $-\frac{1}{2}$ & $\frac{1}{2}$ & $0$ & $0$ & $0$ & $0$ & $0$ & $0$ \\
\jvsb & & $(\mathbbm 1,{\bf 4},\mathbbm 1,\mathbbm 1)$ & $+$ & $-$ & $-\frac{1}{2}$ & $\frac{1}{2}$ & $0$ & $0$ & $0$ & $-15$ & $0$ & $0$ \\
\jvsb & & $(\mathbbm 1,{\bf \bar 4},\mathbbm 1,\mathbbm 1)$ & $+$ & $-$ & $-\frac{1}{2}$ & $\frac{1}{2}$ & $0$ & $0$ & $0$ & $15$ & $0$ & $0$ \\
\hline
\jvsb $(\mathbbm 1,\mathbbm 1,{\bf 8})$& & $(\mathbbm 1,\mathbbm 1,\mathbbm 1,{\bf \bar 4})$ & $+$ & $-$ & $0$ & $0$ & $0$ & $-1$ & $0$ & $0$ & $0$ & $1$ \\
\jvsb & & $(\mathbbm 1,\mathbbm 1,\mathbbm 1,{\bf 4})$ & $-$ & $+$ & $0$ & $0$ & $0$ & $-1$ & $0$ & $0$ & $0$ & $-1$ \\
\hline
\jvsb $(\mathbbm 1,\mathbbm 1,{\bf 8}_s)$& & $(\mathbbm 1,\mathbbm 1,\mathbbm 1,{\bf 4})$ & $-$ & $+$ & $0$ & $0$ & $0$ & $\frac{1}{2}$ & $\frac{3}{2}$ & $0$ & $0$ & $1$ \\
\jvsb & & $(\mathbbm 1,\mathbbm 1,\mathbbm 1,{\bf \bar 4})$ & $+$ & $-$ & $0$ & $0$ & $0$ & $\frac{1}{2}$ & $\frac{3}{2}$ & $0$ & $0$ & $-1$ \\
\hline
\jvsb $(\mathbbm 1,\mathbbm 1,{\bf 8}_c)$& & $(\mathbbm 1,\mathbbm 1,\mathbbm 1,{\bf 6})$ & $-$ & $+$ & $0$ & $0$ & $0$ & $\frac{1}{2}$ & $-\frac{3}{2}$ & $0$ & $0$ & $0$ \\
\jvsb & & $(\mathbbm 1,\mathbbm 1,\mathbbm 1,\mathbbm 1)$ & $+$ & $-$ & $0$ & $0$ & $0$ & $\frac{1}{2}$ & $-\frac{3}{2}$ & $0$ & $0$ & $2$ \\
\jvsb & & $(\mathbbm 1,\mathbbm 1,\mathbbm 1,\mathbbm 1)$ & $+$ & $-$ & $0$ & $0$ & $0$ & $\frac{1}{2}$ & $-\frac{3}{2}$ & $0$ & $0$ & $-2$ \\
\hline
\jvsb $(\mathbbm 1,\mathbbm 1,\mathbbm 1)$&$U_1$ & $(\mathbbm 1,\mathbbm 1,\mathbbm 1,\mathbbm 1)$ & $-$ & $+$ & $\frac{1}{2}$ & $\frac{1}{2}$ & $3$ & $0$ & $0$ & $0$ & $0$ & $0$ \\
\jvsb $(\mathbbm 1,\mathbbm 1,\mathbbm 1)$&$U_2$  & $(\mathbbm 1,\mathbbm 1,\mathbbm 1,\mathbbm 1)$ & $-$ & $+$ & $\frac{1}{2}$ & $\frac{1}{2}$ & $-3$ & $0$ & $0$ & $0$ & $0$ & $0$ \\
\jvsb $(\mathbbm 1,\mathbbm 1,\mathbbm 1)$&$U_3$  & $(\mathbbm 1,\mathbbm 1,\mathbbm 1,\mathbbm 1)$ & $-$ & $+$ & $1$ & $-1$ & $0$ & $0$ & $0$ & $0$ & $0$ & $0$ \\
\jvsb $(\mathbbm 1,\mathbbm 1,\mathbbm 1)$&$U_4$  & $(\mathbbm 1,\mathbbm 1,\mathbbm 1,\mathbbm 1)$ & $-$ & $+$ & $-1$ & $-1$ & $0$ & $0$ & $0$ & $0$ & $0$ & $0$ 
\end{tabular}
\caption{Chiral projection of bulk hypermultiplets multiplets from the untwisted sector to the fixed points $n_2=1$.
 \index{Spectrum!exotic fixed points}
The representations are given with respect to $\SU6\times[\SU3\times\SO8]$ and 
 $\SU2\times\SU4\times[\SU2'\times\SU4']$ in the bulk and at $n_2=1$, respectively.}
\label{tab:n21U}
\end{table}
\begin{table}[htb]
\centering
\scriptsize
\begin{tabular}{c||c|c||c|c||c|c||c|c|c|c|c|c|c|c}
\jvsb Bulk & \multicolumn{2}{c||}{$n_2=1$} & $n_3$ & $\gamma_3'$ &$H_L$ & $H_R$ & 
 $t_1$ & $t_2$ & $t_3$ & $t_4$ & $t_5$ & $t_6^1$ & $t_7$ & $t_8$  \\
 \hline
\hline
 \jvsb $( \mathbbm 1,\mathbbm 1,\mathbbm 1)$ & $(\mathbbm 1,\mathbbm 1,\mathbbm 1,\mathbbm 1)$ & $Y_0^*$  & $0$ & $\frac{2}{3}$ & $+,-+$ & $-,+,-$ & $0$ & $\frac{2}{3}$ & $0$ & $\frac{2}{3}$ & $0$ & $0$ & $0$ & $0$ \\
 \hline
\jvsb $( \mathbbm 1,\mathbbm 1,\mathbbm 1)$ & $(\mathbbm 1,\mathbbm 1,\mathbbm 1,\mathbbm 1)$ & $\bar Y_0^*$   & $0$ & $\frac{2}{3}$ & $-,+,-$ & $+,-,+$ & $0$ & $\frac{2}{3}$ & $0$ & $\frac{2}{3}$ & $0$ & $0$ & $0$ & $0$ \\
\hline
\hline
\jvsb $( \mathbbm 1,\mathbbm 1,\mathbbm 1)$ & $(\mathbbm 1,\mathbbm 1,\mathbbm 1,\mathbbm 1)$ & $Y_1^*$   & $1$ & $\frac{2}{3}$ & $+,-+$ & $-,+,-$ & $-\frac{1}{2}$ & $\frac{1}{6}$ & $1$ & $-\frac{1}{3}$ & $-1$ & $0$ & $0$ & $0$ \\
\hline
\jvsb $( \mathbbm 1,\mathbbm 1,\mathbbm 1)$ & $(\mathbbm 1,\mathbbm 1,\mathbbm 1,\mathbbm 1)$ & $\bar Y_1^*$   & $1$ & $\frac{2}{3}$ & $-,+,-$ & $+,-,+$ & $-\frac{1}{2}$ & $\frac{1}{6}$ & $1$ & $-\frac{1}{3}$ & $-1$ & $0$ & $0$ & $0$ \\
\hline
\hline
\jvsb $( \mathbbm 1,\mathbbm 1,\mathbbm 1)$ & $(\mathbbm 1,\mathbbm 1,\mathbbm 1,\mathbbm 1)$ & $Y_2^*$   & $2$ & $\frac{2}{3}$ & $-,+,-$ & $+,-,+$ & $-\frac{1}{2}$ & $\frac{1}{6}$ & $-1$ & $-\frac{1}{3}$ & $1$ & $0$ & $0$ & $0$ \\
\hline
\jvsb $( \mathbbm 1,\mathbbm 1,\mathbbm 1)$ & $(\mathbbm 1,\mathbbm 1,\mathbbm 1,\mathbbm 1)$ & $\bar Y_2^*$   & $2$ & $\frac{2}{3}$ & $+,-+$ & $-,+,-$ & $-\frac{1}{2}$ & $\frac{1}{6}$ & $-1$ & $-\frac{1}{3}$ & $1$ & $0$ & $0$ & $0$ \\
\end{tabular}
\caption{Projection of  bulk hypermultiplets with oscillator numbers from twisted sectors $T_2$ and $T_4$
 to the fixed points $n_2=1$. 
  \index{Spectrum!exotic fixed points}
  The singlets $Y_{n_3}^*$ have ${\bf \tilde N}=(0,1,0,0)$, ${\bf \tilde N}^*=(0,0,0,0)$,
 the singlets $\bar Y_{n_3}^*$ have ${\bf \tilde N}=(0,0,0,0)$, ${\bf \tilde N}^*=(1,0,0,0)$. The three parities for the chiral multiplet components
 $H_L, H_R$ correspond to $\gamma=0,\frac12,1$. The localization in the $\G2$-plane
 is given by $n_6=0$ for $\gamma=0$, and $n_6=2$ otherwise. }
\label{tab:n21T2T4*}
\end{table}
\begin{center}
\scriptsize
\begin{longtable}[c]{c||c|c||c|c||c|c||c|c|c|c|c|c|c|c}
\hline
\multicolumn{1}{c||}{Bulk} &
\multicolumn{2}{c||}{$n_2=1$} &
\multicolumn{1}{c|}{ $n_3$} &
\multicolumn{1}{c||}{ $\gamma_3'$} &
\multicolumn{1}{c|}{$H_L$} &
\multicolumn{1}{c||}{$H_R$} &
\multicolumn{1}{c|}{$t_1$} &
\multicolumn{1}{c|}{$t_2$} &
\multicolumn{1}{c|}{$t_3$} &
\multicolumn{1}{c|}{$t_4$} &
\multicolumn{1}{c|}{$t_5$\jvsb} &
\multicolumn{1}{c|}{$t_6^1$} &
\multicolumn{1}{c|}{$t_7$} &
\multicolumn{1}{c}{$t_8$} 
\\ \hline \hline
\endfirsthead
\multicolumn{15}{c}%
{{ \tablename\ \thetable{} - continued from previous page}}  \\
\hline 
\multicolumn{1}{c||}{Bulk} &
\multicolumn{2}{c||}{$n_2=1$} &
\multicolumn{1}{c|}{ $n_3$} &
\multicolumn{1}{c||}{ $\gamma_3'$} &
\multicolumn{1}{c|}{$H_L$} &
\multicolumn{1}{c||}{$H_R$} &
\multicolumn{1}{c|}{$t_1$} &
\multicolumn{1}{c|}{$t_2$} &
\multicolumn{1}{c|}{$t_3$} &
\multicolumn{1}{c|}{$t_4$} &
\multicolumn{1}{c|}{$t_5$\jvsb} &
\multicolumn{1}{c|}{$t_6^1$} &
\multicolumn{1}{c|}{$t_7$} &
\multicolumn{1}{c}{$t_8$} 
\\ \hline \hline
\endhead
\endlastfoot
\multicolumn{15}{c}%
{{ \tablename\ \thetable{} - continued on next page}} \\
\endfoot
\jvsb  $({\bf  6}, \mathbbm 1,\mathbbm 1)$& $({\bf 2}, \mathbbm 1,\mathbbm 1,\mathbbm 1)$ &   & $0$ & $1$ & $+,-,+$ & $-,+,-$ & $0$ & $-\frac{1}{3}$ & $-1$ & $\frac{2}{3}$ & $0$ & $10$ & $0$ & $0$ \\
\jvsb  & $(\mathbbm 1,{\bf 4},\mathbbm 1,\mathbbm 1)$ &   & $0$ & $\frac{1}{3}$ & $-,+,-$ & $+,-,+$ & $0$ & $-\frac{1}{3}$ & $1$ & $\frac{2}{3}$ & $0$ & $5$ & $0$ & $0$ \\
\hline
\jvsb  $({\bf  \bar 6}, \mathbbm 1,\mathbbm 1)$& $({\bf 2}, \mathbbm 1,\mathbbm 1,\mathbbm 1)$ &   & $0$ & $\frac{1}{3}$ & $+,-,+$ & $-,+,-$ & $0$ & $-\frac{1}{3}$ & $1$ & $\frac{2}{3}$ & $0$ & $-10$ & $0$ & $0$ \\
\jvsb  & $(\mathbbm 1,{\bf \bar 4},\mathbbm 1,\mathbbm 1)$ &   & $0$ & $1$ & $-,+,-$ & $+,-,+$ & $0$ & $-\frac{1}{3}$ & $-1$ & $\frac{2}{3}$ & $0$ & $-5$ & $0$ & $0$ \\
\hline
\jvsb  $(\mathbbm 1,\mathbbm 1,\mathbbm 1)$ & $(\mathbbm 1,\mathbbm 1,\mathbbm 1,\mathbbm 1)$ & $Y_0$  & $0$ & $\frac{2}{3}$ & $+,-,+$ & $-,+,-$ & $1$ & $-\frac{1}{3}$ & $0$ & $\frac{2}{3}$ & $0$ & $0$ & $0$ & $0$ \\
\hline
\jvsb  $(\mathbbm 1,\mathbbm 1,\mathbbm 1)$ & $(\mathbbm 1,\mathbbm 1,\mathbbm 1,\mathbbm 1)$ & $\bar Y_0$  & $0$ & $\frac{2}{3}$ & $+,-,+$ & $-,+,-$ & $-1$ & $-\frac{1}{3}$ & $0$ & $\frac{2}{3}$ & $0$ & $0$ & $0$ & $0$ \\
\hline
\hline
\jvsb  $( \mathbbm 1, {\bf   3},\mathbbm 1)$& $( \mathbbm 1,\mathbbm 1,{\bf 2},\mathbbm 1)$ &   & $0$ & $\frac{1}{3}$ & $+,-,+$ & $-,+,-$ & $0$ & $\frac{2}{3}$ & $0$ & $-\frac{1}{3}$ & $1$ & $0$ & $1$ & $0$ \\
\jvsb  & $(\mathbbm 1,\mathbbm 1,\mathbbm 1,\mathbbm 1)$ &   & $0$ & $\frac{1}{3}$ & $-,+,-$ & $+,-,+$ & $0$ & $\frac{2}{3}$ & $0$ & $-\frac{1}{3}$ & $1$ & $0$ & $-2$ & $0$ \\
\hline
\jvsb $( \mathbbm 1, {\bf \bar 3},\mathbbm 1)$ & $( \mathbbm 1,\mathbbm 1,{\bf 2},\mathbbm 1)$ &   & $0$ & $1$ & $-,+,-$ & $+,-,+$ & $0$ & $\frac{2}{3}$ & $0$ & $-\frac{1}{3}$ & $-1$ & $0$ & $-1$ & $0$ \\
\jvsb  & $(\mathbbm 1,\mathbbm 1,\mathbbm 1,\mathbbm 1)$ &   & $0$ & $1$ & $+,-,+$ & $-,+,-$ & $0$ & $\frac{2}{3}$ & $0$ & $-\frac{1}{3}$ & $-1$ & $0$ & $2$ & $0$ \\
\hline
\jvsb  $( \mathbbm 1,\mathbbm 1, {\bf 8})$& $(\mathbbm 1,\mathbbm 1,\mathbbm 1,{\bf 4})$ &   & $0$ & $\frac{2}{3}$ & $+,-,+$ & $-,+,-$ & $0$ & $\frac{2}{3}$ & $0$ & $-\frac{1}{3}$ & $0$ & $0$ & $0$ & $-1$ \\
\jvsb  & $(\mathbbm 1,\mathbbm 1,\mathbbm 1,{\bf \bar 4})$ &   & $0$ & $\frac{2}{3}$ & $-,+,-$ & $+,-,+$ & $0$ & $\frac{2}{3}$ & $0$ & $-\frac{1}{3}$ & $0$ & $0$ & $0$ & $1$ \\
\hline
\hline
\jvsb  $({\bf  6}, \mathbbm 1,\mathbbm 1)$& $({\bf 2}, \mathbbm 1,\mathbbm 1,\mathbbm 1)$ &   & $1$ & $\frac{1}{3}$ & $+,-,+$ & $-,+,-$ & $0$ & $-\frac{1}{3}$ & $-1$ & $-\frac{1}{3}$ & $-1$ & $-10$ & $0$ & $0$ \\
\jvsb  & $(\mathbbm 1,{\bf 4},\mathbbm 1,\mathbbm 1)$ &   & $1$ & $\frac{1}{3}$ & $-,+,-$ & $+,-,+$ & $0$ & $-\frac{1}{3}$ & $-1$ & $-\frac{1}{3}$ & $-1$ & $5$ & $0$ & $0$ \\
\hline
\jvsb  $({\bf  \bar6}, \mathbbm 1,\mathbbm 1)$& $({\bf 2}, \mathbbm 1,\mathbbm 1,\mathbbm 1)$ &   & $1$ & $1$ & $+,-,+$ & $-,+,-$ & $\frac{1}{2}$ & $\frac{1}{6}$ & $0$ & $-\frac{1}{3}$ & $-1$ & $10$ & $0$ & $0$ \\
\jvsb  & $(\mathbbm 1,{\bf \bar 4},\mathbbm 1,\mathbbm 1)$ &   & $1$ & $1$ & $-,+,-$ & $+,-,+$ & $\frac{1}{2}$ & $\frac{1}{6}$ & $0$ & $-\frac{1}{3}$ & $-1$ & $-5$ & $0$ & $0$ \\
\hline
\jvsb  $(\mathbbm 1,\mathbbm 1,\mathbbm 1)$ & $(\mathbbm 1,\mathbbm 1,\mathbbm 1,\mathbbm 1)$ & $Y_1$  & $1$ & $\frac{2}{3}$ & $+,-,+$ & $-,+,-$ & $0$ & $\frac{2}{3}$ & $-2$ & $-\frac{1}{3}$ & $-1$ & $0$ & $0$ & $0$ \\
\jvsb  $(\mathbbm 1,\mathbbm 1,\mathbbm 1)$  & $(\mathbbm 1,\mathbbm 1,\mathbbm 1,\mathbbm 1)$ & $\bar Y_1$  & $1$ & $\frac{2}{3}$ & $+,-,+$ & $-,+,-$ & $\frac{1}{2}$ & $-\frac{5}{6}$ & $1$ & $-\frac{1}{3}$ & $-1$ & $0$ & $0$ & $0$ \\
\hline
\hline
\jvsb  $( \mathbbm 1,{\bf  3},\mathbbm 1)$& $( \mathbbm 1,\mathbbm 1,{\bf 2},\mathbbm 1)$ &   & $1$ & $\frac{1}{3}$ & $-,+,-$ & $+,-,+$ & $-\frac{1}{2}$ & $\frac{1}{6}$ & $1$ & $\frac{2}{3}$ & $0$ & $0$ & $1$ & $0$ \\
\jvsb  & $(\mathbbm 1,\mathbbm 1,\mathbbm 1,\mathbbm 1)$ &   & $1$ & $\frac{1}{3}$ & $+,-,+$ & $-,+,-$ & $-\frac{1}{2}$ & $\frac{1}{6}$ & $1$ & $\frac{2}{3}$ & $0$ & $0$ & $-2$ & $0$ \\
\hline
\jvsb $( \mathbbm 1,{\bf  \bar 3},\mathbbm 1)$ & $( \mathbbm 1,\mathbbm 1,{\bf 2},\mathbbm 1)$ &   & $1$ & $1$ & $+,-,+$ & $-,+,-$ & $-\frac{1}{2}$ & $\frac{1}{6}$ & $1$ & $-\frac{1}{3}$ & $1$ & $0$ & $-1$ & $0$ \\
\jvsb  & $(\mathbbm 1,\mathbbm 1,\mathbbm 1,\mathbbm 1)$ &   & $1$ & $1$ & $-,+,-$ & $+,-,+$ & $-\frac{1}{2}$ & $\frac{1}{6}$ & $1$ & $-\frac{1}{3}$ & $1$ & $0$ & $2$ & $0$ \\
\hline
\jvsb $( \mathbbm 1,\mathbbm 1,{\bf  8}_s)$ & $(\mathbbm 1,\mathbbm 1,\mathbbm 1,{\bf 4})$ &   & $1$ & $\frac{2}{3}$ & $+,-,+$ & $-,+,-$ & $-\frac{1}{2}$ & $\frac{1}{6}$ & $1$ & $\frac{1}{6}$ & $\frac{1}{2}$ & $0$ & $0$ & $1$ \\
\jvsb  & $(\mathbbm 1,\mathbbm 1,\mathbbm 1,{\bf \bar 4})$ &   & $1$ & $\frac{2}{3}$ & $-,+,-$ & $+,-,+$ & $-\frac{1}{2}$ & $\frac{1}{6}$ & $1$ & $\frac{1}{6}$ & $\frac{1}{2}$ & $0$ & $0$ & $-1$ \\
\hline
\hline
\jvsb  $({\bf  6}, \mathbbm 1,\mathbbm 1)$ & $({\bf 2}, \mathbbm 1,\mathbbm 1,\mathbbm 1)$ &   & $2$ & $\frac{1}{3}$ & $-,+,-$ & $+,-,+$ & $\frac{1}{2}$ & $\frac{1}{6}$ & $0$ & $-\frac{1}{3}$ & $1$ & $-10$ & $0$ & $0$ \\
\jvsb  & $(\mathbbm 1,{\bf 4},\mathbbm 1,\mathbbm 1)$ &   & $2$ & $\frac{1}{3}$ & $+,-,+$ & $-,+,-$ & $\frac{1}{2}$ & $\frac{1}{6}$ & $0$ & $-\frac{1}{3}$ & $1$ & $5$ & $0$ & $0$ \\
\hline
\jvsb  $({\bf  \bar6}, \mathbbm 1,\mathbbm 1)$ & $({\bf 2}, \mathbbm 1,\mathbbm 1,\mathbbm 1)$ &   & $2$ & $1$ & $-,+,-$ & $+,-,+$ & $0$ & $-\frac{1}{3}$ & $1$ & $-\frac{1}{3}$ & $1$ & $10$ & $0$ & $0$ \\
\jvsb  & $(\mathbbm 1,{\bf \bar 4},\mathbbm 1,\mathbbm 1)$ &   & $2$ & $1$ & $+,-,+$ & $-,+,-$ & $0$ & $-\frac{1}{3}$ & $1$ & $-\frac{1}{3}$ & $1$ & $-5$ & $0$ & $0$ \\
\hline
\jvsb  $( \mathbbm 1,\mathbbm 1,\mathbbm 1)$& $(\mathbbm 1,\mathbbm 1,\mathbbm 1,\mathbbm 1)$ & $Y_2$  & $2$ & $\frac{2}{3}$ & $-,+,-$ & $+,-,+$ & $0$ & $\frac{2}{3}$ & $2$ & $-\frac{1}{3}$ & $1$ & $0$ & $0$ & $0$ \\
\hline
\jvsb $( \mathbbm 1,\mathbbm 1,\mathbbm 1)$ & $(\mathbbm 1,\mathbbm 1,\mathbbm 1,\mathbbm 1)$ & $\bar Y_2$  & $2$ & $\frac{2}{3}$ & $-,+,-$ & $+,-,+$ & $\frac{1}{2}$ & $-\frac{5}{6}$ & $-1$ & $-\frac{1}{3}$ & $1$ & $0$ & $0$ & $0$ \\
\hline
\hline
\jvsb $( \mathbbm 1, {\bf 3},\mathbbm 1)$ & $( \mathbbm 1,\mathbbm 1,{\bf 2},\mathbbm 1)$ &   & $2$ & $\frac{1}{3}$ & $-,+,-$ & $+,-,+$ & $-\frac{1}{2}$ & $\frac{1}{6}$ & $-1$ & $-\frac{1}{3}$ & $-1$ & $0$ & $1$ & $0$ \\
\jvsb  & $(\mathbbm 1,\mathbbm 1,\mathbbm 1,\mathbbm 1)$ &   & $2$ & $\frac{1}{3}$ & $+,-,+$ & $-,+,-$ & $-\frac{1}{2}$ & $\frac{1}{6}$ & $-1$ & $-\frac{1}{3}$ & $-1$ & $0$ & $-2$ & $0$ \\
\hline
\jvsb $( \mathbbm 1, {\bf \bar 3},\mathbbm 1)$ & $( \mathbbm 1,\mathbbm 1,{\bf 2},\mathbbm 1)$ &   & $2$ & $1$ & $-,+,-$ & $+,-,+$ & $-\frac{1}{2}$ & $\frac{1}{6}$ & $-1$ & $\frac{2}{3}$ & $0$ & $0$ & $-1$ & $0$ \\
\jvsb  & $(\mathbbm 1,\mathbbm 1,\mathbbm 1,\mathbbm 1)$ &   & $2$ & $1$ & $+,-,+$ & $-,+,-$ & $-\frac{1}{2}$ & $\frac{1}{6}$ & $-1$ & $\frac{2}{3}$ & $0$ & $0$ & $2$ & $0$ \\
\hline
\jvsb $( \mathbbm 1,\mathbbm 1,{\bf  8}_c)$ & $(\mathbbm 1,\mathbbm 1,\mathbbm 1,{\bf 6})$ &   & $2$ & $\frac{2}{3}$ & $-,+,-$ & $+,-,+$ & $-\frac{1}{2}$ & $\frac{1}{6}$ & $-1$ & $\frac{1}{6}$ & $-\frac{1}{2}$ & $0$ & $0$ & $0$ \\
\jvsb  & $(\mathbbm 1,\mathbbm 1,\mathbbm 1,\mathbbm 1)$ &   & $2$ & $\frac{2}{3}$ & $+,-,+$ & $-,+,-$ & $-\frac{1}{2}$ & $\frac{1}{6}$ & $-1$ & $\frac{1}{6}$ & $-\frac{1}{2}$ & $0$ & $0$ & $2$ \\
\jvsb  & $(\mathbbm 1,\mathbbm 1,\mathbbm 1,\mathbbm 1)$ &   & $2$ & $\frac{2}{3}$ & $+,-,+$ & $-,+,-$ & $-\frac{1}{2}$ & $\frac{1}{6}$ & $-1$ & $\frac{1}{6}$ & $-\frac{1}{2}$ & $0$ & $0$ & $-2$ \\
\caption{Projection of  bulk hypermultiplets from twisted sectors $T_2$ and $T_4$ with vanishing oscillator numbers
 to the fixed points $n_2=1$. For excited states see Table \ref{tab:n21T2T4*}.
  \index{Spectrum!exotic fixed points}
 The representations are given with respect to $\SU6\times[\SU3\times\SO8]$ and 
 $\SU2\times\SU4\times[\SU2'\times\SU4']$ in the bulk and at $n_2=1$, respectively.
 The three parities for the chiral multiplet components
 $H_L, H_R$ correspond to $\gamma=0,\frac12,1$. The localization in the $\G2$-plane
 is given by $n_6=0$ for $\gamma=0$, and $n_6=2$ otherwise.}
\label{tab:n21T2T4}
\end{longtable}
\end{center}

 \section{Zero modes in four dimensions}
 \label{app:4d}

The spectrum of the low-energy effective field theory in four dimensions follows from
solutions of the mass equations (\ref{eq:masslessL}), (\ref{eq:masslessR}) and the
projection conditions (\ref{eq:1ststep}), (\ref{eq:2ndstepS}). Up to the
geometrical $\gamma'$-phases, the result was already published in \cite{bhx06}.
Here we present the list of chiral gauge multiplets of $\SU3\times\SU2$ with charges under
nine linearly independent $\U1$ factors, \index{U1@$\U1$!factors} with generators
\begin{align}
t_1\,\dots,t_5,t_6^0,t_7,t_8,t_Y\,,
\end{align}
as defined in Table \ref{tab:t}. We additionally list the $B-L$ charges and 
the multiplet of the six-dimensional model which contains the zero mode.
In the case of bulk multiplets, we chose to represent the latter
by multiplets with respect to the local gauge group at $n_2=0$, rather
than the bulk gauge group, since we are mainly interested in the effective $\SU5$ GUT at
this fixed point in the main text. All left-handed chiral multiplets which arise from sectors $k=0,2,4$ are listed in
Table \ref{tab:4d}, together with all left-handed chiral multiplets from $k=1,3,5$ at $(n_2,n_2')=(0,0),(1,0)$. Note
that the spectrum of the sectors $k=1,3,5$ at $(n_2,n_2')=(0,1),(1,1)$ is an exact copy of the latter, and
not shown explicitly. The $\U1$ charges with respect to the generators $t_1,\dots,t_5$ can be 
inferred from the Tables  \ref{tab:n2T}, \ref{tab:n20U}, \ref{tab:n20T2T4} and \ref{tab:n21T}.

\begin{center}
\footnotesize
\begin{longtable}[c]{c||c|c|c|c|c|c||c|c|c|c|c||c|c}
\multicolumn{1}{c||}{Multiplet} &
\multicolumn{1}{c|}{$k$} &
\multicolumn{1}{c|}{ $n_2$} &
\multicolumn{1}{c|}{ $n_3$} &
\multicolumn{1}{c|}{$\gamma$} &
\multicolumn{1}{c|}{$\gamma_3'$} &
\multicolumn{1}{c||}{$\gamma_5'$} &
\multicolumn{1}{c|}{$t_6^0$} &
\multicolumn{1}{c|}{$t_7$} &
\multicolumn{1}{c|}{$t_8$} &
\multicolumn{1}{c|}{$t_Y$\jvsb} &
\multicolumn{1}{c||}{$t_{B-L}$} &
\multicolumn{2}{c}{Origin} 
\\ \hline \hline
\endfirsthead
\multicolumn{14}{c}%
{{ \tablename\ \thetable{} - continued from previous page}}  \\
\multicolumn{1}{c||}{Multiplet} &
\multicolumn{1}{c|}{$k$} &
\multicolumn{1}{c|}{ $n_2$} &
\multicolumn{1}{c|}{ $n_3$} &
\multicolumn{1}{c|}{$\gamma$} &
\multicolumn{1}{c|}{$\gamma_3'$} &
\multicolumn{1}{c||}{$\gamma_5'$} &
\multicolumn{1}{c|}{$t_6^0$} &
\multicolumn{1}{c|}{$t_7$} &
\multicolumn{1}{c|}{$t_8$} &
\multicolumn{1}{c|}{$t_Y$\jvsb} &
\multicolumn{1}{c||}{$t_{B-L}$} &
\multicolumn{2}{c}{Origin} 
\\ \hline \hline
\endhead
\endlastfoot
\multicolumn{14}{c}%
{{ \tablename\ \thetable{} - continued on next page}} \\
\endfoot
\jvsb$({\bf 3},{\bf 2},\mathbbm 1,\mathbbm 1)$ & $0$ & $0$ & $0$ & $0$ & $0$ & $0$ & $3$ & $0$ & $0$ & $\frac{1}{6}$ & $\frac{1}{3}$ & $({\bf 10}, \mathbbm 1,\mathbbm 1)$ & $\T_{(4)}$ \\
\jvsb$({\bf \bar 3},\mathbbm 1,\mathbbm 1,\mathbbm 1)$ & $0$ & $0$ & $0$ & $0$ & $0$ & $0$ & $3$ & $0$ & $0$ & $-\frac{2}{3}$ & $-\frac{1}{3}$ & $({\bf 10}, \mathbbm 1,\mathbbm 1)$ & $\T_{(3)}$ \\
\jvsb$(\mathbbm 1,\mathbbm 1,\mathbbm 1,\mathbbm 1)$ & $0$ & $0$ & $0$ & $0$ & $0$ & $0$ & $3$ & $0$ & $0$ & $1$ & $1$ & $({\bf 10}, \mathbbm 1,\mathbbm 1)$ & $\T_{(3)}$ \\
\hline
\jvsb$(\mathbbm 1,{\bf 2},\mathbbm 1,\mathbbm 1)$ & $0$ & $0$ & $0$ & $0$ & $0$ & $0$ & $6$ & $0$ & $0$ & $-\frac{1}{2}$ & $0$ & $({\bf \bar 5},\mathbbm 1, \mathbbm 1)$ & $\Fb$ \\
\jvsb$(\mathbbm 1,{\bf 2},\mathbbm 1,\mathbbm 1)$ & $0$ & $0$ & $0$ & $0$ & $0$ & $0$ & $-6$ & $0$ & $0$ & $\frac{1}{2}$ & $0$ & $({\bf  5},\mathbbm 1, \mathbbm 1)$ & $\F$ \\
\jvsb$(\mathbbm 1,\mathbbm 1,\mathbbm 1,{\bf \bar 4})$ & $0$ & $0$ & $0$ & $0$ & $0$ & $0$ & $0$ & $0$ & $1$ & $0$ & $\frac{1}{2}$ & $(\mathbbm 1, \mathbbm 1,{\bf 8})$ &  \\
\jvsb$(\mathbbm 1,\mathbbm 1,\mathbbm 1,{\bf \bar 4})$ & $0$ & $0$ & $0$ & $0$ & $0$ & $0$ & $0$ & $0$ & $-1$ & $0$ & $\frac{1}{2}$ & $(\mathbbm 1, \mathbbm 1,{\bf 8}_s)$ &  \\
\jvsb$(\mathbbm 1,\mathbbm 1,\mathbbm 1,\mathbbm 1)$ & $0$ & $0$ & $0$ & $0$ & $0$ & $0$ & $0$ & $0$ & $-2$ & $0$ & $0$ & $(\mathbbm 1, \mathbbm 1,{\bf 8}_c)$ &  \\
\jvsb$(\mathbbm 1,\mathbbm 1,\mathbbm 1,\mathbbm 1)$ & $0$ & $0$ & $0$ & $0$ & $0$ & $0$ & $0$ & $0$ & $2$ & $0$ & $0$ & $(\mathbbm 1, \mathbbm 1,{\bf 8}_c)$ &  \\
\jvsb$(\mathbbm 1,\mathbbm 1,\mathbbm 1,\mathbbm 1)$ & $0$ & $0$ & $0$ & $0$ & $0$ & $0$ & $0$ & $0$ & $0$ & $0$ & $0$ & $(\mathbbm 1, \mathbbm 1,\mathbbm 1)$ & $U_1^c$ \\
\hline
\jvsb$({\bf 3},{\bf 2},\mathbbm 1,\mathbbm 1)$ & $1$ & $0$ & $0$ & $0$ & $1$ & $1$ & $\frac{1}{2}$ & $0$ & $0$ & $\frac{1}{6}$ & $\frac{1}{3}$ & $({\bf 10}, \mathbbm 1,\mathbbm 1)$ & $\T_{(1)}$ \\
\jvsb$({\bf \bar 3},\mathbbm 1,\mathbbm 1,\mathbbm 1)$ & $1$ & $0$ & $0$ & $0$ & $1$ & $\frac{1}{2}$ & $\frac{1}{2}$ & $0$ & $0$ & $-\frac{2}{3}$ & $-\frac{1}{3}$ & $({\bf 10}, \mathbbm 1,\mathbbm 1)$ & $\T_{(1)}$ \\
\jvsb$(\mathbbm 1,\mathbbm 1,\mathbbm 1,\mathbbm 1)$ & $1$ & $0$ & $0$ & $0$ & $1$ & $\frac{1}{2}$ & $\frac{1}{2}$ & $0$ & $0$ & $1$ & $1$ & $({\bf 10}, \mathbbm 1,\mathbbm 1)$ & $\T_{(1)}$ \\
\jvsb$({\bf \bar 3},\mathbbm 1,\mathbbm 1,\mathbbm 1)$ & $1$ & $0$ & $0$ & $0$ & $\frac{2}{3}$ & $\frac{1}{2}$ & $-\frac{3}{2}$ & $0$ & $0$ & $\frac{1}{3}$ & $-\frac{1}{3}$ & $({\bf \bar 5},\mathbbm 1, \mathbbm 1)$ & $\Fb_{(1)}$ \\
\jvsb$(\mathbbm 1,{\bf 2},\mathbbm 1,\mathbbm 1)$ & $1$ & $0$ & $0$ & $0$ & $\frac{2}{3}$ & $1$ & $-\frac{3}{2}$ & $0$ & $0$ & $-\frac{1}{2}$ & $-1$ & $({\bf \bar 5},\mathbbm 1, \mathbbm 1)$ & $\Fb_{(1)}$ \\
\jvsb$(\mathbbm 1,\mathbbm 1,\mathbbm 1,\mathbbm 1)$ & $1$ & $0$ & $0$ & $0$ & $\frac{1}{3}$ & $1$ & $\frac{5}{2}$ & $0$ & $0$ & $0$ & $-1$ & $(\mathbbm 1, \mathbbm 1,\mathbbm 1)$ & $N_{(1)}^c$ \\
\hline
\jvsb$(\mathbbm 1,\mathbbm 1,\mathbbm 1,\mathbbm 1)$ & $1$ & $0$ & $0$ & $0$ & $\frac{1}{3}$ & $\frac{1}{2}$ & $\frac{5}{2}$ & $0$ & $0$ & $0$ & $1$ & $(\mathbbm 1, \mathbbm 1,\mathbbm 1)$ &   $S_1$\\
\jvsb$(\mathbbm 1,\mathbbm 1,\mathbbm 1,\mathbbm 1)$ & $1$ & $0$ & $0$ & $0$ & $1$ & $\frac{1}{2}$ & $-\frac{5}{2}$ & $0$ & $0$ & $0$ & $0$ & $(\mathbbm 1, \mathbbm 1,\mathbbm 1)$ &  $S_2$ \\
\jvsb$(\mathbbm 1,\mathbbm 1,\mathbbm 1,\mathbbm 1)$ & $1$ & $0$ & $0$ & $0$ & $\frac{1}{3}$ & $\frac{1}{2}$ & $\frac{5}{2}$ & $0$ & $0$ & $0$ & $1$ & $(\mathbbm 1, \mathbbm 1,\mathbbm 1)$ &  $S_3$ \\
\jvsb$(\mathbbm 1,\mathbbm 1,\mathbbm 1,\mathbbm 1)$ & $1$ & $0$ & $0$ & $0$ & $\frac{1}{3}$ & $\frac{1}{2}$ & $\frac{5}{2}$ & $0$ & $0$ & $0$ & $1$ & $(\mathbbm 1, \mathbbm 1,\mathbbm 1)$ &  $S_4$ \\
\jvsb$(\mathbbm 1,\mathbbm 1,\mathbbm 1,\mathbbm 1)$ & $1$ & $0$ & $0$ & $0$ & $1$ & $1$ & $-\frac{5}{2}$ & $0$ & $0$ & $0$ & $0$ & $(\mathbbm 1, \mathbbm 1,\mathbbm 1)$ & $S_5$ \\
\jvsb$(\mathbbm 1,\mathbbm 1,\mathbbm 1,\mathbbm 1)$ & $1$ & $0$ & $0$ & $0$ & $1$ & $1$ & $-\frac{5}{2}$ & $0$ & $0$ & $0$ & $0$ & $(\mathbbm 1, \mathbbm 1,\mathbbm 1)$ &  $S_6$ \\
\hline
\jvsb$(\mathbbm 1,\mathbbm 1,{\bf 2},\mathbbm 1)$ & $1$ & $0$ & $1$ & $0$ & $1$ & $\frac{1}{2}$ & $\frac{5}{2}$ & $-1$ & $0$ & $0$ & $\frac{1}{3}$ & $(\mathbbm 1, {\bf \bar 3},\mathbbm 1)$ &  \\
\jvsb$(\mathbbm 1,\mathbbm 1,\mathbbm 1,{\bf 6})$ & $1$ & $0$ & $1$ & $0$ & $\frac{1}{3}$ & $1$ & $\frac{5}{2}$ & $0$ & $0$ & $0$ & $0$ & $(\mathbbm 1, \mathbbm 1,{\bf 8}_c)$ &  \\
\jvsb$(\mathbbm 1,\mathbbm 1,\mathbbm 1,\mathbbm 1)$ & $1$ & $0$ & $1$ & $0$ & $1$ & $1$ & $\frac{5}{2}$ & $2$ & $0$ & $0$ & $\frac{1}{3}$ & $(\mathbbm 1, {\bf \bar 3},\mathbbm 1)$ &  \\
\jvsb$(\mathbbm 1,\mathbbm 1,\mathbbm 1,\mathbbm 1)$ & $1$ & $0$ & $1$ & $0$ & $\frac{1}{3}$ & $\frac{1}{2}$ & $\frac{5}{2}$ & $0$ & $0$ & $0$ & $0$ & $(\mathbbm 1, \mathbbm 1,\mathbbm 1)$ & $S_7$  \\
\jvsb$(\mathbbm 1,\mathbbm 1,\mathbbm 1,\mathbbm 1)$ & $1$ & $0$ & $1$ & $0$ & $\frac{1}{3}$ & $\frac{1}{2}$ & $\frac{5}{2}$ & $0$ & $-2$ & $0$ & $0$ & $(\mathbbm 1, \mathbbm 1,{\bf 8}_c)$ &  \\
\jvsb$(\mathbbm 1,\mathbbm 1,\mathbbm 1,\mathbbm 1)$ & $1$ & $0$ & $1$ & $0$ & $\frac{1}{3}$ & $\frac{1}{2}$ & $\frac{5}{2}$ & $0$ & $2$ & $0$ & $0$ & $(\mathbbm 1, \mathbbm 1,{\bf 8}_c)$ &  \\
\hline
\jvsb$(\mathbbm 1,\mathbbm 1,{\bf 2},\mathbbm 1)$ & $1$ & $0$ & $2$ & $0$ & $\frac{2}{3}$ & $1$ & $\frac{5}{2}$ & $1$ & $0$ & $0$ & $-\frac{1}{3}$ & $(\mathbbm 1, {\bf 3},\mathbbm 1)$ &  \\
\jvsb$(\mathbbm 1,\mathbbm 1,\mathbbm 1,\mathbbm 1)$ & $1$ & $0$ & $2$ & $0$ & $\frac{2}{3}$ & $\frac{1}{2}$ & $\frac{5}{2}$ & $-2$ & $0$ & $0$ & $-\frac{1}{3}$ & $(\mathbbm 1, {\bf 3},\mathbbm 1)$ &  \\
\jvsb$(\mathbbm 1,\mathbbm 1,\mathbbm 1,\mathbbm 1)$ & $1$ & $0$ & $2$ & $0$ & $\frac{1}{3}$ & $\frac{1}{2}$ & $\frac{5}{2}$ & $0$ & $0$ & $0$ & $-1$ & $(\mathbbm 1, \mathbbm 1,\mathbbm 1)$ & $S_8$ \\
\hline
\jvsb$(\mathbbm 1,{\bf 2},\mathbbm 1,\mathbbm 1)$ & $1$ & $1$ & $0$ & $0$ & $1$ & $1$ & $0$ & $1$ & $-1$ & $0$ & $-\frac{5}{6}$ & $({\bf 2}, \mathbbm 1,\mathbbm 1,\mathbbm 1)$ &  \\
\jvsb$(\mathbbm 1,\mathbbm 1,\mathbbm 1,\mathbbm 1)$ & $1$ & $1$ & $0$ & $0$ & $\frac{2}{3}$ & $\frac{1}{2}$ & $1$ & $1$ & $-1$ & $-\frac{1}{2}$ & $\frac{1}{6}$ & $(\mathbbm 1,\mathbbm 1,\mathbbm 1,\mathbbm 1)$ &  \\
\jvsb$(\mathbbm 1,\mathbbm 1,\mathbbm 1,\mathbbm 1)$ & $1$ & $1$ & $0$ & $0$ & $\frac{1}{3}$ & $\frac{1}{2}$ & $-1$ & $1$ & $-1$ & $\frac{1}{2}$ & $\frac{1}{6}$ & $(\mathbbm 1,\mathbbm 1,\mathbbm 1,\mathbbm 1)$ &  \\
\hline
\jvsb$(\mathbbm 1,{\bf 2},\mathbbm 1,\mathbbm 1)$ & $1$ & $1$ & $1$ & $0$ & $\frac{1}{3}$ & $\frac{1}{2}$ & $0$ & $-1$ & $1$ & $0$ & $\frac{5}{6}$ & $({\bf 2}, \mathbbm 1,\mathbbm 1,\mathbbm 1)$ &  \\
\jvsb$(\mathbbm 1,\mathbbm 1,\mathbbm 1,\mathbbm 1)$ & $1$ & $1$ & $1$ & $0$ & $1$ & $\frac{1}{2}$ & $1$ & $-1$ & $1$ & $-\frac{1}{2}$ & $-\frac{1}{6}$ & $(\mathbbm 1,\mathbbm 1,\mathbbm 1,\mathbbm 1)$ &  \\
\jvsb$(\mathbbm 1,\mathbbm 1,\mathbbm 1,\mathbbm 1)$ & $1$ & $1$ & $1$ & $0$ & $\frac{2}{3}$ & $\frac{1}{2}$ & $-1$ & $-1$ & $1$ & $\frac{1}{2}$ & $-\frac{1}{6}$ & $(\mathbbm 1,\mathbbm 1,\mathbbm 1,\mathbbm 1)$ &  \\
\hline
\jvsb$(\mathbbm 1,{\bf 2},\mathbbm 1,\mathbbm 1)$ & $1$ & $1$ & $2$ & $0$ & $1$ & $1$ & $0$ & $1$ & $1$ & $0$ & $\frac{1}{6}$ & $({\bf 2}, \mathbbm 1,\mathbbm 1,\mathbbm 1)$ &  \\
\jvsb$(\mathbbm 1,{\bf 2},\mathbbm 1,\mathbbm 1)$ & $1$ & $1$ & $2$ & $0$ & $\frac{1}{3}$ & $\frac{1}{2}$ & $0$ & $-1$ & $-1$ & $0$ & $-\frac{1}{6}$ & $({\bf 2}, \mathbbm 1,\mathbbm 1,\mathbbm 1)$ &  \\
\jvsb$(\mathbbm 1,\mathbbm 1,\mathbbm 1,\mathbbm 1)$ & $1$ & $1$ & $2$ & $0$ & $1$ & $\frac{1}{2}$ & $1$ & $-1$ & $-1$ & $-\frac{1}{2}$ & $-\frac{7}{6}$ & $(\mathbbm 1,\mathbbm 1,\mathbbm 1,\mathbbm 1)$ &  \\
\jvsb$(\mathbbm 1,\mathbbm 1,\mathbbm 1,\mathbbm 1)$ & $1$ & $1$ & $2$ & $0$ & $\frac{2}{3}$ & $1$ & $1$ & $1$ & $1$ & $-\frac{1}{2}$ & $-\frac{5}{6}$ & $(\mathbbm 1,\mathbbm 1,\mathbbm 1,\mathbbm 1)$ &  \\
\jvsb$(\mathbbm 1,\mathbbm 1,\mathbbm 1,\mathbbm 1)$ & $1$ & $1$ & $2$ & $0$ & $\frac{2}{3}$ & $1$ & $-1$ & $-1$ & $-1$ & $\frac{1}{2}$ & $\frac{5}{6}$ & $(\mathbbm 1,\mathbbm 1,\mathbbm 1,\mathbbm 1)$ &  \\
\jvsb$(\mathbbm 1,\mathbbm 1,\mathbbm 1,\mathbbm 1)$ & $1$ & $1$ & $2$ & $0$ & $\frac{1}{3}$ & $\frac{1}{2}$ & $-1$ & $1$ & $1$ & $\frac{1}{2}$ & $\frac{7}{6}$ & $(\mathbbm 1,\mathbbm 1,\mathbbm 1,\mathbbm 1)$ &  \\
\hline
\jvsb$({\bf 3},\mathbbm 1,\mathbbm 1,\mathbbm 1)$ & $2$ & $0$ & $0$ & $\frac{1}{2}$ & $\frac{1}{3}$ & $0$ & $-1$ & $0$ & $0$ & $-\frac{1}{3}$ & $-\frac{2}{3}$ & $({\bf  5},\mathbbm 1, \mathbbm 1)$ & $\F_{0,  \frac12}$ \\
\jvsb$({\bf \bar 3},\mathbbm 1,\mathbbm 1,\mathbbm 1)$ & $2$ & $0$ & $0$ & $\frac{1}{2}$ & $1$ & $0$ & $1$ & $0$ & $0$ & $\frac{1}{3}$ & $\frac{2}{3}$ & $({\bf \bar 5},\mathbbm 1, \mathbbm 1)$ & $\Fb_{0,  \frac12}$ \\
\jvsb$(\mathbbm 1,\mathbbm 1,\mathbbm 1,\mathbbm 1)$ & $2$ & $0$ & $0$ & $\frac{1}{2}$ & $\frac{1}{3}$ & $0$ & $0$ & $-2$ & $0$ & $0$ & $\frac{2}{3}$ & $(\mathbbm 1, {\bf 3},\mathbbm 1)$ &  \\
\jvsb$(\mathbbm 1,\mathbbm 1,{\bf 2},\mathbbm 1)$ & $2$ & $0$ & $0$ & $\frac{1}{2}$ & $1$ & $0$ & $0$ & $-1$ & $0$ & $0$ & $\frac{1}{3}$ & $(\mathbbm 1, {\bf \bar 3},\mathbbm 1)$ &  \\
\jvsb$(\mathbbm 1,\mathbbm 1,\mathbbm 1,{\bf \bar 4})$ & $2$ & $0$ & $0$ & $\frac{1}{2}$ & $\frac{2}{3}$ & $0$ & $0$ & $0$ & $1$ & $0$ & $\frac{1}{2}$ & $(\mathbbm 1, \mathbbm 1,{\bf 8})$ &  \\
\jvsb$(\mathbbm 1,\mathbbm 1,\mathbbm 1,\mathbbm 1)$ & $2$ & $0$ & $0$ & $0$ & $\frac{2}{3}$ & $0$ & $0$ & $0$ & $0$ & $0$ & $1$ & $(\mathbbm 1, \mathbbm 1,\mathbbm 1)$ & $Y_{0,  0}$ \\
\jvsb$(\mathbbm 1,\mathbbm 1,\mathbbm 1,\mathbbm 1)$ & $2$ & $0$ & $0$ & $0$ & $\frac{2}{3}$ & $0$ & $0$ & $0$ & $0$ & $0$ & $-1$ & $(\mathbbm 1, \mathbbm 1,\mathbbm 1)$ &  $\bar Y_{0,  0}$  \\
\jvsb$(\mathbbm 1,\mathbbm 1,\mathbbm 1,\mathbbm 1)$ & $2$ & $0$ & $0$ & $1$ & $\frac{2}{3}$ & $0$ & $0$ & $0$ & $0$ & $0$ & $1$ & $(\mathbbm 1, \mathbbm 1,\mathbbm 1)$ & $Y_{0,  1}$ \\
\jvsb$(\mathbbm 1,\mathbbm 1,\mathbbm 1,\mathbbm 1)$ & $2$ & $0$ & $0$ & $1$ & $\frac{2}{3}$ & $0$ & $0$ & $0$ & $0$ & $0$ & $-1$ & $(\mathbbm 1, \mathbbm 1,\mathbbm 1)$ & $\bar Y_{0,  1}$ \\
\hline
\jvsb$(\mathbbm 1,{\bf 2},\mathbbm 1,\mathbbm 1)$ & $2$ & $0$ & $1$ & $0$ & $\frac{1}{3}$ & $0$ & $-1$ & $0$ & $0$ & $\frac{1}{2}$ & $0$ & $({\bf  5},\mathbbm 1, \mathbbm 1)$ &  $\F_{1,  0}$  \\
\jvsb$(\mathbbm 1,{\bf 2},\mathbbm 1,\mathbbm 1)$ & $2$ & $0$ & $1$ & $0$ & $1$ & $0$ & $1$ & $0$ & $0$ & $-\frac{1}{2}$ & $0$ & $({\bf \bar 5},\mathbbm 1, \mathbbm 1)$ & $\Fb_{1,  0}$ \\
\jvsb$(\mathbbm 1,{\bf 2},\mathbbm 1,\mathbbm 1)$ & $2$ & $0$ & $1$ & $1$ & $\frac{1}{3}$ & $0$ & $-1$ & $0$ & $0$ & $\frac{1}{2}$ & $0$ & $({\bf  5},\mathbbm 1, \mathbbm 1)$ & $\F_{1,  1}$ \\
\jvsb$(\mathbbm 1,{\bf 2},\mathbbm 1,\mathbbm 1)$ & $2$ & $0$ & $1$ & $1$ & $1$ & $0$ & $1$ & $0$ & $0$ & $-\frac{1}{2}$ & $0$ & $({\bf \bar 5},\mathbbm 1, \mathbbm 1)$ & $\Fb_{1,  1}$ \\
\jvsb$(\mathbbm 1,\mathbbm 1,{\bf 2},\mathbbm 1)$ & $2$ & $0$ & $1$ & $\frac{1}{2}$ & $\frac{1}{3}$ & $0$ & $0$ & $1$ & $0$ & $0$ & $-\frac{1}{3}$ & $(\mathbbm 1, {\bf 3},\mathbbm 1)$ &  \\
\jvsb$(\mathbbm 1,\mathbbm 1,\mathbbm 1,\mathbbm 1)$ & $2$ & $0$ & $1$ & $\frac{1}{2}$ & $1$ & $0$ & $0$ & $2$ & $0$ & $0$ & $-\frac{2}{3}$ & $(\mathbbm 1, {\bf \bar 3},\mathbbm 1)$ &  \\
\jvsb$(\mathbbm 1,\mathbbm 1,\mathbbm 1,{\bf 4})$ & $2$ & $0$ & $1$ & $0$ & $\frac{2}{3}$ & $0$ & $0$ & $0$ & $1$ & $0$ & $-\frac{1}{2}$ & $(\mathbbm 1, \mathbbm 1,{\bf 8}_s)$ &  \\
\jvsb$(\mathbbm 1,\mathbbm 1,\mathbbm 1,{\bf 4})$ & $2$ & $0$ & $1$ & $1$ & $\frac{2}{3}$ & $0$ & $0$ & $0$ & $1$ & $0$ & $-\frac{1}{2}$ & $(\mathbbm 1, \mathbbm 1,{\bf 8}_s)$ &  \\
\jvsb$(\mathbbm 1,\mathbbm 1,\mathbbm 1,\mathbbm 1)$ & $2$ & $0$ & $1$ & $\frac{1}{2}$ & $\frac{1}{3}$ & $0$ & $5$ & $0$ & $0$ & $0$ & $0$ & $(\mathbbm 1, \mathbbm 1,\mathbbm 1)$ & $X_{1,  \frac12}$ \\
\jvsb$(\mathbbm 1,\mathbbm 1,\mathbbm 1,\mathbbm 1)$ & $2$ & $0$ & $1$ & $\frac{1}{2}$ & $1$ & $0$ & $-5$ & $0$ & $0$ & $0$ & $0$ & $(\mathbbm 1, \mathbbm 1,\mathbbm 1)$ & $\bar X_{1,  \frac12}$ \\
\jvsb$(\mathbbm 1,\mathbbm 1,\mathbbm 1,\mathbbm 1)$ & $2$ & $0$ & $1$ & $0$ & $\frac{2}{3}$ & $0$ & $0$ & $0$ & $0$ & $0$ & $1$ & $(\mathbbm 1, \mathbbm 1,\mathbbm 1)$ &  $Y_{1,  0}$\\
\jvsb$(\mathbbm 1,\mathbbm 1,\mathbbm 1,\mathbbm 1)$ & $2$ & $0$ & $1$ & $0$ & $\frac{2}{3}$ & $0$ & $0$ & $0$ & $0$ & $0$ & $-1$ & $(\mathbbm 1, \mathbbm 1,\mathbbm 1)$ & $\bar Y_{1,  0}$ \\
\jvsb$(\mathbbm 1,\mathbbm 1,\mathbbm 1,\mathbbm 1)$ & $2$ & $0$ & $1$ & $1$ & $\frac{2}{3}$ & $0$ & $0$ & $0$ & $0$ & $0$ & $1$ & $(\mathbbm 1, \mathbbm 1,\mathbbm 1)$ &$Y_{1,  1}$ \\
\jvsb$(\mathbbm 1,\mathbbm 1,\mathbbm 1,\mathbbm 1)$ & $2$ & $0$ & $1$ & $1$ & $\frac{2}{3}$ & $0$ & $0$ & $0$ & $0$ & $0$ & $-1$ & $(\mathbbm 1, \mathbbm 1,\mathbbm 1)$ & $\bar Y_{1,  1}$ \\
\hline
\jvsb$({\bf \bar 3},\mathbbm 1,\mathbbm 1,\mathbbm 1)$ & $2$ & $0$ & $2$ & $0$ & $1$ & $0$ & $1$ & $0$ & $0$ & $\frac{1}{3}$ & $-\frac{1}{3}$ & $({\bf \bar 5},\mathbbm 1, \mathbbm 1)$ & $\Fb_{2,  0}$ \\
\jvsb$({\bf \bar 3},\mathbbm 1,\mathbbm 1,\mathbbm 1)$ & $2$ & $0$ & $2$ & $1$ & $1$ & $0$ & $1$ & $0$ & $0$ & $\frac{1}{3}$ & $-\frac{1}{3}$ & $({\bf \bar 5},\mathbbm 1, \mathbbm 1)$ & $\Fb_{2,  1}$ \\
\jvsb$(\mathbbm 1,{\bf 2},\mathbbm 1,\mathbbm 1)$ & $2$ & $0$ & $2$ & $\frac{1}{2}$ & $\frac{1}{3}$ & $0$ & $-1$ & $0$ & $0$ & $\frac{1}{2}$ & $1$ & $({\bf  5},\mathbbm 1, \mathbbm 1)$ &  $\F_{2,  \frac12}$\\
\jvsb$(\mathbbm 1,\mathbbm 1,\mathbbm 1,{\bf 6})$ & $2$ & $0$ & $2$ & $\frac{1}{2}$ & $\frac{2}{3}$ & $0$ & $0$ & $0$ & $0$ & $0$ & $0$ & $(\mathbbm 1, \mathbbm 1,{\bf 8}_c)$ &  \\
\jvsb$(\mathbbm 1,\mathbbm 1,\mathbbm 1,\mathbbm 1)$ & $2$ & $0$ & $2$ & $0$ & $1$ & $0$ & $0$ & $2$ & $0$ & $0$ & $\frac{1}{3}$ & $(\mathbbm 1, {\bf \bar 3},\mathbbm 1)$ &  \\
\jvsb$(\mathbbm 1,\mathbbm 1,\mathbbm 1,\mathbbm 1)$ & $2$ & $0$ & $2$ & $0$ & $\frac{1}{3}$ & $0$ & $0$ & $-2$ & $0$ & $0$ & $-\frac{1}{3}$ & $(\mathbbm 1, {\bf 3},\mathbbm 1)$ &  \\
\jvsb$(\mathbbm 1,\mathbbm 1,\mathbbm 1,\mathbbm 1)$ & $2$ & $0$ & $2$ & $1$ & $1$ & $0$ & $0$ & $2$ & $0$ & $0$ & $\frac{1}{3}$ & $(\mathbbm 1, {\bf \bar 3},\mathbbm 1)$ &  \\
\jvsb$(\mathbbm 1,\mathbbm 1,\mathbbm 1,\mathbbm 1)$ & $2$ & $0$ & $2$ & $1$ & $\frac{1}{3}$ & $0$ & $0$ & $-2$ & $0$ & $0$ & $-\frac{1}{3}$ & $(\mathbbm 1, {\bf 3},\mathbbm 1)$ &  \\
\jvsb$(\mathbbm 1,\mathbbm 1,\mathbbm 1,\mathbbm 1)$ & $2$ & $0$ & $2$ & $0$ & $\frac{1}{3}$ & $0$ & $5$ & $0$ & $0$ & $0$ & $1$ & $(\mathbbm 1, \mathbbm 1,\mathbbm 1)$ &  $X_{2,  0}$\\
\jvsb$(\mathbbm 1,\mathbbm 1,\mathbbm 1,\mathbbm 1)$ & $2$ & $0$ & $2$ & $1$ & $\frac{1}{3}$ & $0$ & $5$ & $0$ & $0$ & $0$ & $1$ & $(\mathbbm 1, \mathbbm 1,\mathbbm 1)$ & $X_{2,  1}$ \\
\jvsb$(\mathbbm 1,\mathbbm 1,\mathbbm 1,\mathbbm 1)$ & $2$ & $0$ & $2$ & $\frac{1}{2}$ & $\frac{2}{3}$ & $0$ & $0$ & $0$ & $0$ & $0$ & $0$ & $(\mathbbm 1, \mathbbm 1,\mathbbm 1)$ &  $\bar Y_{2,  \frac12}$\\
\hline
\jvsb$(\mathbbm 1,\mathbbm 1,{\bf 2},\mathbbm 1)$ & $3$ & $0$ & $0$ & $\frac{1}{3}$ & $0$ & $\frac{1}{2}$ & $\frac{5}{2}$ & $1$ & $0$ & $0$ & $-\frac{1}{3}$ & $(\mathbbm 1, {\bf 3},\mathbbm 1)$ &  \\
\jvsb$(\mathbbm 1,\mathbbm 1,{\bf 2},\mathbbm 1)$ & $3$ & $0$ & $0$ & $\frac{1}{3}$ & $0$ & $\frac{1}{2}$ & $-\frac{5}{2}$ & $-1$ & $0$ & $0$ & $\frac{1}{3}$ & $(\mathbbm 1, {\bf \bar 3},\mathbbm 1)$ &  \\
\jvsb$(\mathbbm 1,\mathbbm 1,\mathbbm 1,\mathbbm 1)$ & $3$ & $0$ & $0$ & $\frac{1}{3}$ & $0$ & $1$ & $\frac{5}{2}$ & $-2$ & $0$ & $0$ & $-\frac{1}{3}$ & $(\mathbbm 1, {\bf 3},\mathbbm 1)$ &  \\
\jvsb$(\mathbbm 1,\mathbbm 1,\mathbbm 1,\mathbbm 1)$ & $3$ & $0$ & $0$ & $\frac{1}{3}$ & $0$ & $1$ & $-\frac{5}{2}$ & $2$ & $0$ & $0$ & $\frac{1}{3}$ & $(\mathbbm 1, {\bf \bar 3},\mathbbm 1)$ &  \\
\hline
\jvsb$(\mathbbm 1,\mathbbm 1,\mathbbm 1,\mathbbm 1)$ & $3$ & $1$ & $0$ & $0$ & $0$ & $\frac{1}{2}$ & $1$ & $1$ & $-1$ & $-\frac{1}{2}$ & $-\frac{5}{6}$ & $(\mathbbm 1,\mathbbm 1,\mathbbm 1,\mathbbm 1)$ &  \\
\jvsb$(\mathbbm 1,\mathbbm 1,\mathbbm 1,\mathbbm 1)$ & $3$ & $1$ & $0$ & $0$ & $0$ & $\frac{1}{2}$ & $-1$ & $-1$ & $1$ & $\frac{1}{2}$ & $\frac{5}{6}$ & $(\mathbbm 1,\mathbbm 1,\mathbbm 1,\mathbbm 1)$ &  \\
\jvsb$(\mathbbm 1,\mathbbm 1,\mathbbm 1,\mathbbm 1)$ & $3$ & $1$ & $0$ & $\frac{1}{3}$ & $0$ & $1$ & $1$ & $1$ & $-1$ & $-\frac{1}{2}$ & $\frac{7}{6}$ & $(\mathbbm 1,\mathbbm 1,\mathbbm 1,\mathbbm 1)$ &  \\
\jvsb$(\mathbbm 1,\mathbbm 1,\mathbbm 1,\mathbbm 1)$ & $3$ & $1$ & $0$ & $\frac{1}{3}$ & $0$ & $1$ & $-1$ & $-1$ & $1$ & $\frac{1}{2}$ & $-\frac{7}{6}$ & $(\mathbbm 1,\mathbbm 1,\mathbbm 1,\mathbbm 1)$ &  \\
\jvsb$(\mathbbm 1,\mathbbm 1,\mathbbm 1,\mathbbm 1)$ & $3$ & $1$ & $0$ & $\frac{2}{3}$ & $0$ & $\frac{1}{2}$ & $1$ & $1$ & $-1$ & $-\frac{1}{2}$ & $-\frac{5}{6}$ & $(\mathbbm 1,\mathbbm 1,\mathbbm 1,\mathbbm 1)$ &  \\
\jvsb$(\mathbbm 1,\mathbbm 1,\mathbbm 1,\mathbbm 1)$ & $3$ & $1$ & $0$ & $\frac{2}{3}$ & $0$ & $\frac{1}{2}$ & $-1$ & $-1$ & $1$ & $\frac{1}{2}$ & $\frac{5}{6}$ & $(\mathbbm 1,\mathbbm 1,\mathbbm 1,\mathbbm 1)$ &  \\
\jvsb$(\mathbbm 1,\mathbbm 1,\mathbbm 1,\mathbbm 1)$ & $3$ & $1$ & $0$ & $1$ & $0$ & $\frac{1}{2}$ & $1$ & $1$ & $-1$ & $-\frac{1}{2}$ & $-\frac{5}{6}$ & $(\mathbbm 1,\mathbbm 1,\mathbbm 1,\mathbbm 1)$ &  \\
\jvsb$(\mathbbm 1,\mathbbm 1,\mathbbm 1,\mathbbm 1)$ & $3$ & $1$ & $0$ & $1$ & $0$ & $\frac{1}{2}$ & $-1$ & $-1$ & $1$ & $\frac{1}{2}$ & $\frac{5}{6}$ & $(\mathbbm 1,\mathbbm 1,\mathbbm 1,\mathbbm 1)$ &  \\
\hline
\jvsb$({\bf 3},\mathbbm 1,\mathbbm 1,\mathbbm 1)$ & $4$ & $0$ & $0$ & $0$ & $1$ & $0$ & $-1$ & $0$ & $0$ & $-\frac{1}{3}$ & $-\frac{2}{3}$ & $({\bf  5},\mathbbm 1, \mathbbm 1)$ & $\Fb^c_{0,  0}$ \\
\jvsb$({\bf 3},\mathbbm 1,\mathbbm 1,\mathbbm 1)$ & $4$ & $0$ & $0$ & $1$ & $1$ & $0$ & $-1$ & $0$ & $0$ & $-\frac{1}{3}$ & $-\frac{2}{3}$ & $({\bf  5},\mathbbm 1, \mathbbm 1)$ &  $\Fb^c_{0,  1}$\\
\jvsb$({\bf \bar 3},\mathbbm 1,\mathbbm 1,\mathbbm 1)$ & $4$ & $0$ & $0$ & $0$ & $\frac{1}{3}$ & $0$ & $1$ & $0$ & $0$ & $\frac{1}{3}$ & $\frac{2}{3}$ & $({\bf \bar 5},\mathbbm 1, \mathbbm 1)$ &$\F^c_{0,  0}$  \\
\jvsb$({\bf \bar 3},\mathbbm 1,\mathbbm 1,\mathbbm 1)$ & $4$ & $0$ & $0$ & $1$ & $\frac{1}{3}$ & $0$ & $1$ & $0$ & $0$ & $\frac{1}{3}$ & $\frac{2}{3}$ & $({\bf \bar 5},\mathbbm 1, \mathbbm 1)$ &  $\F^c_{0,  1}$\\
\jvsb$(\mathbbm 1,\mathbbm 1,{\bf 2},\mathbbm 1)$ & $4$ & $0$ & $0$ & $0$ & $1$ & $0$ & $0$ & $1$ & $0$ & $0$ & $-\frac{1}{3}$ & $(\mathbbm 1, {\bf 3},\mathbbm 1)$ &  \\
\jvsb$(\mathbbm 1,\mathbbm 1,{\bf 2},\mathbbm 1)$ & $4$ & $0$ & $0$ & $1$ & $1$ & $0$ & $0$ & $1$ & $0$ & $0$ & $-\frac{1}{3}$ & $(\mathbbm 1, {\bf 3},\mathbbm 1)$ &  \\
\jvsb$(\mathbbm 1,\mathbbm 1,\mathbbm 1,{\bf 4})$ & $4$ & $0$ & $0$ & $0$ & $\frac{2}{3}$ & $0$ & $0$ & $0$ & $-1$ & $0$ & $-\frac{1}{2}$ & $(\mathbbm 1, \mathbbm 1,{\bf 8})$ &  \\
\jvsb$(\mathbbm 1,\mathbbm 1,\mathbbm 1,{\bf 4})$ & $4$ & $0$ & $0$ & $1$ & $\frac{2}{3}$ & $0$ & $0$ & $0$ & $-1$ & $0$ & $-\frac{1}{2}$ & $(\mathbbm 1, \mathbbm 1,{\bf 8})$ &  \\
\jvsb$(\mathbbm 1,\mathbbm 1,\mathbbm 1,\mathbbm 1)$ & $4$ & $0$ & $0$ & $0$ & $\frac{1}{3}$ & $0$ & $0$ & $2$ & $0$ & $0$ & $-\frac{2}{3}$ & $(\mathbbm 1, {\bf \bar 3},\mathbbm 1)$ &  \\
\jvsb$(\mathbbm 1,\mathbbm 1,\mathbbm 1,\mathbbm 1)$ & $4$ & $0$ & $0$ & $1$ & $\frac{1}{3}$ & $0$ & $0$ & $2$ & $0$ & $0$ & $-\frac{2}{3}$ & $(\mathbbm 1, {\bf \bar 3},\mathbbm 1)$ &  \\
\jvsb$(\mathbbm 1,\mathbbm 1,\mathbbm 1,\mathbbm 1)$ & $4$ & $0$ & $0$ & $\frac{1}{2}$ & $\frac{2}{3}$ & $0$ & $0$ & $0$ & $0$ & $0$ & $-1$ & $(\mathbbm 1, \mathbbm 1,\mathbbm 1)$ & $Y_{0,  \frac12}^c$ \\
\jvsb$(\mathbbm 1,\mathbbm 1,\mathbbm 1,\mathbbm 1)$ & $4$ & $0$ & $0$ & $\frac{1}{2}$ & $\frac{2}{3}$ & $0$ & $0$ & $0$ & $0$ & $0$ & $1$ & $(\mathbbm 1, \mathbbm 1,\mathbbm 1)$ &  $\bar Y_{0,  \frac12}^c$\\
\hline
\jvsb$(\mathbbm 1,{\bf 2},\mathbbm 1,\mathbbm 1)$ & $4$ & $0$ & $1$ & $\frac{1}{2}$ & $1$ & $0$ & $-1$ & $0$ & $0$ & $\frac{1}{2}$ & $0$ & $({\bf  5},\mathbbm 1, \mathbbm 1)$ & $\Fb_{1,  \frac12}^c$ \\
\jvsb$(\mathbbm 1,{\bf 2},\mathbbm 1,\mathbbm 1)$ & $4$ & $0$ & $1$ & $\frac{1}{2}$ & $\frac{1}{3}$ & $0$ & $1$ & $0$ & $0$ & $-\frac{1}{2}$ & $0$ & $({\bf \bar 5},\mathbbm 1, \mathbbm 1)$ &  $\F_{1,  \frac12}^c$\\
\jvsb$(\mathbbm 1,\mathbbm 1,{\bf 2},\mathbbm 1)$ & $4$ & $0$ & $1$ & $0$ & $\frac{1}{3}$ & $0$ & $0$ & $-1$ & $0$ & $0$ & $\frac{1}{3}$ & $(\mathbbm 1, {\bf \bar 3},\mathbbm 1)$ &  \\
\jvsb$(\mathbbm 1,\mathbbm 1,{\bf 2},\mathbbm 1)$ & $4$ & $0$ & $1$ & $1$ & $\frac{1}{3}$ & $0$ & $0$ & $-1$ & $0$ & $0$ & $\frac{1}{3}$ & $(\mathbbm 1, {\bf \bar 3},\mathbbm 1)$ &  \\
\jvsb$(\mathbbm 1,\mathbbm 1,\mathbbm 1,{\bf \bar 4})$ & $4$ & $0$ & $1$ & $\frac{1}{2}$ & $\frac{2}{3}$ & $0$ & $0$ & $0$ & $-1$ & $0$ & $\frac{1}{2}$ & $(\mathbbm 1, \mathbbm 1,{\bf 8}_s)$ &  \\
\jvsb$(\mathbbm 1,\mathbbm 1,\mathbbm 1,\mathbbm 1)$ & $4$ & $0$ & $1$ & $0$ & $1$ & $0$ & $0$ & $-2$ & $0$ & $0$ & $\frac{2}{3}$ & $(\mathbbm 1, {\bf 3},\mathbbm 1)$ &  \\
\jvsb$(\mathbbm 1,\mathbbm 1,\mathbbm 1,\mathbbm 1)$ & $4$ & $0$ & $1$ & $1$ & $1$ & $0$ & $0$ & $-2$ & $0$ & $0$ & $\frac{2}{3}$ & $(\mathbbm 1, {\bf 3},\mathbbm 1)$ &  \\
\jvsb$(\mathbbm 1,\mathbbm 1,\mathbbm 1,\mathbbm 1)$ & $4$ & $0$ & $1$ & $0$ & $1$ & $0$ & $5$ & $0$ & $0$ & $0$ & $0$ & $(\mathbbm 1, \mathbbm 1,\mathbbm 1)$ & $\bar X^c_{1,  0}$  \\
\jvsb$(\mathbbm 1,\mathbbm 1,\mathbbm 1,\mathbbm 1)$ & $4$ & $0$ & $1$ & $0$ & $\frac{1}{3}$ & $0$ & $-5$ & $0$ & $0$ & $0$ & $0$ & $(\mathbbm 1, \mathbbm 1,\mathbbm 1)$ & $X^c_{1,  0}$   \\
\jvsb$(\mathbbm 1,\mathbbm 1,\mathbbm 1,\mathbbm 1)$ & $4$ & $0$ & $1$ & $1$ & $1$ & $0$ & $5$ & $0$ & $0$ & $0$ & $0$ & $(\mathbbm 1, \mathbbm 1,\mathbbm 1)$ & $\bar X^c_{1,  1}$  \\
\jvsb$(\mathbbm 1,\mathbbm 1,\mathbbm 1,\mathbbm 1)$ & $4$ & $0$ & $1$ & $1$ & $\frac{1}{3}$ & $0$ & $-5$ & $0$ & $0$ & $0$ & $0$ & $(\mathbbm 1, \mathbbm 1,\mathbbm 1)$ & $ X^c_{1,  1}$  \\
\jvsb$(\mathbbm 1,\mathbbm 1,\mathbbm 1,\mathbbm 1)$ & $4$ & $0$ & $1$ & $\frac{1}{2}$ & $\frac{2}{3}$ & $0$ & $0$ & $0$ & $0$ & $0$ & $-1$ & $(\mathbbm 1, \mathbbm 1,\mathbbm 1)$ & $Y_{1,  \frac12}^c$ \\
\jvsb$(\mathbbm 1,\mathbbm 1,\mathbbm 1,\mathbbm 1)$ & $4$ & $0$ & $1$ & $\frac{1}{2}$ & $\frac{2}{3}$ & $0$ & $0$ & $0$ & $0$ & $0$ & $1$ & $(\mathbbm 1, \mathbbm 1,\mathbbm 1)$ & $\bar Y_{1,  \frac12}^c$ \\
\hline
\jvsb$({\bf 3},\mathbbm 1,\mathbbm 1,\mathbbm 1)$ & $4$ & $0$ & $2$ & $\frac{1}{2}$ & $1$ & $0$ & $-1$ & $0$ & $0$ & $-\frac{1}{3}$ & $\frac{1}{3}$ & $({\bf  5},\mathbbm 1, \mathbbm 1)$ & $\Fb_{2,  \frac12}^c$ \\
\jvsb$(\mathbbm 1,{\bf 2},\mathbbm 1,\mathbbm 1)$ & $4$ & $0$ & $2$ & $0$ & $\frac{1}{3}$ & $0$ & $1$ & $0$ & $0$ & $-\frac{1}{2}$ & $-1$ & $({\bf \bar 5},\mathbbm 1, \mathbbm 1)$ & $\F^c_{2,  0}$ \\
\jvsb$(\mathbbm 1,{\bf 2},\mathbbm 1,\mathbbm 1)$ & $4$ & $0$ & $2$ & $1$ & $\frac{1}{3}$ & $0$ & $1$ & $0$ & $0$ & $-\frac{1}{2}$ & $-1$ & $({\bf \bar 5},\mathbbm 1, \mathbbm 1)$ & $\F^c_{2,  1}$ \\
\jvsb$(\mathbbm 1,\mathbbm 1,\mathbbm 1,{\bf 6})$ & $4$ & $0$ & $2$ & $0$ & $\frac{2}{3}$ & $0$ & $0$ & $0$ & $0$ & $0$ & $0$ & $(\mathbbm 1, \mathbbm 1,{\bf 8}_c)$ &  \\
\jvsb$(\mathbbm 1,\mathbbm 1,\mathbbm 1,{\bf 6})$ & $4$ & $0$ & $2$ & $1$ & $\frac{2}{3}$ & $0$ & $0$ & $0$ & $0$ & $0$ & $0$ & $(\mathbbm 1, \mathbbm 1,{\bf 8}_c)$ &  \\
\jvsb$(\mathbbm 1,\mathbbm 1,\mathbbm 1,\mathbbm 1)$ & $4$ & $0$ & $2$ & $\frac{1}{2}$ & $1$ & $0$ & $0$ & $-2$ & $0$ & $0$ & $-\frac{1}{3}$ & $(\mathbbm 1, {\bf 3},\mathbbm 1)$ &  \\
\jvsb$(\mathbbm 1,\mathbbm 1,\mathbbm 1,\mathbbm 1)$ & $4$ & $0$ & $2$ & $\frac{1}{2}$ & $\frac{1}{3}$ & $0$ & $0$ & $2$ & $0$ & $0$ & $\frac{1}{3}$ & $(\mathbbm 1, {\bf \bar 3},\mathbbm 1)$ &  \\
\jvsb$(\mathbbm 1,\mathbbm 1,\mathbbm 1,\mathbbm 1)$ & $4$ & $0$ & $2$ & $\frac{1}{2}$ & $\frac{1}{3}$ & $0$ & $-5$ & $0$ & $0$ & $0$ & $-1$ & $(\mathbbm 1, \mathbbm 1,\mathbbm 1)$ & $X_{2,  \frac12}^c$ \\
\jvsb$(\mathbbm 1,\mathbbm 1,\mathbbm 1,\mathbbm 1)$ & $4$ & $0$ & $2$ & $0$ & $\frac{2}{3}$ & $0$ & $0$ & $0$ & $0$ & $0$ & $0$ & $(\mathbbm 1, \mathbbm 1,\mathbbm 1)$ & $\bar X_{2,  0}^c$  \\
\jvsb$(\mathbbm 1,\mathbbm 1,\mathbbm 1,\mathbbm 1)$ & $4$ & $0$ & $2$ & $1$ & $\frac{2}{3}$ & $0$ & $0$ & $0$ & $0$ & $0$ & $0$ & $(\mathbbm 1, \mathbbm 1,\mathbbm 1)$ & $\bar X_{2,  1}^c$  \\
\caption{Zero modes in four dimensions and their origin.
 \index{Spectrum!zero modes}}
\label{tab:4d}
\end{longtable}
\end{center}

 \section{Discrete symmetry transformations}
 \label{app:discrete}

In Chapter \ref{cpt:vacuum} three vacua were studied in detail,
\begin{subequations}
\begin{align}
    \mathcal S_0&=\left\{ X_0, \bar X_0^c, U_2, U_4, S_2, S_5 \right\} \,, \\
     	\mathcal S_1&=\mathcal S_0 \cup \left\{
		X_1, \bar X_1, Y_2, S_7
	\right\}\,,\\
	\mathcal S_2&=\mathcal S_0 \cup \left\{ X_1^c,\bar X_1,Y_2^c, \bar Y_2, U_1^c,U_3,S_6,S_7\right\}\,,   
\end{align}
\end{subequations}
with unbroken symmetries \index{Unbroken symmetries}
\begin{subequations}
\begin{align}
 	G_{\rm vac}(\mathcal S_0) &=
	\left[ \tZ{24} \times \Z 6 \right]_R \times
	 \U1_5 \times \Z{120} \,, \\
	 G_{\rm vac}\left(\mathcal S_1\right)&=\left[\Z6\times\tZ2 \right]_R
	 \times\Z{60} \times \Z2\,,\\
	 G_{\rm vac}\left(\mathcal S_2\right)&=\left[ \tZ4 \times \Z2 \right]_R\times \Z{60}\,, 
\end{align}
\end{subequations}
under which a field $\phi$ transforms as \index{Discrete symmetry!transformation phase}
\begin{align}
	\phi &\mapsto e^{2 \pi i \, \eta_{\mathcal S}(\phi)} \phi\,, &
	\eta_{\mathcal S}(\phi)&= \alpha'_{\mathcal S} \cdot {\bf Q}(\phi)+{\bf r}' _{\mathcal S}\cdot  \mathcal K (\phi)\,,
\end{align}
where $\alpha'_{\mathcal S}$ and ${\bf r}_{\mathcal S}'$ can be found in Tables \ref{tab:vacS0}, \ref{tab:vacS1} and \ref{tab:vacS1}
for the vacua $\mathcal S=\mathcal S_0, \mathcal S_1, \mathcal S_2$, respectively.
The phases $\eta_{\mathcal S}(\phi)$ for all fields $\phi$ at the GUT fixed point $(n_2,n_2')=(0,0)$ are
listed in Tables \ref{tab:chns} and \ref{tab:chs}.
These tables confirm that the given symmetries are unbroken in the above vacua.

Furthermore, from the tables one can infer if a coupling of interest can be present in the superpotential
or not. This follows from comparison of the total phase with the vacuum phase with respect
to the unbroken symmetries,
\begin{align}
	 \eta_{\mathcal S_0}(\mathcal K_{\rm vac})&=\left\{\begin{array}{ll}
	\frac56, &\text{for}\, {[\tZ{24}]_R}, \jvsb\\
	\frac13, &\text{for}\, {[\Z{6}]_R}, \jvsb \\
	0, & \text{else},
	\end{array}\right. 
\end{align}
\begin{align}
	 \eta_{\mathcal S_1}(\mathcal K_{\rm vac})&=\left\{\begin{array}{ll}
	\frac13, &\text{for}\, {[\Z{6}]_R}, \jvsb \\
	0, & \text{else},
	\end{array}\right. &
	 \eta_{\mathcal S_2}(\mathcal K_{\rm vac})&=\left\{\begin{array}{ll}
	\frac12, &\text{for}\, {[\tZ{4}]_R}, \jvsb \\
	0, & \text{else}.
	\end{array}\right.	
\end{align}
Couplings may be allowed if the sums of the phases add to the above vacuum values,
up to integers, and for all unbroken symmetries. Otherwise, they are forbidden
to arbitrary order in the singlets. 

Note that the breaking of symmetries is insensitive to overall minus signs. This implies that
singlets $s$ and $s^{-1}$, in the sense of Section~\ref{sec:kernel}, yield the same unbroken
symmetry. In consequence, it is possible that a coupling is allowed from the viewpoint of
the unbroken symmetries, but always involves singlets with negative powers, and is therefore
forbidden for the description of physical interactions. As an example, consider the 
dimension-five proton decay operator
\index{Proton decay!dimension-five} from Equation~(\ref{eq:lsprot5}) in the vacuum $\mathcal S_0$,
\begin{align}
\eta_{\mathcal S_0}\left(\T_{(1)}\T_{(1)}\T_{(1)}\Fb_{(1)}\right)=\eta_{\mathcal S_0}(\mathcal K_{\rm vac})\,.
\end{align}
This term is consistent with the unbroken symmetries of the vacuum. However, an example
for a corresponding coupling reads
\begin{align}
	X_0^{-1} (\bar X_0^c)^2 S_2 S_5 \,\T_{(1)}\T_{(1)}\T_{(1)}\Fb_{(1)}\,,
\end{align}
and in fact it is impossible to find monomials with only positive exponents.
Since the symmetries do not know about the projection rule (\ref{eq:PN})
for physical couplings, they can only  
 state necessary conditions for their existence.


\begin{center}
\footnotesize
\begin{longtable}[c]{c|c|c||c|c|c|c||c c|c c c|c c}
\multicolumn{3}{c||}{} &
\multicolumn{4}{c||}{False vacuum} &
\multicolumn{2}{c|}{Vacuum $\mathcal S_0$} &
\multicolumn{3}{c|}{Vacuum $\mathcal S_1$} &
\multicolumn{2}{c}{Vacuum $\mathcal S_2$}  \jvsb \\
\cline{4-14}
\multicolumn{1}{c|}{Multiplet} &
\multicolumn{1}{c|}{$k$} &
\multicolumn{1}{c||}{ $n_3$} &
\multicolumn{1}{c|}{$R^1$} &
\multicolumn{1}{c|}{$R^2$} &
\multicolumn{1}{c|}{$R^3$} &
\multicolumn{1}{c||}{$t_X$} &
\multicolumn{1}{c}{$[\tZ{24}]_R$} &
\multicolumn{1}{c|}{$[\Z6]_R$} &
\multicolumn{1}{c}{$[\Z6]_R$} &
\multicolumn{1}{c}{$[\tZ2]_R$} &
\multicolumn{1}{c|}{$\Z2$} &
\multicolumn{1}{c}{$[\tZ4]_R$} &
\multicolumn{1}{c}{$[\Z2]_R$} \jvsb
\\ \hline \hline
\endfirsthead
\multicolumn{14}{c}%
{{ \tablename\ \thetable{} - continued from previous page}}  \\
\multicolumn{3}{c||}{} &
\multicolumn{4}{c||}{False vacuum} &
\multicolumn{2}{c|}{Vacuum $\mathcal S_0$} &
\multicolumn{3}{c|}{Vacuum $\mathcal S_1$} &
\multicolumn{2}{c}{Vacuum $\mathcal S_2$}  \jvsb \\
\cline{4-14}
\multicolumn{1}{c|}{Multiplet} &
\multicolumn{1}{c|}{$k$} &
\multicolumn{1}{c||}{ $n_3$} &
\multicolumn{1}{c|}{$R^1$} &
\multicolumn{1}{c|}{$R^2$} &
\multicolumn{1}{c|}{$R^3$} &
\multicolumn{1}{c||}{$t_X$} &
\multicolumn{1}{c}{$[\tZ{24}]_R$} &
\multicolumn{1}{c|}{$[\Z6]_R$} &
\multicolumn{1}{c}{$[\Z6]_R$} &
\multicolumn{1}{c}{$[\tZ2]_R$} &
\multicolumn{1}{c|}{$\Z2$} &
\multicolumn{1}{c}{$[\tZ4]_R$} &
\multicolumn{1}{c}{$[\Z2]_R$} \jvsb
\\ \hline \hline
\endhead
\endlastfoot
\multicolumn{14}{c}%
{{ \tablename\ \thetable{} - continued on next page}} \\
\endfoot
\jvsb$({\bf 10}, \mathbbm 1,\mathbbm 1)$ & $0$ & $0$ & $0$ & $-1$ & $0$ & $\frac{1}{5}$ & $\frac{1}{60}$ & $\frac{2}{3}$ & $\frac{2}{3}$ & $\frac{1}{2}$ & $0$ & $\frac{3}{20}$ & $0$ \\
\jvsb$({\bf 10}, \mathbbm 1,\mathbbm 1)$ & $0$ & $0$ & $-1$ & $0$ & $0$ & $\frac{1}{5}$ & $\frac{1}{60}$ & $\frac{2}{3}$ & $\frac{2}{3}$ & $\frac{1}{2}$ & $0$ & $\frac{3}{20}$ & $0$ \\
\jvsb$({\bf  5},\mathbbm 1, \mathbbm 1)$ & $0$ & $0$ & $0$ & $0$ & $-1$ & $-\frac{2}{5}$ & $\frac{4}{5}$ & $0$ & $0$ & $0$ & $0$ & $\frac{1}{5}$ & $0$ \\
\jvsb$({\bf \bar 5},\mathbbm 1, \mathbbm 1)$ & $0$ & $0$ & $0$ & $0$ & $-1$ & $\frac{2}{5}$ & $\frac{1}{5}$ & $0$ & $0$ & $0$ & $0$ & $\frac{4}{5}$ & $0$ \\
\jvsb$(\mathbbm 1, \mathbbm 1,{\bf 8})$ & $0$ & $0$ & $0$ & $-1$ & $0$ & $\frac{1}{2}$ & $\frac{23}{24}$ & $\frac{1}{3}$ & $\frac{5}{6}$ & $\frac{3}{4}$ & $\frac{1}{2}$ & $\frac{5}{8}$ & $0$ \\
\jvsb$(\mathbbm 1, \mathbbm 1,{\bf 8}_c)$ & $0$ & $0$ & $-1$ & $0$ & $0$ & $0$ & $\frac{13}{16}$ & $0$ & $0$ & $\frac{1}{2}$ & $\frac{1}{2}$ & $\frac{3}{4}$ & $\frac{1}{2}$ \\
\jvsb$(\mathbbm 1, \mathbbm 1,{\bf 8}_s)$ & $0$ & $0$ & $-1$ & $0$ & $0$ & $\frac{1}{2}$ & $\frac{13}{16}$ & $0$ & $\frac{1}{2}$ & $\frac{1}{4}$ & $0$ & $\frac{7}{8}$ & $\frac{1}{2}$ \\
\hline
\jvsb$({\bf 10}, \mathbbm 1,\mathbbm 1)$ & $1$ & $0$ & $-\frac{1}{6}$ & $-\frac{1}{3}$ & $-\frac{1}{2}$ & $\frac{1}{5}$ & $\frac{1}{60}$ & $\frac{2}{3}$ & $\frac{2}{3}$ & $\frac{1}{2}$ & $0$ & $\frac{3}{20}$ & $0$ \\
\jvsb$({\bf \bar 5},\mathbbm 1, \mathbbm 1)$ & $1$ & $0$ & $-\frac{1}{6}$ & $-\frac{1}{3}$ & $-\frac{1}{2}$ & $-\frac{3}{5}$ & $\frac{47}{60}$ & $\frac{1}{3}$ & $\frac{1}{3}$ & $\frac{1}{2}$ & $0$ & $\frac{1}{20}$ & $0$ \\
\hline
\jvsb$(\mathbbm 1, {\bf \bar 3},\mathbbm 1)$ & $1$ & $1$ & $\frac{5}{6}$ & $-\frac{1}{3}$ & $-\frac{1}{2}$ & $\frac{1}{3}$ & $\frac{1}{4}$ & $0$ & $\frac{2}{3}$ & $\frac{1}{2}$ & $0$ & $\frac{7}{12}$ & $0$ \\
\jvsb$(\mathbbm 1, \mathbbm 1,{\bf 8}_c)$ & $1$ & $1$ & $-\frac{1}{6}$ & $-\frac{1}{3}$ & $-\frac{1}{2}$ & $0$ & $\frac{5}{48}$ & $0$ & $0$ & $\frac{1}{2}$ & $\frac{1}{2}$ & $\frac{3}{4}$ & $\frac{1}{2}$ \\
\hline
\jvsb$(\mathbbm 1, {\bf 3},\mathbbm 1)$ & $1$ & $2$ & $-\frac{1}{6}$ & $-\frac{1}{3}$ & $-\frac{1}{2}$ & $-\frac{1}{3}$ & $\frac{11}{12}$ & $0$ & $0$ & $\frac{1}{2}$ & $0$ & $\frac{11}{12}$ & $0$ \\
\hline
\jvsb$({\bf  5},\mathbbm 1, \mathbbm 1)$ & $2$ & $0$ & $-\frac{1}{3}$ & $-\frac{2}{3}$ & $0$ & $-\frac{2}{5}$ & $\frac{4}{5}$ & $0$ & $0$ & $0$ & $0$ & $\frac{1}{5}$ & $0$ \\
\jvsb$({\bf \bar 5},\mathbbm 1, \mathbbm 1)$ & $2$ & $0$ & $-\frac{1}{3}$ & $-\frac{2}{3}$ & $0$ & $\frac{2}{5}$ & $\frac{1}{30}$ & $\frac{1}{3}$ & $\frac{1}{3}$ & $0$ & $0$ & $\frac{3}{10}$ & $0$ \\
\jvsb$(\mathbbm 1, {\bf 3},\mathbbm 1)$ & $2$ & $0$ & $-\frac{1}{3}$ & $-\frac{2}{3}$ & $0$ & $\frac{2}{3}$ & $\frac{23}{24}$ & $\frac{1}{3}$ & $0$ & $0$ & $0$ & $\frac{2}{3}$ & $0$ \\
\jvsb$(\mathbbm 1, {\bf \bar 3},\mathbbm 1)$ & $2$ & $0$ & $-\frac{1}{3}$ & $-\frac{2}{3}$ & $0$ & $\frac{1}{3}$ & $\frac{23}{24}$ & $\frac{1}{3}$ & $\frac{2}{3}$ & $\frac{1}{2}$ & $0$ & $\frac{7}{12}$ & $0$ \\
\jvsb$(\mathbbm 1, \mathbbm 1,{\bf 8})$ & $2$ & $0$ & $-\frac{1}{3}$ & $-\frac{2}{3}$ & $0$ & $\frac{1}{2}$ & $\frac{23}{24}$ & $\frac{1}{3}$ & $\frac{5}{6}$ & $\frac{3}{4}$ & $\frac{1}{2}$ & $\frac{5}{8}$ & $0$ \\
\hline
\jvsb$({\bf  5},\mathbbm 1, \mathbbm 1)$ & $2$ & $1$ & $-\frac{1}{3}$ & $-\frac{2}{3}$ & $0$ & $-\frac{2}{5}$ & $\frac{1}{120}$ & $0$ & $0$ & $0$ & $0$ & $\frac{7}{10}$ & $0$ \\
\jvsb$({\bf \bar 5},\mathbbm 1, \mathbbm 1)$ & $2$ & $1$ & $-\frac{1}{3}$ & $-\frac{2}{3}$ & $0$ & $\frac{2}{5}$ & $\frac{109}{120}$ & $\frac{1}{3}$ & $0$ & $0$ & $0$ & $\frac{4}{5}$ & $0$ \\
\jvsb$(\mathbbm 1, {\bf 3},\mathbbm 1)$ & $2$ & $1$ & $-\frac{1}{3}$ & $-\frac{2}{3}$ & $0$ & $-\frac{1}{3}$ & $\frac{11}{12}$ & $\frac{1}{3}$ & $0$ & $\frac{1}{2}$ & $0$ & $\frac{11}{12}$ & $0$ \\
\jvsb$(\mathbbm 1, {\bf \bar 3},\mathbbm 1)$ & $2$ & $1$ & $-\frac{1}{3}$ & $-\frac{2}{3}$ & $0$ & $-\frac{2}{3}$ & $\frac{23}{24}$ & $\frac{1}{3}$ & $\frac{1}{3}$ & $0$ & $0$ & $\frac{1}{3}$ & $0$ \\
\jvsb$(\mathbbm 1, \mathbbm 1,{\bf 8}_s)$ & $2$ & $1$ & $-\frac{1}{3}$ & $-\frac{2}{3}$ & $0$ & $-\frac{1}{2}$ & $\frac{15}{16}$ & $\frac{1}{3}$ & $\frac{1}{6}$ & $\frac{3}{4}$ & $0$ & $\frac{5}{8}$ & $\frac{1}{2}$ \\
\hline
\jvsb$({\bf  5},\mathbbm 1, \mathbbm 1)$ & $2$ & $2$ & $-\frac{1}{3}$ & $-\frac{2}{3}$ & $0$ & $\frac{3}{5}$ & $\frac{101}{120}$ & $0$ & $0$ & $\frac{1}{2}$ & $0$ & $\frac{19}{20}$ & $0$ \\
\jvsb$({\bf \bar 5},\mathbbm 1, \mathbbm 1)$ & $2$ & $2$ & $-\frac{1}{3}$ & $-\frac{2}{3}$ & $0$ & $-\frac{3}{5}$ & $\frac{109}{120}$ & $\frac{1}{3}$ & $\frac{2}{3}$ & $\frac{1}{2}$ & $0$ & $\frac{11}{20}$ & $0$ \\
\jvsb$(\mathbbm 1, {\bf 3},\mathbbm 1)$ & $2$ & $2$ & $-\frac{1}{3}$ & $-\frac{2}{3}$ & $0$ & $-\frac{1}{3}$ & $\frac{1}{8}$ & $\frac{1}{3}$ & $0$ & $\frac{1}{2}$ & $0$ & $\frac{5}{12}$ & $0$ \\
\jvsb$(\mathbbm 1, {\bf \bar 3},\mathbbm 1)$ & $2$ & $2$ & $-\frac{1}{3}$ & $-\frac{2}{3}$ & $0$ & $\frac{1}{3}$ & $\frac{1}{12}$ & $\frac{1}{3}$ & $0$ & $\frac{1}{2}$ & $0$ & $\frac{1}{12}$ & $0$ \\
\jvsb$(\mathbbm 1, \mathbbm 1,{\bf 8}_c)$ & $2$ & $2$ & $-\frac{1}{3}$ & $-\frac{2}{3}$ & $0$ & $0$ & $\frac{5}{48}$ & $\frac{1}{3}$ & $0$ & $\frac{1}{2}$ & $\frac{1}{2}$ & $\frac{3}{4}$ & $\frac{1}{2}$ \\
\hline
\jvsb$(\mathbbm 1, {\bf 3},\mathbbm 1)$ & $3$ & $0$ & $-\frac{1}{2}$ & $0$ & $-\frac{1}{2}$ & $-\frac{1}{3}$ & $\frac{1}{24}$ & $0$ & $\frac{1}{3}$ & $\frac{1}{2}$ & $0$ & $\frac{5}{12}$ & $0$ \\
\jvsb$(\mathbbm 1, {\bf \bar 3},\mathbbm 1)$ & $3$ & $0$ & $-\frac{1}{2}$ & $0$ & $-\frac{1}{2}$ & $\frac{1}{3}$ & $\frac{19}{24}$ & $0$ & $0$ & $\frac{1}{2}$ & $0$ & $\frac{1}{12}$ & $0$ \\
\hline
\jvsb$({\bf  5},\mathbbm 1, \mathbbm 1)$ & $4$ & $0$ & $-\frac{2}{3}$ & $-\frac{1}{3}$ & $0$ & $-\frac{2}{5}$ & $\frac{4}{5}$ & $0$ & $0$ & $0$ & $0$ & $\frac{1}{5}$ & $0$ \\
\jvsb$({\bf \bar 5},\mathbbm 1, \mathbbm 1)$ & $4$ & $0$ & $-\frac{2}{3}$ & $-\frac{1}{3}$ & $0$ & $\frac{2}{5}$ & $\frac{1}{30}$ & $\frac{1}{3}$ & $\frac{1}{3}$ & $0$ & $0$ & $\frac{3}{10}$ & $0$ \\
\jvsb$(\mathbbm 1, {\bf 3},\mathbbm 1)$ & $4$ & $0$ & $-\frac{2}{3}$ & $-\frac{1}{3}$ & $0$ & $-\frac{1}{3}$ & $\frac{7}{8}$ & $0$ & $\frac{2}{3}$ & $\frac{1}{2}$ & $0$ & $\frac{11}{12}$ & $0$ \\
\jvsb$(\mathbbm 1, {\bf \bar 3},\mathbbm 1)$ & $4$ & $0$ & $-\frac{2}{3}$ & $-\frac{1}{3}$ & $0$ & $-\frac{2}{3}$ & $\frac{7}{8}$ & $0$ & $\frac{1}{3}$ & $0$ & $0$ & $\frac{5}{6}$ & $0$ \\
\jvsb$(\mathbbm 1, \mathbbm 1,{\bf 8})$ & $4$ & $0$ & $-\frac{2}{3}$ & $-\frac{1}{3}$ & $0$ & $-\frac{1}{2}$ & $\frac{7}{8}$ & $0$ & $\frac{1}{2}$ & $\frac{1}{4}$ & $\frac{1}{2}$ & $\frac{7}{8}$ & $0$ \\
\hline
\jvsb$({\bf  5},\mathbbm 1, \mathbbm 1)$ & $4$ & $1$ & $-\frac{2}{3}$ & $-\frac{1}{3}$ & $0$ & $-\frac{2}{5}$ & $\frac{37}{40}$ & $0$ & $\frac{1}{3}$ & $0$ & $0$ & $\frac{7}{10}$ & $0$ \\
\jvsb$({\bf \bar 5},\mathbbm 1, \mathbbm 1)$ & $4$ & $1$ & $-\frac{2}{3}$ & $-\frac{1}{3}$ & $0$ & $\frac{2}{5}$ & $\frac{33}{40}$ & $\frac{1}{3}$ & $\frac{1}{3}$ & $0$ & $0$ & $\frac{4}{5}$ & $0$ \\
\jvsb$(\mathbbm 1, {\bf 3},\mathbbm 1)$ & $4$ & $1$ & $-\frac{2}{3}$ & $-\frac{1}{3}$ & $0$ & $\frac{2}{3}$ & $\frac{7}{8}$ & $0$ & $0$ & $0$ & $0$ & $\frac{1}{6}$ & $0$ \\
\jvsb$(\mathbbm 1, {\bf \bar 3},\mathbbm 1)$ & $4$ & $1$ & $-\frac{2}{3}$ & $-\frac{1}{3}$ & $0$ & $\frac{1}{3}$ & $\frac{11}{12}$ & $0$ & $\frac{1}{3}$ & $\frac{1}{2}$ & $0$ & $\frac{7}{12}$ & $0$ \\
\jvsb$(\mathbbm 1, \mathbbm 1,{\bf 8}_s)$ & $4$ & $1$ & $-\frac{2}{3}$ & $-\frac{1}{3}$ & $0$ & $\frac{1}{2}$ & $\frac{43}{48}$ & $0$ & $\frac{1}{6}$ & $\frac{1}{4}$ & $0$ & $\frac{7}{8}$ & $\frac{1}{2}$ \\
\hline
\jvsb$({\bf  5},\mathbbm 1, \mathbbm 1)$ & $4$ & $2$ & $-\frac{2}{3}$ & $-\frac{1}{3}$ & $0$ & $\frac{3}{5}$ & $\frac{37}{40}$ & $0$ & $\frac{2}{3}$ & $\frac{1}{2}$ & $0$ & $\frac{19}{20}$ & $0$ \\
\jvsb$({\bf \bar 5},\mathbbm 1, \mathbbm 1)$ & $4$ & $2$ & $-\frac{2}{3}$ & $-\frac{1}{3}$ & $0$ & $-\frac{3}{5}$ & $\frac{119}{120}$ & $\frac{1}{3}$ & $\frac{1}{3}$ & $\frac{1}{2}$ & $0$ & $\frac{11}{20}$ & $0$ \\
\jvsb$(\mathbbm 1, {\bf 3},\mathbbm 1)$ & $4$ & $2$ & $-\frac{2}{3}$ & $-\frac{1}{3}$ & $0$ & $-\frac{1}{3}$ & $\frac{3}{4}$ & $0$ & $\frac{1}{3}$ & $\frac{1}{2}$ & $0$ & $\frac{5}{12}$ & $0$ \\
\jvsb$(\mathbbm 1, {\bf \bar 3},\mathbbm 1)$ & $4$ & $2$ & $-\frac{2}{3}$ & $-\frac{1}{3}$ & $0$ & $\frac{1}{3}$ & $\frac{17}{24}$ & $0$ & $\frac{1}{3}$ & $\frac{1}{2}$ & $0$ & $\frac{1}{12}$ & $0$ \\
\jvsb$(\mathbbm 1, \mathbbm 1,{\bf 8}_c)$ & $4$ & $2$ & $-\frac{2}{3}$ & $-\frac{1}{3}$ & $0$ & $0$ & $\frac{35}{48}$ & $0$ & $\frac{1}{3}$ & $\frac{1}{2}$ & $\frac{1}{2}$ & $\frac{3}{4}$ & $\frac{1}{2}$ \\
\caption{Discrete transformation phases $\eta_{\mathcal S}(\phi)$ for non-singlets $\phi$ at $(n_2,n_2')=(0,0)$.
The tilde in $\tZ N$ denotes that the $N$-th application of the generator associated
with the symmetry yields not necessarily an integer, but a phase which can be expressed by other generators.
Note that the vacua $\mathcal S_1, \mathcal S_2$
additionally have an unbroken symmetry whose generator is proportional to $\frac 12 t_X$,
for $\mathcal S_0$ this holds up to $t_5$.
$\mathcal S_0$ has $\U1_5$ as an unbroken continuous symmetry.}
\label{tab:chns}
\end{longtable}
\end{center}

\begin{center}
\footnotesize
\begin{longtable}[c]{c|c|c||c|c|c|c||c c|c c c|c c}
\multicolumn{3}{c||}{} &
\multicolumn{4}{c||}{False vacuum} &
\multicolumn{2}{c|}{Vacuum $\mathcal S_0$} &
\multicolumn{3}{c|}{Vacuum $\mathcal S_1$} &
\multicolumn{2}{c}{Vacuum $\mathcal S_2$}  \jvsb \\
\cline{4-14}
\multicolumn{1}{c|}{Singlet} &
\multicolumn{1}{c|}{$k$} &
\multicolumn{1}{c||}{ $n_3$} &
\multicolumn{1}{c|}{$R^1$} &
\multicolumn{1}{c|}{$R^2$} &
\multicolumn{1}{c|}{$R^3$} &
\multicolumn{1}{c||}{$t_X$} &
\multicolumn{1}{c}{$[\tZ{24}]_R$} &
\multicolumn{1}{c|}{$[\Z6]_R$} &
\multicolumn{1}{c}{$[\Z6]_R$} &
\multicolumn{1}{c}{$[\tZ2]_R$} &
\multicolumn{1}{c|}{$\Z2$} &
\multicolumn{1}{c}{$[\tZ4]_R$} &
\multicolumn{1}{c}{$[\Z2]_R$} \jvsb
\\ \hline \hline
\endfirsthead
\multicolumn{14}{c}%
{{ \tablename\ \thetable{} - continued from previous page}}  \\
\multicolumn{3}{c||}{} &
\multicolumn{4}{c||}{False vacuum} &
\multicolumn{2}{c|}{Vacuum $\mathcal S_0$} &
\multicolumn{3}{c|}{Vacuum $\mathcal S_1$} &
\multicolumn{2}{c}{Vacuum $\mathcal S_2$}  \jvsb \\
\cline{4-14}
\multicolumn{1}{c|}{Singlet} &
\multicolumn{1}{c|}{$k$} &
\multicolumn{1}{c||}{ $n_3$} &
\multicolumn{1}{c|}{$R^1$} &
\multicolumn{1}{c|}{$R^2$} &
\multicolumn{1}{c|}{$R^3$} &
\multicolumn{1}{c||}{$t_X$} &
\multicolumn{1}{c}{$[\tZ{24}]_R$} &
\multicolumn{1}{c|}{$[\Z6]_R$} &
\multicolumn{1}{c}{$[\Z6]_R$} &
\multicolumn{1}{c}{$[\tZ2]_R$} &
\multicolumn{1}{c|}{$\Z2$} &
\multicolumn{1}{c}{$[\tZ4]_R$} &
\multicolumn{1}{c}{$[\Z2]_R$} \jvsb
\\ \hline \hline
\endhead
\endlastfoot
\multicolumn{14}{c}%
{{ \tablename\ \thetable{} - continued on next page}} \\
\endfoot
\jvsb$U_1^c$ & $0$ & $0$ & $0$ & $-1$ & $0$ & $0$ & $\frac{1}{3}$ & $\frac{1}{3}$ & $\frac{1}{3}$ & $0$ & $0$ & $0$ & $0$ \\
\jvsb$U_2$ & $0$ & $0$ & $-1$ & $0$ & $0$ & $2$ & $0$ & $0$ & $0$ & $0$ & $0$ & $0$ & $0$ \\
\jvsb$U_3$ & $0$ & $0$ & $-1$ & $0$ & $0$ & $0$ & $\frac{2}{3}$ & $\frac{2}{3}$ & $\frac{2}{3}$ & $0$ & $0$ & $0$ & $0$ \\
\jvsb$U_4$ & $0$ & $0$ & $-1$ & $0$ & $0$ & $-2$ & $0$ & $0$ & $0$ & $0$ & $0$ & $0$ & $0$ \\
\hline
\jvsb$N^c_{(1)}$ & $1$ & $0$ & $-\frac{1}{6}$ & $-\frac{1}{3}$ & $-\frac{1}{2}$ & $1$ & $\frac{1}{4}$ & $0$ & $0$ & $\frac{1}{2}$ & $0$ & $\frac{1}{4}$ & $0$ \\
\jvsb$S_1$ & $1$ & $0$ & $\frac{5}{6}$ & $-\frac{1}{3}$ & $-\frac{1}{2}$ & $-1$ & $\frac{1}{4}$ & $0$ & $0$ & $\frac{1}{2}$ & $0$ & $\frac{1}{4}$ & $0$ \\
\jvsb$S_2$ & $1$ & $0$ & $\frac{5}{6}$ & $-\frac{1}{3}$ & $-\frac{1}{2}$ & $0$ & $0$ & $0$ & $0$ & $0$ & $0$ & $0$ & $0$ \\
\jvsb$S_3$ & $1$ & $0$ & $-\frac{1}{6}$ & $\frac{2}{3}$ & $-\frac{1}{2}$ & $1$ & $\frac{11}{12}$ & $\frac{2}{3}$ & $\frac{2}{3}$ & $\frac{1}{2}$ & $0$ & $\frac{1}{4}$ & $0$ \\
\jvsb$S_4$ & $1$ & $0$ & $\frac{11}{6}$ & $-\frac{1}{3}$ & $-\frac{1}{2}$ & $1$ & $\frac{1}{4}$ & $0$ & $0$ & $\frac{1}{2}$ & $0$ & $\frac{1}{4}$ & $0$ \\
\jvsb$S_5$ & $1$ & $0$ & $-\frac{1}{6}$ & $\frac{2}{3}$ & $-\frac{1}{2}$ & $0$ & $0$ & $0$ & $0$ & $0$ & $0$ & $0$ & $0$ \\
\jvsb$S_6$ & $1$ & $0$ & $\frac{11}{6}$ & $-\frac{1}{3}$ & $-\frac{1}{2}$ & $0$ & $\frac{1}{3}$ & $\frac{1}{3}$ & $\frac{1}{3}$ & $0$ & $0$ & $0$ & $0$ \\
\hline
\jvsb$S_7$ & $1$ & $1$ & $\frac{5}{6}$ & $-\frac{1}{3}$ & $-\frac{1}{2}$ & $0$ & $\frac{7}{24}$ & $0$ & $0$ & $0$ & $0$ & $0$ & $0$ \\
\hline
\jvsb$S_8$ & $1$ & $2$ & $-\frac{1}{6}$ & $-\frac{1}{3}$ & $-\frac{1}{2}$ & $-1$ & $\frac{23}{24}$ & $0$ & $0$ & $\frac{1}{2}$ & $0$ & $\frac{1}{4}$ & $0$ \\
\hline
\jvsb$X_0$ & $2$ & $0$ & $-\frac{1}{3}$ & $-\frac{2}{3}$ & $0$ & $0$ & $0$ & $0$ & $0$ & $0$ & $0$ & $0$ & $0$ \\
\jvsb$\bar X_0$ & $2$ & $0$ & $-\frac{1}{3}$ & $-\frac{2}{3}$ & $0$ & $0$ & $\frac{5}{6}$ & $\frac{1}{3}$ & $\frac{1}{3}$ & $0$ & $0$ & $\frac{1}{2}$ & $0$ \\
\jvsb$Y_0$ & $2$ & $0$ & $-\frac{1}{3}$ & $-\frac{2}{3}$ & $0$ & $1$ & $\frac{3}{4}$ & $0$ & $0$ & $\frac{1}{2}$ & $0$ & $\frac{3}{4}$ & $0$ \\
\jvsb$\bar Y_0$ & $2$ & $0$ & $-\frac{1}{3}$ & $-\frac{2}{3}$ & $0$ & $-1$ & $\frac{1}{12}$ & $\frac{1}{3}$ & $\frac{1}{3}$ & $\frac{1}{2}$ & $0$ & $\frac{3}{4}$ & $0$ \\
\jvsb$Y_0^*$ & $2$ & $0$ & $-\frac{1}{3}$ & $-\frac{5}{3}$ & $0$ & $1$ & $\frac{11}{12}$ & $\frac{2}{3}$ & $\frac{2}{3}$ & $\frac{1}{2}$ & $0$ & $\frac{1}{4}$ & $0$ \\
\jvsb$\bar Y_0^*$ & $2$ & $0$ & $\frac{2}{3}$ & $-\frac{2}{3}$ & $0$ & $1$ & $\frac{1}{12}$ & $\frac{1}{3}$ & $\frac{1}{3}$ & $\frac{1}{2}$ & $0$ & $\frac{3}{4}$ & $0$ \\
\hline
\jvsb$X_1$ & $2$ & $1$ & $-\frac{1}{3}$ & $-\frac{2}{3}$ & $0$ & $0$ & $\frac{5}{24}$ & $0$ & $0$ & $0$ & $0$ & $\frac{1}{2}$ & $0$ \\
\jvsb$\bar X_1$ & $2$ & $1$ & $-\frac{1}{3}$ & $-\frac{2}{3}$ & $0$ & $0$ & $\frac{17}{24}$ & $\frac{1}{3}$ & $0$ & $0$ & $0$ & $0$ & $0$ \\
\jvsb$Y_1$ & $2$ & $1$ & $-\frac{1}{3}$ & $-\frac{2}{3}$ & $0$ & $1$ & $\frac{1}{8}$ & $\frac{1}{3}$ & $\frac{2}{3}$ & $\frac{1}{2}$ & $0$ & $\frac{3}{4}$ & $0$ \\
\jvsb$\bar Y_1$ & $2$ & $1$ & $-\frac{1}{3}$ & $-\frac{2}{3}$ & $0$ & $-1$ & $\frac{19}{24}$ & $0$ & $\frac{1}{3}$ & $\frac{1}{2}$ & $0$ & $\frac{3}{4}$ & $0$ \\
\jvsb$Y_1^*$ & $2$ & $1$ & $-\frac{1}{3}$ & $-\frac{5}{3}$ & $0$ & $-1$ & $\frac{23}{24}$ & $\frac{2}{3}$ & $0$ & $\frac{1}{2}$ & $0$ & $\frac{1}{4}$ & $0$ \\
\jvsb$\bar Y_1^*$ & $2$ & $1$ & $\frac{2}{3}$ & $-\frac{2}{3}$ & $0$ & $-1$ & $\frac{1}{8}$ & $\frac{1}{3}$ & $\frac{2}{3}$ & $\frac{1}{2}$ & $0$ & $\frac{3}{4}$ & $0$ \\
\hline
\jvsb$X_2$ & $2$ & $2$ & $-\frac{1}{3}$ & $-\frac{2}{3}$ & $0$ & $1$ & $\frac{1}{24}$ & $0$ & $0$ & $\frac{1}{2}$ & $0$ & $\frac{3}{4}$ & $0$ \\
\jvsb$\bar X_2$ & $2$ & $2$ & $-\frac{1}{3}$ & $-\frac{2}{3}$ & $0$ & $-1$ & $\frac{17}{24}$ & $\frac{1}{3}$ & $\frac{2}{3}$ & $\frac{1}{2}$ & $0$ & $\frac{3}{4}$ & $0$ \\
\jvsb$Y_2$ & $2$ & $2$ & $-\frac{1}{3}$ & $-\frac{2}{3}$ & $0$ & $0$ & $\frac{19}{24}$ & $\frac{1}{3}$ & $0$ & $0$ & $0$ & $\frac{1}{2}$ & $0$ \\
\jvsb$\bar Y_2$ & $2$ & $2$ & $-\frac{1}{3}$ & $-\frac{2}{3}$ & $0$ & $0$ & $\frac{23}{24}$ & $0$ & $\frac{2}{3}$ & $0$ & $0$ & $0$ & $0$ \\
\jvsb$X_2^*$ & $2$ & $2$ & $-\frac{1}{3}$ & $-\frac{5}{3}$ & $0$ & $0$ & $\frac{1}{8}$ & $\frac{2}{3}$ & $\frac{1}{3}$ & $0$ & $0$ & $\frac{1}{2}$ & $0$ \\
\jvsb$\bar X_2^*$ & $2$ & $2$ & $\frac{2}{3}$ & $-\frac{2}{3}$ & $0$ & $0$ & $\frac{7}{24}$ & $\frac{1}{3}$ & $0$ & $0$ & $0$ & $0$ & $0$ \\
\hline
\jvsb$X_0^c$ & $4$ & $0$ & $-\frac{2}{3}$ & $-\frac{1}{3}$ & $0$ & $0$ & $\frac{5}{6}$ & $\frac{1}{3}$ & $\frac{1}{3}$ & $0$ & $0$ & $\frac{1}{2}$ & $0$ \\
\jvsb$\bar X_0^c$ & $4$ & $0$ & $-\frac{2}{3}$ & $-\frac{1}{3}$ & $0$ & $0$ & $0$ & $0$ & $0$ & $0$ & $0$ & $0$ & $0$ \\
\jvsb$Y_0^c$ & $4$ & $0$ & $-\frac{2}{3}$ & $-\frac{1}{3}$ & $0$ & $-1$ & $\frac{1}{12}$ & $\frac{1}{3}$ & $\frac{1}{3}$ & $\frac{1}{2}$ & $0$ & $\frac{3}{4}$ & $0$ \\
\jvsb$\bar Y_0^c$ & $4$ & $0$ & $-\frac{2}{3}$ & $-\frac{1}{3}$ & $0$ & $1$ & $\frac{3}{4}$ & $0$ & $0$ & $\frac{1}{2}$ & $0$ & $\frac{3}{4}$ & $0$ \\
\jvsb$Y_0^{*c}$ & $4$ & $0$ & $-\frac{5}{3}$ & $-\frac{1}{3}$ & $0$ & $-1$ & $\frac{3}{4}$ & $0$ & $0$ & $\frac{1}{2}$ & $0$ & $\frac{3}{4}$ & $0$ \\
\jvsb$\bar Y_0^*c$ & $4$ & $0$ & $-\frac{2}{3}$ & $\frac{2}{3}$ & $0$ & $-1$ & $\frac{11}{12}$ & $\frac{2}{3}$ & $\frac{2}{3}$ & $\frac{1}{2}$ & $0$ & $\frac{1}{4}$ & $0$ \\
\hline
\jvsb$X_1^c$ & $4$ & $1$ & $-\frac{2}{3}$ & $-\frac{1}{3}$ & $0$ & $0$ & $\frac{5}{8}$ & $\frac{1}{3}$ & $\frac{1}{3}$ & $0$ & $0$ & $0$ & $0$ \\
\jvsb$\bar X_1^c$ & $4$ & $1$ & $-\frac{2}{3}$ & $-\frac{1}{3}$ & $0$ & $0$ & $\frac{1}{8}$ & $0$ & $\frac{1}{3}$ & $0$ & $0$ & $\frac{1}{2}$ & $0$ \\
\jvsb$Y_1^c$ & $4$ & $1$ & $-\frac{2}{3}$ & $-\frac{1}{3}$ & $0$ & $-1$ & $\frac{17}{24}$ & $0$ & $\frac{2}{3}$ & $\frac{1}{2}$ & $0$ & $\frac{3}{4}$ & $0$ \\
\jvsb$\bar Y_1^c$ & $4$ & $1$ & $-\frac{2}{3}$ & $-\frac{1}{3}$ & $0$ & $1$ & $\frac{1}{24}$ & $\frac{1}{3}$ & $0$ & $\frac{1}{2}$ & $0$ & $\frac{3}{4}$ & $0$ \\
\jvsb$Y_1^{*c}$ & $4$ & $1$ & $-\frac{5}{3}$ & $-\frac{1}{3}$ & $0$ & $1$ & $\frac{17}{24}$ & $0$ & $\frac{2}{3}$ & $\frac{1}{2}$ & $0$ & $\frac{3}{4}$ & $0$ \\
\jvsb$\bar Y_1^{*c}$ & $4$ & $1$ & $-\frac{2}{3}$ & $\frac{2}{3}$ & $0$ & $1$ & $\frac{7}{8}$ & $\frac{2}{3}$ & $\frac{1}{3}$ & $\frac{1}{2}$ & $0$ & $\frac{1}{4}$ & $0$ \\
\hline
\jvsb$X_2^c$ & $4$ & $2$ & $-\frac{2}{3}$ & $-\frac{1}{3}$ & $0$ & $-1$ & $\frac{19}{24}$ & $\frac{1}{3}$ & $\frac{1}{3}$ & $\frac{1}{2}$ & $0$ & $\frac{3}{4}$ & $0$ \\
\jvsb$\bar X_2^c$ & $4$ & $2$ & $-\frac{2}{3}$ & $-\frac{1}{3}$ & $0$ & $1$ & $\frac{1}{8}$ & $0$ & $\frac{2}{3}$ & $\frac{1}{2}$ & $0$ & $\frac{3}{4}$ & $0$ \\
\jvsb$Y_2^c$ & $4$ & $2$ & $-\frac{2}{3}$ & $-\frac{1}{3}$ & $0$ & $0$ & $\frac{1}{24}$ & $0$ & $\frac{1}{3}$ & $0$ & $0$ & $0$ & $0$ \\
\jvsb$\bar Y_2^c$ & $4$ & $2$ & $-\frac{2}{3}$ & $-\frac{1}{3}$ & $0$ & $0$ & $\frac{7}{8}$ & $\frac{1}{3}$ & $\frac{2}{3}$ & $0$ & $0$ & $\frac{1}{2}$ & $0$ \\
\jvsb$Y_2^{*c}$ & $4$ & $2$ & $-\frac{5}{3}$ & $-\frac{1}{3}$ & $0$ & $0$ & $\frac{13}{24}$ & $0$ & $\frac{1}{3}$ & $0$ & $0$ & $\frac{1}{2}$ & $0$ \\
\jvsb$\bar Y_2^{*c}$ & $4$ & $2$ & $-\frac{2}{3}$ & $\frac{2}{3}$ & $0$ & $0$ & $\frac{17}{24}$ & $\frac{2}{3}$ & $0$ & $0$ & $0$ & $0$ & $0$ \\
\caption{Discrete transformation phases $\eta_{\mathcal S}(s)$ for singlets $s$ at $(n_2,n_2')=(0,0)$.
The tilde in $\tZ N$ denotes that the $N$-th application of the generator associated
with the symmetry yields not necessarily an integer, but a phase which can be expressed by other generators.
Note that the vacua $\mathcal S_1, \mathcal S_2$
additionally have an unbroken symmetry whose generator is proportional to $\frac 12 t_X$,
for $\mathcal S_0$ this holds up to $t_5$.
$\mathcal S_0$ has $\U1_5$ as an unbroken continuous symmetry.}
\label{tab:chs}
\end{longtable}
\end{center}

\cleardoublepage

\cleardoublepage
\addcontentsline{toc}{section}{Index}
\printindex

\end{document}